\documentclass{article}
\usepackage{fancybox}
\usepackage[dvips]{graphicx}
\usepackage{cite}  
\makeatletter
\@addtoreset{equation}{section}

\makeatother

\begin{document}

\begin{flushright}

hep-th/0401120

Jan 2004

KUNS-1893
\end{flushright}

\begin{center}
 {\Large \bf Matrix models and the gravitational interaction} \\
\vspace{5mm}
 Takehiro Azuma \\
\vspace{5mm}
\textit{Department of Physics, Kyoto University, Kyoto 606-8502,
  Japan} \\
 \verb| azuma@gauge.scphys.kyoto-u.ac.jp| \\
\vspace{5mm}
 A Dissertation in candidacy for the degree of Doctor of Philosophy
\end{center}

 \begin{abstract}
 The superstring theory is now regarded as the most promising candidate
 for the unification of the Standard Model and gravity, and this field 
 has been rigorously investigated. However, we
  have seen a setback of the perturbative analysis of the superstring
  theory, because it has so many vacua that we have no way to determine
  which are the true ones. In order to remedy this problem and
  thus for the superstring theory to have a power to predict our real
  four-dimensional world, we need the nonperturbative formulation of
  the superstring theory.

  In the late 1990's, our understanding of the nonperturbative aspects 
  of the superstring theory have been greatly deepened. Especially, 
  the large-$N$ ($N$ is the size of the matrices) reduced models have been
  proposed as the nonperturbative 
  formulation of the superstring theory. One of the most promising candidates
  is the IIB matrix model, which is defined by the dimensional
  reduction of the ten-dimensional ${\cal N}=1$ super-Yang-Mills
  theory to zero dimension. It has been conjectured that this model
  resurrects the behavior of the string theory at the large-$N$ limit.
  There have been a lot of interesting
  discoveries of the IIB matrix model, such as the dynamical
  generation of the four-dimensional spacetime and the interpretation of
  the diffeomorphism invariance.

  On the other hand, there are a lot of problems to surmount, if a large-$N$
  reduced model is to be an eligible framework to unify the
  gravitational interaction. Firstly, it is still an enigma how we can 
  realize the local Lorentz invariant matrix model. In addition, we
  need to understand how we can describe the curved spacetime more
  manifestly, in terms of a large-$N$ reduced model.

  This thesis discusses several attempts to address these
  issues concerning the gravitational interaction. This thesis is
  based on the following works\cite{0102168,0204078,0209057,0401038}. 
  
 \end{abstract}

\newpage

\tableofcontents

\newpage
\section{Introduction}
\setcounter{page}{1}
  One of the main themes in the elementary particle
  physics is unification. There are four kinds of interaction in the
  nature; the weak interaction, the strong interaction, the
  electromagnetic interaction and the gravitational interaction. 
  In 1967, Glashow, Weinberg and Salam succeeded in the unification of 
  the weak and electromagnetic interaction in terms of the $SU(2)
  \times U(1)$ gauge group. The ensuing success is the advent of the
  "Standard Model", described by the $SU(3) \times SU(2) \times U(1)$
  gauge group. The Standard Model is striking in the sense that it
  describes all the experimental phenomena by setting the 18
  parameters in the model. 

  However, there are two serious drawbacks to the Standard
  Model. First is that we cannot explain theoretically how these 18
  parameters are fixed. Namely, we have to rely on the experimental
  data in setting these parameters. Second is that the Standard Model
  does not unify the gravitational interaction.

  Now, the superstring theory is regarded as the most promising
  candidate to resolve these two drawbacks. The superstring theory
  regards not a zero-dimensional point but a one-dimensional string
  with the length $l_{p} = 10^{-33}$cm as the fundamental object. Here,  
  $l_{p}$ is called "the Planck Scale", which is the fundamental scale in 
  the superstring theory. The superstring theory incorporates not only 
  the gauge particles but also the gravitons in its oscillation
  modes. There is an infinite tower of the mass level in the oscillation
  modes. The massless modes of the open string, which has two ends, are
  regarded as the gauge particle. On the other hand, the massless
  modes of the closed string, which constitutes a closed circle like a 
  rubber band, gives 
  the spin-2 graviton. In this sense, the superstring theory is
  thought to include not only the Standard Model but also the
  gravitational theory in the low-energy limit, where the massive
  modes of the superstring theory are ignored.  The superstring theory
  is fascinating in that it incorporates no free parameters. It is in
  contrast to the Standard Model, which suffers the problem of a
  plethora of free parameters. The superstring theory has a
  possibility to explain the gauge group of the Standard Model, the
  number of the generations of the quarks, the mass of the Higgs
  particle, quarks or leptons, the dimensionality of our spacetime
  $\cdots$, from a theory without any parameter.

  We have seen the so-called "first string boom" in the early
  1980's, in which the perturbative aspects of the superstring theory
  have been elucidated. The striking discovery in 1984 is that the
  superstring theory is free from the divergence of the gravitational
  energy. Physicists have been so far puzzled by the UV divergence in
  the quantization of the gravitational interaction. This conundrum
  has been resolved by the superstring theory which alleviates this
  divergence. Moreover, it has also been known that there are only
  five kinds of superstring theory well-defined in the ten
  dimensions. They are the type I, type IIA, type IIB, $SO(32)$ heterotic
  and $E_{8} \times E_{8}$ heterotic superstring theory.
  Especially, the $E_{8} \times E_{8}$ heterotic superstring theory
  is propitious for the immersion of the Standard Model in the
  superstring theory. The exceptional Lie algebra $E_{8}$ includes 
  the Lie algebra $SU(3) \times SU(2) \times U(1)$, which is the gauge 
  symmetry of the Standard Model, as its subalgebra.

  The superstring theory is defined in the ten-dimensional spacetime,
  but this is not so problematic because it is a theory of
  gravity. The spacetime is regarded as being given not a priori but
  dynamically from the classical solution of the theory. It is now
  believed that the  superfluous six dimensions are compactified at an
  extremely small size at the incunabula of the
  universe. Mathematically, the six superfluous dimensions are
  compactified by the Calabi-Yau manifold.   

  However, the superstring theory suffers a serious setback. Since the 
  superstring theory is defined only perturbatively, there is no
  telling which is the true vacuum. While there are only five
  well-defined superstring theories, there are a myriad of ways to
  compactify the ten-dimensional superstring theory into the four
  dimensions and thus we are overwhelmed by the infinite vacua.
  This means that the superstring theory does not have the ability to
  predict our four-dimensional world, and we cannot tell whether the
  superstring theory dynamically generates the Standard Model.
  In order to surmount this difficulty, we definitely need the
  constructive definition (i.e. the definition without the
  perturbation).

  The so-called "first string boom" was over at the end of the
  1980's. However, there were important discoveries at this period
  about the noncritical string theory and the bosonic matrix model
  (the extensive review can be found in \cite{9304011,9306153}).
  Distler and Kawai\cite{DDK} succeeded in the quantization of the
  non-critical string theory up to one dimension via the conformal
  gauge. In addition, there is a striking relation between the bosonic 
  string theory and the matrix model, proposed by
  F. David\cite{David}. For the simplest zero-dimensional non-critical 
  string, the corresponding matrix model is a one-matrix model
  described by a simple $N \times N$ hermitian matrix. The
  correspondence has been found for the central charge $c = 1 -
  \frac{6}{m(m+1)}$ (for $m=2,3,4,\cdots$), in which the string
  corresponds to the multi-matrix model.  The path integral of the
  worldsheet for all genera has been approximated by the random
  triangulation of the worldsheet. This has led us to identify the
  Feynman rule with that of the one-matrix model. In this way, we have 
  seen the formal correspondence between the matrix model and the
  string theory. Even more crucial is that the matrix model
  can be solved nonperturbatively. Brezin and
  Kazakov\cite{brezinkazakov} solved the matrix model by means of the
  orthogonal polynomial method. They extracted the nonperturbative
  aspects of the bosonic matrix model via the differential equation
  called Painlev\`{e} equation. They calculated the parameter
  called the string 
  susceptibility, and found that this agrees with the calculation of
  Distler and Kawai\cite{DDK} for the noncritical string.

  We have failed in extending this idea to the "super"string,
  and this progress is limited to the bosonic string.  This is due to
  the same difficulty as we face in putting the chiral fermion onto
  the lattice. Thus, these discoveries, per se, do not give a
  technical clue to the constructive definition of the superstring
  theory. The 'state-of-the-art' matrix models (such as the IIB matrix 
  model) do not inherit the same techniques as the old matrix model.
  Nevertheless, these discoveries serve as an important
  touchstone for the legitimacy of the large-$N$ reduced models as the 
  constructive definition of the string theory.

  In the late 1990's we have seen the so-called "second string
  boom", in which we grasped the nonperturbative aspects of the
  superstring theory. During the second string boom, Polchinski
  discovered the soliton-like object called the D-brane. The discovery of
  the D-brane led to the idea of the two kinds of the duality of the
  string theory. First is the T-duality, which is the duality between
  the large and small radius of the compactification. Second is the
  S-duality, which relates the strong coupling and the weak
  coupling. It is now believed that the five 
  ten-dimensional superstring theories are related with one another
  through the T/S duality. This is an illuminating discovery, in the sense
  that it is reminiscent of the more fundamental theory whose certain
  limit might reproduce these five superstring theories.
  
  Another big progress is the
  proposal  for the constructive definition of the superstring theory
  via the large-$N$ reduced model.  Here, the 'large-$N$ reduced
  model' means the model defined by the $N \times N$ hermitian matrices.
  In 1996, Banks, Fischler, Shenker
  and Susskind proposed the matrix model defined by the dimensional
  reduction of the ten-dimensional super Yang-Mills theory to one
  dimension $-$ the so-called BFSS model\cite{9610043}. 
  This matrix model is more related to the type IIA superstring
  theory, and the effect of the type IIA supergravity is induced by
  the one-loop effect.  

  Other attempts are the IIB matrix model (the IKKT
  model)\cite{9612115} and the matrix string theory\cite{9703030},
  which are defined by the dimensional reduction of the
  ten-dimensional super Yang-Mills theory to the zero and two
  dimensions, respectively. Especially, the IIB matrix model is now
  regarded as one of the most promising candidates for the constructive 
  definition of the superstring theory (this model is extensively
  reviewed in \cite{9908038,shino}).   As its nomenclature shows,
  this model is deeply related to the type IIB superstring theory. 
  The simplest and the most direct correspondence is that the IIB
  matrix model is defined by the matrix regularization of the
  Green-Schwarz action of the type IIB superstring theory; namely by
  replacing the Poisson bracket with the commutator of the $N \times
  N$ matrix. In addition, it has been shown\cite{9705128} that the
  Wilson loops satisfy the string field equations of motion for the
  type IIB superstring in the light-cone gauge.
  The prominent feature of this model is that it does not
  include any free parameter. The overall coefficient which is an
  artifact of the coupling constant before being reduced is
  absorbed by a simple field redefinition. 

  There have been hitherto a lot of progresses with respect to the IIB
  matrix model. First is the relation to the gravitational
  interaction. The most fundamental feature is the spacetime ${\cal
  N}=2$ supersymmetry\cite{9612115} when we regard the eigenvalues of
  the bosonic matrices as the spacetime coordinate. This urges us to
  interpret the  eigenvalues as the spacetime coordinate. In addition, 
  the graviton and dilaton exchange has been calculated in
  \cite{9612115}. The interpretation of the general coordinate
  invariance is proposed in \cite{9903217}, in which the permutation
  invariance of the eigenvalues is identified with the general
  coordinate invariance of the low-energy effective action.
  
  Second is the relation to the noncommutative geometry, which has
  been widely investigated since Seiberg and Witten elucidated the
  relationship with the superstring theory\cite{9908142}. By expanding 
  the IIB matrix model around the classical background, it reproduces
  the noncommutative Yang-Mills theory\cite{9908141}.

  Third is the dynamical generation of the four-dimensional
  spacetime. In \cite{9802085}, it has been found that the Hausdorff
  dimension of the eigenvalue distribution is four. In \cite{0111102}, 
  Nishimura and Sugino proposed the breakdown of the Lorentz symmetry
  from $SO(10)$ to $SO(4)$ by means of the third order of the Gaussian
  expansion. Their analysis has been extended to the seventh order in
  \cite{0204240,0211272}. 

  The discovery of the IIB matrix model implies the trilateral
  relationship of the three important notions in the elementary
  particle physics; the superstring theory, the quantum field theory and
  the matrix model. The discovery of the relationship between the
  one-matrix model and the bosonic string theory is another good
  example to relate these notions. In the course of the 
  "second string boom" we have other important discoveries for this
  trilateral relationships. In 1997, Maldacena advocated the duality
  between the type IIB superstring theory on the $AdS_{5} \times
  S^{5}$ spacetime and the ${\cal N}=4$, four-dimensional super
  Yang-Mills theory. This is the so-called "AdS/CFT
  correspondence"\cite{9711200,9802150,9905111}. This correspondence
  has been conjectured through the coincident symmetries of these two
  theories. Since the proposal of Maldacena, the AdS/CFT correspondence 
  has been rigorously investigated by many authors (among them is the
  author's work\cite{0106063}, while we do not delve into this work in
  this thesis). Another important
  discovery is the so-called Dijkgraaf-Vafa
  duality\cite{0206255,0207111,0208048}. This is a duality between the
  ${\cal N}=1$ four-dimensional super Yang-Mills theory and the
  one-matrix model, which used to attract much attention in relation to
  the bosonic string theory in the early 1990's. This discovery
  reinstated the one-matrix model into many authors' attention.
  At first\cite{0206255,0207111,0208048}, this relationship is
  understood with the intervention of the
  topological string theory. However, Cachazo, Douglas, Seiberg and
  Witten\cite{0211170,0301006,0303207} explained this duality
  directly through the super Yang-Mills theory without topological string
  theory. They elucidated that these two theories comply with the
  same Schwinger-Dyson equation, employing the notion of the chiral
  ring and the Konishi anomaly\cite{konishi,konishishizuya}. Moreover, 
  in \cite{0303210,0312026}, the direct relationship between the large-$N$
  reduction and the Dijkgraaf-Vafa duality has been found.
  In this way, in the course of the "second string boom", many
  interesting discoveries, including the IIB matrix model, have been
  unraveled. In the future, we will surely have a synergy in the
  studies of these three germane areas.

  While the IIB matrix model has a lot of interesting
  properties, there is a plenty of room to investigate. The more manifest 
  correspondence with the gravitational interaction is definitely a
  {\$}64,000 question. If the IIB matrix model or its extensions are to be
  the eligible frameworks to unify the gravitational interaction, we
  must take seriously the relation to the gravitational interaction.
  While several important evidences have been found, there are a lot
  of questions to clarify. It has been an enigma how we can describe
  the local Lorentz symmetry in terms of the large-$N$ reduced
  models. In addition, the definition of the IIB matrix model relies
  heavily on the flat background. Especially, the IIB matrix model
  incorporates the background of the flat space, not the curved
  space. This prohibits the perturbation around the curved space
  background.

  In this thesis, we discuss some of the
  attempts\cite{0102168,0204078,0209057,0401038} to address these
  issues. This thesis is organized as follows.
  In Section 2, we give  a brief review of the IIB matrix
  model. Especially, we focus on the development of the IIB matrix
  model in relation to the gravitational interaction. Moreover, we
  have a careful look at the alterations of the IIB matrix model
  that incorporate a curved-space background, in relation to some of
  the author's works\cite{0209057,0401038}.

  In Section 3, we review the author's works about the supermatrix
  model\cite{0102168,0209057}, based on the $osp(1|32,R)$ super Lie
  algebra.  The studies of the supermatrix models
  give a rich perspective for the generalization of the IIB matrix
  model in relation to the gravitational interaction. We start with
  investigating the correspondence of the nongauged $osp(1|32,R)$
  model  with the IIB  matrix model, paying attention to the
  supersymmetry. We next consider the so-called "gauged version",
  namely the model whose Lorentz symmetry and the gauge symmetry are
  mixed together. This is an important idea in realizing the matrix
  model equipped with the local Lorentz symmetry, because the
  eigenvalues of the bosonic matrices are identified with the
  spacetime in the IIB matrix model.
  We next go back to the nongauged version of the $osp(1|32,R)$ 
  supermatrix model including the mass term\cite{0209057}.
  It has been known that the IIB matrix model with the tachyonic mass
  term, whose review we give in Section 2, incorporates the 
  curved-space background because the trivial commutative background
  is destabilized. We focus on the similarity of such models and the
  massive supermatrix model.

  In Section 4, we elaborate on the idea of the "gauged matrix model"
  without the supermatrix model\cite{0204078}. Especially, we allocate 
  the odd-rank matrices of the ten dimensions for the matter field,
  and the even-rank matrices to the parameters of the local Lorentz
  transformation. When we mix the Lorentz symmetry and the gauge
  symmetry, the model must inevitably include the higher-rank fields
  in order for the action to be invariant and for the algebra of the
  symmetry to close with respect to the commutator. This is a good
  news in that the rank-3 fields are identified with the spin
  connection in the curved space. We consider the explicit
  identification of the matrices with the differential operator, and
  clarify that the bosonic part actually reduces to the Einstein
  gravity in the low-energy limit. We also discuss the structure of
  the ${\cal N}=2$ supersymmetry.

  In Section 5, we study the stability of the fuzzy sphere background
  in the quantum sense. While it is easy to discuss the
  stability of the curved-space background (all we have to do is just
  to plug the solution into the action), it is by no means easy to
  discuss the stability in the quantum sense. In this section, we
  address this question via the heat bath algorithm of the Monte Carlo 
  simulation. We focus on the simplest case; namely the bosonic
  three-dimensional IIB matrix model with the Chern-Simons term. We
  find that this bosonic matrix model has a first-order phase
  transition between the Yang-Mills phase (in which the quantum effect 
  is large and the fuzzy sphere classical solution is unstable) and
  the fuzzy sphere phase (in which the model is subject to the meager
  quantum effect and the fuzzy sphere is stable),
  as we change the radius of the fuzzy-sphere solution
  (equivalently, the coefficient of the Chern-Simons term).
  Moreover, we find that this model has a one-loop exactness in the
  fuzzy sphere phase in the large-$N$ limit. The latter result is
  especially exciting, in the sense that this helps the analysis of
  the dynamical generation of the gauge group.

  Section 6 is devoted to the conclusion and the future outlook of
  the work.

  In Appendix A, we summarize the notation of this thesis. We give the 
  detailed notation of the gamma matrices and the supermatrices. In
  addition, we give a proof of the miscellaneous formulae of
  the gamma matrices.

  In Appendix B, we give the calculation of the
  Seeley-de-Witt coefficients, which plays an important role in
  Section 4. 

  In Appendix C, we give the detailed recipe for the heat bath algorithm of
  the matrix model. We start with the quadratic one-matrix model in
  full detail, because this gives the key idea of the simulation of
  the IIB matrix model with/without the Chern-Simons term.
  We nextly discuss the simulation of the quartic one-matrix model.
  Finally, we discuss the heat bath algorithm for the IIB matrix model 
  with/without the Chern-Simons term.

 \section{Brief review of the IIB matrix model}
  In this section, we give a brief review of the IIB matrix
  model\cite{9612115}. As we have mentioned in the introduction, the
  IIB matrix model is regarded as the most promising candidate for the 
  constructive definition of the superstring theory. This model is
  also called "the IKKT model", where IKKT is the acronym for the
  authors of \cite{9612115}; Ishibashi, Kawai, Kitazawa and Tsuchiya.

 \subsection{Definition and symmetry of the IIB matrix model}
  The IIB matrix model is defined by the dimensional reduction of the
  ten-dimensional ${\cal N}=1$ super Yang-Mills theory to zero
  dimension as
  \begin{eqnarray}
  S = - \frac{1}{g^{2}} Tr \left( \frac{1}{4} \sum_{\mu,\nu=0}^{9}
  [A_{\mu}, A_{\nu}][A^{\mu}, A^{\nu}] + \frac{1}{2} \sum_{\mu=0}^{9}
  {\bar \psi} \Gamma^{\mu} [A_{\mu}, \psi] \right). \label{AZM31IKKT}
  \end{eqnarray}
  Here, $A_{\mu}$ are the ten-dimensional bosonic vector, and $\psi$
  is the ten-dimensional Majorana-Weyl (hence sixteen-component)
  fermion. Both $A_{\mu}$ and $\psi$ are promoted to the $N \times N$
  hermitian matrices. This model incorporates the $SO(9,1)$ Lorentz
  symmetry and the $SU(N)$ gauge symmetry. This model is a totally
  reduced model in that it is reduced to zero dimension.
  The prominent feature of this model is that it has no free
  parameter. The overall coefficient $g$ can be trivially absorbed
  into the fields by the following redefinition.
  \begin{eqnarray}
   A_{\mu} \to g^{\frac{1}{2}} A_{\mu}, \hspace{2mm}
   \psi \to g^{\frac{3}{4}} \psi. \label{redef}
  \end{eqnarray}
  The coefficient $g$ is an artifact of the coupling constant
  of the super Yang-Mills theory before being reduced, and is nothing
  but a scaling parameter.

  The path integral of the IIB matrix model is given by
   \begin{eqnarray}
     Z = \int dA d\psi e^{-S_{(E)}}, \label{pathintegral}
   \end{eqnarray}
  where $S_{(E)}$ is defined in the ten-dimensional Euclidean space by
  the Wick rotation of $A_{0}$ and $\Gamma^{0}$ in the action
  (\ref{AZM31IKKT}). 
  The convergence of the path integral of the IIB matrix model is not
  trivial, because the gauge group $SU(N)$ is not
  compact. However, the convergence of the path integral is discussed
  in the paper \cite{0101071,0103159} (the review is found in the
  Ph.D. thesis of Austing\cite{0108128}). The authors corroborated
  that the path integral converges for\footnote{$d$ is the
  dimension of the IIB matrix model, and the model (\ref{AZM31IKKT})
  is defined for $d=10$. More explicitly, the $d$-dimensional
  supersymmetric IIB matrix model means the following action
  \begin{eqnarray}
    S = - \frac{1}{g^{2}} Tr \left( \frac{1}{4} \sum_{\mu,\nu=0}^{d-1}
  [A_{\mu}, A_{\nu}][A^{\mu}, A^{\nu}] + \frac{1}{2} \sum_{\mu=0}^{d-1}
  {\bar \psi} \Gamma^{\mu} [A_{\mu}, \psi] \right), \nonumber
  \end{eqnarray}
  which is well-defined for $d=3,4,6,10$.
  The explicit definition of the $d$-dimensional bosonic IIB matrix
  model is given by
  \begin{eqnarray}
    S = - \frac{1}{g^{2}} Tr \left( \frac{1}{4} \sum_{\mu,\nu=0}^{d-1}
  [A_{\mu}, A_{\nu}][A^{\mu}, A^{\nu}] \right). \nonumber
  \end{eqnarray} } $d=4,6,10$.
  Only for the bosonic part, the path integral is shown to converge
  for $d \geq d_{c}$. Here, $d_{c}=5$ for the $SU(2)$ gauge group,
  $d_{c}=4$ for the $SU(3)$ and $d_{c}=3$ for all the other simple Lie 
  group. 

  \subsubsection{Relation to the type IIB superstring theory}
  The IIB matrix model has a deep relation to the type IIB superstring.
  Here, we discern that the IIB matrix model is obtained by the matrix 
  regularization of the Green-Schwarz action of the type IIB
  superstring theory. The Green-Schwarz formalism is defined so that
  the spacetime supersymmetry should be more manifest. The bosonic
  Nambu-Goto action is given by
  \begin{eqnarray}
  S_{b} = - \frac{1}{4 \pi \alpha'} \int d^{2} \sigma \sqrt{-h} h^{\alpha
  \beta} \partial_{\alpha} X^{\mu} \partial_{\beta}
  X_{\mu}. \label{nambugoto} 
  \end{eqnarray}
 Here, $h_{\alpha \beta}$ is a metric on the worldsheet, and has a
 Lorentz signature $\eta^{\alpha \beta} = \textrm{diag}(-,+)$.
 A naive guess for a supersymmetric extension is to replace
 $\partial_{\alpha} X_{\mu}$ with 
  \begin{eqnarray}
  \Pi_{\alpha}^{\mu} =
 \partial_{\alpha} X^{\mu} - i ({\bar \theta}^{1} \Gamma^{\mu}
 \partial_{\alpha} \theta^{1} - {\bar \theta}^{2} \Gamma^{\mu}
 \partial_{\alpha} \theta^{2}).
  \end{eqnarray}
 $\Pi_{\alpha}^{\mu}$ is trivially invariant under the supersymmetry
 transformation 
  \begin{eqnarray}
   \delta_{S} \theta^{A} = \epsilon^{A}, \hspace{2mm}
   \delta_{S}  X^{\mu} = i ({\bar \epsilon}^{1} \Gamma^{\mu}
   \theta^{1} - {\bar \epsilon}^{2} \Gamma^{\mu}
   \theta^{2}). \label{SUSY}
  \end{eqnarray}
    Here, $A$ runs over $1,2$, and this is  an index for the supersymmetry. 
   However, the naive replacement of $\partial_{\alpha} X_{\mu}$ with
   $\Pi^{\mu}_{\alpha}$ for the action (\ref{nambugoto}) is not
   enough, because it does not have a $\kappa$ symmetry and hence it has
   twice as many degrees of freedom as it should. Instead, we give the 
   following action:
  \begin{eqnarray}
   S_{GS} = \frac{-1}{4 \pi \alpha'} \int d^{2} \sigma 
   \left( \sqrt{-h} h^{\alpha \beta} \Pi^{\mu}_{\alpha} \Pi_{\beta
   \mu} + 2i \epsilon^{\alpha \beta} \partial_{\alpha} X^{\mu}
     ({\bar \theta^{1}} \Gamma_{\mu} \partial_{\beta} \theta^{1}
   +  {\bar \theta^{2}} \Gamma_{\mu} \partial_{\beta} \theta^{2})
   + 2 \epsilon^{\alpha \beta} ({\bar \theta^{1}} \Gamma^{\mu}
   \partial_{\alpha} \theta^{1})({\bar \theta^{2}} \Gamma_{\mu}
   \partial_{\alpha} \theta^{2}) \right). \nonumber 
 \end{eqnarray}
  In this section, we define the rank-2 epsilon tensor as
  $\epsilon^{01}=1$ (and hence $\epsilon_{01}=-1$). 
  This action is defined for the $d=10$ dimensional spacetime, and the 
  index $\mu$ runs over $0,1,2,\cdots,9$.
  The supersymmetry transformation of this action is 
  \begin{eqnarray}
  \delta_{S} S_{GS} = \frac{-1}{2 \pi \alpha'} \int d^{2} \sigma
    \epsilon^{\alpha \beta} 
    [ ({\bar \epsilon^{1}} \Gamma^{\mu} \partial_{\alpha} \theta^{1})
      ({\bar \theta^{1}} \Gamma_{\mu} \partial_{\beta} \theta^{1})
   - ({\bar \epsilon^{2}} \Gamma^{\mu} \partial_{\alpha} \theta^{2})
      ({\bar \theta^{2}} \Gamma_{\mu} \partial_{\beta} \theta^{2})],
  \label{susygs}
  \end{eqnarray}
 where we drop the surface terms. The term (\ref{susygs}) is shown to
 vanish (i.e. the action $S_{GS}$ is supersymmetry invariant) by
 noting the following rewriting: 
  \begin{eqnarray}
   A &=& \epsilon^{\alpha \beta} {\bar \epsilon} \Gamma^{\mu}
   \partial_{\alpha} \theta {\bar \theta} \Gamma_{\mu}
   \partial_{\beta} \theta 
      =   {\bar \epsilon} \Gamma^{\mu}
   {\dot \theta} {\bar \theta} \Gamma_{\mu}
   \theta' -   {\bar \epsilon} \Gamma^{\mu}
   \theta' {\bar \theta} \Gamma_{\mu}{\dot  \theta}.
  \end{eqnarray}
 Here, $\sigma_{0} = \tau$ and $\sigma_{1} = \sigma$. 
  ${\dot \theta} = \frac{\partial \theta}{\partial \tau}$ and $
 \theta' = \frac{\partial \theta}{\partial \sigma}$. $A$ can be
 further written as
  \begin{eqnarray}
  A =  \frac{2}{3} [ {\bar \epsilon} \Gamma^{\mu} {\dot \theta} {\bar 
  \theta} \Gamma_{\mu} \theta' +
   {\bar \epsilon} \Gamma^{\mu} \theta' {\dot {\bar 
  \theta}} \Gamma_{\mu} \theta 
  + {\bar \epsilon} \Gamma^{\mu} \theta {\bar 
  \theta}' \Gamma_{\mu} {\dot \theta}]
  + \frac{1}{3} \frac{\partial}{\partial \tau}[ {\bar \epsilon}
  \Gamma^{\mu} \theta {\bar \theta}\Gamma_{\mu} \theta']
  -  \frac{1}{3} \frac{\partial}{\partial \sigma}[ {\bar \epsilon}
  \Gamma^{\mu} \theta {\bar \theta}\Gamma_{\mu} {\dot \theta}].
  \end{eqnarray}
 The first term vanishes due to the Fierz identity 
  \begin{eqnarray}
 {\bar \epsilon} \Gamma^{\mu} \psi_{[1} {\bar \psi}_{2} \Gamma_{\mu}
 \psi_{3]} = 0. \label{cycleferm}
 \end{eqnarray}
 We delegate the
 proof of this Fierz identity to Appendix \ref{sec-fierz2}. The second and
 third terms vanish because they are nothing but surface terms.

  This model also incorporates the symmetry named the $\kappa$ symmetry:
  \begin{eqnarray}
  & & \delta_{\kappa} \theta^{A} = 2 \Gamma^{\mu} \Pi^{\alpha}_{\mu}
   \kappa^{A \alpha}, \hspace{2mm}
   \delta_{\kappa} X^{\mu} = i {\bar \theta^{1}} \Gamma^{\mu}
   \delta_{\kappa} \theta^{1} - i {\bar \theta^{2}} \Gamma^{\mu}
   \delta_{\kappa} \theta^{\mu}, \nonumber \\
  & & \delta_{\kappa} (\sqrt{-h} h^{\alpha \beta}) = - 16i \sqrt{-h}
   ( \partial_{\gamma} {\bar \theta^{1}} \kappa^{1\beta} P_{-}^{\alpha 
   \gamma} - \partial_{\gamma} {\bar \theta^{2}} \kappa^{2 \beta}
   P_{+}^{\alpha \gamma} ). \label{kappa}
  \end{eqnarray}
  Here, $P_{\pm}^{\alpha \beta}$ is a projection operator defined by
  \begin{eqnarray}
   P_{\pm}^{\alpha \beta} = \frac{1}{2} (h^{\alpha \beta} \pm
   \frac{\epsilon^{\alpha \beta}}{\sqrt{-h}}). \label{proj}
  \end{eqnarray}
  This satisfies the following properties
   \begin{eqnarray}
    P_{\pm}^{\alpha \beta} h_{\beta \gamma} P_{\pm}^{\gamma \delta} =
    P_{\pm}^{\alpha \delta}, \hspace{2mm}
    P_{\pm}^{\alpha \beta} h_{\beta \gamma} P_{\mp}^{\gamma \delta} =
    0, \hspace{2mm}
    P_{\pm}^{\alpha \beta} P_{\pm}^{\gamma \delta} = 
    P_{\pm}^{\gamma \beta} P_{\pm}^{\alpha \delta},
   \end{eqnarray}
  which follow from the inversion rule
    \begin{eqnarray}
     h^{\alpha \beta} = \frac{1}{h} \left( \begin{array}{cc} h_{11} &
     - h_{01} \\ - h_{10} & h_{00} \end{array} \right). 
    \end{eqnarray}
  Here, the parameters of the $\kappa$-symmetry is subject to the
  following projection rule:
   \begin{eqnarray}
   P^{\alpha \beta}_{-} \kappa^{1}_{\beta} = \kappa^{1 \alpha},
   \hspace{2mm}
   P^{\alpha \beta}_{+} \kappa^{2}_{\beta} = \kappa^{2
   \alpha}. \label{projection} 
   \end{eqnarray}

  The $\kappa$-symmetry transformation varies the action as
  \begin{eqnarray}
   \delta_{\kappa}S_{GS} &=& \frac{-1}{2 \pi \alpha'} \int d^{2} \sigma
   \epsilon^{\alpha \beta}
   \left(   [(\partial_{\beta} {\bar \theta}^{1}) \Gamma^{\mu}
             (\partial_{\alpha} \theta^{1})]
            [{\bar \theta}^{1} \Gamma_{\mu} \delta_{\kappa} \theta^{1}]
          + 2[(\partial_{\alpha} {\bar \theta}^{1}) \Gamma^{\mu}
              \delta_{\kappa} \theta^{1}]
             [{\bar \theta}^{1} \Gamma_{\mu} \partial_{\beta}
   \theta^{1}] \right. \nonumber \\
    & & \hspace{20mm} \left.
        - [(\partial_{\beta} {\bar \theta}^{2}) \Gamma^{\mu}
             (\partial_{\alpha} \theta^{2})]
            [{\bar \theta}^{2} \Gamma_{\mu} \delta_{\kappa} \theta^{2}]
        - 2[(\partial_{\alpha} {\bar \theta}^{2}) \Gamma^{\mu}
              \delta_{\kappa} \theta^{2}]
             [{\bar \theta}^{2} \Gamma_{\mu} \partial_{\beta}
   \theta^{2}] \right), \label{kappatrans}
  \end{eqnarray}
  up to the surface terms that vanish in the integral. We can show
  that the Green-Schwarz action is invariant under the
  $\kappa$-symmetry with the help of the identity (\ref{cycleferm}).  

  Nextly, we explain how the matrix regularization of the
  Green-Schwarz action of the type IIB superstring theory leads to the 
  IIB matrix model.
  For the type IIB superstring theory, the chirality of the two
  spinors $\theta^{1}$ and $\theta^{2}$ is identical. Therefore, we
  set $\theta^{1} = \theta^{2} = \theta$. This simplifies the term 
  $\Pi_{\alpha}^{\mu}$ as $\Pi_{\alpha}^{\mu} = \partial_{\alpha}
  X^{\mu}$. Then, the action is simplified as
   \begin{eqnarray}
   S_{GS} = \frac{-1}{4 \pi \alpha'} \int d^{2} \sigma
    \left( h^{\alpha \beta} \sqrt{-h} \partial_{\alpha} X^{\mu} 
       \partial_{\beta} X_{\mu} + 4i \epsilon^{\alpha \beta}
    \partial_{\alpha} X^{\mu} {\bar \theta} \Gamma_{\mu}
    \partial_{\beta} \theta \right). \label{gs2}
   \end{eqnarray}
  Integrating out $h_{\alpha \beta}$, we obtain
   \begin{eqnarray}
   S_{GS} &=& \frac{-1}{2 \pi \alpha'} \int d^{2} \sigma
    \left( \sqrt{- \det (\partial_{\alpha} X^{\mu} \partial_{\beta}
    X_{\mu})} + 2i \epsilon^{\alpha \beta}
    \partial_{\alpha} X^{\mu} {\bar \theta} \Gamma_{\mu}
    \partial_{\beta} \theta \right) \nonumber \\
   &=&  \frac{-1}{2 \pi \alpha'} \int d^{2} \sigma
    \left( \sqrt{-  \frac{1}{2} (\epsilon^{\alpha \beta}
    \partial_{\alpha} X^{\mu} 
    \partial_{\beta} X_{\mu})^{2}} + 2i \epsilon^{\alpha \beta}
    \partial_{\alpha} X^{\mu} {\bar \theta} \Gamma_{\mu}
    \partial_{\beta} \theta \right). \label{gs3}
   \end{eqnarray}
  In the last equality, we note that this determinant can be
    rewritten as
  \begin{eqnarray}
    \det (\partial_{\alpha} X^{\mu} \partial_{\beta}
    X_{\mu}) 
  &=& (\partial_{0} X^{\mu} \partial_{0} X_{\mu})
    (\partial_{1} X^{\nu} \partial_{1} X_{\nu})
 -  (\partial_{0} X^{\mu} \partial_{1} X_{\mu})
    (\partial_{1} X^{\nu} \partial_{0} X_{\nu}) \nonumber \\
  &=& \frac{1}{2} (\epsilon^{\alpha \beta} \partial_{\alpha} X^{\mu}
    \partial_{\beta} X_{\mu})^{2}.
  \end{eqnarray}

  We introduce the Schild action and discern that this reduces to
  the Green-Schwarz action. The Schild action is defined by
  \begin{eqnarray}
   S_{\textrm{Schild}} =
   - \int d^{2} \sigma \left( \sqrt{-h_{s}} a \left( - \frac{1}{4N} 
   \{ X_{\mu}, X_{\nu} \}^{2} + \frac{i}{2} {\bar \psi} \Gamma^{\mu}
   \{ X_{\mu}, \psi \} \right) +  \sqrt{-h_{s}} bN \right). \label{schild}
  \end{eqnarray}
  Here, we introduce the scalar density $\sqrt{-h_{s}}$ as an independent
  variable. The Poisson bracket $\{ q,p \}$ is defined by
   \begin{eqnarray}
    \{ q, p \} = \frac{1}{\sqrt{-h_{s}}} \epsilon^{\alpha \beta}
    \partial_{\alpha} q \partial_{\beta} p. \label{poisson}
   \end{eqnarray}
  Varying this Schild action with respect to $\sqrt{-h_{s}}$, we obtain
   \begin{eqnarray}
     \sqrt{-h_{s}} = \frac{1}{2N} \sqrt{\frac{a}{b}} \sqrt{-(
     \epsilon^{\alpha \beta} \partial_{\alpha} X^{\mu}
     \partial_{\beta} X_{\mu} )^{2}}. 
   \end{eqnarray}
   This scalar density is clearly identical to the determinant in the
   Green-Schwarz action $\sqrt{-h}$ up to a constant.
   Plugging this into the action, we obtain
   \begin{eqnarray}
    S_{\textrm{Schild}} = - \int d^{2} \sigma \left( \sqrt{ab}
    \sqrt{-(\epsilon^{\alpha \beta} 
    \partial_{\alpha} X^{\mu} \partial_{\beta} X_{\mu})^{2}}
    + \frac{ia}{2} \epsilon^{\alpha \beta} {\bar \psi} \Gamma^{\mu}
   \partial_{\alpha} X_{\mu} \partial_{\beta} \psi \right),
   \end{eqnarray}
 which reduces to the Green-Schwarz action by setting $(a,b) =
 (\frac{2}{\pi \alpha'}, \frac{1}{16 \pi \alpha'})$. 
 We discuss the matrix regularization of the Schild action
 (\ref{schild}). We perform the following replacement:
  \begin{eqnarray}
   - i \{ , \} = N [ , ], \hspace{2mm}
  \int d^{2} \sigma = \frac{1}{N} Tr. \label{matrixregularization}
  \end{eqnarray}
  By setting $a = \frac{1}{g^{2}}$ and dropping the term $\sqrt{-h_{s}}
  b$, the Schild action arrives at the IIB matrix model.

  We next discuss the supersymmetry of the Green-Schwarz action. 
  We identify the two combinations of the parameter of the
  original supersymmetry and the $\kappa$ symmetry, to obtain a new
  supersymmetry transformation. To this end, we relate the parameters
  of the supersymmetry and the $\kappa$ symmetry as
  \begin{eqnarray}
   \kappa^{1 \alpha} = - \frac{1}{4} \Gamma^{\mu} (\epsilon^{1} -
   \epsilon^{2}) \Pi_{\beta \mu} P_{-}^{\alpha \beta}, \hspace{2mm}
   \kappa^{2 \alpha} =   \frac{1}{4} \Gamma^{\mu} (\epsilon^{1} -
   \epsilon^{2}) \Pi_{\beta \mu} P_{+}^{\alpha \beta}. \label{param}
  \end{eqnarray}
  We take the following gauge for the Schild action of the type IIB
  superstring theory:
  \begin{eqnarray}
   \theta^{1} = \theta^{2} (= \theta), \hspace{2mm}
   \Pi^{\mu}_{\alpha} = \partial_{\alpha} X^{\mu}, \hspace{2mm}
   \frac{1}{\sqrt{-h_{s}}} \epsilon^{\alpha \beta} \Pi^{\mu}_{\alpha}
   \Pi^{\nu}_{\beta} = \{ X^{\mu}, X^{\nu} \}.
  \end{eqnarray}
 Thus, the fermionic supersymmetry transformation is written as
  \begin{eqnarray}
   \delta_{\kappa^{1}} \theta^{1}
  =  \frac{1}{2} \left( 1 - \frac{1}{2} \Gamma^{\mu \nu} \{ X_{\mu},
  X_{\nu} \} \right) 
   (\epsilon^{1} - \epsilon^{2}), \hspace{2mm}
  \delta_{\kappa^{2}} \theta^{2}
  = \frac{1}{2} \left( 1 + \frac{1}{2} \Gamma^{\mu \nu} \{ X_{\mu},
  X_{\nu} \} \right) (\epsilon^{1} - \epsilon^{2}). \label{susyret1}
  \end{eqnarray}
  Thus, the $\kappa$ symmetry transformation for $X^{\mu}$ is
   \begin{eqnarray}
    \delta_{\kappa} X^{\mu} = - i {\bar \theta} \Gamma^{\mu}
    (\epsilon_{1} - \epsilon_{2}). \label{susycoordinate2}
   \end{eqnarray}
  We introduce the following redefinition of the supersymmetry
  parameter:
  \begin{eqnarray}
    \chi  = 2 (\epsilon^{1} - \epsilon^{2}), \hspace{2mm}
    \chi' = \frac{(\epsilon^{1} + \epsilon^{2})}{2}. 
  \end{eqnarray}
  The combination of the supersymmetry and the $\kappa$ symmetry is
  thus given by
  \begin{eqnarray}
   (\delta_{\epsilon} + \delta_{\kappa}) X^{\mu}
 = i {\bar \chi} \Gamma^{\mu} \theta, \hspace{2mm}
  (\delta_{\epsilon} + \delta_{\kappa}) \theta
 = \chi' + \frac{1}{8} \Gamma^{\mu \nu} \{ X_{\mu}, X_{\nu} \} \chi.
  \label{susyeclectic} 
  \end{eqnarray}

 \subsubsection{Supersymmetry}
 Now, let us have a careful look at the supersymmetry of the IIB
  matrix model. We have 
  formulated the supersymmetry of the  Schild action of the type IIB
  superstring. In the IIB matrix model we consider that the
  matrix-regularized version of that supersymmetry is inherited. The
  ${\cal N}=2$ supersymmetry of the IIB matrix model is then 
  distinguished by
   \begin{itemize}
    \item{homogeneous: $\delta^{(1)}_{\epsilon} \psi = \frac{i}{2}
    [A_{\mu} , A_{\nu}] \Gamma^{\mu\nu} \epsilon, \hspace{2mm} 
    \delta^{(1)}_{\epsilon} A_{\mu} = i {\bar  \epsilon} \Gamma_{\mu}
    \psi$. \\ 
    The feature of the
    {\it 'homogeneous supersymmetry'} is that this supersymmetry transformation depends
    on the matter fields $A_{\mu}$ and $\psi$. And this supersymmetry
    transformation vanishes if there is no matter field.}
   \item{inhomogeneous: $\delta^{(2)}_{\xi} \psi = \xi, \hspace{2mm}
       \delta^{(2)}_{\xi}  A_{\mu} = 0$. \\
       The feature of the {\it
       'inhomogeneous supersymmetry'} is that the translation  survives without
       the matter fields.} 
   \end{itemize}
  The commutators of these supersymmetries give the following important
  results:
   \begin{eqnarray}
    &(1)& [\delta^{(1)}_{\epsilon_{1}}, \delta^{(1)}_{\epsilon_{2}}]
    A_{\mu} = 0 ,\hspace{2mm} [\delta^{(1)}_{\epsilon_{1}},
    \delta^{(1)}_{\epsilon_{2}} ] \psi = 0,
    \label{AZM31SUSY1com} \\
    &(2)& [\delta^{(2)}_{\xi_{1}}, \delta^{(2)}_{\xi_{2}}] A_{\mu} =
    0, \hspace{2mm} [\delta^{(2)}_{\xi_{1}}, \delta^{(2)}_{\xi_{2}}] = 
    0, \label{AZM31SUSY2com} \\
    &(3)& [\delta^{(1)}_{\epsilon}, \delta^{(2)}_{\xi}] A_{\mu} = - i
    {\bar \epsilon} \Gamma_{\mu} \xi, \hspace{2mm}
    [\delta^{(1)}_{\epsilon}, \delta^{(2)}_{\xi}] \psi = 0.
    \label{AZM31SUSY3com} 
   \end{eqnarray}
  These properties can be verified by taking the difference of
 the two supersymmetry transformations.
  \begin{enumerate}
   \item{This is the most complicated to compute. For the gauge field, 
       we should consider the following transformation
      \begin{eqnarray}
          A_{\mu} &\stackrel{\delta^{(1)}_{\epsilon_{2}}}{\rightarrow}& 
          A_{\mu} + i \epsilon_{2} \Gamma_{\mu} \psi
          \stackrel{\delta^{(1)}_{\epsilon_{1}}}{\rightarrow} A_{\mu} +
          i ({\bar \epsilon_{1}} + {\bar \epsilon_{2}}) \Gamma_{\mu}
          \psi - \frac{1}{2} {\bar \epsilon_{2}} \Gamma_{\mu} [A_{\nu},
          A_{\rho}] \Gamma^{\nu\rho} \epsilon_{1},   \\
          A_{\mu} &\stackrel{\delta^{(1)}_{\epsilon_{1}}}{\rightarrow}& 
          A_{\mu} + i \epsilon_{1} \Gamma_{\mu} \psi
          \stackrel{\delta^{(1)}_{\epsilon_{2}}}{\rightarrow} A_{\mu} +
          i ({\bar \epsilon_{1}} + {\bar \epsilon_{2}}) \Gamma_{\mu}
          \psi - \frac{1}{2} {\bar \epsilon_{1}} \Gamma_{\mu} [A_{\nu},
          A_{\rho}] \Gamma^{\nu\rho} \epsilon_{2}.
      \end{eqnarray}
      Then, the commutator is
     \begin{eqnarray}
       [\delta^{(1)}_{\epsilon_{1}}, \delta^{(1)}_{\epsilon_{2}}]
       A_{\mu} =  - \frac{1}{2} {\bar \epsilon_{2}} \Gamma_{\mu} [A_{\nu},
          A_{\rho}] \Gamma^{\nu\rho} \epsilon_{1}  + \frac{1}{2} {\bar
       \epsilon_{1}} \Gamma_{\mu} [A_{\nu}, A_{\rho}] \Gamma^{\nu\rho}
       \epsilon_{2}.         
     \end{eqnarray}
    Utilizing the formula $\Gamma^{\mu} \Gamma^{\nu\rho} =
    \Gamma^{\mu\nu\rho} + 
    \eta^{\mu\nu} \Gamma^{\rho} - \eta^{\mu\rho} \Gamma^{\nu}$, we obtain
    \begin{eqnarray}
      [\delta^{(1)}_{\epsilon_{1}}, \delta^{(1)}_{\epsilon_{2}}]
    A_{\mu} = 2 {\bar \epsilon_{1}} \Gamma^{\rho} \epsilon_{2} [A_{\mu},
    A_{\rho}]. \label{AZM31SUSY11bos}
    \end{eqnarray}
   For the fermions, we have only to repeat the similar procedure:
    \begin{eqnarray}
     \psi &\stackrel{\delta^{(1)}_{\epsilon_{2}}}{\to}& \psi + \frac{i}{2}
     [A_{\mu}, A_{\nu}] \Gamma^{\mu\nu} \epsilon_{2}
     \stackrel{\delta^{(1)}_{\epsilon_{1}}}{\to} \psi + \frac{i}{2}
     [A_{\mu}, A_{\nu}] \Gamma^{\mu\nu} (\epsilon_{1} + \epsilon_{2}) -
     [A_{\mu}, {\bar \epsilon_{1}} \Gamma_{\nu} \psi ] \Gamma^{\mu\nu}
     \epsilon_{2}, \\
     \psi &\stackrel{\delta^{(1)}_{\epsilon_{1}}}{\to}& \psi + \frac{i}{2}
     [A_{\mu}, A_{\nu}] \Gamma^{\mu\nu} \epsilon_{1}
     \stackrel{\delta^{(1)}_{\epsilon_{2}}}{\to} \psi + \frac{i}{2}
     [A_{\mu}, A_{\nu}] \Gamma^{\mu\nu} (\epsilon_{1} + \epsilon_{2}) -
     [A_{\mu}, {\bar \epsilon_{2}} \Gamma_{\nu} \psi ] \Gamma^{\mu\nu}
     \epsilon_{1}.
    \end{eqnarray}    
    Using the formula of the Fierz transformation, 
     \begin{eqnarray}
      {\bar \epsilon_{1}} \Gamma_{\nu} \psi \Gamma^{\mu\nu} \epsilon_{2} =
      {\bar \epsilon_{1}} \Gamma^{\mu} \epsilon_{2} \psi - \frac{7}{16}
      {\bar \epsilon_{1}} \Gamma^{\rho} \epsilon_{2} \Gamma_{\rho}
      \Gamma^{\mu} \psi - \frac{1}{16 \times 5!} {\bar \epsilon_{1}}
      \Gamma^{\rho_{1} \cdots \rho_{5}} \epsilon_{2} \Gamma_{\rho_{1} \cdots
      \rho_{5}} \Gamma^{\mu} \psi + (\textrm{rank 3 term}), \nonumber \\
     \label{AZM31fierz}
     \end{eqnarray} 
  whose proof we present in Appendix \ref{sec-fierz},
  we verify that the commutator of the supersymmetry transformation is 
    \begin{eqnarray}
      [\delta^{(1)}_{\epsilon_{1}},
    \delta^{(1)}_{\epsilon_{2}} ] \psi = 2 [\psi,   {\bar
    \epsilon_{1}} \Gamma^{\rho} \epsilon_{2} A_{\rho} ]. \label{AZM31aho}
    \label{AZM31SUSY11ferm} 
    \end{eqnarray}
  Here, we utilize the equation of motion
       \begin{eqnarray}
     \frac{d S}{d {\bar \psi}} = - \frac{1}{g^{2}} \Gamma^{\mu} 
      [A_{\mu}, \psi] = 0,
       \end{eqnarray}
   in order to eliminate the second and the third terms of the Fierz
   identity (\ref{AZM31fierz}).

   Next, we note that the commutators of the supersymmetry transformation
   (\ref{AZM31SUSY11bos}) and 
   (\ref{AZM31SUSY11ferm}) vanish up to the gauge transformation. The
   gauge transformation of the IIB model is to multiply the unitary
   matrix $i \alpha \in SU(N)$. The gauge transformation is expressed in
   the infinitesimal form as follows :
  \begin{eqnarray}
   A_{\mu}, \psi \rightarrow A_{\mu}+i[A_{\mu}, \alpha], \hspace{2mm} \psi
   + i [\psi, \alpha]. \label{AZM31gaugetr}
  \end{eqnarray}
   The supersymmetry transformation  (\ref{AZM31SUSY11bos}) and
   (\ref{AZM31SUSY11ferm}) can be gauged away by the gauge parameter
   $\alpha = 2 {\bar \epsilon_{1}} \Gamma^{\rho} \epsilon_{2} A_{\rho}$.  We 
   now complete the proof of (\ref{AZM31SUSY1com}) up to the gauge
   transformation. 
      }
   \item{This is trivial because the supersymmetry $\delta^{(2)}_{\xi}$
       involves only a constant.}
    \item{This can be proven by taking the difference of these two
        transformations  
       \begin{eqnarray}
        A_{\mu} &\stackrel{ \delta^{(2)}_{\xi}}{\to}&  A_{\mu}
       \stackrel{\delta^{(1)}_{\epsilon}}{\to} A_{\mu} +  i {\bar \epsilon}
      \Gamma_{\mu} \psi \textrm{, whereas }
       A_{\mu}\stackrel{\delta^{(1)}_{\epsilon}}{\to} A_{\mu} +  i
       {\bar \epsilon}  \Gamma_{\mu} \psi  \stackrel
       { \delta^{(2)}_{\xi}}{\to}  A_{\mu} +  i{\bar \epsilon}
       \Gamma_{\mu} (\psi + \xi),  \nonumber  \\
      \psi  &\stackrel{ \delta^{(2)}_{\xi}}{\to}& \psi + \xi
       \stackrel{\delta^{(1)}_{\epsilon}}{\to}  \psi + \xi +  \frac{i}{2}
      \Gamma^{\mu\nu} [ A_{\mu} , A_{\nu} ] \epsilon 
       \textrm{, whereas } \nonumber \\
   \psi &\stackrel{\delta^{(1)}_{\epsilon}}{\to}&   \psi +  \frac{i}{2}
      \Gamma^{\mu\nu} [ A_{\mu} , A_{\nu} ] \epsilon 
    \stackrel{ \delta^{(2)}_{\xi}}{\to} \psi + \xi +  \frac{i}{2}
      \Gamma^{\mu\nu} [ A_{\mu} , A_{\nu} ] \epsilon.
    \end{eqnarray}
  }
  \end{enumerate}
  This completes the proof of the commutation relation of the supersymmetry
  transformation. When we take the linear combination of the
  homogeneous and the inhomogeneous supersymmetry as
  \begin{eqnarray}
   {\tilde \delta}^{(1)}_{\epsilon} = \delta^{(1)}_{\epsilon} +
   \delta^{(2)}_{\epsilon}, \hspace{2mm}
   {\tilde \delta}^{(2)}_{\epsilon} = i (\delta^{(1)}_{\epsilon} -
   \delta^{(2)}_{\epsilon}),
  \end{eqnarray}
  the commutator is written as
  \begin{eqnarray}
   [{\tilde \delta}^{(\alpha)}_{\epsilon}, {\tilde
   \delta}^{(\beta)}_{\xi}] \psi = 0, \hspace{2mm}
   [{\tilde \delta}^{(\alpha)}_{\epsilon}, {\tilde
   \delta}^{(\beta)}_{\xi}] A_{\mu} = - 2i {\bar \epsilon}
   \Gamma_{\mu} \xi \delta^{\alpha \beta}, \label{finalcc} 
  \end{eqnarray}
 where $\alpha,\beta$ run over $1,2$. The commutation relation
 (\ref{finalcc}) has a serious consequence. When we regard the
 eigenvalues of the bosonic matrices $A_{\mu}$ as the spacetime
 coordinate, the IIB matrix model carries a spacetime ${\cal N}=2$
 supersymmetry. Namely, the commutator of the supersymmetry
 transformations gives a translation of the spacetime by
 $a_{\mu} = - 2i {\bar \epsilon} \Gamma_{\mu} \xi$.
 This urges us to interpret the spacetime as emerging from the
 eigenvalues of $A_{\mu}$. 
 This is a crucial property in order for this model to 
 include the gravitational interaction. If the IIB matrix model has a
 massless particle, it must have a spin-2 graviton.
 We have a good reason to reduce not the four or six-dimensional, but
 the ten-dimensional ${\cal N}=1$ super Yang-Mills theory. This model
 has the 32 maximal supersymmetry only if we reduce the
 ten-dimensional super Yang-Mills theory. The maximal supersymmetry 
 is an essential aspect for the theory of gravity.

 \subsection{Interaction between the BPS object}
  We next review the interpretation of the BPS objects (such as the
  D-brane) in terms of the IIB matrix model. We start with considering 
  the classical equation of motion of the IIB matrix model:
  \begin{eqnarray}
   [A_{\nu}, [A_{\mu}, A_{\nu}]] = 0. \label{IIBeom}
  \end{eqnarray}
  Here, we set the fermion $\psi$ to zero. The corresponding equation
  of motion for the type IIB superstring theory is given by
  \begin{eqnarray}
   \{ X_{\nu}, \{ X^{\mu}, X^{\nu} \} \} = 0. \label{IIBeomschild}
  \end{eqnarray}
  In terms of the type IIB superstring theory, this has a solution
  representing the D1-brane:
  \begin{eqnarray}
   X_{0} = T \tau, \hspace{2mm}
   X_{1} = \frac{L}{2\pi} \sigma, \hspace{2mm}
   X_{2} = X_{3} = \cdots = X_{9} = 0, \label{d1brane}
  \end{eqnarray}
 where $T$ and $L$ are the compactification radii of $X^{0}$ and
 $X^{1}$, respectively. The parameter $\tau$ and $\sigma$ take values
  $0 \leq \tau \leq 1$ and $0 \leq \sigma \leq 2 \pi$. The Poisson
 bracket is given by
  \begin{eqnarray}
   \{ X_{0}, X_{1} \} = \epsilon^{01} \partial_{0} X_{0} \partial_{1}
   X_{1} = T \times \frac{L}{2 \pi} =  \frac{TL}{2\pi}.
  \end{eqnarray}
  The corresponding commutator for the matrix regularization is
  \begin{eqnarray}
  - i [A_{0}, A_{1}] =  \frac{TL}{2 \pi N}
 \Leftrightarrow \{ X_{0}, X_{1} \} = \frac{TL}{2\pi}.
  \end{eqnarray}
  This commutation relation is realized only for the infinite-size
  matrices (we can see that this is impossible for the finite-size
  matrices by using the cyclic rule of the trace $Tr$).
  Namely, we have the following solution:
   \begin{eqnarray}
     A_{0} = \frac{T}{\sqrt{2 \pi N}} q, \hspace{2mm}
     A_{1} = \frac{L}{\sqrt{2 \pi N}} p, \hspace{2mm}
     A_{2} = A_{3} = \cdots = A_{9} = 0. \label{d1branematrix}
   \end{eqnarray}
 where $q,p$ are the infinite-size canonical pair satisfying
 $[q,p]=i$.

 We discern that the solution (\ref{d1branematrix}) is a BPS saturated 
 state in the following
 way. We substitute the solution (\ref{d1branematrix}), along with 
 $A_{2} = A_{3} = \cdots = A_{9} = 0$ and $\psi = 0$, into the
 supersymmetry transformation in the previous section to obtain
  \begin{eqnarray}
  & & \delta_{\epsilon}^{(1)} \psi = - \frac{TL}{4 \pi N}
   \Gamma^{01} \epsilon, \hspace{2mm}
   \delta_{\epsilon}^{(1)} A_{\mu} = 0, \nonumber \\
  & & \delta_{\xi}^{(2)} \psi = \xi, \hspace{2mm}
    \delta_{\xi}^{(2)} A_{\mu} = 0. \label{susyd1brane}
  \end{eqnarray}
  The homogeneous supersymmetry can be clearly canceled by the
  inhomogeneous supersymmetry, when we properly choose the parameter
  $\xi$. In this sense, the D1-brane solution of the IIB matrix model
  (\ref{d1branematrix}) actually preserves half of the
  supersymmetry, and thus is a BPS saturated object.

  By the same token, we can establish the solutions of the IIB matrix
  model that correspond to the $D3,D5, \cdots$ brane, by taking the
  similar canonical pairs as
  \begin{eqnarray}
   [X_{2}, X_{3}] \propto {\bf 1}_{N \times N}, \hspace{2mm}
   [X_{4}, X_{5}] \propto {\bf 1}_{N \times N}, \cdots. \label{d3579brane}
  \end{eqnarray}
  This nicely corresponds to the fact that the type IIB superstring
  theory accepts the D1,3,5,$\cdots$-branes.

  We next see that the bosonic matrices $A_{\mu}$ of the IIB matrix
  model describe the multi-body system. We can easily build the
  solution for the two D1-branes located at $x_{2} = \pm \frac{b}{2}$
  by taking the following block-diagonal matrices. 
 \begin{eqnarray}
  & & A_{0} = \left( \begin{array}{cc} \frac{T}{\sqrt{2 \pi n_{1}}} q & 0 \\
   0 & \frac{T}{\sqrt{2 \pi n_{2}}} q' \end{array} \right), \hspace{2mm}
     A_{1} = \left( \begin{array}{cc} \frac{T}{\sqrt{2 \pi n_{1}}} p & 0 \\
   0 & \frac{T}{\sqrt{2 \pi n_{2}}} p' \end{array} \right), \hspace{2mm}
     A_{2} = \left( \begin{array}{cc} \frac{b}{2} & 0 \\
   0 & - \frac{b}{2} \end{array} \right), \nonumber \\
  & & A_{3} = \cdots = A_{9} = 0, \label{d1-d1brane}
  \end{eqnarray}
 
 where $n_{1,2}$ is the size of each block, and $q,q',p,p'$ satisfy
 $[q,p] = [q',p'] = i$. 
  
 By the same token, we can build the $D1-{\bar D1}$ brane system
 through the following classical solution.
   \begin{eqnarray}
  & & A_{0} = \left( \begin{array}{cc} \frac{T}{\sqrt{2 \pi n_{1}}} q & 0 \\
   0 & \frac{T}{\sqrt{2 \pi n_{2}}} q' \end{array} \right), \hspace{2mm}
   A_{1} = \left( \begin{array}{cc} \frac{T}{\sqrt{2 \pi n_{1}}} p & 0 \\
   0 & - \frac{T}{\sqrt{2 \pi n_{2}}} p' \end{array} \right), \hspace{2mm}
   A_{2} = \left( \begin{array}{cc} \frac{b}{2} & 0 \\
   0 & - \frac{b}{2} \end{array} \right), \nonumber \\
 & & A_{3} = \cdots = A_{9} = 0,  \label{d1-antid1brane}
  \end{eqnarray}
   \begin{figure}[htbp]
    \begin{center}
    \scalebox{.5}{\includegraphics{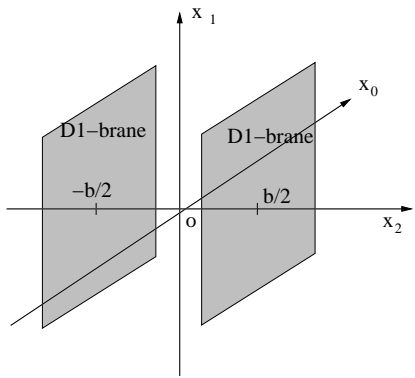} }
    \end{center}
   \caption{The two parallel D1-branes, represented by the classical
     solution of the IIB matrix model (\ref{d1-d1brane}).} 
  \label{d1brane-fig}
  \end{figure} 
    In this way, the IIB matrix model describes not only the single
  D-objects , but also the multi-body system by allocating the plural
  matters on the block-diagonal components. In this sense, there is a
  great difference from the action of the D-instanton action, which
  appears to be the same action. Unlike the D-instanton action, the
  IIB matrix model accepts the multi-body system and thus is a
  second-quantized action.

  Not only do the matrices of the IIB matrix model $A_{\mu}$
  accommodate the multi-body system but also the interaction is
  embedded in the same matrices. Namely, the off-diagonal part can be
  interpreted as representing the interaction of each block.
    \begin{figure}[htbp]
   \begin{center}
    \scalebox{.4}{\includegraphics{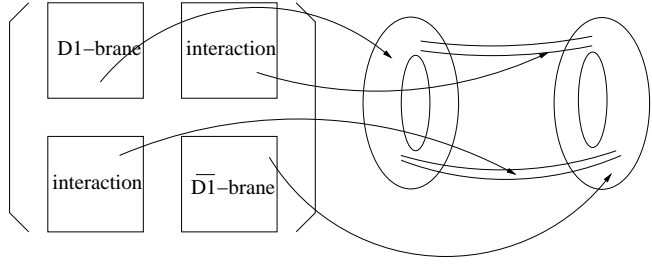} }
   \end{center}
   \caption{The off-diagonal parts represent the
     interaction in the IIB matrix model.} 
  \label{interaction}
  \end{figure}

  We next investigate the interaction between the D-objects by the
  one-loop calculation. To this end, we derive the one-loop effective
  action. We separate the matrices with the classical solution and the 
  fluctuation as
  \begin{eqnarray}
   A_{\mu} = p_{\mu} + a_{\mu}, \hspace{2mm}
   \psi = \chi + \varphi. \label{classical-fluctuation}
  \end{eqnarray}

  Here, $p_{\mu}$ and $\chi$ are the classical solution, and $a_{\mu}$
  and $\varphi$ are the fluctuation. Since the IIB matrix model is
  invariant under the $SU(N)$ gauge transformation (\ref{AZM31gaugetr}),
  we need the gauge fixing. Here, we adopt the background gauge
  \begin{eqnarray}
   -i [p_{\mu}, A^{\mu}] = 0. \label{backgroundgauge}
  \end{eqnarray}
  The corresponding gauge-fixing term  $S_{\textrm{g.f.}}$ and the
  ghost term $S_{\textrm{ghost}}$ are 
  respectively
  \begin{eqnarray}
   S_{\textrm{g.f.}} = - \frac{1}{2} Tr [p_{\mu}, A^{\mu}]^{2},
   \hspace{2mm}
  S_{\textrm{ghost}} = - Tr ([p_{\mu}, b][p^{\mu},c]), \label{gaugefixingterm} 
  \end{eqnarray}
 where $c$ and $b$ are the ghosts and the anti-ghosts respectively.
 In the following, we set the parameter $g$ to 1. Then, we extract the 
 quadratic term of the fluctuation for the total action
 $S^{\textrm{(total)}} = S + S_{\textrm{g.f.}} + S_{\textrm{ghost}}$ as
  \begin{eqnarray}
   S_{2}^{\textrm{(total)}} &=& - Tr \left( \frac{1}{2} [p_{\mu},
   a_{\nu}]^{2} + [p_{\mu}, p_{\nu}][a^{\mu}, a^{\nu}] 
   + \frac{1}{2} {\bar \varphi} \Gamma^{\mu} [p_{\mu}, \varphi]
   + [p_{\mu}, b][p^{\mu}, c] \right) \nonumber \\
  &=& Tr \left( \frac{1}{2} a^{\mu} (P^{2} \delta_{\mu \nu} - 2i
   F_{\mu \nu}) a^{\nu} + \frac{1}{2} {\bar \varphi}
   \Gamma^{\mu} P_{\mu} \varphi - b P^{2} c \right). \label{one-loop}
  \end{eqnarray}
 Here, we introduce the following adjoint operators
  \begin{eqnarray}
   P_{\mu} X = (\textrm{ad}p_{\mu}) X = [p_{\mu},X], \hspace{2mm} 
   F_{\mu \nu} X = (\textrm{ad} (-i[p_{\mu}, p_{\nu}])) X
             = -i [P_{\mu}, P_{\nu}] X, \label{adjoint}
  \end{eqnarray}
 where the last equality for $F_{\mu \nu}$ follows from the Jacobi
 identity of the commutators. Integrating the fluctuation, we obtain
 the following effective action.
  \begin{eqnarray}
   W &=& - \log \int da db dc d \varphi \exp (-S_{2}^{\textrm{(total)}})
     \nonumber \\
     &=& \frac{1}{2} Tr \left( tr_{10 \times 10} \log (P^{2}
     \delta_{\mu \nu} - 2i F_{\mu \nu}) 
   - \log P^{2} - \frac{1}{4} tr_{16 \times 16} \log \left( P^{2}+ \frac{i}{2}
     F_{\mu \nu} \Gamma^{\mu \nu} \frac{1+\Gamma^{\sharp}}{2} \right)
     \right).   \label{effectiveaction}
  \end{eqnarray}
 Here, the trace $Tr$ is for the $N \times N$ matrices, while the
 trace $tr_{10 \times 10}$ and $tr_{16 \times 16}$ are respectively
 for the ten-dimensional vector indices and the ten-dimensional gamma
 matrices ($\frac{1+\Gamma^{\sharp}}{2}$ is the Weyl projector, and
 effectively the size is not 32 but 16). 

 We evaluate the interaction between the diagonal blocks using the
 one-loop effective action (\ref{effectiveaction}), and elucidate that 
 the IIB matrix model includes the gravitational interaction. 
 We consider the backgrounds having a block-diagonal form:
  \begin{eqnarray}
   A_{\mu} = p_{\mu} = \textrm{diag}(p^{(1)}_{\mu}, p^{(2)}_{\mu},
   \cdots ). \label{bdbg}
  \end{eqnarray}
 Here, each block $p^{(i)}_{\mu}$ represents the separate D-object and 
 is an $n_{i} \times n_{i}$ matrix. We decompose the classical
 background between the trace and the traceless part as
  \begin{eqnarray}
   p^{(i)}_{\mu} = d^{(i)}_{\mu} 1_{n_{i} \times n_{i}} + {\tilde
   p}^{(i)}_{\mu}.
  \end{eqnarray}
 Since the eigenvalues of the bosonic matrices are regarded as the
 spacetime coordinate in the IIB matrix model, the trace
 $d^{(i)}_{\mu}$ should be identified with the collective coordinate
 of the D-objects. We assume that the classical D-objects are
 separated with each other so that the distances $d^{(i)}_{\mu} -
 d^{(j)}_{\mu}$ are sufficiently large.

 The adjoint operator $P_{\mu}$ operates on the $(i,j)$ components of
 the matrix $X^{(i,j)}$ as
  \begin{eqnarray}
   (P_{\mu} X)^{(i,j)} = (d^{(i)}_{\mu} - d^{(j)}_{\mu}) X^{(i,j)}
   + {\tilde p}^{(i)}_{\mu} X^{(i,j)}
  - X^{(i,j)} {\tilde p}^{(j)}_{\mu}
  = ((d^{(i)}_{\mu} - d^{(j)}_{\mu}) + P_{L,\mu}^{(i,j)} +
   P_{R,\mu}^{(i,j)} ) X^{(i,j)}, 
  \end{eqnarray}
 where $P_{L,\mu}^{(i,j)} X^{(i,j)}={\tilde p}^{(i)}_{\mu} X^{(i,j)}$
 and $P_{R,\mu}^{(i,j)} =  - X^{(i,j)} {\tilde p}^{(j)}_{\mu}$ represent the
 operation of $P_{\mu}$ from the left and right, respectively.
 Similarly, we simplify the notation of the operation of the
 adjoint operator $F_{\mu \nu}$. The background field strength is
 rewritten as
  \begin{eqnarray}
  f_{\mu \nu} = \textrm{diag} ( 
  i[p^{(1)}_{\mu}, p^{(1)}_{\nu}], i [p^{(2)}_{\mu},
  p^{(2)}_{\nu}], \cdots)
  =   \textrm{diag} ( 
  {\tilde f}^{(1)}_{\mu \nu}, {\tilde f}^{(2)}_{\mu \nu}, \cdots).
  \end{eqnarray}
 The operation of $F_{\mu \nu}$ can be likewise decomposed into the
 left and right operation as
  \begin{eqnarray}
  (F_{\mu \nu} X)^{(i,j)} = {\tilde f}^{(i)}_{\mu \nu} X^{(i,j)}
  - X^{(i,j)} {\tilde f}^{(j)}_{\mu \nu} 
  = (F^{(i,j)}_{L, \mu \nu} + F^{(i,j)}_{R, \mu \nu}) X^{(i,j)}. 
  \end{eqnarray}
 Since the left and right operation are independent, we have
  \begin{eqnarray}
  Tr O = \sum_{i,j=1}^{n} Tr O_{L}^{(i,j)} Tr O_{R}^{(i,j)}.
  \end{eqnarray}
  We obtain the one-loop interaction for these D-objects. To this end, 
  we expand the effective action (\ref{effectiveaction}) with respect
  to the power of $F_{\mu}$. The first bosonic term is expanded as
  \begin{eqnarray}
 & & \frac{1}{2} Tr \left( tr_{10 \times 10} \log (P^{2} \delta_{\mu
 \nu} - 2i F_{\mu \nu}) \right) \nonumber \\ 
 &=& Tr \left( 5 \log P^{2} 
   + \frac{1}{P^{2}} F_{\mu \nu} \frac{1}{P^{2}} F_{\nu \mu}
     - \frac{4i}{3} \frac{1}{P^{2}} F_{\mu \nu} \frac{1}{P^{2}} F_{\nu 
     \rho} \frac{1}{P^{2}} F_{\rho \mu} 
   - 2 \frac{1}{P^{2}} F_{\mu \nu} \frac{1}{P^{2}} F_{\nu \rho}
     \frac{1}{P^{2}} F_{\rho \chi} \frac{1}{P^{2}} F_{\chi \mu}
     \right) + {\cal O}(P^{-10}). \nonumber 
  \end{eqnarray}
 Whereas, the third fermionic term of (\ref{effectiveaction}) is
 expanded as
  \begin{eqnarray}
  & & - \frac{1}{4} Tr \left( tr_{16 \times 16} \log \left( P^{2}+ \frac{i}{2}
     F_{\mu \nu} \Gamma^{\mu \nu} \frac{1+\Gamma^{\sharp}}{2} \right)
     \right) \nonumber \\
  &=& Tr \left( -4 \log P^{2} 
   - \frac{1}{P^{2}} F_{\mu \nu} \frac{1}{P^{2}} F_{\nu \mu}
   + \frac{4i}{3} \frac{1}{P^{2}} F_{\mu \nu} \frac{1}{P^{2}} F_{\nu 
     \rho} \frac{1}{P^{2}} F_{\rho \mu} 
   +  \frac{1}{P^{2}} F_{\mu \nu} \frac{1}{P^{2}} F_{\nu \rho}
     \frac{1}{P^{2}} F_{\rho \chi} \frac{1}{P^{2}} F_{\chi \mu}
     \right. \nonumber \\ 
  & & + \frac{1}{4} \frac{1}{P^{2}} F_{\mu \nu}  \frac{1}{P^{2}}
     F_{\rho \chi} \frac{1}{P^{2}} F_{\mu \nu} \frac{1}{P^{2}} F_{\rho 
     \chi} 
    + \frac{1}{2}  \frac{1}{P^{2}} F_{\mu \nu}  \frac{1}{P^{2}} F_{\mu 
     \nu} \frac{1}{P^{2}} F_{\rho \chi} \frac{1}{P^{2}} F_{\rho \chi}
     \nonumber \\
  & & \left.
   - 2  \frac{1}{P^{2}} F_{\mu \nu} \frac{1}{P^{2}} F_{\mu \rho}
     \frac{1}{P^{2}} F_{\chi \nu} \frac{1}{P^{2}} F_{\chi \rho} \right) 
   + {\cal O}(P^{-10}). \nonumber
  \end{eqnarray}

In this derivation, we utilize the following formulae of the gamma matrices:
  \begin{eqnarray}
 & & tr (\Gamma^{\mu_{1} \mu_{2}} \Gamma^{\nu_{1} \nu_{2}})
 = -2 tr( \eta^{[\mu_{1} [\nu_{1}} \eta^{\mu_{2}]\nu_{2}]}) \nonumber
 \\
 & & tr (\Gamma^{\mu_{1} \mu_{2}} \Gamma^{\nu_{1} \nu_{2}}
 \Gamma^{\rho_{1} \rho_{2}})
  =  8 tr (\eta^{[\mu_{1} [\nu_{1}} \eta^{\mu_{2}][\rho_{1}}
 \eta^{\nu_{2}] \rho_{2}]}), \nonumber \\
 & & tr (\Gamma^{\mu_{1} \mu_{2}} \Gamma^{\nu_{1} \nu_{2}}
 \Gamma^{\rho_{1} \rho_{2}} \Gamma^{\chi_{1} \chi_{2}}) 
  = tr(4 \eta^{[\mu_{1} [\nu_{1}} \eta^{\mu_{2}] \nu_{2}]} 
         \eta^{[\rho_{1} [\chi_{1}} \eta^{\rho_{2}] \chi_{2}]}
    + 8 \eta^{[\mu_{1} [\rho_{1}} \eta^{\mu_{2}] \rho_{2}]}
        \eta^{[\nu_{1} [\chi_{1}} \eta^{\nu_{2}] \chi_{2}]} \nonumber \\
 & & \hspace{40mm} 
    - 16 \eta^{[\mu_{1} [\nu_{1}} \eta^{[\rho_{1} [\chi_{1}}
 \eta^{\mu_{2}] \chi_{2}]} \eta^{\nu_{2}] \chi_{2}]} 
    - 16 \eta^{[\mu_{1} [\rho_{1}} \eta^{\mu_{2}] [\chi_{1}}
         \eta^{[\nu_{1} \chi_{2}]} \chi^{\nu_{2}] \chi_{2}]} \nonumber \\
 & & \hspace{40mm}
    + 16 \eta^{[\mu_{1} [\nu_{1}} \eta^{[\rho_{1} [\chi_{1}}
 \chi^{\mu_{2}] \chi_{2}]} \eta^{\nu_{2}] \rho_{2}]}).   
  \nonumber
  \end{eqnarray}

 Along with the second ghost term $-Tr \log P^{2}$,
 we discern that the effect of the boson
 and the fermion cancels up to the ${\cal O}(P^{-6})$ order. This
 cancellation is ascribed to the supersymmetry of the
 IIB matrix model. Due to this cancellation, the leading effect of the 
 one-loop interaction is of the order ${\cal O}(P^{-8})$. Namely, the
 interaction complies with the power law $\sim \frac{1}{r^{d-2}}$
 (where $r$ is the distance of the two D-objects, and the
 dimensionality is $d=10$). 
 In this sense, we regard this one-loop interaction as the
 gravitational interaction. This is an important evidence that the IIB 
 matrix model describes the gravitational interaction, and this is a
 consequence of the supersymmetry. Especially, the
 contribution of the $(i,j)$ block is expressed using the
 notation for the left and right operation as
  
 {\footnotesize
 \begin{eqnarray}
   W^{(i,j)} &=& \frac{n_{j}}{4(d^{(i)} - d^{(j)})^{8}} Tr \left(- 4
   ({\tilde f}^{(i)}_{\mu \nu} {\tilde f}^{(i)}_{\nu \rho}  {\tilde
   f}^{(i)}_{\rho \chi}  {\tilde f}^{(i)}_{\chi \mu} )
   - 8 ({\tilde f}^{(i)}_{\mu \nu} {\tilde f}^{(i)}_{\rho
   \chi}  {\tilde f}^{(i)}_{\mu \chi}  {\tilde f}^{(i)}_{\rho \nu} ) 
   + 2  ({\tilde f}^{(i)}_{\mu \nu} {\tilde f}^{(i)}_{\mu
   \nu}  {\tilde f}^{(i)}_{\rho \chi}  {\tilde f}^{(i)}_{\rho \chi} )
   +   ({\tilde f}^{(i)}_{\mu \nu} {\tilde f}^{(i)}_{\rho
   \chi}  {\tilde f}^{(i)}_{\mu \nu}  {\tilde f}^{(i)}_{\rho \chi} ) ) 
   \right) \nonumber \\  
  &+&  \frac{n_{i}}{4(d^{(i)} - d^{(j)})^{8}} Tr \left(- 4
   ({\tilde f}^{(j)}_{\mu \nu} {\tilde f}^{(j)}_{\nu \rho}  {\tilde
   f}^{(j)}_{\rho \chi}  {\tilde f}^{(j)}_{\chi \mu} )
   - 8 ({\tilde f}^{(j)}_{\mu \nu} {\tilde f}^{(j)}_{\rho
   \chi}  {\tilde f}^{(j)}_{\mu \chi}  {\tilde f}^{(j)}_{\rho \nu} ) 
   + 2  ({\tilde f}^{(j)}_{\mu \nu} {\tilde f}^{(j)}_{\mu
   \nu}  {\tilde f}^{(j)}_{\rho \chi}  {\tilde f}^{(j)}_{\rho \chi} )
   +   ({\tilde f}^{(j)}_{\mu \nu} {\tilde f}^{(j)}_{\rho
   \chi}  {\tilde f}^{(j)}_{\mu \nu}  {\tilde f}^{(j)}_{\rho \chi} ) ) 
   \right) \nonumber \\
  &+&  \frac{1}{4(d^{(i)} - d^{(j)})^{8}} \left(
    -48 Tr( {\tilde f}^{(i)}_{\mu \nu}{\tilde f}^{(i)}_{\nu \rho})
        Tr( {\tilde f}^{(j)}_{\mu \chi}{\tilde f}^{(j)}_{\chi \rho})
    + 6 Tr( {\tilde f}^{(i)}_{\mu \nu}{\tilde f}^{(i)}_{\mu \nu})
        Tr( {\tilde f}^{(j)}_{\rho \chi}{\tilde f}^{(j)}_{\rho \chi})
   \right) + {\cal O}((d^{(i)} - d^{(j)})^{-10}). \label{gravitondilaton}
  \end{eqnarray} } 
  The tensor structure indicates that the first and second term of the 
  last line of (\ref{gravitondilaton}) represent the graviton and
  dilaton exchange respectively. This result augurs very well for the
  IIB matrix model to be a bona fide framework for the gravitation
  interaction. This argument is limited to the background gauge, and
  it is an interesting future work to extend this argument in a gauge
  independent way.

  \subsection{Interpretation of the diffeomorphism invariance}
 In this subsection, we review the interpretation of the
 diffeomorphism invariance on the target space for the IIB matrix
 model. In \cite{9903217}, it has been pointed out that the
 invariance under the permutation of the eigenvalues $S_{N}$ is interpreted as
 the diffeomorphism invariance of the low-energy action. Since the
 permutation invariance is of course a subgroup of the $SU(N)$
 symmetry, this means that both the diffeomorphism invariance and the
 gauge invariance emerge from the $SU(N)$ invariance! This is a
 surprising aspect of the IIB matrix model, in that the unification of
 these two symmetries has been achieved in a natural way.
 
 In discussing the dynamics of the spacetime, we expand the IIB matrix 
 model around the diagonal background
  \begin{eqnarray}
   p_{\mu} = \textrm{diag}(x_{\mu}^{1}, x_{\mu}^{2}, \cdots,
   x_{\mu}^{N}), \hspace{2mm}
   \chi = \textrm{diag} (\chi^{1}, \chi^{2}, \cdots,
   \chi^{N}), \label{genericmoduli} 
  \end{eqnarray}
 where these satisfy the constraint $\sum_{i=1}^{N}
 x_{\mu}^{i} = 0$
 and $\sum_{i=1}^{N} \chi^{i} = 0$, because they belong to the $SU(N)$ 
 gauge group. We delegate the details to the references
 \cite{9802085,9903217,9908038,shino}, but after we integrate the
 fluctuation at the one-loop level and further the fermion zero mode
 $\chi$, we finally obtain the following interesting result:
  \begin{eqnarray}
   \int dX d\chi \exp (-S_{\textrm{eff}}^{\textrm{one-loop}}[x,\chi] )
  = \sum_{G:\textrm{graph}} \int dX W[X;G] \label{diffeoevi1}.
  \end{eqnarray}
 Here, $G$ denotes the graphs that connects the eigenvalues $X$, and
 $W$ is the Boltzmann weight for the graph $G$ and the configuration
 of the eigenvalues $X$. An explicit calculation also indicates that
 the dependence on the two different eigenvalues $x^{i}$ and $x^{j}$
 contributes at the order $(x^{i}-x^{j})^{-12}$, when $x^{i}$ and
 $x^{j}$ are connected by the graph $G$. 

 The important property of the integral (\ref{diffeoevi1}) is that it
 is invariant under the permutation of the eigenvalues
 $S_{N}$. We note that this is of course not the case with each graph
 $G$, because the way to connect the eigenvalues by the graphs
 impairs the $S_{N}$ invariance. Nevertheless, this symmetry is
 retrieved when we take a summation for all the graphs.
 This phenomenon is somewhat reminiscent of the dynamical
 triangulation approach to the quantum gravity, in which the
 diffeomorphism invariance is retrieved by summing over all the
 triangulation. 

 Here, we see how the permutation invariance of the eigenvalues leads
 to the diffeomorphism invariance. In a sense, this is a very natural
 interpretation, since we identify the eigenvalues with the spacetime
 coordinate. To elaborate on this viewpoint, we consider the scalar
 field $\phi^{i}$ propagating in the distributed eigenvalues.
 Namely, we consider the following effective action as an example;
  \begin{eqnarray}
  S = \sum_{i,j} \frac{(\phi^{i} - \phi^{j})^{2}}{2} f(x^{i} - x^{j})
  + \sum_{i} m (\phi^{i})^{2}. \label{toyscalar}
  \end{eqnarray}
  $f(x)$ is a function damping fast sufficiently at infinity. We
  introduce the density function $\rho(x) = \sum_{i}
  \delta^{(10)}(x-x^{i})$. Along with the continuous function
  $\phi(x)$ that satisfies $\phi^{i} = \phi(x^{i})$, the action
  (\ref{toyscalar}) is rewritten as
  \begin{eqnarray}
  S = \int dx dy \langle \rho(x) \rho(y) \rangle \frac{(\phi(x) -
  \phi(y))^{2}}{2} f(x-y) +  m \int dx \langle \rho(x) \rangle
  \phi(x)^{2}. \label{toyscalar2}
  \end{eqnarray}
 Here, we take an average with respect to the eigenvalue configuration 
 $X$ and the graphs $G$. In order to see clearly the correspondence
 between the gravitational background, we further rewrite the action as
  \begin{eqnarray}
   S &=& \frac{1}{2} \int dx \langle \rho(x) \rangle \left[ \int dy
   \langle \rho(y) \rangle (x-y)_{\mu} (x-y)_{\nu} f(x-y) (1+c(x,y))
   \right]
  \partial^{\mu} \phi(x) \partial^{\nu} \phi(x) \nonumber \\
  &+& m \int dx \langle \rho(x) \rangle \phi^{2}(x) + \cdots. 
  \end{eqnarray}
 Here, we normalize the density correlation as $\langle \rho(x)
 \rho(y) \rangle = \langle \rho(x) \rangle \langle \rho(y) \rangle (1
 + c(x,y))$. This urges us to identify the eigenvalue density with the 
 gravitational background as
  \begin{eqnarray}
   \sqrt{g} e^{-\Phi(x)} \sim \langle \rho(x) \rangle, \hspace{2mm}
   g_{\mu \nu} (x) =  \int dy \langle \rho(y) \rangle (x-y)_{\mu}
   (x-y)_{\nu} f(x-y) (1+c(x,y)). \label{backgroudmetric}
  \end{eqnarray}
 
 Nextly, we see how the diffeomorphism invariance is realized. Since
 the IIB matrix model itself is $SU(N)$ invariant, the model itself is 
 of course invariant under its subgroup $S_{N}$. Under the permutation
 $x^{i} \to x^{\sigma(i)}$ for $\sigma \in S_{N}$, the fields $\phi^{i}$ 
 transform to $\phi^{\sigma(i)}$. We extend the permutation of the
 eigenvalue to the transformation of the continuous transformation
 $x \to \xi(x)$, where $\xi(x)$ is a continuous function such that
 $\xi(x^{i}) = x^{\sigma(i)}$. It is trivial that $\phi(x)$ is subject
 to the general coordinate transformation as a scalar field.
 In addition, the background metric given in (\ref{backgroudmetric})
 also receives a transformation as a rank-2 tensor, when the function
 $f(x)$ decreases rapidly and has a support only near the origin $x=0$.
 In this sense, we can fairly interpret the $S_{N}$ invariance as the
 general coordinate invariance of the low-energy limit. 
 
 While the IIB matrix model has only a flat noncommutative background
 and depends heavily on the flat metric, we have an interpretation for
 the curved space background. A nontrivial background is induced
 dynamically through the condensation of the graviton.

 \subsection{Alternative totally reduced models} \label{alternativemodel}
  In this subsection, we review the attempts for the generalization of 
  the IIB matrix model. The IIB matrix model augurs well for the 
  unification of the gravity. The ${\cal N}=2$ supersymmetry is a
  fundamental object for the theory of gravity. In the preceding two
  sections, we have reviewed some of the evidences for the
  gravitational interaction: the graviton-dilaton exchange in the
  one-loop calculation and the interpretation of the diffeomorphism
  invariance. 
  However, the IIB matrix model suffers a serious drawback in that it
  has only the flat noncommutative background. However, if a matrix
  model is to be a bona fide framework to describe the
  gravitational interaction, it should accommodate a curved space
  background in a natural way. In this section, we introduce several
  generalizations of the IIB matrix model, so that they should
  incorporate a curved space background.

  \subsubsection{The IIB matrix model with the Chern-Simons term}
  \label{sec0101102}
  Firstly, we introduce the matrix model with the Chern-Simons term,
  defined by the following action\cite{0101102}:
  \begin{eqnarray}
   S = \frac{1}{g^{2}} Tr \left( - \frac{1}{4}[A_{\mu}, A_{\nu}][A^{\mu},
   A^{\nu}] + \frac{2i \alpha}{3} \epsilon_{\mu \nu \rho} A^{\mu}
   A^{\nu} A^{\rho}  
   + \frac{1}{2} {\bar \psi} \sigma^{\mu} [A_{\mu}, \psi]
   \right). \label{IIBfs} 
  \end{eqnarray}
  Here, the indices $\mu, \nu, \rho, \cdots$ run over $1,2,3$ and this model is
  defined on the three-dimensional Euclidean space. $A_{\mu}$ are the
  three-dimensional bosonic vectors and $\psi$ is the
  three-dimensional Majorana fermion. This is also a totally reduced
  model like the IIB matrix model, and $A_{\mu}$ and $\psi$ are promoted 
  to the $N \times N$ hermitian matrices. $\sigma_{\mu}$ are the Pauli
  matrices and in the following $\sigma_{\mu \nu}$ denotes
  $\sigma_{\mu \nu}  = \frac{1}{2}[\sigma_{\mu}, \sigma_{\nu}] = i
  \epsilon_{\mu \nu \rho} \sigma_{\rho}$. 
 This model has the $SO(3)$ rotational symmetry and the
  $SU(N)$ gauge symmetry. 

  In \cite{0309264}, Tomino performed the calculation of the path
  integral for the $N=2$ case as a toy model, and elucidated that the
  path integral converges for the $N=2$ case (while the bosonic model
  (\ref{verydefinition}) which we analyze in Sec. 5 does
  diverge). This is in contrast to 
  the IIB matrix model without the Chern-Simons term, in which the
  path integral diverges for $d=3$ and arbitrary $N$. 
  Austing and Wheater in \cite{0310170} corroborated that adding the
  Chern-Simons term does not affect the convergence as long as the
  original path integral (without the Chern-Simons term) converges
  absolutely.

  The interesting property of this matrix model is that it
  incorporates the classical solution of the fuzzy sphere. Its
  classical equation of motion is given by
  \begin{eqnarray}
   [A_{\mu}, [A_{\mu}, A_{\nu}]] + i \alpha \epsilon_{\nu \rho \chi}
   [A_{\rho}, A_{\chi}] = 0. \label{fseq}
  \end{eqnarray}
  The fuzzy-sphere solution is given by the $N$-dimensional 
  irreducible representation of the $SU(2)$ Lie algebra $L_{\mu}$:
  \begin{eqnarray}
   A_{\mu} = \alpha L_{\mu}, \textrm{ where } [L_{\mu}, L_{\nu}] = i
   \epsilon_{\mu \nu \rho} L_{\rho}. \label{fssolution}
  \end{eqnarray}
  This is regarded as the sphere, firstly because it satisfies the
  relation
  \begin{eqnarray}
   A_{1}^{2} + A_{2}^{2} + A_{3}^{2} = R^{2} {\bf 1}_{N \times N}
   = \alpha^{2} \frac{N^{2}-1}{4} {\bf 1}_{N \times N}. \label{radius}
  \end{eqnarray}
 The relation (\ref{radius}) determines the radius of the sphere as
 $R^{2} =  \alpha^{2} \frac{N^{2}-1}{4}$. In addition, the
 eigenvalues of $A_{\mu}$ are distributed sphere-like. In this sense,
 this classical solution is interpreted as the sphere in the
 three-dimensional spacetime. That this matrix model has the $S^{2}$
 fuzzy sphere classical solution is a great advantage. While the
 $S^{2}$ fuzzy sphere is nothing but a simple manifold, it is of
 significance as a prototype of the curved space background, which is
 a fundamental principle of the general relativity.

 This matrix model also incorporates the ${\cal N}=2$ supersymmetry.
 Like the IIB matrix model, this model also has the homogeneous and
 the inhomogeneous supersymmetry.
  \begin{eqnarray}
  & & \textrm{homogeneous: } \delta^{(1)}_{\epsilon} A_{\mu} = i {\bar
   \epsilon} \sigma_{\mu} \psi, \hspace{2mm}
   \delta^{(1)}_{\epsilon} \psi = \frac{i}{2} ([A_{\mu}, A_{\nu}] - i
   \alpha \epsilon_{\mu \nu \rho} A_{\rho}) \sigma^{\mu \nu} \epsilon,
   \label{homofsIIB} \\
  & & \textrm{inhomogeneous: } \delta^{(2)}_{\epsilon} A_{\mu} = 0,
   \hspace{2mm} 
   \delta^{(2)}_{\epsilon} \psi = \xi. \label{inhomofsIIB}
  \end{eqnarray}
  This model has the similar commutation relation to that of the IIB
  matrix model:
  \begin{eqnarray}
 & &  [\delta^{(1)}_{\epsilon_{1}}, \delta^{(1)}_{\epsilon_{2}}] A_{\mu} =
   0, \hspace{2mm}
   [\delta^{(1)}_{\epsilon_{1}}, \delta^{(1)}_{\epsilon_{2}}] \psi =
   0, \label{susyfsIIBcom1} \\
 & & [\delta^{(2)}_{\xi_{1}}, \delta^{(2)}_{\xi_{2}}] A_{\mu} =
   0, \hspace{2mm}
   [\delta^{(2)}_{\xi_{1}}, \delta^{(2)}_{\xi_{2}}] \psi =
   0, \label{susyfsIIBcom2} \\
 & &  [\delta^{(1)}_{\epsilon}, \delta^{(2)}_{\xi}] A_{\mu} =
   - i {\bar \epsilon} \sigma_{\mu} \xi, \hspace{2mm}
   [\delta^{(1)}_{\epsilon}, \delta^{(2)}_{\xi}] \psi =
   0. \label{susyfsIIBcom3}
  \end{eqnarray}
  The proof of these commutation relations goes in the same way as in
  the IIB matrix model, and we give only the proof of
  (\ref{susyfsIIBcom1}). 
  Using the formula $\sigma_{\mu} \sigma_{\nu \rho} = \delta_{\mu \nu}
 \sigma_{\rho} - \delta_{\mu \rho} \sigma_{\nu}$, 
  we obtain the commutation relation
  \begin{eqnarray}
   [\delta^{(1)}_{\epsilon_{1}}, \delta^{(1)}_{\epsilon_{2}}] A_{\mu}
 = i [A_{\mu}, \lambda] + \epsilon_{\mu \nu \rho} \theta_{\nu}
 A_{\rho}, \label{step001}
  \end{eqnarray}
 where $\lambda = 2 i ({\bar \epsilon_{2}} \sigma_{\mu} \epsilon_{1}) A_{\mu}$
 and $\theta_{\mu} = 2 i ({\bar \epsilon_{2}} \sigma_{\mu} \epsilon_{1})$.
 For the fermion, we obtain the commutator by taking the difference
 likewise:
  \begin{eqnarray}
    [\delta^{(1)}_{\epsilon_{1}}, \delta^{(1)}_{\epsilon_{2}}] \psi
 = \psi - \left([{\bar \epsilon}_{1} \sigma_{\mu} \psi, A_{\nu}]
 \sigma_{\mu \nu} \epsilon_{2} + \alpha ({\bar \epsilon}_{1}
 \sigma_{\rho} \psi) \sigma_{\rho} \epsilon_{2} \right)
  + (\epsilon_{1} \leftrightarrow \epsilon_{2}).   
  \end{eqnarray}
 We use the following Fierz transformation whose detailed proof we
 give in Appendix \ref{sec-fierz}:
  \begin{eqnarray}
  ({\bar \epsilon}_{1} \sigma_{\mu} \psi) \sigma_{\mu \nu}
  \epsilon_{2}
 = - ({\bar \epsilon}_{1} \sigma_{\nu} \epsilon_{2}) \psi + 
  \textrm{(rank-0 terms)},
  \hspace{2mm} 
  ({\bar \epsilon}_{1} \sigma_{\mu} \psi) \sigma_{\mu} \epsilon_{2}
 = \frac{1}{2} ({\bar \epsilon}_{1} \sigma_{\mu} \epsilon_{2})
 \sigma_{\mu} \psi + \textrm{(rank-0 terms)}. \label{fierz3dim} 
  \end{eqnarray}
 In (\ref{fierz3dim}), we omit the rank-0 contribution, since we are
 interested in the commutator $[\delta^{(1)}_{\epsilon_{1}},
 \delta^{(1)}_{\epsilon_{2}}] \psi$. 
 Then, we obtain the commutation relation
  \begin{eqnarray}
   [\delta^{(1)}_{\epsilon_{1}}, \delta^{(1)}_{\epsilon_{2}}] \psi 
 = i [\psi, \lambda] - i \sigma_{\mu} \frac{\theta_{\mu}}{2}
 \psi. \label{step002}  
  \end{eqnarray}
 The difference from the IIB matrix model is that the commutation
 relation (\ref{step002}) is satisfied without using the classical
 equation of motion.
 The relations (\ref{step001}) and (\ref{step002}) indicate that the
 commutators vanish up to the $SO(3)$ rotation by $\theta_{\mu}$, as
 well as the gauge transformation by $\lambda$. The appearance of the
 $SO(3)$ symmetry is a novelty of this matrix model.

 We likewise take the linear combination of the supersymmetry as
  \begin{eqnarray}
   {\tilde \delta}^{(1)} = \delta^{(1)} + \delta^{(2)}, \hspace{2mm}
   {\tilde \delta}^{(2)} = i (\delta^{(1)} - \delta^{(2)}).
  \end{eqnarray}
 Their commutation relation gives the translation of the bosonic
 matrices $A_{\mu}$ as
  \begin{eqnarray}
    [{\tilde \delta}^{(\alpha)}_{\epsilon}, {\tilde
   \delta}^{(\beta)}_{\xi}] \psi = 0, \hspace{2mm}
   [{\tilde \delta}^{(\alpha)}_{\epsilon}, {\tilde
   \delta}^{(\beta)}_{\xi}] A_{\mu} = - 2i ({\bar \epsilon} \sigma_{\mu}
   \xi) \delta^{\alpha \beta}. \label{susyfsIIBtrans}
  \end{eqnarray}
 The fuzzy-sphere classical solution is a BPS saturated state. Plugging 
 the classical solution (\ref{fssolution}) and $\psi = 0$ into the
 homogeneous transformation (\ref{homofsIIB}), we immediately find that 
 the homogeneous supersymmetry vanishes. Whereas, the inhomogeneous
 supersymmetry (\ref{inhomofsIIB}) survives. In this sense, the
 fuzzy-sphere solution (\ref{fssolution}) preserves half of the
 supersymmetry. 

 We next discuss the expansion of this matrix model around the
 fuzzy sphere solution (\ref{fssolution}). In the case of the IIB
 matrix model, we derive the noncommutative Yang-Mills theory by the
 expansion around the flat noncommutative
 background\cite{9908141}. This derivation of the noncommutative
 Yang-Mills theory is quite natural, because the matrices, per se, are 
 noncommutative objects. The studies of the noncommutative field theory 
 have become explosively popular since Seiberg and
 Witten\cite{9908142} elucidated the relation to the superstring
 theory. In the noncommutative spacetime, we impose the noncommutativity 
 on the spacetime as $[x_{\mu}, x_{\nu}] = i c_{\mu \nu}$. This
 relationship is reminiscent of the canonical commutation relation of
 the space and the momentum in the quantum mechanics. Just as the
 spacetime and momentum have the uncertainty principle, the
 noncommutative spacetime incorporates the uncertainty of the
 spacetime. In the context of the superstring theory, the
 noncommutativity of the spacetime is introduced by turning on the B
 field\cite{9908142}, which gives the noncommutativity
  \begin{eqnarray}
   c_{\mu \nu} = 2 \pi \alpha' \left(\frac{1}{g+2 \pi \alpha' B}
   \right)_{A} 
   = - (2 \pi \alpha') \left( \frac{1}{g + 2 \pi \alpha' B} B
   \frac{1}{g - 2 \pi \alpha' B} \right)_{\mu \nu}.  
  \end{eqnarray}
  This leads to the uncertainty of the spacetime at the string scale ${\cal
  O}(\sqrt{\alpha'}) \sim {\cal O}(l_{s})$.

  When we expand the matrix model (\ref{IIBfs}) around the fuzzy
  sphere background (\ref{fssolution}), we obtain a noncommutative
  Yang-Mills theory on the $S^{2}$ sphere. The expansion around the
  sphere is given by
  \begin{eqnarray}
   A_{\mu} = \alpha L_{\mu} + \alpha R {\hat a}_{\mu},
  \end{eqnarray}
 where ${\hat a}_{\mu}$ is the fluctuation. The fluctuation is
 expanded by the noncommutative spherical harmonics as
  \begin{eqnarray}
   {\hat a}_{\mu} = \sum_{l=0}^{N-1} \sum_{m=-l}^{l} a_{\mu lm} {\hat
   Y}_{lm}, \hspace{2mm}
   {\hat \psi} = \sum_{l=0}^{N-1} \sum_{m=-l}^{l} \psi_{lm} {\hat
   Y}_{lm}, \label{ncsh}
  \end{eqnarray}
 where the noncommutative spherical harmonics ${\hat Y}_{lm}$ is
 given by
  \begin{eqnarray}
   {\hat Y}_{lm} = R^{-l} \sum_{a} f^{lm}_{a_{1} \cdots a_{l}}
   {\hat x}^{a_{1}} {\hat x}^{a_{2}} \cdots {\hat x}^{a_{l}}.
  \end{eqnarray}
  The indices $a_{\mu}$ run over $1,2,3$, and $f^{lm}_{a_{1} \cdots
  a_{l}}$ is a symmetric and traceless tensor. We can understand that
  this is a $(2l+1)$-dimensional tensor as follows. Firstly, when we
  allocate $1,2,3$ to $a_{1}, \cdots, a_{l}$ in a symmetric way, we
  have ${_{l+2}}C_{2} = \frac{(l+2)(l+1)}{2}$ ways. On the other hand, 
  the tracelessness condition implies that $f^{lm}_{a a a_{1} a_{2}
  \cdots a_{l-2}}$ vanishes. This condition stems from the constraint
  $({\hat x}^{a})^{2} =  {\bf 1}_{N \times N}$.
  This eliminates ${_{l}}C_{2} = \frac{l(l-1)}{2}$ ways. Therefore,
  the dimensionality of this tensor  is $\frac{(l+2)(l+1)}{2} -
  \frac{l(l-1)}{2} = 2l+1$. Its normalization is defined by
   \begin{eqnarray}
    \frac{1}{N} Tr( {\hat Y}_{lm} {\hat Y}^{\dagger}_{l'm'}) =
    \delta_{ll'} \delta_{mm'}. \label{normalmatfs}
   \end{eqnarray}

  This of course corresponds to the expansion in terms of the c-number
  spherical harmonics 
  \begin{eqnarray}
 & &  a_{\mu} (\Omega) = \sum_{l=0}^{\infty} \sum_{m=-l}^{l} a_{\mu lm} Y_{lm}
   (\Omega), \hspace{2mm}
    \psi = \sum_{l=0}^{\infty} \sum_{m=-l}^{l} \psi_{lm} Y_{lm}
   (\Omega), \textrm{ where } \\
 & & Y_{lm}(\Omega) = R^{-l} \sum_{a} f^{lm}_{a_{1} \cdots a_{l}} x^{a_{1}} 
   \cdots x^{a_{l}}, \hspace{2mm}
  \int \frac{d \Omega}{4 \pi} Y_{lm} Y^{\ast}_{l'm'} = 
  \frac{1}{4 \pi} \int^{2 \pi}_{0} d \varphi \int^{\pi}_{0} \sin
   \theta d \theta Y_{lm} Y^{\ast}_{l'm'}
     = \delta_{ll'} \delta_{mm'}. \nonumber \\
  \end{eqnarray}
 Now, we can build the mapping rule between the matrices and the
 functions on the $S^{2}$ sphere. The rule is listed below:
   \begin{eqnarray}
    & & {\hat a}_{\mu} = \sum_{l=0}^{N-1} \sum_{m=-l}^{l} a_{\mu lm}
    {\hat Y}_{lm} \to 
     a_{\mu} (\Omega) = \sum_{l=0}^{\infty} \sum_{m=-l}^{l} a_{\mu lm}
    Y_{lm} (\Omega), \label{mappingfunc} \\
  & &  \frac{1}{N} Tr \to \int \frac{d \Omega}{4 \pi},
    \label{mappingtrace} \\
  & &  {\hat a} {\hat b} \to a(\Omega) \star b(\Omega). \label{mappingprod}
   \end{eqnarray} 
  We need to look more closely at the star product
  (\ref{mappingprod}). Firstly, the correspondence (\ref{mappingfunc}) 
  is rewritten, using the orthogonality condition (\ref{normalmatfs}),
  as
 \begin{eqnarray}
   {\hat a}_{\mu} \to a_{\mu} (\Omega) = \frac{1}{N}
   \sum_{l=0}^{N-1} \sum_{m=-l}^{l} Tr ({\hat a}_{\mu} {\hat
   Y}^{\dagger}_{lm}) Y_{lm} (\Omega).
 \end{eqnarray}
  Here, we require that the maximum value of $l$ is $N-1$. This leads us 
  to define the star product as follows. Here, we require that $l+l'$
  should not exceed $N-1$.
  \begin{eqnarray}
   a \star b (\Omega) &=& \frac{1}{N} \sum_{l=0}^{N-1} \sum_{m=-l}^{l}
   Tr ({\hat a}{\hat b} {\hat Y}^{\dagger}_{lm}) Y_{lm} (\Omega)
   \nonumber \\
 &=& \frac{1}{N} \sum_{lm} \sum_{l'm'} \sum_{l''m''} \int d \Omega' d
   \Omega'' Y^{\ast}_{l'm'} (\Omega') Y^{\ast}_{l''m''} (\Omega)
   a(\Omega') b(\Omega'') Y_{lm} (\Omega)
   Tr ({\hat Y}_{l'm'} {\hat Y}_{l''m''} {\hat
   Y}^{\dagger}_{lm}). \nonumber \\
  \end{eqnarray}
 Then, the action (\ref{IIBfs}) is mapped as follows:
  \begin{eqnarray}
   S &=& - \frac{\alpha^{4} R^{4}}{4g^{2}} Tr({\hat F}_{\mu \nu} {\hat
   F}_{\mu \nu}) - \frac{\alpha^{4}}{6g^{2}} Tr L_{\mu}^{2}
   - \frac{i}{2g^{2}} \alpha^{4} R^{3} \epsilon_{\mu \nu \rho}
   Tr \left( \frac{1}{R} [L_{\mu}, {\hat a}_{\nu}] {\hat a}_{\rho} +
   \frac{1}{3} {\hat a}_{\mu} [{\hat a}_{\nu}, {\hat a}_{\rho}] - \frac{i}{2
   R} \epsilon_{\nu \rho \chi} {\hat a}_{\mu} {\hat a}_{\chi}
   \right) \nonumber \\  
   &+& \frac{\alpha}{2g^{2}} Tr {\bar {\hat \psi}} \sigma^{\mu} [L_{\mu} +
   R {\hat a}_{\mu}, \psi] \nonumber \\
  &=& - \frac{\alpha^{4} (N^{3} - N)}{24 g^{2}} 
      - \frac{\alpha^{4} R^{4} N}{4g^{2}} \int \frac{d \Omega}{4
   \pi} \left\{ (F_{\mu \nu} F_{\mu \nu}) + \frac{2i}{R}
   \epsilon_{\mu \nu \rho}
   \left( \frac{1}{R} (L_{\mu} a_{\nu}) a_{\rho} + \frac{1}{3} a_{\mu}
   [a_{\nu}, a_{\rho}] - \frac{i}{2 R} \epsilon_{\nu \rho \chi}
   a_{\mu} a_{\chi} \right) \right\}_{\star} \nonumber \\
  &+& \frac{\alpha R N}{2g^{2}} \int \frac{d \Omega}{4 \pi} \left(
   \frac{1}{R} {\bar \psi} \sigma^{\mu} L_{\mu} \psi 
   + {\bar \psi} \sigma^{\mu} [a_{\mu}, \psi]
   \right)_{\star}. \label{ncymfs} 
  \end{eqnarray}
 where we define the field strength ${\hat F}_{\mu \nu}$ as
  \begin{eqnarray}
   {\hat F}_{\mu \nu} = \frac{1}{R} ([L_{\mu}, a_{\nu}] - [L_{\nu},
   a_{\mu}]) - 
   \frac{i}{R} \epsilon_{\mu \nu \rho} a_{\rho} + [a_{\mu},
   a_{\nu}]. \label{fsfs} 
  \end{eqnarray}
 The subscript $_{\star}$ indicates that the product is defined by the 
 noncommutative star product. Therefore, the commutator survives even
 for the $U(1)$ abelian gauge group of the Yang-Mills theory.

 This gives the relationship between the coupling constant of the
 Yang-Mills theory $g_{YM}$ and the {\it artifact} of the coupling
 constant of the Yang-Mill theory before being reduced as
  \begin{eqnarray}
   g^{2}_{YM} = \frac{4 \pi g^{2}}{\alpha^{4} R^{2} N}.
  \end{eqnarray}
 We go on to the symmetry of this Yang-Mills theory. Firstly, the
 $SU(N)$ symmetry is interpreted as the gauge symmetry. For the
 element of the $SU(N)$ Lie {\it group} $U$, it is expanded as
 $U = \exp(i \lambda) \sim 1 + i \lambda + \cdots$, where $\lambda$ is 
 an element of the $su(N)$ Lie {\it algebra}. The gauge transformation 
 for $A_{\mu}$ is given by
  \begin{eqnarray}
    \delta A_{\mu} = [A_{\mu}, i \lambda] = [\alpha L_{\mu} + \alpha R
    {\hat a}_{\mu}, i \lambda].
  \end{eqnarray}
 This transforms the fluctuation as
 \begin{eqnarray}
   \delta {\hat a_{\mu}} = \frac{i}{R} \alpha [L_{\mu}, \lambda] 
                + i [{\hat a}_{\mu}, \lambda]. \label{gaugefs}
 \end{eqnarray}
 This is immediately translated in terms of the Yang-Mills theory as
  \begin{eqnarray}
   \delta a_{\mu} = \frac{i}{R} L_{\mu} \lambda +  i [{\hat a}_{\mu},
   \lambda]_{\star}. \label{gaugefsnc}
  \end{eqnarray}
 Secondly, we note that this noncommutative Yang-Mills theory
 preserves the following ${\cal N}=1$ supersymmetry.
  \begin{eqnarray}
   \delta_{S} a_{\mu} = \frac{i}{\alpha R} {\bar \epsilon} \sigma_{\mu} 
   \psi, \hspace{2mm}
   \delta_{S} \psi = \frac{i \alpha^{2} R^{2}}{2} F_{\mu \nu}
   \sigma^{\mu \nu} \epsilon.
  \end{eqnarray}
  
  We have hitherto discussed the perturbation around the irreducible
  representation of the fuzzy sphere (\ref{fssolution}). In this case, 
  we obtain a $U(1)$ abelian noncommutative Yang-Mills theory.
  The matrix model (\ref{IIBfs}) also accommodates the fuzzy
  sphere solution of the reducible representation 
  \begin{eqnarray}
   A_{\mu} = \alpha (L_{\mu} \otimes {\bf 1}_{k \times k})
         = \left( \begin{array}{cccc} \alpha L_{\mu} & & & \\
      & \alpha L_{\mu} & & \\ & & \ddots & \\
      & & & \alpha L_{\mu} \end{array} \right).
  \end{eqnarray}
  When $L_{\mu}$ is the $n \times n$ representation of the $SU(2)$ Lie
  algebra, the size of the matrices $N$ satisfies $N=nk$. The
  perturbation around this reducible representation gives the $U(k)$
  nonabelian theory.

 \subsubsection{Other attempts for higher-dimensional fuzzy sphere
  solution} \label{kimurayhigher}
  We next review the extension of the above discussion to the
  higher-dimensional fuzzy sphere classical solution. To this end,
  there are two strategies. First is the addition of the
  higher-dimensional Chern-Simons term, and second is the addition of
  the tachyonic mass term.

  The first attempt is achieved by defining the following
  action\cite{0204256,0301055} :
  \begin{eqnarray}
   S = - \frac{1}{g^{2}} \left( \frac{1}{4} Tr [A_{\mu}, A_{\nu}]^{2} + 
         \frac{\lambda}{2k+1}
         \epsilon_{\mu_{1} \cdots \mu_{2k+1}} Tr A_{\mu_{1}}
         A_{\mu_{2}} \cdots A_{\mu_{2k+1}} \right). \label{IIBfsp}
 \end{eqnarray}
  Here, we define this matrix model in the $d=(2k+1)$-dimensional
  Euclidean space. The indices $\mu, \nu, \cdots$ run over
  $1,2,\cdots, (2k+1)$. The supersymmetric Yang-Mills theory can be defined 
  only for the $d=3,4,6,10$ case, and the supersymmetric matrix model
  cannot be defined except for the three-dimensional case as we have
  reviewed before. Hence, we focus on the bosonic matrix model in the
  following. Except for the $k=1$ (three-dimensional) case, the
  convergence of the path integral is not yet manifestly proven.
  Nevertheless, we surmise that the fuzzy-sphere classical solution
  might be a metastable state, in that it is barriered by a high
  potential. We attribute this metastability to the similar reasoning
  to the $\phi^{4}$ matrix model. As we explain in  Appendix
  \ref{simulation},  the $\phi^{4}$ matrix model retains the
  metastability near the origin despite its unbounded-below potential.
  
  The illuminating feature of this higher-dimensional matrix model is
  that it incorporates the classical solution of the
  $S^{2k}$ fuzzy sphere. The equation of motion is given by
  \begin{eqnarray}
   [A_{\nu}, [A_{\mu}, A_{\nu}]] + \lambda \epsilon_{\mu
   \nu_{1} \cdots \nu_{2k}} A_{\nu_{1}} A_{\nu_{2}} \cdots
   A_{\nu_{2k}} = 0. \label{eomfsp}
  \end{eqnarray}
  It has the following classical solution of the $S^{2k}$
  fuzzy sphere\cite{9712105,0105006,0207111}:
  \begin{eqnarray}
  A_{\mu} &=& \frac{\alpha}{2} G^{(2k)}_{\mu}, \textrm{ where }
  \nonumber \\
  G^{(2k)}_{\mu} &=& (\Gamma^{(2k)}_{\mu} \otimes {\bf 1}_{2^{k} \times
  2^{k}}  \otimes \cdots
  \otimes {\bf 1}_{2^{k} \times 2^{k}})_{\textrm{sym}} + \cdots +
  ( {\bf 1}_{2^{k} \times 2^{k}} \otimes {\bf 1}_{2^{k} \times 2^{k}}
  \otimes \cdots \otimes 
  \Gamma^{(2k)}_{\mu})_{\textrm{sym}}. \nonumber \\ \label{fspsolution} 
  \end{eqnarray}

  We define the symmetric tensor product more explicitly, for $k=1, 2$
  case (namely, for the $S^{2}$ and $S^{4}$ fuzzy sphere). 
  We denote the matrix element using the orthonormal state $|i \rangle$ 
  and $|j \rangle$ as $(A)_{ij} = \langle i| A |j \rangle$. The usual tensor 
  product (two-fold product, for brevity) is defined by
  \begin{eqnarray}
   \langle i_{1},i_{2} | A_{1} \otimes A_{2} | j_{1},j_{2} \rangle
 = \langle i_{1} | A_{1} | j_{1} \rangle \langle i_{2} | A_{2} | j_{2}
 \rangle, \textrm{ where }
  |j_{1}, j_{2} \rangle = |j_{1} \rangle |j_{2}
 \rangle. \label{ordinarytensor} 
  \end{eqnarray}
 On the other hand, the symmetric tensor product is defined as
  \begin{eqnarray}
  _{\textrm{sym}}\langle i_{1}, i_{2} | A \otimes B | j_{1} j_{2}
  \rangle_{\textrm{sym}}, \textrm{ where }
  | j_{1}, j_{2} \rangle_{\textrm{sym}}
 = \frac{1}{n} ( |j_{1} \rangle |j_{2} \rangle + |j_{2} \rangle |j_{1}
  \rangle). \label{symmetrictensor}
  \end{eqnarray}
  The coefficient $n$ is determined, so that the symmetrized state may 
  retain the orthonormality. Here, we give a more explicit definition
  for the two-fold tensor product of the Paulian matrices
  ($k=1$ case). For the two-fold tensor product, the symmetrized
  orthonormal state is given by
  \begin{eqnarray}
   |1,1 \rangle = |1 \rangle |1 \rangle, \hspace{2mm}
   |2,2 \rangle = |2 \rangle |2 \rangle, \hspace{2mm}
   |1,2 \rangle = |2,1 \rangle = \frac{1}{\sqrt{2}}
    (|1 \rangle |2 \rangle + |2 \rangle |1 \rangle). 
  \end{eqnarray}
 This difference comes from the fact that the states $|1,1 \rangle$
 and $|2,2 \rangle$ are symmetrized ab ovo and hence that we do not
 need symmetrization or normalization. We elaborate on these
 definitions in terms of explicit representation in terms of
 matrices. The usual tensor product (\ref{ordinarytensor}) is
 rewritten as
  \begin{eqnarray}
    A \otimes B = \left( \begin{array}{c|cccc}
   & |j_{1}=1, j_{2}=1 \rangle  & |j_{1}=2, j_{2}=1 \rangle &
   |j_{1}=1, j_{2}=2 \rangle & |j_{1}=2, j_{2}=2 \rangle \\ \hline 
 \langle i_{1}=1, i_{2}=1 | & a_{11} b_{11} & a_{12} b_{11} & a_{11} b_{12} &
   a_{12} b_{12} \\ 
 \langle i_{1}=2, i_{2}=1 |  & a_{21} b_{11} & a_{22} b_{11} & a_{21}
   b_{12} & a_{22}  b_{12} \\  
 \langle i_{1}=1, i_{2}=2 | &  a_{11} b_{21} & a_{12} b_{21} & a_{11} b_{22} &
   a_{12} b_{22} \\ 
 \langle i_{1}=2, i_{2}=2 | & a_{21} b_{21} & a_{22} b_{21} & a_{21} b_{22} &
   a_{22} b_{22} \end{array} \right). \nonumber \\ \label{ordinarytensor2}
  \end{eqnarray}
 On the other hand, its symmetrization is rewritten as
  \begin{eqnarray}
 (A \otimes B)_{\textrm{sym}} 
 = \left( \begin{array}{c|ccc}
 &  |1,1 \rangle_{\textrm{sym}} & |1,2 \rangle_{\textrm{sym}} & |2,2
 \rangle_{\textrm{sym}} \\ \hline 
 _{\textrm{sym}}\langle 1,1| & a_{11} b_{11} & \frac{(a_{12} b_{11} + a_{11} b_{12})}{\sqrt{2}}  & a_{12} b_{12} \\ 
 _{\textrm{sym}}\langle 1,2| & \frac{(a_{21} b_{11} + a_{11}
 b_{21})}{\sqrt{2}}  & \frac{(a_{22} b_{11} + a_{21} b_{12} +
 a_{12} b_{21} + a_{11} b_{22})}{(\sqrt{2})^{2}}  & \frac{(a_{22} b_{12} +
 a_{12} b_{22})}{\sqrt{2}} \\
  _{\textrm{sym}}\langle 2,2| & a_{21} b_{21} & \frac{(a_{22} b_{21} +
 a_{21} b_{22})}{\sqrt{2}} & a_{22} b_{22} \end{array}
 \right). \nonumber \\ \label{symmetrictensor2}
  \end{eqnarray} 
  The coefficient $\lambda$ is related to the radius parameter of the
  fuzzy sphere $\alpha$ as
  \begin{eqnarray}
   \lambda = (\frac{\alpha}{2})^{3-2k} \frac{8k}{m_{k}},
   \label{lambdaalpha} 
  \end{eqnarray}
  where the coefficient $m_{k}$ is given later in the self-dual
  relation relation (\ref{fsprel5}).

  The $S^{2k}$ fuzzy sphere is mathematically more involved than
  the $S^{2}$ fuzzy sphere. We introduce the properties of the
  higher-dimensional fuzzy sphere solution. 
  Firstly, the gamma matrices $\Gamma^{(2k)}_{\mu}$ is defined by the
  Wick rotation of those for the Minkowski spacetime given in Appendix 
  \ref{notation}. The definition is given by
   \begin{eqnarray}
 & & \Gamma^{(2)}_{p} = \sigma_{p}, \hspace{2mm} (p=1,2,3), \hspace{2mm}
   \Gamma^{(2k)}_{p} = \Gamma^{(2k-2)}_{p} \otimes \sigma_{3} 
   = \left( \begin{array}{cc} \Gamma^{(2k)}_{p} & 0 \\ 0 & -
   \Gamma^{(2k)}_{p} \end{array} \right), \nonumber \\
 & & \Gamma^{(2k)}_{2k} = {\bf 1}_{2^{k-1} \times 2^{k-1}} \otimes
   \sigma_{2} = \left( \begin{array}{cc} 0 & -i {\bf 1}_{2^{k-1} \times
   2^{k-1}}  \\ i {\bf 1}_{2^{k-1} \times 2^{k-1}} & 0 \end{array}
   \right), \nonumber \\
 & &   \Gamma^{(2k)}_{2k+1} = {\bf 1}_{2^{k-1} \times 2^{k-1}} \otimes
   \sigma_{1} = \left( \begin{array}{cc} 0 &  {\bf 1}_{2^{k-1} \times
   2^{k-1}}  \\ {\bf 1}_{2^{k-1} \times 2^{k-1}} & 0 \end{array}
   \right), \label{gammawick}
   \end{eqnarray}
 where $p$ runs over $p=1,2,\cdots, 2k-1$.
 Namely, we identify $i \Gamma_{0}$ in the Minkowski space with
 $\Gamma^{(2k)}_{2k}$ in the Euclidean spacetime.

   The size of the representation (\ref{fspsolution}) is given in
  \cite{0105006} as
  \begin{eqnarray}
  & & N_{1} = (n+1), \hspace{2mm}
   N_{2} = \frac{(n+1)(n+2)(n+3)}{6}, \hspace{2mm}
   N_{3} = \frac{(n+1)(n+2)(n+3)^{2} (n+4)(n+5)}{360}, \nonumber \\
  & & N_{4} = \frac{(n+1)(n+2)(n+3)^{2}(n+4)^{2}
   (n+5)^{2}(n+6)(n+7)}{302400}. \label{sizefsp}
  \end{eqnarray}
  We see a grave difference between the $k=1$ (three-dimensional) case
  and otherwise. For $k=1$, the representation (\ref{fspsolution}) can 
  be defined for an arbitrary integer-size matrices (of course, except 
  for 1). However, this never holds true for $k \geq 2$ case.
  For example, the representation (\ref{fspsolution}) is possible only 
  for a limited integer size; namely $N_{2} = 4, 10, 20, \cdots$.

 The important property is that $G^{(2k)}_{\mu}$ does not generically
 close with respect to the commutator for $k \geq 2$. This is also a
 tremendous difference between the three-dimensional case and the
 higher-dimensional case. Here, we denote $G^{(2k)}_{\mu
 \nu} = [G^{(2k)}_{\mu}, G^{(2k)}_{\nu}]$. These satisfy the following 
 relations:
  \begin{eqnarray}
 & &  G^{(2k)}_{\mu} G^{(2k)}_{\mu} = n(n+2k) {\bf 1}_{N_{k} \times
   N_{k}}, \label{fsprel1} \\
 & &  G^{(2k)}_{\mu \nu} G^{(2k)}_{\mu \nu} = -8kn(n+2k) {\bf
   1}_{N_{k} \times N_{k}}, \label{fsprel2} \\
 & & [G^{(2k)}_{\mu \nu}, G^{(2k)}_{\rho}] = 4(-\delta_{\mu \rho}
   G^{(2k)}_{\nu} + \delta_{\nu \rho} G^{(2k)}_{\mu} ),
   \label{fsprel3} \\
 & & [G^{(2k)}_{\mu \nu}, G^{(2k)}_{\rho \chi}] = 
  4 (\delta_{\nu \rho} G^{(2k)}_{\mu \chi} + \delta_{\mu \chi}
  G^{(2k)}_{\nu \rho} - \delta_{\mu \chi} G^{(2k)}_{\nu \rho} - \delta_{\nu
  \rho} G^{(2k)}_{\mu \chi}). \label{fsprel4}
  \end{eqnarray}
 In addition, $G^{(2k)}_{\mu}$ satisfies the following self-dual
 relation.
  \begin{eqnarray}
 & & \epsilon_{\mu \nu_{1} \cdots \nu_{2k}} G^{(2k)}_{\nu_{1}}
    G^{(2k)}_{\nu_{2}} \cdots G^{(2k)}_{\nu_{2k}} 
  = m_{k} G_{\mu}, \textrm{ where } \label{fsprel5} \\
 & & m_{1} = 2i, \hspace{2mm} m_{2} = 8(n+2), \hspace{2mm}
     m_{k} = - 2ki (n+2k-2) m_{k-1}. \label{selfdualcoef}
  \end{eqnarray}

 Before verifying these properties, we start with convincing ourselves 
 that the notion of the higher-dimensional fuzzy sphere solution is a
 generalization of the $SU(2)$ Lie algebra. To this end, we focus on
 the relation between the $k=1$ (three-dimensional) case and the
 $SU(2)$ Lie algebra.

 Firstly, the $n$-fold symmetric tensor product for $k=1$ is identical 
 to the $(n+1)$-dimensional irreducible representation of the $SU(2)$
 Lie algebra. We explicitly give the matrix element of the $n=2$ case as
  \begin{eqnarray}
   G^{(2)}_{1} &=& [(\sigma_{1} \otimes {\bf 1}_{2 \times 2}) + ({\bf
   1}_{2 \times 2} \otimes \sigma_{1})]_{\textrm{sym}} \nonumber \\
   &=&  \left( \begin{array}{c|cc|c} 0 & 1 & 0 & 0 \\ \hline
    1 & 0 & 0 & 0 \\ 0 & 0 & 0 & 1 \\ \hline
   0 & 0 & 1 & 0 \end{array} \right)_{\textrm{sym}}
  +  \left( \begin{array}{c|cc|c} 0 & 0 & 1 & 0 \\ \hline
    0 & 0 & 0 & 1 \\ 1 & 0 & 0 & 0 \\ \hline
   0 & 1 & 0 & 0 \end{array} \right)_{\textrm{sym}}
   = \sqrt{2} \left( \begin{array}{ccc} 0 & 1 & 0 \\ 1 & 0 & 1 \\ 0 &
   1 & 0 \end{array} \right), \nonumber \\
  G^{(2)}_{2} &=& [(\sigma_{2} \otimes {\bf 1}_{2 \times 2}) + ({\bf
   1}_{2 \times 2} \otimes \sigma_{2})]_{\textrm{sym}} \nonumber \\
   &=& \left( \begin{array}{c|cc|c} 0 & -i & 0 & 0 \\ \hline
    i & 0 & 0 & 0 \\ 0 & 0 & 0 & -i \\ \hline
   0 & 0 & i & 0 \end{array} \right)_{\textrm{sym}}
  +  \left( \begin{array}{c|cc|c} 0 & 0 & -i & 0 \\ \hline
    0 & 0 & 0 & -i \\ i & 0 & 0 & 0 \\ \hline
   0 & i & 0 & 0 \end{array} \right)_{\textrm{sym}} 
  = \sqrt{2} \left( \begin{array}{ccc} 0 & -i & 0 \\ i & 0 & -i \\ 0 &
   i & 0 \end{array} \right), \nonumber \\
 G^{(2)}_{3} &=& [(\sigma_{3} \otimes {\bf 1}_{2 \times 2}) + ({\bf
   1}_{2 \times 2} \otimes \sigma_{3})]_{\textrm{sym}} \nonumber \\
  &=& \left( \begin{array}{c|cc|c} 1 & 0 & 0 & 0 \\ \hline
    0 & -1 & 0 & 0 \\ 0 & 0 & 1 & 0 \\ \hline
   0 & 0 & 0 & -1 \end{array} \right)_{\textrm{sym}}
  +  \left( \begin{array}{c|cc|c} 1 & 0 & 0 & 0 \\ \hline
    0 & 1 & 0 & 0 \\ 0 & 0 & -1 & 0 \\ \hline
   0 & 0 & 0 & -1 \end{array} \right)_{\textrm{sym}}
   = 2 \left( \begin{array}{ccc} 1 & 0 & 0 \\ 0 & 0 & 0 \\ 0 &
   0 & -1 \end{array} \right).  \label{n2k1}
  \end{eqnarray} 
 These are clearly identical to the $3 \times 3$
 representation of the $SU(2)$ Lie algebra. 
 In addition, the 2-fold tensor product obeys the following important
 relation:
 \begin{eqnarray}
  \sum_{i=1}^{3} (\sigma_{\mu} \otimes \sigma_{\mu})_{\textrm{sym}}
  &=&  \left( \begin{array}{c|cc|c} 0 & 0 & 0 & 1 \\ \hline
   0 & 0 & 1 & 0 \\ 0 & 1 & 0 & 0 \\ \hline
   1 & 0 & 0 & 0 \end{array} \right)_{\textrm{sym}}
  +  \left( \begin{array}{c|cc|c} 0 & 0 & 0 & -1 \\ \hline
   0 & 0 & 1 & 0 \\ 0 & 1 & 0 & 0 \\ \hline
   -1 & 0 & 0 & 0 \end{array} \right)_{\textrm{sym}}
  +  \left( \begin{array}{c|cc|c} 1 & 0 & 0 & 0 \\ \hline
   0 & -1 & 0 & 0 \\ 0 & 0 & -1 & 0 \\ \hline
   0 & 0 & 0 & 1 \end{array} \right)_{\textrm{sym}} \nonumber \\
  &=&  \left( \begin{array}{ccc} 1 & 0 & 0 \\
    0 & 1 & 0 \\ 0 & 0 & 1 \end{array} \right)
   = {\bf 1}_{3 \times 3}. \label{basiclaw}
 \end{eqnarray} 
 Utilizing this relation, the radius of the fuzzy sphere is derived as 
 follows:
  \begin{eqnarray}
   (G^{(2)}_{\mu})^{2} &=& \underbrace{(\sigma_{\mu}^{2} \otimes {\bf 1}
   \otimes \cdots \otimes {\bf 1})_{\textrm{sym}}
   + \cdots + ({\bf 1} \otimes {\bf 1}
   \otimes  \cdots \otimes
   \sigma_{\mu}^{2})_{\textrm{sym}}}_{\textrm{$n$ terms}} 
  + \underbrace{(\sigma_{\mu} \otimes \sigma_{\mu} \otimes {\bf 1} \otimes 
   \cdots \otimes {\bf 1})_{\textrm{sym}} + \cdots}_{\textrm{$n(n-1)$
   terms}} \nonumber \\
  &=& (3n + n(n-1)) {\bf 1}_{N_{1} \times N_{1}} 
   =  (N_{1}^{2} - 1) {\bf 1}_{N_{1} \times N_{1}}, \label{casimir}
  \end{eqnarray}
 where we used (\ref{basiclaw}) in the second equality and recall that 
 $N_{1} = n+1$. This proves the relation (\ref{fsprel1}) for $k=1$.
 This is identical to the Casimir of the $J_{\mu}$, which satisfies
 $J_{\mu}^{2} = \frac{N^{2}-1}{4}$. Here, the relation between
 $G^{(2)}_{\mu}$ and $J_{\mu}$ is given by $\frac{1}{2} G^{(2)}_{\mu} =
 J_{\mu}$. 

 The self-dual relation (\ref{fsprel5}) clearly reduces to the
 commutation relation of the $SU(2)$ Lie algebra:
  \begin{eqnarray}
   \epsilon_{\mu \nu \rho} G^{(2)}_{\nu} G^{(2)}_{\rho}
 = \frac{1}{2} \epsilon_{\mu \nu \rho} [G^{(2)}_{\nu}, G^{(2)}_{\rho}]
 = \frac{1}{2} \epsilon_{\mu \nu \rho} (2i \epsilon_{\nu \rho \chi}
 G^{(2)}_{\chi})  = 2i G^{(2)}_{\mu}.
  \end{eqnarray}

 We next go into the proof of the relations (\ref{fsprel1}) $\sim$
 (\ref{fsprel5}). We immediately discern that (\ref{fsprel3}) and
 (\ref{fsprel4}) are verified from the commutation relations of the
 gamma matrices. (\ref{fsprel1}) can be substantiated by repeating the same 
 argument as for (\ref{casimir}) in the $k=1$ case. In promoting to
 the higher-dimensional case, we note that the similar relation to
 (\ref{basiclaw}) holds true of the general case; namely
  \begin{eqnarray}
   \sum_{\mu=1}^{2k+1} (\Gamma^{(2k)}_{\mu} \otimes
   \Gamma^{(2k)}_{\mu})_{\textrm{sym}} = {\bf 1}. \label{basiclaw2}
  \end{eqnarray}
  For $k=2$, this relation can be verified as
  \begin{eqnarray}
  & & \sum_{\mu=1}^{5} (\Gamma^{(2k)}_{\mu} \otimes
   \Gamma^{(2k)}_{\mu})_{\textrm{sym}} \nonumber \\
  &=& \left( \sum_{\mu=1}^{3} ((\sigma_{\mu} \otimes
   \sigma_{3}) \otimes  (\sigma_{\mu} \otimes
   \sigma_{3}) )_{\textrm{sym}} \right)
   + (( {\bf 1} \otimes \sigma_{2}) \otimes  ( {\bf 1} \otimes
   \sigma_{2}) )_{\textrm{sym}} + (( {\bf 1} \otimes \sigma_{1})
   \otimes  ( {\bf 1} \otimes \sigma_{1}) )_{\textrm{sym}}
   = {\bf 1}, \nonumber
  \end{eqnarray}
 by using (\ref{basiclaw}) twice. We can verify the higher-$k$ case
 recursively in the same way. Once we prove (\ref{basiclaw2}), the
 proof of (\ref{fsprel1}) goes in the same way as for (\ref{casimir}).

 The relation (\ref{fsprel2}) is verified by the following
 calculation:
  \begin{eqnarray}
   G^{(2k)}_{\mu \nu} G^{(2k)}_{\mu \nu}
 &=& [(2 \Gamma^{\mu \nu} \otimes {\bf 1} \otimes \cdots \otimes {\bf
 1})_{\textrm{sym}} + \cdots ]^{2} 
 \nonumber \\
 &=&  \underbrace{(4 \Gamma^{\mu \nu} \Gamma^{\mu \nu} \otimes {\bf 1} 
 \otimes \cdots \otimes {\bf 1})_{\textrm{sym}}}_{\textrm{$n$ terms}}
  + \underbrace{4 ((\Gamma^{\mu} \Gamma^{\nu} - \delta^{\mu \nu})
 \otimes (\Gamma^{\mu} \Gamma^{\nu} - \delta^{\mu \nu})
 \otimes {\bf 1} \otimes \cdots \otimes {\bf 1})_{\textrm{sym}} +
 \cdots }_{\textrm{$n(n-1)$ terms}} \nonumber \\
  &=& 4 ( - n2k(2k+1) - n(n-1)2k) {\bf 1} 
   =  - 8kn(n+2k) {\bf 1}.
  \end{eqnarray}

 Lastly, we prove the relation (\ref{fsprel5}). The brutal-force proof 
 up to $k=4$ is given in \cite{0209057}. However, the Gordian knot is
 cut in one stroke in \cite{0301055} by noting the following reduction 
 law. We consider the algebra at the north pole of the fuzzy sphere;
 $G^{(2k)}_{2k+1}  \sim n$. Then, $G^{(2k)}_{ab}$ are the generator of 
 the $SO(2k)$ rotation around the north pole, where the indices
 $a,b,\cdots$ run over $1,2,\cdots,2k$. For $p, q, \cdots =
 1,2,\cdots 2k-1$, the elements for the $S^{2k}$ fuzzy sphere can be
 identified with those of the $S^{2k-2}$ fuzzy sphere;
  \begin{eqnarray}
   i G^{(2k)}_{p 2k} \sim G^{(2k-2)}_{p}, \hspace{2mm}
   G^{(2k)}_{pq} \sim G^{(2k-2)}_{pq}. \label{identification} 
  \end{eqnarray}
 This gives the recursive relation of the coefficient $m_{k}$ as
  \begin{eqnarray}
   m_{k} G^{(2k)}_{2k+1} &=& n m_{k} = \epsilon_{\mu_{1} \cdots
   \mu_{2k} (2k+1)} G^{(2k)}_{\mu_{1}} \cdots G^{(2k)}_{\mu_{2k}}
   = 2k \epsilon_{\mu_{1} \cdots
   \mu_{2k-1}(2k) (2k+1)} G^{(2k)}_{\mu_{1} \mu_{2}}
       \cdots G^{(2k)}_{\mu_{2k-1}2k} \nonumber \\
  &=& - 2ki \epsilon_{p_{1} \cdots p_{2k-1}}
   G^{(2k-2)}_{p_{1}} \cdots G^{(2k-2)}_{p_{2k-1}}
   = -2ki m_{k-1} (G^{(2k-2)}_{p_{2k-1}})^{2}
   = -2ki m_{k-1} n(n+2k-2). \nonumber \\ \label{recursive} 
  \end{eqnarray}
  This shows that the coefficient $m_{k}$ obeys the recursion relation
 (\ref{selfdualcoef}).

 This completes the relations (\ref{fsprel1}) $\sim$ (\ref{fsprel5}).
 We can see that (\ref{fspsolution}) represents the sphere simply due
 to the relation (\ref{fsprel1}). This gives the radius of the
 $S^{2k}$ fuzzy sphere as
  \begin{eqnarray}
   R^{2} = \frac{\alpha^{2} n(n+2k)}{4}.
  \end{eqnarray}
  The relation (\ref{lambdaalpha}) has a serious consequence for the
  radius. For the $k=1$ case, the radius is proportional to the
  coefficient $\lambda$. However, for $k \geq 2$ case, the radius
  becomes smaller for the larger coefficient $\lambda$.

 Nextly, we point out that  the $S^{2k-2}$ fuzzy sphere is 
 attached to the $S^{2k}$ fuzzy sphere at each point. 
 This property is understood by considering the algebra on the north
 pole $G^{(2k)}_{2k+1} \sim n$. As we mentioned before,
 $G^{(2k)}_{ab}$ are the generators of the $SO(2k)$ rotation.
 Since identification (\ref{identification}) gives the algebra of the
 $S^{2k-2}$ sphere, we see that the $S^{2k-2}$ sphere is attached on
 the north pole. While we see
 only the north pole, it is clear that this holds true of all the
 points of the $S^{2k}$ sphere due to the rotational symmetry.
 Due to the identification (\ref{identification}), the radius of the
 $S^{2k-2}$ fuzzy sphere attached to the $S^{2k}$ sphere is clearly
 given by 
  \begin{eqnarray}
   R_{S^{2k-2}}^{2} = \frac{\alpha^{2} n(n+2k-2)}{4}.
   \label{radiusattached}
  \end{eqnarray}
 This property can be understood from the viewpoint of the stabilizer
 group. The stabilizer group of the $S^{2k}$ fuzzy sphere is
 $SO(2k+1)/U(k)$. This is identical to
  \begin{eqnarray}
    SO(2k+1)/U(k) &\sim& (SO(2k+1)/SO(2k)) \times (SO(2k)/U(k))
    \nonumber \\
    &\sim& (SO(2k+1)/SO(2k)) \times (SO(2k-1)/U(k-1)).  \label{stabilizer}
  \end{eqnarray}
 This solidifies the fact that the $S^{2k-2}$ sphere is attached on
 the $S^{2k}$ sphere. 

 We see another grave difference between the $S^{2}$ and the
 higher-dimensional $S^{2k}$ fuzzy sphere with respect to the
 classical stability. The classical energy is easily calculated by
 plugging the fuzzy sphere solution $S^{2k}$ into the action as
  \begin{eqnarray}
   S = \frac{kn(n+2k) N_{k} \alpha^{4}}{2g^{2}} \left( \frac{1}{4} -
   \frac{1}{2k+1} \right). \label{fspclassical}
  \end{eqnarray}
 In this derivation, we use the relation (\ref{fsprel1}) and
 (\ref{fsprel2}), as well as the self-dual relation (\ref{fsprel5}).
 The classical energy (\ref{fspclassical}) is negative for the
 three-dimensional ($k=1$) case, and thus energetically favored than
 the flat background $A_{\mu} = \textrm{diag}(x^{1}_{\mu}, \cdots,
 x^{N}_{\mu})$, 
 for which $S=0$. However, the reverse is the case with the $k \geq 2$ 
 case. In this case, the fuzzy sphere solution is {\it less}
 energetically favored.

 We have seen the alternative matrix model incorporating the
 higher-dimensional fuzzy sphere solution and the properties of the
 solution. Another choice is to add the tachyonic mass term to the
 bosonic part of the IIB matrix model as\cite{0103192}
  \begin{eqnarray}
   S = - \frac{1}{g^{2}} Tr \left( \frac{1}{4} [A_{\mu}, A_{\nu}]^{2}
       + \lambda A_{\mu}^{2} \right). \label{IIBtmass}
  \end{eqnarray}
 This model is defined by the general $d$-dimensional Euclidean
 space (here, the indices $\mu, \nu, \cdots$ run over $1,2,\cdots,
 d$). The convergence of the path integral of this matrix model is not 
 clear, while it is balanced by the mass term and the quartic term.
 This comes from the diagonal part of the matrices $A_{\mu}$. The
 diagonal part of $A_{\mu}$ of course contributes to the tachyonic
 mass term. Whereas, the quartic term comprises the commutator, so
 that the diagonal part is canceled. This is a serious drawback,
 compared with the matrix model with the Chern-Simons term. 

 The equation of motion 
  \begin{eqnarray}
   [A_{\nu}, [A_{\mu}, A_{\nu}]] + 2 \lambda A_{\mu} = 0
   \label{eomIIBtmass} 
  \end{eqnarray}
 gives a variety of the curved-space classical solution. Due to the
 commutation relation (\ref{fsprel3}) and (\ref{fsprel4}), (for $d
 \geq 2k+1$) it accommodates the $S^{2k}$ fuzzy sphere solution 
  \begin{eqnarray} 
  A _{\mu} = \frac{\alpha}{2} G_{\mu},
  \end{eqnarray}
 where $\alpha$ and the mass $\lambda$ is now related as 
 \begin{eqnarray}
  \lambda = k \alpha^{2}.
  \end{eqnarray}
 The classical energy of the fuzzy sphere solution $A_{\mu} =
 \frac{\alpha}{2} G_{\mu}$ for the action (\ref{eomIIBtmass}) is  
  \begin{eqnarray}
    S = - \frac{kn(n+2k)N_{k} \alpha^{4}}{8g^{2}}. \label{fstmassclassical}
  \end{eqnarray}

 This matrix model model has a wider class of the non-trivial
 classical background than the matrix model with the Chern-Simons
 term. While we delegate the explicit definition to the paper
 \cite{0103192}, this matrix model accommodates the classical solution 
 of the fuzzy torus. This property is not shared with the matrix model 
 with the Chern-Simons term. In this sense, the matrix model with a
 tachyonic mass term is an interesting attempt for a richer curved
 space background of the matrix model.

 \subsection{Other related studies}
 In this brief review, we do not exhaust all the recent developments of 
 the IIB matrix model. Instead, we touch on the other important
 studies of the IIB matrix model to conclude this review.
 Firstly, in \cite{9802085}, the reasoning for the IIB matrix model to 
 induce the four-dimensional spacetime dynamically has been found.
 The novelty is that they related the branched polymer and the
 eigenvalue distribution of the IIB matrix model. In this way, they
 found that the Hausdorff dimension of the set of the eigenvalues of
 the matrices $A_{\mu}$ is four. This is a great discovery, in that
 the IIB matrix model implies the dynamical generation of our real
 four-dimensional world. This issue is reviewed in detail in
 \cite{9908038,shino}. 

 Another proposal for the dynamical generation of the four-dimensional 
 spacetime is proposed in \cite{0111102}. In this paper, they
 discussed the Gaussian expansion of the IIB matrix model, by taking
 the quadratic action to be the "test action". They conducted the
 third-order calculation of the Gaussian expansion, and found that the 
 $SO(10)$ Lorentz symmetry breaks down to the $SO(4)$ symmetry.
 This also serves as an important evidence for the generation of the
 four-dimensional spacetime. In \cite{0204240,0211272}, they refined
 the calculation by means of the improved Taylor expansion and the
 notion of the 2PI (two-particle irreducible) Feynman diagram.
 They substantiated the breakdown of the $SO(10)$ Lorentz symmetry to
 the $SO(4)$ up to the seventh order of the Gaussian expansion.

 The IIB matrix model is a successful proposal for the constructive
 definition of the superstring theory, and possesses many exciting
 properties which solidify the confidence that it is an authentic
 constructive definition.
 
 \section{Studies of the $osp(1|32,R)$ supermatrix model}
 In this section, we discuss the matrix model based on the
 $osp(1|32,R)$ super Lie algebra, as a natural extension of the IIB
 matrix model. Firstly, L. Smolin\cite{0002009} proposed the matrix
 model based on the $osp(1|32,R)$ super Lie algebra as an M-theory
 matrix model. Smolin discussed the possibility for this matrix model
 to induce both the BFSS model\cite{9610043} and the IIB matrix
 model\cite{9612115}.

 As we have briefly reviewed in the previous section,
 the IIB matrix model enjoys a lot of successful aspects. And, we have 
 also discussed some of the attempts for the extensions of the IIB
 matrix model for a more manifest formulation of the gravitational
 interaction. Here, we introduce the "supermatrix models", which are
 defined by the "super Lie algebra". There are a lot of illuminating
 features to the supermatrix model. Firstly, the models include both the 
 bosons and the fermions in a unified way, since the bosons and the
 fermions are embedded in the same multiplet. Secondly, as we will see 
 more closely in the definition, the supermatrix models accommodate
 the higher-rank tensor fields, as well as the rank-1 bosonic fields
 $A_{\mu}$. Especially, this opens the door to the generalization of
 the spin connection in terms of the large-$N$ reduced models. 
 If we are to describe a gravitational interaction more manifestly,
 we should take this question more seriously.
 Thirdly, the action is described by the cubic term of the
 supermatrix. This might be a natural aspect for the generalization of 
 the superstring theory, because the cubic interaction is the
 fundamental one in the sense that the four-point interaction is identical to
 the connection of the two cubic interactions due to the conformal
 invariance. Moreover, this model might be solvable in relation to the 
 Chern-Simons theory. The Chern-Simons theory is known to be exactly
 solvable by means of the Jones Polynomial in the knot
 theory\cite{Jones}. If we find a correspondence between the cubic
 matrix model and Witten's technique of solving the Chern-Simons
 theory, we may obtain a way to solve the superstring theory exactly
 in the distant future.
 This section is based on the author's works \cite{0102168,0209057}. 

 \subsection{Definition of the $osp(1|32,R)$ super Lie algebra}
   Before entering the investigation of the superstring action, we
  settle the definition of a super Lie algebra. A super Lie algebra
  is an algebra of supermatrix in which both bosonic matrices and the
  fermionic matrices are embedded in one matrix.
  Supermatrices  possess many properties different from (ordinary)
  matrices, and these properties and the notations are summarized in
  detail in Appendix. \ref{AZMA091441}.  
  
 Let us start with the definition of the $osp(1|32,R)$ super Lie
 algebra.
 \begin{eqnarray} 
  &\clubsuit& \textrm{ If } M \in osp(1|32,R),  \textrm{ then } {^{T}M }
  G + GM = 0 
  \textrm{ for }  G =  \left( \begin{array}{cc}  \Gamma^{0}  & 0  \\  0
  & i   \end{array} \right). \nonumber \\
  &\clubsuit& M \textrm{ is traceless with respect to 33 } \times
  \textrm{ 33 supermatrix.} \nonumber \\
  &\clubsuit& M \textrm{ is a real supermatrix in that } M^{\ast} =
  M. \nonumber 
 \end{eqnarray} 
   We confirm that, for the first condition,  $osp(1|32,R)$ forms a
  closed super Lie algebra. Suppose matrices $M_{1}$  
  and $M_{2}$ satisfy the condition 
  \begin{eqnarray} {^{T} M_{1}} G + G M_{1} = 0, \hspace{4mm}
    {^{T} M_{2}} G + G M_{2} = 0. \label{AZMspade1}\end{eqnarray}
  If $osp(1|32,R)$ is to be a closed super Lie algebra, we call for the
  following  condition:
   \begin{eqnarray} {^{T}([M_{1} , M_{2}])} G + G [M_{1} , M_{2}]
  = 0. \label{AZMspade2} \end{eqnarray}
  Multiplying $G^{-1}$ on both (\ref{AZMspade1}) and (\ref{AZMspade2}) from 
  the left, we respectively obtain
   \begin{eqnarray} G^{-1} {^{T} M_{k}} G + M_{k} = 0,
  \label{AZMspade1+} \end{eqnarray}
  \begin{eqnarray} G^{-1} {^{T} ([M_{1} , M_{2}])} G +  [M_{1} , M_{2}] = 0,
  \label{AZMspade2+} \end{eqnarray}
 where $k=1,2$. The proof that $osp(1|32,R)$ is a closed super Lie
 algebra is equivalent to deriving (\ref{AZMspade2+}) utilizing
 (\ref{AZMspade1+}):
 \begin{eqnarray} 
  \textrm{(\ref{AZMspade2+})} = [ G^{-1} {^{T}M_{2}} G , G^{-1}
 {^{T}M_{1}} G] + [ M_{1} , M_{2}] \stackrel{(\ref{AZMspade1+})}{=} 
 [-M_{2} , -M_{1}] + [M_{1} , M_{2}] = 0.
   \end{eqnarray} 
  This statement, per se, can be satisfied whatever the matrix $G$ may
  be so long as $G$ has an inverse matrix. 

  Here comes one question:
 {\it Why do we define a metric $G$ as } $G
  = \left( \begin{array}{cc} \Gamma^{0} & 0 \\ 0 & i \end{array}
  \right)$ ?
  This stems from the requirement that $M$ should be a real
  matrix; namely $M^{\ast} = ({^{T}M})^{\dagger} = M$. Let us
  consider the consistency between this reality condition and the
  very definition of $osp(1|32,R)$ super Lie algebra. Take a hermitian
  conjugate of the definition ${^{T}M} G + GM=0$. This gives 
   \begin{eqnarray}
    G^{\dagger} ({^{T}M})^{\dagger} + M^{\dagger} G^{\dagger} = 0.
   \end{eqnarray}  
  Utilizing the properties introduced in the
  Appendix. \ref{AZCaho}, and the reality condition, this is
  rewritten as 
   \begin{eqnarray}
   0 =  G^{\dagger} ({^{T}M})^{\dagger} + M^{\dagger} G^{\dagger}
   \stackrel{M^{\ast} = ({^{T}M})^{\dagger}}{=} G^{\dagger} M^{\ast} +
   M^{\dagger} G^{\dagger} \stackrel{M^{\dagger} = {^{T}(M^{\ast})}
   }{=} G^{\dagger} M^{\ast} + {^{T} (M^{\ast})} G^{\dagger}
   \stackrel{M=M^{\ast}}{=} G^{\dagger} M + {^{T}M} G^{\dagger}.
  \nonumber \\  \label{AZM3ccdef} \end{eqnarray}
  In order for the relationship (\ref{AZM3ccdef}) to be consistent
  with the very definition of $osp(1|32,R)$, we must call for a
  condition $G^{\dagger} = G$. 
 And the starting point of $osp(1|32,R)$ is well-defined if we take a
 metric as $G = \left( \begin{array}{cc} \Gamma^{0} & 0 \\ 0 & i
 \end{array} \right)$.\\
  
 The next issue is to investigate the explicit form of this
 super Lie algebra. This is expressed by 
  \begin{eqnarray}
   \textrm{ If } M \in osp(1|32), \textrm{ then } M = \left
   ( \begin{array}{cc} m  & \psi \\   i {\bar \psi} & 0 \end{array}
   \right). \label{AZMsu1-16-16}
  \end{eqnarray}
 \begin{itemize}
  \item{ $m$ contains only the terms of rank 1,2,5. In other
      words,  $m$ is expressed as 
     \begin{eqnarray}
     m = u_{A_{1}} \Gamma^{A_{1}} + \frac{1}{2!} u_{A_{1} A_{2}}
      \Gamma^{A_{1} A_{2}} + \frac{1}{5!} u_{A_{1} \cdots
      A_{5} } \Gamma^{A_{1} \cdots A_{5} }. \label{11dimexpansion}
     \end{eqnarray}
   } 
  \item{Here ${\bar \psi}$ denotes $\psi^{\dagger} 
      \Gamma^{0}$. However, this is equivalent to ${^{T} \psi}
      \Gamma^{0}$, because we are now considering a real super Lie algebra.}
  \end{itemize}
 The proof of this restriction goes as follows.  Let $M$ be the 
 element of $osp(1|32,R)$ and $M =  \left( \begin{array}{cc} m  & \psi
 \\ i {\bar \phi} & v  \end{array} \right)$:
 \begin{itemize}
  \item{ ${\bar \phi}$ is defined as ${\bar \phi} = \psi^{\dagger}
      \Gamma^{0} = {^{T} \phi} \Gamma^{0}$ .}
  \item{ Because $M$ is a real supermatrix, $m$, $v$ and $\psi$ are
      real, while $i {\bar \phi}$ is pure imaginary (and hence ${\bar
      \phi}$ is real).} 
 \end{itemize} 

 By substituting this formula into the very definition, we obtain 
  \begin{eqnarray}
   {^{T}M } G + GM &=&  \left( \begin{array}{cc} {^{T} m}  & - i {^{T}
 {\bar \phi}} \\ {^{T} \psi} & {^{T}v}  \end{array} \right) 
 \left( \begin{array}{cc}  \Gamma^{0}  & 0  \\  0  & i   \end{array}
 \right)  +  \left( \begin{array}{cc}  \Gamma^{0}  & 0  \\  0  & i
 \end{array}  \right) \left( \begin{array}{cc} m  & \psi
 \\ i {\bar \phi} & v  \end{array} \right) \nonumber \\
  &=& \left( \begin{array}{cc}
 {^{T}m} \Gamma^{0} + \Gamma^{0} m  & {^{T}{\bar \phi}} + \Gamma^{0}
 \psi \\ {^{T} \psi} \Gamma^{0} - {\bar \phi} & 2iv  \end{array}
 \right) = 0. \nonumber 
  \end{eqnarray}
  We immediately obtain the relationship between two fermionic fields
  $\psi$ and $\phi$ from this definition: 
  \begin{eqnarray}
     {^{T}{\bar \phi}} + \Gamma^{0}\psi = {^{T} \Gamma^{0}} \phi +
     \Gamma^{0} \psi = - \Gamma^{0} \phi + \Gamma^{0} \psi = 0.
  \label{AZMdine}  \end{eqnarray}
 By multiplying $\Gamma^{0}$ on the both hand sides from the left , we 
 obtain the relationship between $\psi$ and $\phi$. We immediately note that
 ${^{T} \psi} \Gamma^{0} - {\bar \phi} = 0$ is equivalent to the
 equation (\ref{AZMdine}). 

 The constraint on $v$ is trivial,  and $v$ must vanish. On the other
 hand, the constraint on $m$ is worth a careful investigation.  $m$ is
 imposed on the constraint 
 \begin{eqnarray} 
 {^{T}m} \Gamma^{0} + \Gamma^{0} m = 0. \label{AZ41sp32}
 \end{eqnarray}
 This is the very definition of the  Lie algebra called
 $sp(32)$, and this statement indicates that the bosonic $32 \times
 32$ matrix of the $osp(1|32,R)$ super Lie algebra must belong to the
 $sp(32)$ Lie  algebra. In analyzing
 this supermatrix model, it is more convenient to decompose the
 bosonic part $m \in sp(32)$ in terms of the basis of arbitrary $32 \times 32$
 matrices\footnote{Actually this is a  $32 \times 32 = 1024$
 dimensional basis, because the dimension of this basis is
 $1 + {_{11}C_{1}} + \cdots {_{11}C_{5}} = 1 + 11 + 55 + 165 + 330 +
 462 = \frac{1}{2}(1+1)^{11} = 1024$. }.
 Namely, we decompose the bosonic part $m$ in terms of the
 eleven-dimensional gamma matrices  ${\bf 1}_{32 \times 32}$,
 $\Gamma^{A}$, $\Gamma^{A_{1} A_{2}}$, $\Gamma^{A_{1} A_{2}
 A_{3}}$, $\Gamma^{A_{1} \cdots A_{4}}$ and $\Gamma^{A_{1}
 \cdots A_{5}}$, rather than to obey the expression in the paper
 \cite{0002009}. 

 The relationship (\ref{AZ41sp32}) determines what rank of the 11
 dimensional gamma matrices survive. 
  Suppose $m \in sp(32)$ are expressed in terms of the gamma matrices: 
   \begin{eqnarray}
    m = u {\bf 1} + u_{A} \Gamma^{A} + \frac{1}{2!} u_{A_{1}
    A_{2}} \Gamma^{A_{1} A_{2}} + \frac{1}{3!} u_{A_{1}
    A_{2} A_{3}} \Gamma^{A_{1} A_{2} A_{3}} + \frac{1}{4!}
    u_{A_{1} \cdots A_{4}} \Gamma^{A_{1} \cdots A_{4}} +
    \frac{1}{5!} u_{A_{1} \cdots A_{5}} \Gamma^{A_{1} \cdots
    A_{5}}. \label{AZ41assumption}
   \end{eqnarray}
 The  condition (\ref{AZ41sp32}) is rewritten as 
  \begin{eqnarray}
 m = - (\Gamma^{0})^{-1} ({^{T}m}) \Gamma^{0} = \Gamma^{0} ({^{T}m})
 \Gamma^{0}. \label{AZ41constm2}
  \end{eqnarray}
 Then, performing the following computation for $k=0, 1, \cdots, 5$,
 we obtain 
  \begin{eqnarray}
    \Gamma^{0} ({^{T}\Gamma^{A_{1} \cdots A_{k}}}) \Gamma^{0}
  =  (-1)^{\frac{(k+2)(k-1)}{2}} \Gamma^{A_{1} \cdots A_{k}}
  = \left\{ \begin{array}{ll} + \Gamma^{A_{1} \cdots A_{k}} &
     (\textrm{ for } k=1,2,5), \\
     - \Gamma^{A_{1} \cdots A_{k}} & (\textrm{ for }
  k=0,3,4), \end{array} \right.
    \label{AZ41kth} 
  \end{eqnarray}
 where $k$ is the rank of the gamma matrices.
 Combining this result with the constraint of the $sp(32)$ Lie algebra
 (\ref{AZ41constm2}), we discern that only the gamma matrices of
 rank 1, 2 and 5 survive.  We are thus finished with verifying the
 explicit form of the $osp(1|32,R)$ super Lie algebra.

 \subsection{$osp(1|32,R)$ (nongauged) cubic matrix model} \label{osp132r}
 \subsubsection{Action of the cubic supermatrix model}
 We first discuss the nongauged action of the $osp(1|32,R)$ super Lie
 algebra (we postpone the meaning of the "(non)gauged action" later).
 We start with the following action proposed by
 L. Smolin\cite{0002009}:
  \begin{eqnarray}
   S &=& \frac{i}{g^{2}} Tr \sum_{Q,R=1}^{33} \left( 
  \left( \sum_{p=1}^{32} M_{P}^{Q} [M_{Q}^{R}, M_{R}^{P}] \right)
  - M_{33}^{Q} [M_{Q}^{R}, M_{R}^{33}] \right) \nonumber \\
    &=& \frac{i}{g^{2}} \sum_{a,b,c=1}^{N^{2}} str_{33 \times
  33}(M^{a} M^{b} M^{c}) Tr(t^{a}[t^{b},t^{c}]) 
  = - \frac{f_{abc}}{2g^{2}} str(M^{a}M^{b}M^{c}). \label{AZM42cubic2}  
  \end{eqnarray}
 Here, $M$ is a supermatrix belonging to the $osp(1|32,R)$ super Lie
 algebra. The indices $P,Q,R,\cdots$ and $p,q,r,\cdots$ respectively
 run over $P, Q, R, \cdots = 1, 2, \cdots, 32, 33$ and $p, q, r,
 \cdots, = 1, 2, \cdots, 32$. These indices are respectively allocated 
 for the whole $33 \times 33$ $osp(1|32,R)$ supermatrix, and the $32 \times 32$
 bosonic part. Here, we promote each component of the $osp(1|32,R)$
 super Lie algebra to the $u(N)$ hermitian matrices. In the following, 
 the lowercase trace and the uppercase trace are for the $32 \times 32 
 (33 \times 33)$ matrices and the $N \times N$ matrices, respectively.
 $f_{abc}$ is a structure constant of the $u(N)$ Lie algebra, and $i$
 is necessitated in the overall definition of the supermatrix model in 
 order to render the action real.

  This model apparently possesses the following pathological
  properties \cite{0002009}. The first is that this theory is not
  bounded from above or below. This pathology stems from the fact that 
  the action is {\it cubic}, and can be seen by a naive
  field redefinition $M \rightarrow M \lambda$. The action is
  rewritten as  $S \to \lambda^{3} S$,
  and we can set this action to be $S \to \pm \infty$ by setting the
  parameter to be $\lambda \to \pm \infty$. This means that the path
  integral of the theory $Z = \int e^{-S}$ does not converge. However, 
  note that this pathology is shared by general relativity. 
  This pathological property of general relativity can be seen by Weyl
  transformation
  \begin{eqnarray}
   S_{einstein} = \int d^{d} x \sqrt{g} R = \int d^{d} x e^{(\frac{d}{2} -1)
   \omega} \sqrt{g} ( R - (d-1) \nabla^{2} \omega -
   \frac{(d-2)(d-1)}{4} \partial_{\mu} \omega \partial^{\mu} \omega).
  \nonumber \end{eqnarray}
  Although this pathology of Smolin's proposal is not a happy aspect,
  this may be regarded as a good news because this may indicate that
   Smolin's proposals may include general relativity by taking a due
   limit. The second problem is that this theory possesses no explicit 
   time coordinate. This is again a property shared by the general
   relativity. We can introduce a time coordinate by expanding the
   theory around a certain background. Once a time coordinate is
   introduced, we can construct a Hamiltonian of this theory. 
 
   We mention the symmetries of this matrix model. Firstly, this model 
   naturally has the $osp(1|32,R)$ symmetry, which is the extension of 
   the $SO(9,1)$ Lorentz symmetry. We note that this symmetry is
   possible only when we define the theory in terms of the supertrace, 
   which is defined as $strM = \sum_{p=1}^{32} M_{p}^{p} - M_{33}$.
   The cyclic rule of the supermatrix holds true not for the ordinary
   trace but for the supertrace. This property  can be verified by the
   following argument. We consider the following two arbitrary matrices 
  $Z_{1} =  \left( \begin{array}{cc} A_{11} &
  \alpha_{12} \\ \alpha^{T}_{21} & a_{22} \end{array} \right)$,
  $Z_{2} = \left( \begin{array}{cc} B_{11} & \beta_{12} \\ 
  \beta^{T}_{21} & b_{22} \end{array} \right)$, 
   where $A_{11}$ and $B_{11}$ are 32 $\times$ 32 bosonic matrices,
  $\alpha_{12}$ , $\alpha_{21}$ , $\beta_{12}$ and $\beta_{21}$ are
  32-component fermionic vectors, and $a_{22}$ and $b_{22}$ are
  c-numbers. We consider the supertrace of
  $Z_{1} Z_{2}$ and $Z_{2} Z_{1}$, which are calculated as 
  $str(Z_{1} Z_{2}) = tr(A_{11} B_{11} +
  \alpha_{12} \beta^{T}_{21} ) - (\alpha^{T}_{21} \beta_{12} + a_{22}
  b_{22})$, and $str(Z_{2} Z_{1}) = tr(A_{11} B_{11} +
  \beta_{12} \alpha^{T}_{21} ) - (\beta^{T}_{21} \alpha_{12} + a_{22}
  b_{22})$. 
  Since $\alpha$ and $\beta$ are fermionic quantities, the sign flips
  if we change the order. Therefore, we establish that it is {\it not an
  ordinary trace but a supertrace} that the cyclic rule  $str(Z_{1}
  Z_{2}) = str(Z_{2} Z_{1})$ holds true of the supermatrix.

  We next discuss its gauge symmetry. This model is invariant under
  the following $U(N)$ transformation:
  \begin{eqnarray}
   \delta M = i [({\bf 1}_{33 \times 33} \otimes u), M], \textrm{ where }
   i u \in u(N).
  \end{eqnarray} 
  In this symmetry, we do not mix the $osp(1|32,R)$ transformation and
  the $u(N)$ transformation. We have a closer look at the mixed
  symmetry in an attempt to realize the local Lorentz symmetry, and we 
  call such models "gauged models". On the other hand, we call the
  action (\ref{AZM42cubic2}) "nongauged", because we do not mix these
  two symmetry. This "nongauged symmetry" is close to the symmetry of
  the IIB matrix model, in which the $SO(9,1)$ Lorentz symmetry and
  the $U(N)$ gauge symmetry are separate. 

  In addition, this model incorporates the trivial shift of the
  supermatrix 
  \begin{eqnarray}
   \delta M = c {\bf 1}_{N \times N}, \textrm{ where } c \in osp(1|32,R).
  \end{eqnarray}

  The action (\ref{AZM42cubic2}) is rewritten in terms of the explicit
  expression of the $osp(1|32,R)$ super Lie algebra
  $M = \left( \begin{array}{cc} m & \psi \\ i {\bar \psi} & 0
  \end{array} \right)$ as
  \begin{eqnarray}
    S = - \frac{f_{abc}}{2g^{2}} ( tr(m^{a} m^{b} m^{c}) - 3i {\bar
     \psi^{a}} m^{b} \psi^{c})  
     = \frac{i}{g^{2}} Tr( {m_{p}}^{q} [ {m_{q}}^{r} , {m_{r}}^{p}
     ] ) - 3i {\bar \psi} [m, \psi] ).  
   \label{AZ42actioncomp} 
  \end{eqnarray}
  
  We consider the reduction of the model to the ten-dimensions. 
  To this end, we first rewrite the action (\ref{AZM42cubic2}) in
  terms of the expansion for the eleven-dimensional gamma matrices
  as (\ref{11dimexpansion}).
  In order to see the correspondence between the ten-dimensional IIB
  matrix model, we perform a reduction to the ten dimensions. To this
  end, we specialize the tenth direction $x^{10}=x^{\sharp}$, and we retain
  the $SO(9,1)$ rotational symmetry. We define the ten-dimensional
  fields by the following chiral decomposition of the bosonic matrix $m$:
  \begin{eqnarray}
   m &=& W \Gamma^{\sharp} + A_{\mu}^{(+)} \Gamma^{\mu} \frac{1 +
   \Gamma^{\sharp}}{2} + A_{\mu}^{(-)} \Gamma^{\mu} \frac{1 -
   \Gamma^{\sharp}}{2} 
   + \frac{1}{2} C_{\mu \nu} \Gamma^{\mu \nu}
   + \frac{1}{4!} H_{\mu_{1} \cdots \mu_{4}}\Gamma^{\mu_{1} \cdots
   \mu_{4} \sharp} \nonumber \\ 
   &+& \frac{1}{5!} \left( I_{\mu_{1} \cdots \mu_{5}}^{(+)}
     \Gamma^{\mu_{1} \cdots \mu_{5}} (1 + \Gamma^{(+)})
   + I_{\mu_{1} \cdots \mu_{5}}^{(-)}
     \Gamma^{\mu_{1} \cdots \mu_{5}} (1 + \Gamma^{(-)}) \right). 
  \label{chiraldecom}
  \end{eqnarray}
  Here, we also recall the notation of the left(right)-handed fermion:
  \begin{eqnarray}
   \psi_{L} = \frac{1 + \Gamma^{\sharp}}{2} \psi, \hspace{2mm}
   \psi_{R} = \frac{1 - \Gamma^{\sharp}}{2} \psi, \nonumber
  \end{eqnarray}
 The coefficients $ I_{\mu_{1} \cdots \mu_{5}}^{(+)}$ and
 $I_{\mu_{1} \cdots \mu_{5}}^{(-)}$ are self-dual to each
 other. Namely, they satisfy the relation
  \begin{eqnarray}
   I^{(\pm)}_{\mu_{1} \cdots \mu_{5}} = \frac{-1}{5!}
   I^{(\mp)}_{\mu_{6} \cdots \mu_{10}} \epsilon^{\mu_{1} \cdots
   \mu_{10} \sharp}. \label{selfdualrank5}
  \end{eqnarray}

  The action is thus expanded as
   \begin{eqnarray}
  S_{b} &=& \frac{i}{g^{2}} Tr_{N \times N} ( -96[A^{(+)}_{\mu_{1}} ,
  A^{(-)}_{\mu_{2}} ] C^{\mu_{1} \mu_{2}} - 96 W [A^{(+) \mu} ,
  A^{(-)}_{\mu}] +  
  \frac{4}{5} W [ I^{(+)}_{\mu_{1} \cdots \mu_{5}} , I^{(-) \mu_{1} \cdots
  \mu_{5}}] \nonumber \\
  &-& 4 H_{\mu_{1} \cdots \mu_{4}} ( [ A^{(+)}_{\mu_{5}} , I^{(-) \mu_{1}
  \cdots \mu_{5}}] - [ A^{(-)}_{\mu_{5}} , I^{(+) \mu_{1} \cdots \mu_{5}}]  )
  - 8 C_{\mu_{1} \mu_{2}} [ {I^{(+)\mu_{1}}}_{\mu_{3} \cdots \mu_{6}} , I^{(-)
  \mu_{2} \cdots \mu_{6}}]  \nonumber \\
  &+&  \frac{8}{3} {H^{\nu \rho}}_{\mu_{1} \mu_{2}} ( [{I^{(+)}}_{\nu
  \rho  \mu_{3} \mu_{4} \mu_{5}} ,  I^{(-) \mu_{1} \cdots \mu_{5}}] -  
  [{I^{(-)}}_{\nu \rho \mu_{3} \mu_{4} \mu_{5}} ,  I^{(+) \mu_{1} \cdots
  \mu_{5}}]  ) \nonumber \\
  &+& 32 [ {C^{\mu_{1}}}_{\mu_{2}} , C_{\mu_{1} \mu_{3}}] C^{\mu_{2}
  \mu_{3}} - 16 C_{\mu_{1} \mu_{2}} [ {H^{\mu_{1}}}_{\mu_{3} \mu_{4}
  \mu_{5}} , H^{\mu_{2} \cdots \mu_{5}}] 
  + \frac{1}{27} H_{\mu_{1} \cdots \mu_{4}} [ {H^{\nu}}_{\mu_{5} \cdots
  \mu_{7}} , H_{\nu \mu_{8} \mu_{9} \mu_{10}}] \epsilon^{\mu_{1}
  \cdots \mu_{10} \sharp} ), \nonumber \\ \label{AZM52actionpmb} \\
 S_{f} &=& \frac{i}{g^{2}} Tr_{N \times N} ( 
   -3i (- {\bar \psi_{L}}  [ W , \psi_{R}] +  {\bar \psi_{R}} [ W ,
  \psi_{L}] ) 
  - 3i  ({\bar \psi_{L}} \Gamma^{\mu} [ A^{(+)}_{\mu} , \psi_{L}] 
     + {\bar \psi_{R}} \Gamma^{\mu} [ A^{(-)}_{\mu} , \psi_{R}] ) \nonumber \\ 
   &-&\frac{3i}{2!} ( {\bar \psi_{L}} \Gamma^{\mu_{1} \mu_{2}}
  [ C_{\mu_{1} \mu_{2}}, \psi_{R}] + {\bar \psi_{R}} \Gamma^{\mu_{1}
  \mu_{2}} [ C_{\mu_{1} \mu_{2}}, \psi_{L}] ) \nonumber \\ 
  &-& \frac{3i}{4!} ( - {\bar \psi_{L}} \Gamma^{\mu_{1} \mu_{2} \mu_{3}
  \mu_{4}} [ H_{\mu_{1} \mu_{2} \mu_{3} \mu_{4}} ,   \psi_{R}] +
  {\bar \psi_{R}} \Gamma^{\mu_{1} \mu_{2} \mu_{3} \mu_{4}}
  [ H_{\mu_{1} \mu_{2} \mu_{3} \mu_{4}}, \psi_{L}] ) \nonumber \\ 
   &-& \frac{3i}{5!} ( 2 {\bar \psi_{L}} \Gamma^{\mu_{1} \mu_{2} \mu_{3}
  \mu_{4} \mu_{5}}  
 [ {I^{(+)}}_{\mu_{1} \mu_{2} \mu_{3} \mu_{4} \mu_{5}} , \psi_{L}] 
    + 2 {\bar \psi_{R}} \Gamma^{\mu_{1} \mu_{2} \mu_{3} \mu_{4} \mu_{5}} 
 [ {I^{(-)}}_{\mu_{1} \mu_{2} \mu_{3} \mu_{4} \mu_{5}} ,\psi_{R}] ) ),
  \label{AZM52actionpmf} 
 \end{eqnarray}
 where $S_{b}$ and $S_{f}$ are the bosonic and fermionic parts of the
 action $S$, respectively. The whole action is $S = S_{b} + S_{f}$. 
 The computation of this action is lengthy, and we delegate the proof to
 Appendix B.2. of \cite{0103003}. 

 \subsubsection{Identification of the supersymmetry with the
 IIB matrix model}
 We next discuss the identification of the supersymmetry with that of
 the IIB matrix model. This model has two kinds of supersymmetry
 transformation, corresponding to the $osp(1|32,R)$ rotational
 symmetry and the translation symmetry.
  \begin{itemize}
   \item{The former supersymmetry corresponding to the $osp(1|32,R)$
       rotation is called the "homogeneous supersymmetry", in the same
       sense as for the IIB matrix model. This is defined by the
       rotation by the supercharge
      $Q = \left( \begin{array}{cc} 0 & \chi \\ i {\bar \chi} & 0
       \end{array} \right)$. Namely, the transformation of the
       matrices $M$ is given by
      \begin{eqnarray}
       \delta^{(1)}_{\chi} M = [Q,M] = [\left( \begin{array}{cc} 0 & \chi \\
       i {\bar \chi} & 0 \end{array} \right), 
      \left( \begin{array}{cc} m & \psi \\ i {\bar \psi} & 0
       \end{array} \right) ]
       = \left( \begin{array}{cc} i (\chi {\bar \psi} - \psi {\bar
     \chi}) & - m \chi \\ i {\bar \chi} m & 0 \end{array} \right).
      \label{homogeneousosp} 
      \end{eqnarray}
       }
  \item{The latter supersymmetry corresponds to the translation
      symmetry. This is called the "inhomogeneous supersymmetry", and is
      defined by the simple translation of the fermion
      $\delta^{(2)}_{\xi} \psi = \xi$.}
  \end{itemize}
 In this way, the $osp(1|32,R)$ supermatrix model has $32 + 32= 64$
 supercharges. This is twice as much as the IIB matrix model. This
 simply leads us to speculate that the double structures of the
 IIB matrix model are embedded in this supermatrix model. We elaborate 
 on this point, by scrutinizing the commutation relation.
 Especially, we elucidate that there is a nice correspondence with the 
 supersymmetry of the IIB matrix model, with the chiral decomposition 
 (\ref{chiraldecom}). Namely, we identify $A^{(\pm)}_{\mu}$ with the
 vector fields of the IIB matrix model, and consider its supersymmetry 
 transformation.

 Since the fields $A^{(\pm)}$ are given by\footnote{We recall that
 $\frac{1}{32} tr \Gamma_{A} \Gamma^{A} = 1$ while $\frac{1}{32} tr
 \Gamma_{AB} \Gamma^{AB} = -1$. Here, we do not take a summation with
 respect to the duplicate indices $A,B$.}
  \begin{eqnarray}
   A^{(\pm)}_{\mu} = \frac{1}{32} tr(m \Gamma_{\mu}) \mp \frac{1}{32}
   tr(m \Gamma_{\mu \sharp}), 
  \end{eqnarray}
 the homogeneous transformation of the vector fields $A^{(\pm)}_{\mu}$ 
 and $\psi$ is computed as
  \begin{eqnarray}
   \delta^{(1)}_{\chi} A^{(+)}_{\mu} &=& \frac{1}{32} tr( (\delta^{(1)}_{\chi} 
   m ) \Gamma_{\mu} ) + \frac{-1}{32} tr( (\delta^{(1)}_{\chi}
   m)\Gamma_{\mu \sharp} )  =  \frac{i}{8} {\bar \chi_{R}}
   \Gamma_{\mu} \psi_{R}, \label{AZ43homoa+} \\
  \delta^{(1)}_{\chi} A^{(-)}_{\mu} &=& \frac{1}{32} tr( (\delta^{(1)}_{\chi} 
   m ) \Gamma_{\mu} ) - \frac{-1}{32} tr( (\delta^{(1)}_{\chi}
   m)\Gamma_{\mu \sharp} ) 
  =   \frac{i}{8} {\bar \chi_{L}} \Gamma_{\mu} \psi_{L},  
   \label{AZ43homoa-} \\
  \delta^{(1)}_{\chi} \psi &=&  - m \chi.  \label{AZ43homopsi}
  \end{eqnarray}
  The transformation of the vector fields $A^{(\pm)}_{\mu}$ is
  beautifully reminiscent of the homogeneous transformation of the 
  IIB matrix model. In this sense, we see a correspondence between the 
  chirality of the vector fields and the fermions as
  \begin{eqnarray}
    A^{(+)}_{\mu} \Leftrightarrow \psi_{R}, \hspace{2mm}
    A^{(-)}_{\mu} \Leftrightarrow \psi_{L}. \label{correspondence1}
  \end{eqnarray}
  The chirally-decomposed fermions $\psi_{L,R}$ clearly have 16
  components, which coincides the number of the supercharges for the
  IIB matrix model. In this sense, this identification augurs very
  well for the reproduction of the supersymmetry of the IIB matrix
  model. On the other hand, the homogeneous transformation of the
  fermion is given by (\ref{AZ43homoa-}), and is devoid of the
  commutator of the vector fields like the IIB matrix model.
  This is a downside in the identification of the supersymmetry
  transformation with that of the IIB matrix model.

  Nextly, we have a close look at the commutation relation of the
  supersymmetry transformation. We wish to find the similar structure
  of the commutation relations as that of the IIB matrix model.
  We consider the following commutation relations:
  \begin{eqnarray}
   (1) [\delta^{(1)}_{\chi}, \delta^{(1)}_{\epsilon}] A^{(\pm)}_{\mu},
   \hspace{2mm} 
   (2) [\delta^{(2)}_{\chi}, \delta^{(2)}_{\epsilon}] A^{(\pm)}_{\mu},
   \hspace{2mm} 
   (3) [\delta^{(1)}_{\chi}, \delta^{(2)}_{\epsilon}] A^{(\pm)}_{\mu},
   \label{AZ43namco}
  \end{eqnarray}
  It is trivial that the commutator (2) vanishes, because the
  inhomogeneous supersymmetry is nothing but a translation of the
  fermion. Here, we scrutinize the rest of the commutators one by one.
 
  \paragraph{Supersymmetry transformation $(3)  [\delta^{(1)}_{\chi},
  \delta^{(2)}_{\epsilon} ] A^{(\pm)}_{\mu}$} \hspace{0mm}

  In the IIB matrix model, only the commutator (\ref{AZM31SUSY3com})
  remains nonvanishing. We attempt to
  extract the similar structure to the commutation relation
  (\ref{AZM31SUSY3com}). The commutator of the supersymmetry
  transformation for each $m$ and $\psi$ is given by
    \begin{eqnarray}
    [ \delta^{(1)}_{\chi}, \delta^{(2)}_{\epsilon} ] m = -i 
    ( \chi {\bar \epsilon} - \epsilon {\bar \chi} ), \hspace{3mm}
    [ \delta^{(1)}_{\chi}, \delta^{(2)}_{\epsilon} ] \psi = 0,
  \label{AZ43SUSYcom69}
   \end{eqnarray}
  which can be derived in the same way as
  (\ref{AZM31SUSY3com}). Namely, we take the difference of the
  following paths:
   \begin{eqnarray}
   & & m \stackrel{\delta^{(2)}_{\epsilon}}{\to} m 
         \stackrel{\delta^{(1)}_{\chi}}{\to} m + i (\chi {\bar \psi} -
         \psi {\bar \chi} ), \textrm{ whereas } m
         \stackrel{\delta^{(1)}_{\chi}}{\to} m + i (\chi {\bar
         \psi} - \psi {\bar \chi} )  
         \stackrel{\delta^{(2)}_{\epsilon}}{\to} m + i \chi 
         ({\bar \psi} + {\bar \epsilon})  - i (\psi + \epsilon) {\bar
         \chi}, \nonumber \\
  & &\psi \stackrel{\delta^{(2)}_{\epsilon}}{\to} \psi +
           \epsilon  \stackrel{\delta^{(1)}_{\chi}}{\to} \psi +
           \epsilon - m \chi, \textrm{ whereas } \psi
           \stackrel{\delta^{(1)}_{\chi}}{\to} \psi - m \chi
           \stackrel{\delta^{(2)}_{\epsilon}}{\to}  \psi +
           \epsilon - m \chi. \nonumber
   \end{eqnarray}
  The commutator of the supersymmetry transformation for each
  $A^{(\pm)}_{\mu}$ can be extracted as
   \begin{eqnarray}
  & &  [\delta^{(1)}_{\chi}, \delta^{(2)}_{\epsilon}] A^{(+)}_{\mu} =
    \frac{1}{32} tr(( [\delta^{(1)}_{\chi}, \delta^{(2)}_{\epsilon}] m)
    \Gamma_{\mu}) + \frac{-1}{32} tr(( [\delta^{(1)}_{\chi},
    \delta^{(2)}_{\epsilon}] m) \Gamma_{\mu \sharp}) =
   \frac{i}{8} {\bar \epsilon}_{R} \Gamma_{\mu} 
    \chi_{R}, \label{AZM43SUSYcom12+} \\
  & &  [\delta^{(1)}_{\chi}, \delta^{(2)}_{\epsilon}] A^{(-)}_{\mu} =
    \frac{1}{32} tr (( [\delta^{(1)}_{\chi}, \delta^{(2)}_{\epsilon}]m)
    \Gamma_{\mu}) - \frac{-1}{32} tr(( [\delta^{(1)}_{\chi},
    \delta^{(2)}_{\epsilon}] m) \Gamma_{\mu \sharp}) =
    \frac{i}{8} {\bar \epsilon}_{L} \Gamma_{\mu} \chi_{L}.
    \label{AZM43SUSYcom12-} 
   \end{eqnarray}
   Extracting the specific chirality of the SUSY parameters, we obtain 
  the nonvanishing commutators as
   \begin{eqnarray}
   & & [ \delta^{(1)}_{\chi_{L}}, \delta^{(2)}_{\epsilon_{L}} ]
   A^{(+)}_{\mu} = 0 , \hspace{15mm}
     [ \delta^{(1)}_{\chi_{R}}, \delta^{(2)}_{\epsilon_{R}} ]
   A^{(+)}_{\mu} = \frac{i}{8} {\bar \epsilon}_{R} \Gamma_{\mu} \chi_{R},
 \nonumber \\
  & &  [\delta^{(1)}_{\chi_{L}}, \delta^{(2)}_{\epsilon_{L}} ]
   A^{(-)}_{\mu} = \frac{i}{8} {\bar \epsilon}_{L} \Gamma_{\mu} \chi_{L},
   \hspace{3mm} [\delta^{(1)}_{\chi_{R}}, \delta^{(2)}_{\epsilon_{R}} ]
   A^{(-)}_{\mu} = 0.  \label{AZ43SUSYmainres}
   \end{eqnarray}
  The commutators with different chirality of the two supersymmetry
  parameters clearly vanish:
   \begin{eqnarray}
       [\delta^{(1)}_{\chi_{L}}, \delta^{(2)}_{\epsilon_{R}} ]
       A^{(\pm)}_{\mu} =  [\delta^{(1)}_{\chi_{R}},
       \delta^{(2)}_{\epsilon_{L}} ]  A^{(\pm)}_{\mu} = 0.
   \end{eqnarray}

  These combinations of the supersymmetry transformations clearly
  resemble the structure of the IIB matrix model. The
  correspondence between the vector fields and the chirality of the
  fermionic fields matches that obtained by the correspondence of the
  homogeneous transformation itself. 

  \paragraph{Supersymmetry transformation $(1) [\delta^{(1)}_{\chi},
  \delta^{(1)}_{\epsilon} ] A^{(\pm)}_{\mu} $} \hspace{0mm}

   In the case of the IIB matrix model, this supersymmetry
   transformation vanishes on shell up to the gauge transformation, as
   we have seen in (\ref{AZM31SUSY1com}). We want to find the same
   vanishing in our cubic model, if we are to identify the
   supersymmetry transformation.  
  However, it turns out that this commutator does not vanish.
 In investigating this commutation relation, it is easier
 to utilize the  following identity:
      \begin{eqnarray}
     [ \delta^{(1)}_{\chi}, \delta^{(1)}_{\epsilon} ]
       M = [ Q_{\chi}, [Q_{\epsilon} , M]] -  [ Q_{\epsilon},
       [Q_{\chi} , M]] = [ [Q_{\chi}, Q_{\epsilon}], M]. \label{AZ43comm1964}
    \end{eqnarray}
   This commutator of the
       SUSY transformation is thus obtained by
        \begin{eqnarray}
          [ \delta^{(1)}_{\chi}, \delta^{(1)}_{\epsilon} ] M =
          [ \left( \begin{array}{cc} i (\chi {\bar \epsilon} -
          \epsilon {\bar \chi} ) & 0 \\ 0 & 0 \end{array} \right),
          \left( \begin{array}{cc} m & \psi \\ i {\bar \psi} & 0
          \end{array} \right) ] = \left( \begin{array}{cc} [i(\chi
          {\bar \epsilon} - \epsilon {\bar \chi}), m] & i (\chi
          {\bar \epsilon} - \epsilon {\bar \chi}) \psi \\ -i {\bar
          \psi} (\chi {\bar \epsilon} - \epsilon {\bar \chi}) &  0
          \end{array} \right). \label{AZM43SUSY1coma}
        \end{eqnarray}      
    This relation reveals that the commutators of the two
    homogeneous transformation are
     \begin{eqnarray}
    & &  [ \delta^{(1)}_{\chi}, \delta^{(1)}_{\epsilon} ] m = i [(\chi
          {\bar \epsilon} - \epsilon {\bar \chi}) ,m ], \label{AZ43hhm} \\
    & &   [ \delta^{(1)}_{\chi}, \delta^{(1)}_{\epsilon} ] \psi = i (\chi
          {\bar \epsilon} - \epsilon {\bar \chi}) \psi. \label{AZ43hhpsi}
     \end{eqnarray}
  We have seen in
   (\ref{AZ43SUSYmainres}) which chiralities of the supersymmetry parameters
   correspond to the vector fields $A^{(\pm)}_{\mu}$. According to this
   correspondence, we examine the commutation relation with respect 
   to the following cases. 

    We first investigate the commutation relation  
    $ [ \delta^{(1)}_{\chi_{L}}, \delta^{(1)}_{\epsilon_{L}}
          ]$. We examine
          this commutation relation of this SUSY transformation:
          \begin{eqnarray}
      [ \delta^{(1)}_{\chi_{L}}, \delta^{(1)}_{\epsilon_{L}}
          ] A^{(-)}_{\mu} = \frac{1}{32} tr( 
          ( [ \delta^{(1)}_{\chi_{L}}, \delta^{(1)}_{\epsilon_{L}}] m) 
          \Gamma_{\mu} )- \frac{-1}{32} tr( ( [ \delta^{(1)}_{\chi_{L}},
          \delta^{(1)}_{\epsilon_{L}} ] m) \Gamma_{\mu \sharp} )
       = \frac{i}{8} ( {\bar \chi_{L}}
          [m , \Gamma_{\mu}] \epsilon_{L} ),  \label{AZM43SUSY11a-}
        \end{eqnarray}
    Here, we recall that the commutator $[m,\Gamma_{\mu}]$ belong to the
    $sp(32)$ Lie algebra (namely, only the rank-1,2,5 contributions remain).
     Unfortunately, this commutation relation does
        not vanish exactly in the ten-dimensional reduction.
     Now, we have a close look at which components survive or vanish:
   Since the chirality of the supersymmetry parameter is identical,
  the contribution of the odd-rank fields of $m$ (with respect 
    to the ten-dimensional indices) clearly vanishes. 
  On the other hand, the contribution of the even-rank fields $W,
    C_{\mu_{1} \mu_{2}}, H_{\mu_{1} \cdots \mu_{4}}$ survives.
  For example, the contribution of the rank-0 field $W$ is obtained as
  \begin{eqnarray}
   {\bar \chi_{L}} [ W \Gamma_{\sharp}, \Gamma_{\mu} ]
             \epsilon_{L} = {\bar \chi_{L}} W \Gamma_{\sharp} \Gamma_{\mu} 
             \epsilon_{L} (\neq 0). 
  \end{eqnarray}
    We next investigate the commutation relation of other
      chirality.  $ [ \delta^{(1)}_{\chi_{R}},
      \delta^{(1)}_{\epsilon_{R}}]$. The corresponding field is now
      $A^{(+)}_{\mu}$, and the commutation relation is now 
         \begin{eqnarray}
          [ \delta^{(1)}_{\chi_{R}}, \delta^{(1)}_{\epsilon_{R}}
          ] A^{(+)}_{\mu} = \frac{1}{32} tr( 
          ( [ \delta^{(1)}_{\chi_{R}}, \delta^{(1)}_{\epsilon_{R}}] m) 
          \Gamma_{\mu} ) + \frac{-1}{32} tr( ( [ \delta^{(1)}_{\chi_{R}},
          \delta^{(1)}_{\epsilon_{R}} ] m) \Gamma_{\mu \sharp} ) = 
          \frac{i}{8} ( {\bar \chi_{R}} 
          [m , \Gamma_{\mu}] \epsilon_{R} ),  \label{AZM43SUSY11a+}
        \end{eqnarray}
     by completely computing in the same fashion as in
     (\ref{AZM43SUSY11a-}), and we note that the contribution of the
     even-rank fields $W, C_{\mu_{1} \mu_{2}}, H_{\mu_{1} \cdots \mu_{4}}$
     remains nonvanishing.

  Even the worse news comes in the commutator for the different
  chirality of the supersymmetry parameter. The commutator for the
  rank-1 vector field is given by
     \begin{eqnarray}
  & & [\delta^{(1)}_{\chi_{L}}, \delta^{(1)}_{\epsilon_{R}}] A^{(+)}_{\mu} = 
      \frac{1}{32} tr(([\delta^{(1)}_{\chi_{L}},
      \delta^{(1)}_{\epsilon_{R}}] m) \Gamma_{\mu} ) + \frac{-1}{32} tr(
      ( [\delta^{(1)}_{\chi_{L}}, \delta^{(1)}_{\epsilon_{R}}] m)
      \Gamma_{\mu \sharp} )
      =  \frac{i}{16} ({\bar \chi}_{L} m \Gamma_{\mu} 
      \epsilon_{R} + {\bar \epsilon}_{R} \Gamma_{\mu} m \chi_{L} ),
      \nonumber \\ \label{AZ43SUSY11lra+}  \\ 
  & & [\delta^{(1)}_{\chi_{L}}, \delta^{(1)}_{\epsilon_{R}}]
      A^{(-)}_{\mu} =   \frac{1}{32} tr( ([\delta^{(1)}_{\chi_{L}},
      \delta^{(1)}_{\epsilon_{R}}] m ) \Gamma_{\mu} ) - \frac{-1}{32}
      tr(( [\delta^{(1)}_{\chi_{L}}, \delta^{(1)}_{\epsilon_{R}}] m)
      \Gamma_{\mu \sharp} )  = -  \frac{i}{16} ({\bar
      \epsilon}_{R} m \Gamma_{\mu} 
      \chi_{L} + {\bar \chi}_{L} \Gamma_{\mu} m \epsilon_{R}
      ). \nonumber \\ \label{AZ43SUSY11lra-}
    \end{eqnarray}
     The disastrous fact is that these commutators include the fields
  $A^{(+)}_{\mu}$ and $A^{(-)}_{\mu}$ as
  \begin{eqnarray}
  [\delta^{(1)}_{\chi_{L}}, \delta^{(1)}_{\epsilon_{R}}] A^{(+)}_{\mu}
    =  - \frac{i}{8} {\bar \chi}_{L} 
    A^{(+)}_{\nu} {\Gamma_{\mu}}^{\nu} \epsilon_{R} + \cdots,
    \hspace{2mm} 
  [\delta^{(1)}_{\chi_{L}}, \delta^{(1)}_{\epsilon_{R}}]
    A^{(-)}_{\mu} = - \frac{i}{8} {\bar \chi}_{L} 
     A^{(-)}_{\nu} {\Gamma_{\mu}}^{\nu} \epsilon_{R} + \cdots. 
    \label{AZ43lrmix-} 
  \end{eqnarray}
   These commutation relations reveal that the two-fold supersymmetry is
  not independent of each other, but are connected by not the impurity 
  $W, C_{\mu_{1} \mu_{2}}$ and $H_{\mu_{1} \cdots \mu_{4}}$, but the fields
  $A^{(\pm)}_{\mu}$. This is an  unfavorable situation in the analysis
  of the supersymmetry transformation of this cubic model. 

  \paragraph{Summary of the supersymmetry structure} \hspace{0mm}

  We recapitulate what we have obtained from the above argument of the
  supersymmetry. Firstly, we have noted that this supermatrix model
  incorporates the two-fold structure of the supersymmetry of the IIB
  matrix model. This is predictable from the number of the
  supercharges. This model has the homogeneous supersymmetry and the
  inhomogeneous supersymmetry coming from the $osp(1|32,R)$
  rotation and the translation symmetry, respectively. Therefore, this 
  model has 64 supercharges, which is twice as many as the IIB matrix
  model. The IIB-like structure of the supersymmetry is extracted by
  the chiral decomposition of the rank-1 field and the supercharges, and
  we have seen the correspondence (\ref{correspondence1}). 
  While we have reproduced the commutation relation
  (\ref{AZM31SUSY1com}) and (\ref{AZM31SUSY2com}) from the commutator 
  $[\delta^{(1)}_{\chi}, \delta^{(2)}_{\epsilon}]$ and
  $[\delta^{(2)}_{\chi}, \delta^{(2)}_{\epsilon}]$, the commutator
  $[\delta^{(1)}_{\chi}, \delta^{(1)}_{\epsilon}]$ does not trivially
  vanish, unlike the IIB matrix model. Moreover, the commutation
  relation (\ref{AZ43lrmix-}) unravels that these two supersymmetries
  are not independent of each other. Its physical interpretation will
  be reported elsewhere.

 \subsubsection{Induction of the IIB matrix model}
  We have seen the correspondence of the supersymmetry with the IIB
  matrix model, and we have elucidated the correspondence of the
  chirality between the rank-1 vector fields and the supersymmetry
  parameter as (\ref{correspondence1}). This leads us to the idea that 
  we may be able to induce the IIB matrix model if we integrate out
  certain fields. To this end, we retain the field $A^{(+)}_{\mu}$ and 
  $\psi_{R}$, and integrate out the other fields. 
  However, the solid discussion for the extraction of the IIB matrix
  model is an onerous question, and we contend ourselves with a
  hand-waving argument. This supermatrix model possesses no time
  spacetime derivative from the outset, as is true of the IIB matrix
  model. Its classical equation of motion 
  $\partial_{A} \frac{\partial S}{ \partial (\partial_{A} X)} -
    \frac{\partial S}{\partial X} = 0$ clearly incorporates the
  following noncommutative background
   \begin{eqnarray}
    (A^{(+)}_{\mu})_{\textrm{b.g.}} = {\hat p}_{\mu}, \hspace{2mm} 
  \textrm{(other fields)}=0, \label{mcsupermat} 
   \end{eqnarray}
  where ${\hat p}_{\mu}$ satisfies the canonical commutation relation 
  $[{\hat p}_{\mu}, {\hat p}_{\nu}] = i c_{\mu \nu}$.  When we
  denote the fluctuation as $A^{(+)}_{\mu} = {\hat p}_{\mu} +
  a^{(+)}_{\mu}$, the commutator is mapped as
  \begin{eqnarray}
    [ A^{(+)}_{\mu}, X ] = - i \partial_{\mu} X + [a^{(+)}_{\mu},
    X]. \nonumber 
  \end{eqnarray} 
  Then, the bosonic and the fermionic terms of the action are expanded as
  \begin{eqnarray}
     S_{b} &=& \frac{1}{g^{2}} Tr_{N \times N} ( -96(\partial_{\mu_{1}}
     A^{(-)}_{\mu_{2}} ) C^{\mu_{1} \mu_{2}} + 96 ( \partial_{\mu} W )
     A^{(-) \mu} + 4 (\partial_{\mu_{1}} H_{\mu_{2} \cdots \mu_{5}} ) I^{(-)
     \mu_{1} \cdots \mu_{5}} ) \label{AZ44partial} \\
 &+&
  \frac{i}{g^{2}} Tr_{N \times N} ( -96[a^{(+)}_{\mu_{1}} ,
  A^{(-)}_{\mu_{2}} ] C^{\mu_{1} \mu_{2}} - 96 W [a^{(+) \mu} ,
     A^{(-)}_{\mu}] +  
  \frac{4}{5} W [ I^{(+)}_{\mu_{1} \cdots \mu_{5}} , I^{(-) \mu_{1} \cdots
  \mu_{5}}] \nonumber \\
  &+& 4 ( [ a^{(+)}_{\mu_{1}} , H_{\mu_{2} \cdots \mu_{5}}] I^{(-) \mu_{1}
  \cdots \mu_{5}} - [ A^{(-)}_{\mu_{1}} , H_{\mu_{2} \cdots \mu_{5}}] I^{(+)
  \mu_{1} \cdots \mu_{5}}  ) 
    - 8 C_{\mu_{1} \mu_{2}} [ {I^{(+)\mu_{1}}}_{\mu_{3} \cdots
     \mu_{6}} , I^{(-) \mu_{2} \cdots \mu_{6}}]  \nonumber \\
  &+&  \frac{8}{3} {H^{\nu \rho}}_{\mu_{1} \mu_{2}} ( [{I^{(+)}}_{\nu
     \rho \mu_{3} \mu_{4} \mu_{5}} ,  I^{(-) \mu_{1} \cdots \mu_{5}}]
     - [{I^{(-)}}_{\nu \rho  \mu_{3} \mu_{4} \mu_{5}} ,  I^{(+) \mu_{1} \cdots
  \mu_{5}}]  ) \nonumber \\
  &+& 32 [ {C^{\mu_{1}}}_{\mu_{2}} , C_{\mu_{1} \mu_{3}}] C^{\mu_{2}
     \mu_{3}} - 16  C_{\mu_{1} \mu_{2}} [ {H^{\mu_{1}}}_{\mu_{3}
     \mu_{4} \mu_{5}} , H^{\mu_{2} \cdots \mu_{5}}]  
  + \frac{1}{27} H_{\mu_{1} \cdots \mu_{4}} [ {H^{\rho}}_{\mu_{5} \cdots
  \mu_{7}} , H_{\rho \mu_{8} \mu_{9} \mu_{10}}] \epsilon^{\mu_{1}
     \cdots \mu_{10} \sharp} ), \nonumber \\ \label{AZ44others} \\ 
  S_{f} &=& \frac{1}{g^{2}} Tr_{N \times N} ( - 3i  {\bar
     \psi_{L}} \Gamma^{\mu} \partial_{\mu} \psi_{L} )
     \label{AZ44fermpartial} \\ 
   &+& \frac{i}{g^{2}} Tr_{N \times N} (-3i ( - {\bar \psi_{L}} [ W ,
     \psi_{R}] +  {\bar  \psi_{R}} [ W , \psi_{L}] )  
   - 3i ( {\bar \psi_{L}} \Gamma^{\mu} [ a^{(+)}_{\mu} , \psi_{L}] 
     + {\bar \psi_{R}} \Gamma^{\mu} [ A^{(-)}_{\mu} , \psi_{R}] ) \nonumber \\ 
   &-&\frac{3i}{2!} ( {\bar \psi_{L}} \Gamma^{\mu_{1} \mu_{2}}
     [ C_{\mu_{1} \mu_{2}} , \psi_{R}] + {\bar \psi_{R}}
     \Gamma^{\mu_{1} \mu_{2}} [ C_{\mu_{1} \mu_{2}} , \psi_{L}] )
     \nonumber \\
   &-& \frac{3i}{4!} ( - {\bar \psi_{L}} \Gamma^{\mu_{1} \mu_{2}
     \mu_{3} \mu_{4}} [ H_{\mu_{1} \mu_{2} \mu_{3} \mu_{4}} ,
     \psi_{R}] +  {\bar \psi_{R}} 
  \Gamma^{\mu_{1} \mu_{2} \mu_{3} \mu_{4}} [ H_{\mu_{1} \mu_{2}
     \mu_{3} \mu_{4}},  \psi_{L}] ) \nonumber \\ 
   &-& \frac{3i}{5!} ( 2 {\bar \psi_{L}} \Gamma^{\mu_{1} \mu_{2} \mu_{3}
     \mu_{4} \mu_{5}}  
 [ {I^{(+)}}_{\mu_{1} \mu_{2} \mu_{3} \mu_{4} \mu_{5}} , \psi_{L}] 
   +  2 {\bar \psi_{R}} \Gamma^{\mu_{1} \mu_{2} \mu_{3} \mu_{4} \mu_{5}} 
 [ {I^{(-)}}_{\mu_{1} \mu_{2} \mu_{3} \mu_{4} \mu_{5}} ,\psi_{R}] ) ).
     \label{AZ44fermother} 
  \end{eqnarray}
   The cubic action, per se, does not include a quartic term of vector
  fields in the action. However, we can interpret that the bosonic
  term is induced by the fermionic term of the the IIB matrix model. The idea
  that the theory consisting of  fermionic fields and
  a Dirac operator induces the Einstein gravity and the Yang-Mills
  theory has long been suggested. The proposal of Connes and
  Chamseddine is one of these 
  suggestions of the induced gravity \cite{9606001}. Based on these
  suggestions, we find it a natural idea that 
  the bosonic term of the IIB matrix model is induced from its fermionic
  term. And we hypothesize that the whole IIB model should be induced
  only by the fermionic field. 

  Our goal is thus to find the fermionic terms in this cubic matrix model
  to be identified with that of the IIB matrix model. 
  We have seen in the previous section the correspondence between the
  vector fields and the fermionic fields according to the
  identification of ${\cal N}=2$ supersymmetry with that of the IIB
  matrix model. And the fermionic terms to be identified with that if
  the IIB matrix model is 
    \begin{eqnarray}
     {\bar \psi_{R}} \Gamma^{\mu} A^{(+)}_{\mu} \psi_{R}
     \stackrel{or}{\Leftrightarrow} 
     {\bar \psi_{L}} \Gamma^{\mu} A^{(-)}_{\mu} \psi_{L}. \label{AZ44goal} 
    \end{eqnarray}
  If either of these terms exists in the action, this can be
  identified with the fermionic term of the IIB matrix model.  However, the
  kinetic term of the fermionic fields 
  (\ref{AZ44fermpartial}), per se, does not include such terms as
  (\ref{AZ44goal}), and does not serve to induce the IIB matrix
  model due to the discrepancy of the correspondence of the vector
  fields and the chirality of the fermions.
  In order to remedy this situation, we consider inducing the terms
  (\ref{AZ44goal}) by the multi-loop effect. The action (\ref{AZ44partial})
  $\sim$ (\ref{AZ44fermother}) just tells us that there is no terms to 
  be identified with the IIB matrix model {\it at a tree level}. Since the idea
  of 'inducing the IIB matrix model' is to construct the bosonic term
  by the one-loop 
  effect of the fermionic Lagrangian, the idea of 'induced theory' is
  a notion based on the multi-loop effect of the perturbative
  theory. In this sense, there is no problem if the fermionic term is
  induced by the multi-loop effect.

  We can read off the Feynman rule from the above action. The typical
  propagators and the vertices are given below:
   \begin{figure}[htbp]
   \begin{center}
    \scalebox{.4}{\includegraphics{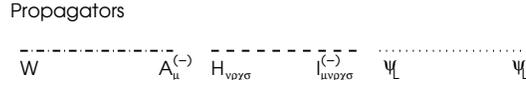} }
   \end{center}
    \caption{Propagators of this cubic supermatrix model.}
   \label{feynprop}
  \end{figure}
   \begin{figure}[htbp]
   \begin{center}
    \scalebox{.4}{\includegraphics{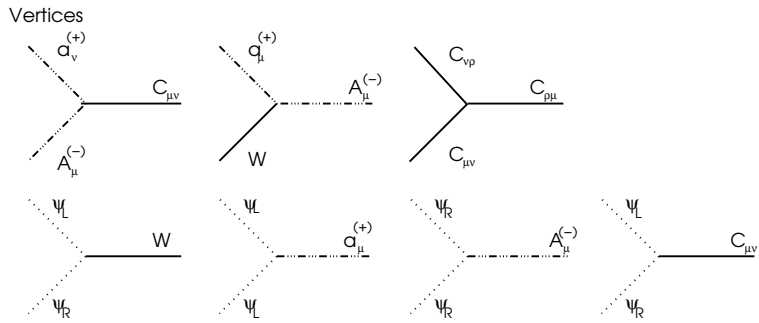} }
   \end{center}
    \caption{Vertices of this cubic supermatrix model.}
   \label{feynvertex}
  \end{figure}
   Our goal is to build a fermion vertex (\ref{AZ44goal}) by means of
  the multi-loop effect. In order to build such a term, the above
  Feynman diagrams do not suffice. Especially we are lacking the
  propagators of the fields. It is thus necessary to build an induced
  propagators by means of the multi-loop effect of the existing
  Feynman rule. These induced propagators are constructed by the
  following loop effect.
   \begin{figure}[htbp]
   \begin{center}
    \scalebox{.4}{\includegraphics{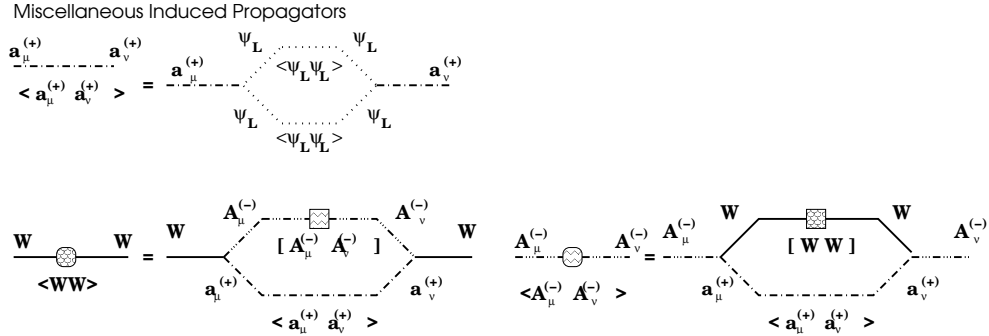} }
   \end{center}
    \caption{The induced propagators of the bosonic fields}
   \label{propwwaa}
  \end{figure}
 \begin{itemize}
  \item{$\langle a^{(+)}_{\mu} a^{(+)}_{\nu} \rangle$ : It is easy to
      construct this induced propagator. All we have to do is to
      connect the fermions using the existing fermion propagators
      $\langle \psi_{L} \psi_{L} \rangle$. }
  \item{$\langle WW \rangle$ and $\langle A^{(-)}_{\mu} A^{(-)}_{\nu}
      \rangle$ : The construction of these propagators is a difficult
      problem. Unfortunately, it is impossible to construct these
      propagators perturbatively, and we delegate the disproof of
      their existence in Appendix B.4. of \cite{0103003}.
      However, this is not the end of the 
      story. Even if we fail to induce these propagators
      perturbatively, we have a choice to induce these propagators by
      means of the nonperturbative effect. These propagators may be
      induced by the following recursive structure. 
  \begin{figure}[htbp]
   \begin{center}
    \scalebox{.4}{\includegraphics{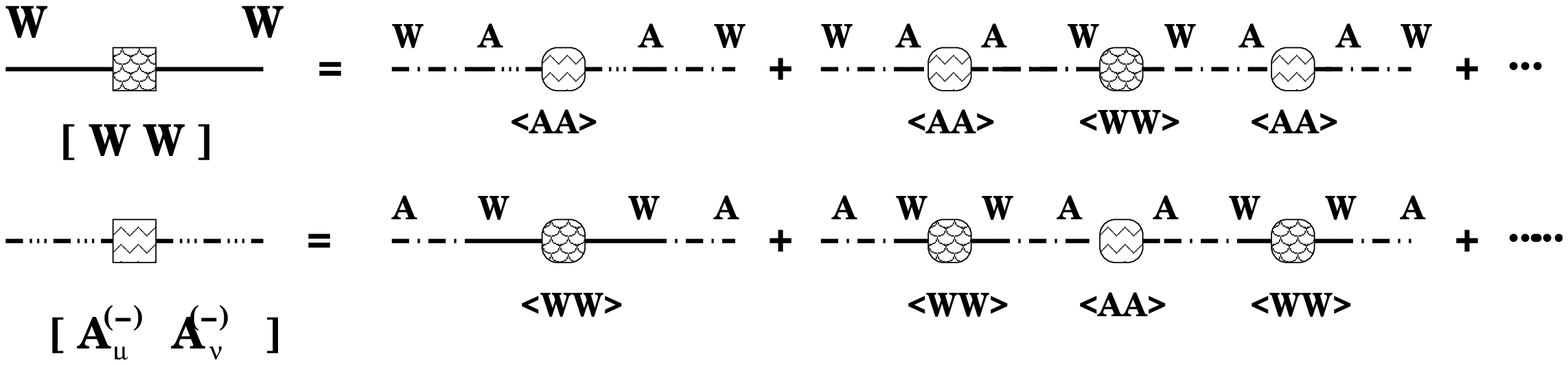} }
   \end{center}
    \caption{The induced propagators of the bosonic fields}
   \label{inducedwwaa}
  \end{figure}
      This structure is reminiscent of the self-consistency condition
      of Nambu-Jona-Lasino model. And when we consider the
      nonperturbative\footnote{Here, we mean the word 'nonperturbative' 
      by the effect not stemming from the multi-loop effect of
      Feynman diagram.} effect, there is no particular conservation
      law which prohibits the existence of the propagators $\langle WW
      \rangle$ or $\langle A^{(-)}_{\mu} A^{(-)}_{\nu} \rangle$.  
      Throughout our discussion, we assume the existence of these
      propagators. } 
 \end{itemize}
  Now we are ready to construct the induced propagators of this cubic
  model. In constructing the vertices of the fermionic terms
  (\ref{AZ44goal}).  The answer is now easy, and the IIB-like vertex
  is constructed by the following procedure.
   \begin{figure}[htbp]
   \begin{center}
    \scalebox{.4}{\includegraphics{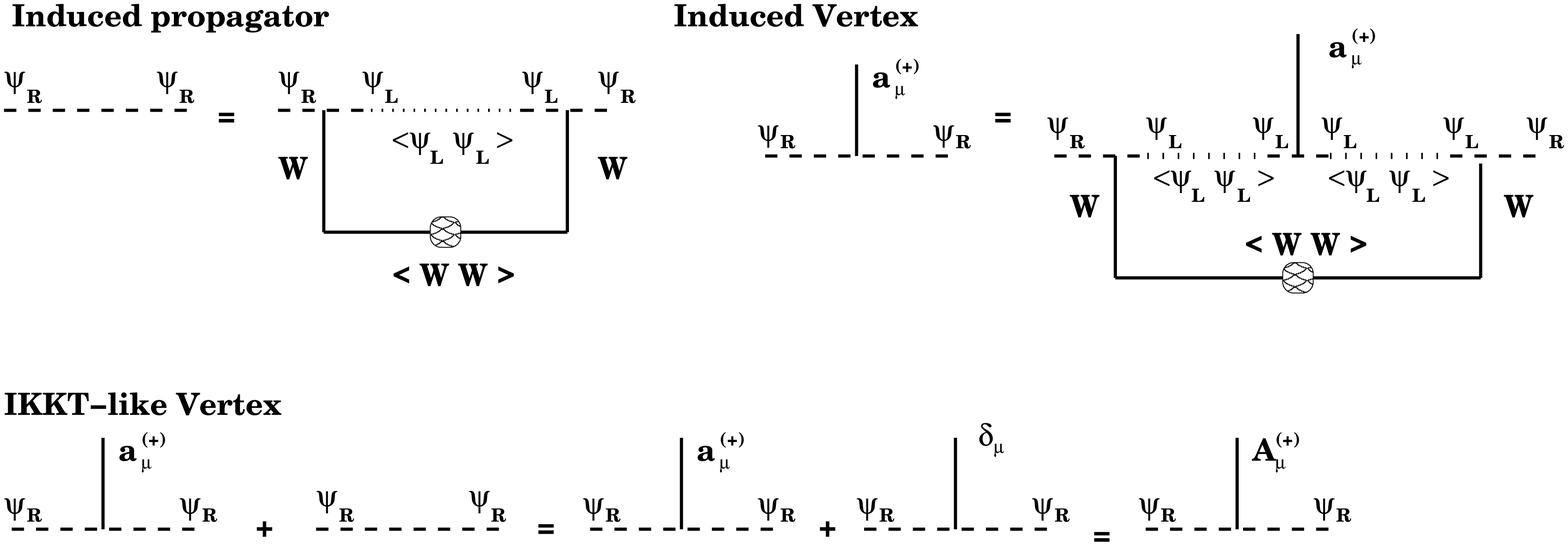} }
   \end{center}
    \caption{The induced IKKT-like vertex ${\bar \psi} \Gamma^{\mu}
      A^{(+)}_{\mu} \psi$}
   \label{truefinalanswer}
  \end{figure}
  \begin{itemize}
   \item{To construct the induced vertex corresponding to
       (\ref{AZ44goal}), we must first construct two objects. One is
       the propagator $\langle \psi_{R} \psi_{R} \rangle$. This is
       easily constructed once we admit the existence of the
       propagator $\langle WW \rangle$. }
   \item{ Another object is the vertex ${\bar \psi_{R}} \Gamma^{\mu}
       a^{(+)}_{\mu} \psi_{R}$. It is also easy to construct this vertex 
       utilizing the induced propagator $\langle WW \rangle$.}
   \item{ These two objects serve to induce our desired term ${\bar
         \psi_{R}} \Gamma^{\mu} A^{(+)}_{\mu} \psi_{R}$. Note that
         this induced 
       propagator indicates that the kinetic term ${\bar \psi_{R}}
       \Gamma^{\mu} \partial_{\mu} \psi_{R}$. This is diagrammatically
       regarded as the vertex of $\psi_{R}$, $\psi_{R}$ and
       $\partial_{\mu} \sim {\hat p}_{\mu}$, where ${\hat p}_{\mu}$, around
         which we have expanded the theory. Therefore, the sum of
         these two objects is  regarded as  
       \begin{eqnarray}
       \langle \psi_{R} \psi_{R} \rangle + {\bar \psi_{R}} \Gamma^{\mu}
       a^{(+)}_{\mu} \psi_{R} = -i {\bar \psi_{R}} \partial_{\mu} \psi_{R} 
       + {\bar \psi_{R}} \Gamma^{\mu} a^{(+)}_{\mu} \psi_{R} =
       {\bar \psi_{R}} \Gamma^{\mu} A^{(+)}_{\mu} \psi_{R}.
       \end{eqnarray} }
  \end{itemize}
  We thus build the vertex for the bosonic field
  $A^{(+)}_{\mu}$ and the corresponding fermionic field
  $\psi_{R}$. This fermionic operator serves to derive the bosonic
  quartic terms of the IIB matrix model. In this way, we expect that
  the IIB matrix model is induced from the $osp(1|32,R)$ supermatrix
  model.

 \subsection{$gl(1|32,R) \otimes gl(N,R)$ gauged cubic matrix model}
  \label{sec0102168l}
  We next investigate the gauged version of the supermatrix model.
  Originally, L. Smolin\cite{0002009} touched on the suggestion to
  enhance the symmetry of the matrix model, by regarding the tensor
  product as not for the Lie group but for the Lie algebra.
  This attempt is fascinating, in that it broadens the symmetry
  greatly. Smolin's proposal turns out to be
  essential to build a matrix model with local Lorentz invariance. 
  To elaborate on this point, let us think about the extension of the
  IIB matrix model to the local Lorentz invariant model.
  In the IIB matrix model, we regard the eigenvalues of the matrices
  as the spacetime coordinate. Therefore, in order to render the
  Lorentz symmetry "local", it is imperative that the parameter of the 
  Lorentz transformation should be dependent on the $u(N)$ element in 
  a nontrivial way\cite{0102168,0204078}. Then, we introduce the
  so-called "gauged matrix model", in order to build such a matrix
  model.

  In the IIB matrix model, the $SO(9,1)$ Lorentz symmetry and the
  $SU(N)$ symmetry are totally decoupled. Now, let $\xi$ and $u$ be
  the generator of the $SO(9,1)$ and the $U(N)$ symmetry,
  respectively (i.e. $\xi \in so(9,1)$ and $u \in u(N)$).
  The symmetry is a tensor product of the group; namely they are
  decoupled in terms of the Lie algebra as
  \begin{eqnarray}
   \exp( \xi \otimes {\bf 1} + {\bf 1} \otimes u) = e^{\xi} \otimes
   e^{u}. \label{symIIB}
  \end{eqnarray}
  This holds of the nongauged version of the supermatrix model.
  
  On the other hand, in the gauged model, we enhance the symmetry
  drastically by mixing these two symmetry. Namely, the element of the 
  transformation group is not $\exp( \xi \otimes {\bf 1} + {\bf 1}
  \otimes u)$ but 
   \begin{eqnarray}
   \exp( \xi \otimes u). \label{symgauged}
   \end{eqnarray}
  In this way, we build a Lorentz symmetry dependent on the $u(N)$
  symmetry. We define a generator with the tensor product of the 
  two Lie algebra. However, we need a caution in defining a tensor
  product of the Lie algebra. Suppose ${\cal A}, {\cal B}$ are two
  different Lie algebra whose bases are $\{ a_{i} \}$ and $\{ b_{i}
  \}$, respectively. Generally, the space ${\cal A} \otimes {\cal B}$, 
  which is spanned by the basis $a_{i} \otimes b_{j}$, does not close
  with respect to the commutator
   \begin{eqnarray}
    [a_{1} \otimes b_{1}, a_{2} \otimes b_{2}] 
 = \frac{1}{2} \left( [a_{1},a_{2}] \otimes \{ b_{1}, b_{2} \}
   + \{a_{1}, a_{2} \} \otimes [b_{1}, b_{2}] \right),
  \textrm{ for } a_{1},a_{2} \in A, \hspace{2mm}
    b_{1}, b_{2} \in B. \label{commutatorgauged}
   \end{eqnarray}
  Rather, we define the tensor product 
  by ${\cal A} \check{\otimes} {\cal B}$, which is the smallest Lie
  algebra  that includes ${\cal A} \otimes {\cal B}$ as a
  subset\footnote{To illustrate this idea, we introduce the following
  simple case. The Lie algebra $su(6)$ is known as the tensor product
  of the Lie algebras $su(3)$ and $su(2)$: 
   \begin{eqnarray}
     su(6) = su(3) \check{\otimes} su(2). \nonumber
   \end{eqnarray}
  This fact is discerned as follows. Let $\lambda^{a}$ and
  $\sigma^{i}$ be the basis of the Lie algebra $su(3)$ and $su(2)$,
  respectively. $su(3)$ and $su(2)$ are 8 and 3 dimensional Lie
  algebras respectively, and the indices run $a = 1, \cdots, 8$ and $i 
  = 1, 2, 3$. The tensor product $su(3) \check{\otimes} su(2)$
  consists of the following elements.
   \begin{itemize}
    \item{ $\lambda^{a} \otimes \sigma^{i}$: These are the elements of 
        the tensor product as a set of matrices: $su(3) \otimes
        su(2)$. This set, per se, does not constitute a closed Lie
        algebra.
         } 
   \item{ $\lambda^{a} \otimes {\bf 1}$ and ${\bf 1} \otimes
       {\sigma^{i}}$: These are the generators of the
       group $SU(3) \times SU(2)$.}
   \end{itemize}   
  These elements are known to constitute the algebra of $su(6)$. 
  This example shows in a pedagogical way how the notion of 'gauged
  theory' enhances the gauge symmetry. While the Lie algebra of the
  gauge group $SU(3) \times SU(2)$ is a $8+3=11$ dimensional algebra,
  the tensor product $su(3) \check{\otimes} su(2)$ is a $8+3+ 8 
  \times 3 =35$
  dimensional Lie algebra. This is the structure of the enhancement of 
  the gauge symmetry. Note that the similar enhancement of the gauge
  symmetry is seen  in Smolin's proposal\cite{0002009,0006137}.}.

  In the large-$N$ reduced models, there are several ways to interpret
  the spacetime coordinate. In the twisted reduced
  models\cite{TEK,TEK2}, the matrices $A_{\mu}$ represent the
  covariant derivative in the spacetime. On the other hand, in the IIB 
  matrix model, the eigenvalues of the matrices $A_{\mu}$ are regarded 
  as the spacetime coordinate. Both relations are linked by the
  expansion around the flat noncommutative background
  \begin{eqnarray}
   A_{\mu} = {\hat p}_{\mu} + a_{\mu}, \textrm{ where }
  [{\hat p}_{\mu}, {\hat p}_{\nu}] = i c_{\mu \nu}.
  \end{eqnarray}
  The IIB matrix model reduces to the non-commutative Yang-Mills
  theory after this reduction. The fermionic term of the IIB matrix
  model $\frac{-1}{2g^{2}} Tr {\bar \psi} \Gamma^{\mu} [A_{\mu},
  \psi]$ reduced to the action of the fermion in the flat space
  \begin{eqnarray}
   \int d^{d} x {\bar \psi}(x) i \Gamma^{\mu} (\partial_{\mu} \psi(x)
   + [a_{\mu}(x), \psi(x)] ). \label{YMclassical}
  \end{eqnarray}
  This reminds us of the correspondence between the differential
  operators and the matrices. 
  
   With this idea in mind, we attempt to formulate a matrix model with 
  the local Lorentz invariance, in which the matrices $A_{\mu}$ look
  like differential operators.
  The action should reduce in the classical low-energy
  limit to the fermionic action on the curved spacetime  
  \begin{eqnarray}
   S_{F} = \int d^{d} x e(x) {\bar \psi}(x) i \Gamma^{\mu}
       {e_{\mu}}^{i}(x) \left(  
       \partial_{i} \psi(x) + [A_{i}(x), \psi(x)] +
       \frac{1}{4} \Gamma^{\nu \rho} 
       \omega_{i \nu \rho} (x) \psi(x) \right). \label{AZ11ll}
  \end{eqnarray}
  Here, the indices $\mu, \nu, \rho, \cdots$ and $i, j, k, 
  \cdots$ both run over $0, 1, \cdots, d-1$.  The former and the
  latter denote the indices of the Minkowskian and the curved
  spacetime, respectively. In considering the correspondence with the
  matrix model, we need to absorb $e(x)$ into 
  the definition of the fermionic field, since it is not $\int
  d^{d} x e(x)$ but $\int d^{d} x$ that corresponds to the trace of
  the large $N$ matrices.  And we regard the "spinor root density"  
  \begin{eqnarray}
   \Psi(x) = e^{\frac{1}{2}} (x) \psi (x) \label{AZ1rootdensity}
  \end{eqnarray}
 as the fundamental quantity. Then, the action is rewritten as
  \begin{eqnarray}
   S_{F} &=& \int d^{d} x {\bar \Psi} (x) e^{\frac{1}{2}} (x) 
    i \Gamma^{\mu} {e_{\mu}}^{i}(x) \left\{ 
      \partial_{i} (e^{-\frac{1}{2}} (x) \Psi(x) )
   + [A_{i} (x), e^{- \frac{1}{2}} (x) \Psi(x)] \right. \nonumber \\
 & & \hspace{15mm} \left. 
   + \frac{1}{4} \Gamma^{\nu \rho}
       \omega_{i \nu \rho} (x) e^{-\frac{1}{2}} (x) \Psi(x) \right\}
    \nonumber \\
    &=& \int d^{d} x \left\{ 
     {\bar \Psi} (x) i \Gamma^{\mu} \left[ {e_{\mu}}^{i} (x)
    \partial_{i} 
      + \frac{1}{2} {e_{\rho}}^{i} (x) \omega_{i \rho \mu} (x)
      + {e_{\mu}}^{i} (x) e^{\frac{1}{2}} (x) (\partial_{i}
    e^{-\frac{1}{2}} (x) ) \right] \Psi (x)  \right. \nonumber \\
    & & \hspace{15mm} \left. + i {\bar \Psi} (x) \Gamma^{\mu}
    {e_{\mu}}^{i} (x) [A_{i}(x), \Psi (x)] 
     + \frac{i}{4} {\bar \Psi} (x) \Gamma^{\mu_{1} \mu_{2} \mu_{3}}
    {e_{[\mu_{1}}}^{i} (x) \omega_{i \mu_{2} \mu_{3}]} (x) \Psi (x)
    \right\} \nonumber \\
   &=& \int d^{d} x \left\{ {\bar \Psi} (x) i \Gamma^{\mu}
    {e_{\mu}}^{i} (x)  
       (\partial_{i} \Psi(x) + [A_{i} (x), {\bar \Psi} (x) ] )
     + \frac{i}{4} {\bar \Psi} (x) \Gamma^{\mu_{1} \mu_{2} \mu_{3}}
       {e_{[\mu_{1}}}^{i} (x) \omega_{i \mu_{2} \mu_{3}]} (x) \Psi
    (x) \right\}, \nonumber \\
     \label{AZ1bonafidefermion}
   \end{eqnarray}
  where we have utilized in the last equality the fact that the
  fermionic field $\Psi (x)$  is Majorana, which leads to the
  cancellation ${\bar \Psi}(x) \Gamma^{\mu} \Psi(x) = 0$.
  The corresponding matrix model is formulated as
  \begin{eqnarray}
   S'_{F} = - \frac{1}{2} Tr {\bar \psi} \Gamma^{\mu} [A_{\mu}, \psi]
     -  \frac{i}{2} Tr {\bar \psi} \Gamma^{\mu_{1} \mu_{2} \mu_{3}}
      \{ A_{\mu_{1} \mu_{2} \mu_{3}}, \psi \}.
   \label{AZ11naivell}
  \end{eqnarray}
  In promoting the action (\ref{AZ1bonafidefermion}) to the matrix
  model, we have identified the covariant derivative with
  the commutator with the rank-1 matrix $A_{\mu}$. 
  The rank-3 term is a naive product of the spin connection and
  the fermion, and it is natural to promote the product to the
  anticommutator of the large $N$ hermitian matrices\footnote{For the
  readers' convenience, we summarize the commutation 
  relations of the hermitian and anti-hermitian
  operators. Let ${\bf H}$ and ${\bf A}$ be the set of the hermitian and
  anti-hermitian operators, respectively. When $h, h_{1}, h_{2} \in
  {\bf H}$ and $a, a_{1}, a_{2} \in {\bf A}$, their commutation
  relations are as follows: 
  \begin{eqnarray}
  [h_{1}, h_{2}] \in {\bf A}, \hspace{2mm}
  [h, a] \in {\bf H}, \hspace{2mm}
  [a_{1}, a_{2}] \in {\bf A},  \hspace{2mm}
  \{ h_{1}, h_{2} \} \in {\bf H},  \hspace{2mm}
  \{ h, a \} \in {\bf A}, \hspace{2mm}
  \{ a_{1}, a_{2} \} \in {\bf H}. \nonumber
 \end{eqnarray}
  We prove only the first relation, because the others are easily
  derived in the same way:
  \begin{eqnarray}
   [h_{1}, h_{2}]^{\dagger} = (h_{1} h_{2} - h_{2} h_{1})^{\dagger}
  = h_{2}^{\dagger} h_{1}^{\dagger} - h_{1}^{\dagger} h_{2}^{\dagger}
  = h_{2} h_{1} - h_{1} h_{2} = - [h_{1}, h_{2}]. \nonumber
  \end{eqnarray}}.
  In the rank-3 term, the pure imaginary number $i$ is necessary so
  that the action should be hermitian.
  The action (\ref{AZ11naivell}) is an addition of the
  rank-3 term to the fermionic term of the action of IIB matrix
  model. Especially when $\psi$ is a Majorana fermion,   
  (\ref{AZ11naivell}) is equivalent to the following action:   
   \begin{eqnarray}
     S''_{F} =  - Tr {\bar \psi} (\Gamma^{\mu} A_{\mu} + i
   \Gamma^{\mu_{1} \mu_{2} 
   \mu_{3}} A_{\mu_{1} \mu_{2} \mu_{3}} ) \psi.  \label{AZ11naivell2}
   \end{eqnarray}
  The equivalence between (\ref{AZ11naivell}) and (\ref{AZ11naivell2}) 
  is verified as
   \begin{eqnarray}
     S'_{F} &=& - \frac{1}{2} Tr {\bar \psi} \Gamma^{\mu} [A_{\mu}, \psi]
     -  \frac{i}{2} Tr {\bar \psi} \Gamma^{\mu_{1} \mu_{2} \mu_{3}}
      \{ A_{\mu_{1} \mu_{2} \mu_{3}}, \psi \} \nonumber \\
     &=& - \frac{1}{2} {\bar \psi}^{A} \Gamma^{\mu} A_{\mu}^{B} \psi^{C} 
       Tr(t^{A} [t^{B},t^{C}])
       - \frac{i}{2} {\bar \psi}^{A} \Gamma^{\mu_{1} \mu_{2} \mu_{3}}
         A^{B}_{\mu_{1} \mu_{2} \mu_{3}} \psi^{C}
       Tr(t^{A} \{ t^{B}, t^{C} \}) \nonumber \\
     &=& - Tr(t^{A} t^{B} t^{C})
       \left( \frac{1}{2} ({\bar \psi}^{A} \Gamma^{\mu} A_{\mu}^{B}
     \psi^{C} - {\bar \psi}^{C} \Gamma^{\mu} A_{\mu}^{B} \psi^{A} )
     \right. \nonumber \\
     & & \left. \hspace{20mm} 
      + \frac{i}{2} ({\bar \psi}^{A} \Gamma^{\mu_{1} \mu_{2} \mu_{3}}
         A^{B}_{\mu_{1} \mu_{2} \mu_{3}} \psi^{C} + {\bar \psi}^{C}
     \Gamma^{\mu_{1} \mu_{2} \mu_{3}}  A^{B}_{\mu_{1} \mu_{2} \mu_{3}}
     \psi^{A} ) \right) \nonumber \\
   &=&  - Tr {\bar \psi} (\Gamma^{\mu} A_{\mu} + i \Gamma^{\mu_{1} \mu_{2}
   \mu_{3}} A_{\mu_{1} \mu_{2} \mu_{3}} ) \psi. \nonumber  
   \end{eqnarray}

  Let us next consider the local Lorentz transformation of this
  matrix model. Originally the local Lorentz transformation of
  the fermionic field $\delta \psi (x) = \frac{1}{4} \Gamma^{\mu_{1}
  \mu_{2}} \varepsilon_{\mu_{1} \mu_{2}} (x) \psi(x)$ should be
  promoted to the anticommutator of the hermitian matrices as 
  \begin{eqnarray}
  \delta \psi \stackrel{?}{=} \frac{1}{4} \Gamma^{\mu_{1} \mu_{2} } 
  \{ \varepsilon_{\mu_{1} \mu_{2}},  \psi \}, \label{ll?}
  \end{eqnarray} 
  in order to retain the hermiticity.
  However, it is a very onerous problem to  find an action invariant
  under this transformation. The conundrum stems from the  {\it
  noncommutativity of the matrices}. The local Lorentz 
  transformation of the fermion (\ref{ll?}) with respect to the action
  (\ref{AZ11naivell2}) entails the terms
   \begin{eqnarray}
   \delta S''_{F} = - \frac{1}{4} Tr {\bar \psi} \Gamma^{\mu} A_{\mu}
   \Gamma^{\nu_{1} \nu_{2}} \psi \varepsilon_{\nu_{1} \nu_{2}} + \cdots. 
   \end{eqnarray}
  It is extremely difficult to absorb this transformation via the
  local Lorentz transformation of the bosonic fields $A_{\mu}$.
 
  This obstacle forces us to abstain the hermiticity of the
  matrices. We instead promote this transformation 
  to the matrix version as
  \begin{eqnarray}
   \delta \psi = \frac{1}{4} \Gamma^{\mu_{1} \mu_{2}} \varepsilon_{\mu_{1}
   \mu_{2}}  \psi, \label{AZ1alt}
  \end{eqnarray}
 and we take the action (\ref{AZ11naivell2}).
 Since the fermion is no longer hermitian,  the actions
 (\ref{AZ11naivell}) and (\ref{AZ11naivell2}) are no longer
 equivalent. There are two prices we
 must pay for this alteration. One is that the product $A_{\mu} \psi$ does
 not directly correspond to the covariant derivative. The other is
 that the matrix model (\ref{AZ11naivell2}) is no longer invariant
 under the translation $\delta A_{\mu} = c_{\mu} {\bf 1}$, which is a
 spacetime translation in the original interpretation of the IIB
 matrix model.

 The local Lorentz transformation of the action (\ref{AZ11naivell2}) is  
  \begin{eqnarray}
   \delta S''_{F} = \frac{1}{4} Tr {\bar \psi} \left[ \Gamma^{\mu} A_{\mu} +
   i \Gamma^{\mu_{1} \mu_{2} \mu_{3}} A_{\mu_{1} \mu_{2} \mu_{3}},
   \Gamma^{\nu_{1} \nu_{2}} \varepsilon_{\nu_{1} \nu_{2}} \right] \psi.
  \end{eqnarray}
  Since we are now considering the local Lorentz
  transformation and their space-time dependent parameters are promoted 
  to $u(N)$ matrices, the infinitesimal parameters  $\varepsilon_{\mu \nu}$
  are $u(N)$ matrices. Then, the commutator 
  \begin{eqnarray}
   [ i \Gamma^{\mu_{1} \mu_{2} \mu_{3}} A_{\mu_{1} \mu_{2} \mu_{3}},
  \Gamma^{\nu_{1} \nu_{2}} \varepsilon_{\nu_{1} \mu_{2}} ] 
   = \frac{i}{2} \underbrace{[ \Gamma^{\mu_{1} \mu_{2} \mu_{3}},
  \Gamma^{\nu_{1} \nu_{2}}]}_{\textrm{rank-3}}
       \{  A_{\mu_{1} \mu_{2} \mu_{3}},  \varepsilon_{\nu_{1} \nu_{2}} \}
   +  \frac{i}{2} \underbrace{\{  \Gamma^{\mu_{1} \mu_{2} \mu_{3}},
  \Gamma^{\nu_{1} \nu_{2}} \} }_{\textrm{rank-1, 5}}
      [  A_{\mu_{1} \mu_{2} \mu_{3}},  \varepsilon_{\nu_{1} \nu_{2}} ]
  \nonumber \\ \label{higheroddgauged}
  \end{eqnarray}
  must include the rank-5 gamma matrices, and the action
  (\ref{AZ11naivell2}) is not invariant 
  under the local Lorentz transformation.   

  The similar thing holds true of the generator of the local Lorentz
  transformation (\ref{AZ1alt}). In order for the algebra to close,
  only  the rank-2 terms are not sufficient, because the commutator  
  \begin{eqnarray}
   [\Gamma^{\mu_{1} \mu_{2}} \varepsilon_{\mu_{1} \mu_{2}}, \Gamma^{\nu_{1}
   \nu_{2}} \varepsilon'_{\nu_{1} \nu_{2}} ]
 =  \frac{1}{2} \underbrace{[ \Gamma^{\mu_{1} \mu_{2}}, \Gamma^{\nu_{1}
   \nu_{2}} ]}_{\textrm{rank-2}}
     \{  \varepsilon_{\mu_{1} \mu_{2}},  \varepsilon'_{\nu_{1} \nu_{2}} \}
  + \frac{1}{2} \underbrace{\{ \Gamma^{\mu_{1} \mu_{2}}, \Gamma^{\nu_{1}
   \nu_{2}} \} }_{\textrm{rank-0, 4}}
     [  \varepsilon_{\mu_{1} \mu_{2}},  \varepsilon'_{\nu_{1} \nu_{2}} ]
  \label{higherevengauged} 
 \end{eqnarray}
  includes the rank-4 gamma matrices.

  Let us recapitulate what we have obtained through this
  observation. 
   \begin{itemize}
   \item{Since the information of the spacetime is encoded in the
        bosonic matrices $A_{\mu}$, the parameter of the local Lorentz 
        transformation must depend on the $u(N)$ element in a
        nontrivial manner. This leads us to consider the "gauged"
        action.} 
    \item{The correspondence of the matrices $A_{\mu}$ and the
        covariant derivative motivates us to construct a matrix model
        in which the matrices $A_{\mu}$ looks like a differential
        operators.} 
    \item{It is difficult to retain the hermiticity of the fermionic
        field, when we build a local Lorentz invariant action. This
        forces us to abstain the hermiticity of the fields,
        despite the following drawbacks.
       \begin{itemize}
        \item{The term ${\bar \psi} \Gamma^{\mu} A_{\mu} \psi$ no
            longer corresponds to the covariant derivative
            ${\bar \psi}(x) (\partial_{\mu} \psi(x)+ [a_{\mu}(x),
            \psi(x)])$.}
        \item{The matrix model (\ref{AZ11naivell2}) is no longer invariant
 under the translation $\delta A_{\mu} = c_{\mu} {\bf 1}$, which is a
 spacetime translation in the original interpretation of the IIB
 matrix model} 
       \end{itemize}}
     \item{The "gauged" symmetry necessitates the higher-rank fields and
         the local Lorentz transformation parameters.}
   \end{itemize}

  This observation leads us to regard the "gauged" version of the
  supermatrix model as a mesmerizing candidate for the local Lorentz
  invariant extension of the IIB matrix model. To this end, we
  consider the $gl(1|32,R)$ super Lie algebra, which is the analytic
  continuation of the $u(1|16,16)$, the complexification of the
  $osp(1|32,R)$. In addition, we mix the $gl(1|32,R)$ symmetry and the 
  $gl(N)$ gauge symmetry. 

  This observation is also elaborated in another angle without the
  supermatrix models in Section \ref{manifestgr}.

 \subsubsection{Definition of the $u(1|16,16)$ and $gl(1|32,R)$ super Lie
  algebra} 
  The $gl(1|32,R)$ super Lie algebra is a cousin of the $u(1|16,16)$
  super Lie algebra. More accurately, $gl(1|32,R)$ is obtained by the
  analytic continuation of the imaginary part of the $u(1|16,16)$.
  Therefore, we start with introducing the $u(1|16,16)$ super Lie
  algebra.

  The very definition of this super Lie algebra is that 
 \begin{eqnarray} 
  \textrm{ If } M \in u(1|16,16),  \textrm{ then } M^{\dagger} G + GM = 0
  \textrm{ for }  G =  \left( \begin{array}{cc}  \Gamma^{0}  & 0  \\  0
  & i   \end{array} \right).
 \end{eqnarray} 
  That this is a complexification of $osp(1|32,R)$ can be seen from
  the following aspect. Unlike the $osp(1|32,R)$ super Lie algebra, we
  do not restrict $M$ to be a real supermatrix, where the reality of
  the supermatrix is defined as $M = M^{\ast} = ({^{T}
  M})^{\dagger}$. Therefore, we must replace the {\it transpose} by
  the {\it hermitian conjugate}. Note that in the real version the
  complex conjugate is equivalent to the transpose according to the
  property in Appendix. \ref{AZMA091441}. 

  We can confirm that the super Lie algebra $u(1|16,16)$ actually
   closes in totally the same fashion as in $osp(1|32,R)$. The
   legitimacy of the metric $G$ also stems from the same logic as in
   $osp(1|32,R)$.  
   We specify the element of $u(1|16,16)$ group according to the
 above definition.  The result is that 
  \begin{eqnarray}
   \textrm{ If } M \in u(1|16,16) , \textrm{ then } M = 
   \left( \begin{array}{cc} m  & \psi \\   i {\bar \psi} & v \end{array}
   \right). \label{AZ51su11616}
  \end{eqnarray}
 \begin{itemize}
  \item{ $v$ is restricted to be a pure imaginary number.} 
  \item{ $u_{A_{1}}$ ,$u_{A_{1} A_{2}}$ and
      $u_{A_{1} \cdots A_{5}}$ are real numbers, while $u$, $u_{A_{1}
      A_{2} A_{3}}$ and $u_{A_{1} \cdots A_{4}}$ are
      pure imaginary numbers.} 
 \end{itemize}
 This result is derived from the very definition of the super Lie
 algebra $u(1|16,16)$. The complex conjugate of supermatrices is defined
 in Appendix. \ref{AZMA091441}:
  \begin{eqnarray} M^{\dagger} G + GM &=&  \left( \begin{array}{cc}
 m^{\dagger}   & \phi \\   \psi^{\dagger} & v^{\dagger} \end{array}
 \right)  \left( \begin{array}{cc} \Gamma^{0}   & 0 \\ 0 & i
 \end{array} \right) +  \left( \begin{array}{cc} \Gamma^{0}   & 0 \\ 0 & i
 \end{array} \right)  \left( \begin{array}{cc} m   & \psi \\
 \phi^{\dagger} & v \end{array} \right) \nonumber \\
  &=& \left( \begin{array}{cc}
 m^{\dagger} \Gamma^{0} + \Gamma^{0} m  & i \phi + \Gamma^{0} \psi \\
 \psi^{\dagger} \Gamma^{0} + i \phi^{\dagger}  & i( v + v^{\dagger})
 \end{array} \right) = 0. \nonumber \\ \label{AZ51su116162}
 \end{eqnarray}
  \begin{itemize}
  \item{Since $v + v^{\dagger} = 0$, (\ref{AZ51su116162}) immediately
      constrains $v$ to be a pure imaginary number.}
  \item{We first investigate the constraint of the bosonic matrix
      $m$. These are decomposed as (\ref{AZ41assumption}), like the
      $osp(1|32,R)$ super Lie algebra.
     We use the relationship of the gamma matrices
     $\Gamma^{0} {^{T} \Gamma^{\mu_{1} \cdots \mu_{k}} } \Gamma^{0} =
     (-1)^{\frac{(k+2)(k-1)}{2}} \Gamma^{\mu_{1} \cdots \mu_{k}}$.
    Therefore,  $(u_{A_{1} \cdots A_{k}})^{\ast} =
      (-1)^{\frac{(k+2)(k-1)}{2}} u_{A_{1} \cdots A_{k}}$. This is
      real for $k=1,2,5$ and pure imaginary for $k=0,3,4$.}
  \item{We investigate the relationship between two fermions $\psi$ and
  $\phi^{\dagger}$ utilizing the result  $ i \phi + \Gamma^{0} \psi=0$: 
 \begin{eqnarray}
  \phi^{\dagger} = (i \Gamma^{0} \psi)^{\dagger} = (-i)
  \psi^{\dagger} ({^{T} \Gamma^{0}})  = (-i) \psi^{\dagger}
  (-\Gamma^{0})  = i {\bar \psi}.
 \end{eqnarray}
 We can verify that this is consistent with the condition
 ${\psi}^{\dagger} \Gamma^{0} + i {\phi}^{\dagger} = 0 $.  }
  \end{itemize}
 We are thus finished with the determination of the elements of
 $u(1|16,16)$ super Lie algebra. 

  The important property of $u(1|16,16)$ super Lie algebra is
  that these can be uniquely decomposed into the direct sum of two
  different representations of $osp(1|32,R)$. We introduce two
  different representations of $osp(1|32,R)$ super Lie algebra
   \begin{eqnarray}
   &\clubsuit& {\cal H} = \{ M = 
   \left( \begin{array}{cc} m_{h} & \psi_{h} \\ i {\bar \psi_{h}} & 0
   \end{array} \right) | m_{h} = u_{A_{1}} \Gamma^{A_{1}} +
   \frac{1}{2!} u_{A_{1} A_{2}} \Gamma^{A_{1} A_{2}} +
   \frac{1}{5!} u_{A_{1} \cdots A_{5}} \Gamma^{A_{1} \cdots
   A_{5}}, \nonumber \\
  & &  \hspace{45mm} u_{A_{1}}, u_{A_{1} A_{2}}, u_{A_{1}
   \cdots A_{5}}, \psi_{h} \in {\cal R}  \}, \label{hdef} \\
  &\clubsuit&  {\cal A}' = \{ M = 
   \left( \begin{array}{cc} m_{a} & i \psi_{a} \\ {\bar \psi_{a}}
   & iv \end{array} \right) | m_{a} = u + \frac{1}{3!} u_{A_{1}
   A_{2} A_{3}} \Gamma^{A_{1} A_{2} A_{3}} + \frac{1}{4!}
   u_{A_{1} \cdots A_{4}} \Gamma^{A_{1} \cdots
   A_{4}}, \nonumber \\
  & & \hspace{45mm} u, u_{A_{1} A_{2} A_{3}}, u_{A_{1} \cdots
   A_{4}}, i \psi_{a}, iv \in (\textrm{pure imaginary}) \}. \label{aprimedef} 
   \end{eqnarray}

  And let $H$ and $A'$ be the element of ${\cal H}$ and ${\cal A}'$
  respectively. Because these are real (pure imaginary), these
  elements  respectively satisfy $H^{\dagger} = {^{T} H'}$ and
  $A'^{\dagger} = - {^{T} A'}$. Therefore, these elements satisfy the
  following property
   \begin{eqnarray}
     {^{T} H} G + G H = 0, \textrm{ for } H \in {\cal H}, \hspace{3mm} 
     {^{T} A}' G - G A' = 0, \textrm{ for } A' \in {\cal A}'.
     \label{AZ51hasu}
   \end{eqnarray}
  The set ${\cal H}$ is, by definition, the $osp(1|32,R)$ super Lie
  algebra itself.
  We investigate an important property of the subset of these two
  subalgebras. The commutation and anti-commutation relations are
  properties of grave importance in getting acquainted with their
  algebras.
    \begin{eqnarray}
    & & (1) [H_{1} , H_{2}  ] \in {\cal H}, \hspace{3mm}
        (2) [H     , A'     ] \in {\cal A}',  \hspace{3mm}
        (3) [A'_{1}, A'_{2} ] \in {\cal H}, \nonumber \\
    & & (4)\{H_{1} , H_{2} \} \in {\cal A}', \hspace{3mm}
        (5)\{H     , A'    \} \in {\cal H}, \hspace{3mm}
        (6)\{A'_{1}, A'_{2}\} \in {\cal A}'. \label{AZ51comhasu}
   \end{eqnarray} 
   where $H, H_{1}, H_{2} \in {\cal H}$ and $A', A'_{1}, A'_{2} \in {\cal
   A'}$. \\
    Here, we only give the proof of the first property, because the
   others can be easily verified likewise.
  \begin{eqnarray}
   {^{T}[H_{1}, H_{2}]} G 
   &=& {^{T}H_{2}} {^{T}H_{1}} G -{^{T}H_{1}} {^{T}H_{2}} G
   = {^{T} H_{2}}(- G H_{1}) - {^{T}H_{1}} (-G H_{2})
   = G H_{2} H_{1} - G H_{1} H_{2} \nonumber \\
   &=& -  G [H_{1}, H_{2}].
  \end{eqnarray}

  Utilizing these relations, we can discern that
  ${\cal A}'$, as well as ${\cal H}'$ is the representations of $osp(1|32,R)$
  and also  that  the algebra ${\cal A}'$ is a
         representation of $osp(1|32,R)$ super Lie algebra by the
         commutation relation  $(2) [H, A' ] \in {\cal
         A}'$ for $H \in {\cal H}$ and $A' \in {\cal A}'$. This
         commutation relation states that $A'$ remain in the super 
         Lie algebra ${\cal A}'$ after the 
         infinitesimal translation by the elements $H \in {\cal
         H}$. In this sense, we can understand that ${\cal A}'$ is
         another representation of $osp(1|32,R)$.

  The introduction of these two representations of $osp(1|32,R)$
  teaches us the relationship of $osp(1|32,R)$ and $u(1|16,16)$
  super Lie algebras. ${\cal H} ( = osp(1|32,R))$ is a real part of
  $u(1|16,16)$ Lie algebra, while ${\cal A}'$ is its imaginary
  part. It is clear that the elements of $u(1|16,16)$ can be
  uniquely decomposed into the direct sum of ${\cal H}$ and ${\cal
  A}'$.
  \begin{eqnarray}
    u(1|16,16) \equiv {\cal H} \oplus {\cal A}',
  \end{eqnarray}
 where $\oplus$ denotes the direct sum of two sets.

 Nextly, we introduce the $gl(1|32,R)$ super Lie algebra, especially
 paying attention to the relation with the $u(1|16,16)$.
  The definition of $gl(1|32,R)$ super Lie algebra is, per se, simple:
   \begin{eqnarray}
     &\clubsuit&  \textrm{If } M \in gl(1|32,R), \hspace{3mm} \textrm
     { then } M = \left( \begin{array}{cc} m & \psi \\ i {\bar \phi} & 
     v \end{array} \right). \label{gl132rdef}
   \end{eqnarray}
  \begin{itemize}
   \item{ $m$ is an element of the Lie algebra $gl(32,R)$, id est, $m$
       is allowed to be {\it an arbitrary $32 \times 32$ bosonic
       matrix}. Decomposing this by the gamma matrices, this can be
       expressed by $m = \sum_{k=0}^{5} \frac{1}{k!} u_{A_{1} \cdots
       A_{k}} \Gamma^{A_{1} \cdots A_{k}}$, where the coefficients
       $u_{A_{1} \cdots A_{k}}$ are all real.}
   \item{ $\psi$ and $\phi$ are independent fermionic vectors. Each of 
       them possesses 32 components, and the components are fermionic
       real number.}
   \item{ $v$ is also a real number.}
  \end{itemize}
  The definition of $gl(1|32,R)$ states nothing. This definition just
  states that an arbitrary real $33 \times 33$ supermatrix is an eligible
  member of the super Lie algebra $gl(1|32,R)$. Although this
  definition does not give any restriction to the elements, the
  correspondence with the complex group $u(1|16,16)$ is an
  interesting aspect of $gl(1|32,R)$ super Lie algebra. Since $\psi$
  and $\phi$ are independent fermionic vectors, these can be rewritten 
  as 
   \begin{eqnarray}
    \psi = \psi_{1} + \psi_{2}, \hspace{2mm} \phi = \psi_{1} - \psi_{2}.
   \end{eqnarray}
  And the bosonic $32 \times 32$ matrices are separated by $m = m_{1}
  + m_{2}$, where $m_{1}$ is rank-1,2,5 and $m_{2}$ is rank-0,3,4.
  Then, we define the sets ${\cal A}$ as follows.
  \begin{eqnarray}
  &\clubsuit&  {\cal A} = \{ M = 
   \left( \begin{array}{cc} m_{2} & \psi_{2} \\  -i {\bar \psi_{2}}
   & v \end{array} \right) | m_{2} = u + \frac{1}{3!} u_{A_{1}
   A_{2} A_{3}} \Gamma^{A_{1} A_{2} A_{3}} + \frac{1}{4!}
   u_{A_{1} \cdots A_{4}} \Gamma^{A_{1} \cdots A_{4}}, \nonumber \\
  & &  \hspace{45mm} u, u_{A_{1} A_{2} A_{3}}, u_{A_{1} \cdots 
   A_{4}}, \psi_{2}, v \in {\cal R} \}. \label{adef}
   \nonumber 
  \end{eqnarray}
  The super Lie algebra $gl(1|32,R)$ is clearly the direct sum of
  these two super Lie algebras 
   \begin{eqnarray}
    gl(1|32,R) = {\cal H} \oplus {\cal A}.
   \end{eqnarray}

  These two subalgebras are also the two different representations of
  $osp(1|32,R)$ super Lie algebra. ${\cal H}$ is $osp(1|32,R)$ itself, 
  and the same super Lie algebra as in introduced in $u(1|16,16)$.
  On the other hand, the  subalgebra ${\cal A}$ is ${\cal
  A}' = i {\cal A}$.\footnote{ This means that, if $A \in {\cal A}$,
  then $i A \in {\cal A}'$.} 
  And the elements of these subalgebras readily satisfy
   \begin{eqnarray}
   {^{T} H} G + G H = 0, \textrm{ for } H \in {\cal H},
   \hspace{3mm} {^{T} A} G - GA = 0, \textrm{ for } A \in {\cal A}.   
   \label{AZ51ha}
   \end{eqnarray}
  And it is clear that these two subalgebras obey
  totally the same commutation relations as those of  ${\cal H}$ and ${\cal 
  A}'$: 
   \begin{eqnarray}
    & & (1) [H_{1}, H_{2} ] \in {\cal H}, \hspace{3mm}
        (2) [H    , A     ] \in {\cal A},  \hspace{3mm}
        (3) [A_{1}, A_{2} ] \in {\cal H}, \nonumber \\
    & & (4)\{H_{1}, H_{2}\} \in {\cal A}, \hspace{3mm}
        (5)\{H    , A    \} \in {\cal H}, \hspace{3mm}
        (6)\{A_{1}, A_{2}\} \in {\cal A}, \label{AZB1comha}
   \end{eqnarray} 
  where $H, H_{1}, H_{2} \in {\cal H}$ and $A, A_{1}, A_{2} \in {\cal
  A}$. The proof is completely the same as that of
  (\ref{AZ51comhasu}), and we  do not repeat it.  
  The commutation relation $[{\cal H}, {\cal A}] \in {\cal A}$
  indicates that ${\cal A}$ is a representation of $osp(1|32,R)$ super 
  Lie algebra. \\

  Now, the relationship of the three super Lie algebras $osp(1|32,R)$,
  $u(1|16,16)$ and $gl(1|32,R)$ is clear. We have seen that both
  $u(1|16,16)$ and $gl(1|32,R)$ are represented by the direct sum
  of two different representations of $osp(1|32,R)$:
   \begin{eqnarray}
    u(1|16,16) = {\cal H} \oplus {\cal A}' ,\hspace{3mm} 
    gl(1|32,R)   = {\cal H} \oplus {\cal A}. \label{AZ51lovetriangle}
   \end{eqnarray}
  The relationship between ${\cal A}$ and ${\cal A}'$ is 
   \begin{eqnarray}
    {\cal A}' = i {\cal A} \Rightarrow \textrm{ If } A \in {\cal A},
    \textrm{ then } i A \in {\cal A}'.
   \end{eqnarray}
  In this sense, we can regard $gl(1|32,R)$ super Lie algebra as {\it
  the analytic continuation} of $u(1|16,16)$. Although we adopt a
  matrix theory with the gauge symmetry $gl(1|32,R)$ unlike Smolin's
  original proposal \cite{0006137}, we note that $gl(1|32,R)$ is a
  cousin of the original $u(1|16,16)$.

  Nextly, we explain why we have to introduce the complexification of
  the $osp(1|32,R)$ from the outset. 
  It turns out that the simple tensor product $osp(1|32,R) \otimes
  u(N)$ does not close with respect to the commutator in mixing these
  two symmetries. Using the relation (\ref{AZ51comhasu}), we obtain
  the following commutation relation:
  \begin{eqnarray}
    [({\cal H} \otimes {\bf H}), ({\cal H} \otimes {\bf H})]
  = ( \{ {\cal H}, {\cal H} \} \otimes [  {\bf H}, {\bf H}])
  + ( [  {\cal H}, {\cal H} ]  \otimes \{ {\bf H}, {\bf H} \} )
  = ({\cal A'} \otimes {\bf A}) + ({\cal H} \otimes {\bf H}).
  \end{eqnarray} 
  Therefore, the tensor product $osp(1|32,R) \otimes u(N) = {\cal H}
  \otimes {\bf H}$ is not a closed Lie algebra. In order to build a
  closed Lie algebra, we are urged to include ${\cal A}' \otimes {\bf
  A}$. This satisfies the commutation relation
  \begin{eqnarray}
   [({\cal A}' \otimes {\bf A}), ({\cal A}' \otimes {\bf A})]
  = ( \{ {\cal A}', {\cal A}' \} \otimes [  {\bf A}, {\bf A}])
  + ( [  {\cal A}', {\cal A}' ]  \otimes \{ {\bf A}, {\bf A} \} )
  = ({\cal A'} \otimes {\bf A}) + ({\cal H} \otimes {\bf H}).
  \end{eqnarray}
  Therefore, we establish the smallest closed Lie algebra that include 
  $osp(1|32,R) \otimes u(N)$ as a subset as
  \begin{eqnarray}
   osp(1|32,R) \check{\otimes} u(N) = ({\cal H} \otimes {\bf H})
 + ({\cal A}' \otimes {\bf A}). \label{u11616necessary}
  \end{eqnarray}
  This is analogous to the situation in which we had to introduce the
  higher-odd-rank fields in the action and the higher-even-rank fields 
  as a parameter of the local Lorentz transformation once we mix the
  symmetries, as we have seen in (\ref{higheroddgauged}) and
  (\ref{higherevengauged}). 

  Here, instead of promoting each element to a hermitian matrix, we
  can make a closed Lie algebra by restricting them to real matrices.
  It is clear that $({\cal H} + {\cal A}) \otimes gl(N,R) = gl(1|32,R) 
  \otimes gl(N,R)$ forms another closed algebra. In this case, we
  embed the spacetime into real matrices, instead of hermitian
  matrices.

  \subsubsection{Action of $gl(1|32,R) \otimes gl(N,R)$ supermatrix model}
    The next job is to investigate the action of this theory. The basic
  idea is similar to $osp(1|32,R)$ non-gauged cubic matrix model, and
  we proceed rather quickly. The action is 
  \begin{eqnarray}
   S &=& \frac{1}{g^{2}} Tr_{N \times N} \sum_{Q,R=1}^{33}
        ((\sum_{p=1}^{32}  {M_{p}}^{Q} {M_{Q}}^{R}
        {M_{R}}^{p})  -{M_{33}}^{Q} {M_{Q}}^{R}
        {M_{R}}^{33} ) = \frac{1}{g^{2}} Tr_{N \times N}( Str_{33
        \times 33} M^{3}) \nonumber \\
    &=& \frac{1}{g^{2}} \sum_{a,b,c=1}^{N^{2}} Str(M^{a} M^{b} M^{c}
     ) Tr(T^{a} T^{b} T^{c}). \label{AZ54action}
  \end{eqnarray}
 $M$ is now a multiplet of $gl(1|32,R)$ super Lie algebra,
        with each component promoted to the element of $gl(N,R)$ Lie
        algebra. The indices
        $P, Q, R, \cdots$ runs $P, Q, R, \cdots = 1, \cdots, 33$,
        while $p, q, r, \cdots = 1, \cdots, 32$.

 We have promoted the $33 \times 33$ matrix $M$ to a large
        $33N \times 33N$ matrices. However, the structure of the promotion is
      completely different from $osp(1|32,R)$ model. The gauge
      transformation is with respect to not the separate $gl(1|32,R)$
      and $gl(N,R)$, but the tensor product of the Lie algebra
        $gl(1|32,R) \otimes gl(N,R)$. This 
      drastically enhances the gauge symmetry of the theory.
   The gauge transformation is thus $M \Rightarrow M + [u, M]$ for an
        arbitrary element of $ u \in gl(1|32,R) \otimes
     gl(N,R)$.

  We can express this supermatrix model in terms of the basis of the
  $gl(N,R)$. Here, $M_{P}^{Q}$ is expanded as
  $M_{P}^{Q} = \sum_{a=1}^{N} (M^{a})_{P}^{Q} T^{a}$. 
  Then, the trace is written as
  \begin{eqnarray}
   Tr(T^{a} T^{b} T^{c}) = \frac{1}{2} Tr(T^{a}[T^{b},T^{c}]) +
   \frac{1}{2} Tr(T^{a} \{ T^{b}, T^{c} \})
  = \frac{1}{4} (f_{abc} + d_{abc}).
  \end{eqnarray}
  Here, the structure constant $f_{abc}$ and $d_{abc}$ are real.
  Using this decomposition, we can rewrite the action
  (\ref{AZ54action}) as
  \begin{eqnarray}
   S = \frac{1}{4 g^{2}} (f_{abc} + d_{abc} ) Str(M^{a}
   M^{b} M^{c} ). \label{AZ53colormatrix}  
   \end{eqnarray}
 This supermatrix model can be expressed using its explicit expression
 (\ref{gl132rdef}) as
  \begin{eqnarray}
   S = \frac{1}{g^{2}} Tr ( tr(m^{3}) - 3i {\bar \phi} m \psi - 3i {\bar 
 \phi} \psi v - v^{3}). \label{AZ53action}
  \end{eqnarray}

 \subsubsection{Wigner-In\"{o}n\"{u} contraction and supersymmetry}
 In considering the relation to the IIB matrix model, we need to
 reduce the model to the ten dimensions. However, this matrix model
 has a grave difference from the $osp(1|32,R)$ nongauged
 supermatrix. Since the action (\ref{AZ54action}) is no longer based
 on the commutator for the $gl(N|R)$ matrices, it is not invariant
 under the inhomogeneous supersymmetry, namely the translation of the
 fermion. Nevertheless, the matrix model (\ref{AZ54action})
 accommodates $32+32=64$ supercharges. This leads us to speculate that 
 this model may also have the two-fold supersymmetry structure of the
 IIB-like supersymmetry.

 Here, we reduce the model to the ten dimensions by the
 Wigner-In\"{o}n\"{u} contraction, in order to extract the translation 
 symmetry. Just as we perceive the earth as a flat
 two-dimensional space because the earth is much bigger than we are, we
  consider the physics in apparently 'ten-dimensional' space because of 
  the large radius of the hyperboloid.
 To this end, we add the linear term of $M$ to the action
 (\ref{AZ54action}) and start from
    \begin{eqnarray}
    S = \frac{1}{3} Tr_{N \times N} Str( M_{t}^{3}) - R^{2} Tr_{N \times
    N} Str M_{t}. \label{AZ54veryaction}
   \end{eqnarray}
  By adding the linear term, the matrix model (\ref{AZ54veryaction})
  incorporates the classical solution\footnote{This
  classical solution is impossible in the original $u(1|16,16)$
  gauged theory, because the (33,33) component $v$ is restricted to
  be an anti-hermitian matrix. If we are to consider the
  Wigner-In{\"o}n{\"u} contraction, one way is to consider the quintic 
  action $S_{u(1|16,16)} = \frac{1}{5} Str(M^{5}_{t}) - R^{4} Str
  M_{t}$. Then, the classical solution $\langle M \rangle = 
  \left( \begin{array}{cc} R  \Gamma^{\sharp} \otimes {\bf 1}_{N
  \times N} & 0 \\ 0 & i R \otimes {\bf 1}_{N \times N} \end{array}
  \right)$ is possible because $ i R \otimes {\bf 1}_{N \times N}$ is now an
  anti-hermitian matrix. Another caution is that the gauge group must
  be not $u(1|16,16) \otimes su(N,C)$ but $u(1|16,16) \otimes
  u(N,C)$. If the gauge group is $u(1|16,16) \otimes su(N,C)$, the
  linear term in (\ref{AZ54veryaction}) vanishes because the
  generators are $Tr(T^{a})=0$, and the Wigner In{\"o}n{\"u}
  contraction is impossible from the beginning.}. 
  \begin{eqnarray}
    \langle M \rangle = \left( \begin{array}{cc} R \Gamma^{\sharp}
  \otimes {\bf 1}_{N \times N} & 0 \\ 0 & R \otimes {\bf 1}_{N \times N}
  \end{array} \right), 
  \end{eqnarray}
  The expansion around this classical solution amounts to the
 Wigner-In\"{o}n\"{u} contraction. We separate
  the original matrix    between the classical solution and the
  fluctuation  and the classical solution as 
   \begin{eqnarray}
    M_{t} = \langle M \rangle + M = \left( \begin{array}{cc} R
    \Gamma^{\sharp} & 0 \\ 0 & R \end{array} \right) + \left
    ( \begin{array}{cc} m & \psi \\ i {\bar \phi} & v \end{array}
    \right).
   \end{eqnarray}
  Then, the action is 
   \begin{eqnarray}
    I = \frac{1}{3} tr( (m + R \Gamma^{\sharp})^{3}) - i ( {\bar \phi} 
    m \psi + {\bar \phi} \psi v + R {\bar \phi} (1 + \Gamma^{\sharp} ) 
    \psi ) - \frac{(v+R)^{3}}{3} - R^{2} ( tr(m + R \Gamma^{\sharp}) - 
    v).
   \end{eqnarray}
    We ignore the terms of ${\cal O}(R^{3})$, because this is just a constant.
   The terms of ${\cal O}(R^{2})$ vanish, because this is a
    linear term with respect to the fluctuation. Then, the action is
    expressed as follows: 
   \begin{eqnarray}
    S = R( tr(m^{2} \Gamma^{\sharp}) - v^{2} - i {\bar \phi} (1 +
    \Gamma^{\sharp}) \psi)  + \frac{1}{3} tr(m^{3}) -
    \frac{v^{3}}{3} -i ({\bar  \phi} m {\psi} + v {\bar \phi} \psi ). 
   \end{eqnarray}
  We divide the bosonic fluctuation $m$ into $m = m_{e} + m_{o}$,
  where $m_{e}$ and $m_{o}$ consist of the even-rank and the odd-rank
  components in terms of the ten dimensions. Namely, they should be
  explicitly expressed as
   \begin{eqnarray}
    m_{e} &=& Z {\bf 1} + W \Gamma^{\sharp} + \frac{1}{2}
             ( C_{\mu_{1} \mu_{2}} \Gamma^{\mu_{1} \mu_{2}} +
             D_{\mu_{1} \mu_{2}} 
             \Gamma^{\mu_{1} \mu_{2} \sharp} ) + \frac{1}{4!} ( G_{\mu_{1}
             \cdots \mu_{4}} \Gamma^{\mu_{1} \cdots \mu_{4}} +
             H_{\mu_{1} \cdots \mu_{4}} \Gamma^{\mu_{1} \cdots \mu_{4}
             \sharp}), \\
    m_{o} &=& \frac{1}{2} ( A^{(+)}_{\mu} \Gamma^{\mu} ( 1 +
           \Gamma^{\sharp}) + A^{(-)}_{\mu} \Gamma^{\mu} (1-
           \Gamma^{\sharp}) ) \nonumber \\
       &+&  \frac{1}{2 \times 3!} (E^{(+)}_{\mu_{1} \mu_{2} \mu_{3}}
           \Gamma^{\mu_{1} \mu_{2} \mu_{3}} (1 + \Gamma^{\sharp} ) +
           E^{(-)}_{\mu_{1} \mu_{2} \mu_{3}} \Gamma^{\mu_{1} \mu_{2}
             \mu_{3}} (1 - \Gamma^{\sharp} ) ) \nonumber \\
           &+& \frac{1}{5!} (I^{(+)}_{\mu_{1} \cdots \mu_{5}}
           \Gamma^{\mu_{1} \cdots \mu_{5}} (1 + \Gamma^{\sharp}) +
           I^{(-)}_{\mu_{1} \cdots \mu_{5}} \Gamma^{\mu_{1} \cdots \mu_{5}} (1
           - \Gamma^{\sharp}) ). 
   \end{eqnarray}
  Later, we further decompose $m_{o}$ as $m_{o} = m^{(+)}_{o} +
  m^{(-)}_{o}$ according to the $(\pm)$, namely the plus or minus
  chirality, in the above decomposition.
  The fermions are divided according to its chirality.
   The action is then written as follows.
   \begin{eqnarray}
    S &=& R( tr(m^{2}_{e} \Gamma^{\sharp}) - v^{2} - 2i {\bar \phi}_{R}
    \psi_{L} ) + tr (\frac{1}{3} m^{3}_{e} + m_{e} m^{2}_{o} )
    \nonumber \\
    &-& i ( {\bar \phi}_{R} (m_{e} + v) \psi_{L}  + {\bar \phi}_{L}
    (m_{e} + v) \psi_{R} + {\bar \phi}_{L} m_{o} \psi_{L} + {\bar
    \phi}_{R} m_{o} \psi_{R} ) - \frac{1}{3} v^{3}. \label{AZ55cocoro} 
   \end{eqnarray}
  We integrate out the fields of order ${\cal
  O}(R)$ and consider the effective theory. In order to cope with the
  cubic term $tr(m_{e}^{3})$, we consider the following rescaling:
   \begin{eqnarray}
  & &   m_{t} = R \Gamma^{\sharp} + m = R \Gamma^{\sharp} +
    R^{-\frac{1}{2}} m'_{e} + R^{\frac{1}{4}} m'_{o}, \hspace{3mm}
    v_{t} = R + v = R + R^{-\frac{1}{2}} v', \nonumber \\
  & & \psi = \psi_{L} + \psi_{R} = R^{-\frac{1}{2}} \psi'_{L} +
    R^{\frac{1}{4}} \psi'_{R} , \hspace{3mm} {\bar \phi} = {\bar
    \phi}_{L} + {\bar \phi}_{R} = 
    R^{\frac{1}{4}} {\bar \phi}'_{L} + R^{-\frac{1}{2}} {\bar
    \phi}'_{R}.
   \end{eqnarray}
  Following this rescaling, $tr(m'^{3}_{e})$ and ${\bar \phi}'_{R}
  (m'_{e} + v') \psi_{L}$ are excluded, because this is rescaled as
  ${\cal O}(R^{- \frac{3}{2}})$. This theory is thus rescaled to be
   \begin{eqnarray}
    S = ( tr(m'^{2}_{e} \Gamma^{\sharp}) - v'^{2} + tr(m'_{e}
    m'^{2}_{o}) ) -i ( 2 {\bar \phi}'_{R} \psi'_{L} + {\bar \phi}'_{L}
    (m'_{e} + v') \psi'_{R} + {\bar \psi}'_{L} m'_{o} \psi'_{L} +
    {\bar \phi}'_{R} m'_{o} \psi'_{R} ).
   \end{eqnarray}
  We integrate out the fields $m'_{e}$, $\psi'_{L}$ and ${\bar
  \phi}'_{R}$ by Gaussian integration. Completing this action square,
  we obtain 
   \begin{eqnarray}
   S &=& tr(  \{ m'_{e} + \frac{1}{2} ( m'^{2}_{o}
   \Gamma^{\sharp} + i (\psi'_{R} {\bar \phi}'_{L} ) \Gamma^{\sharp})
   \}^{2} \Gamma^{\sharp} ) - \frac{1}{4} tr( \{ m'^{2}_{o} +
  i (\psi'_{R} {\bar \phi}'_{L} ) \}^{2} \Gamma^{\sharp} ) \nonumber \\
    &-& (v' + \frac{i}{2} ({\bar \phi}_{L} \psi_{R}) )^{2} -
   \frac{1}{4} ({\bar \phi}'_{L} \psi_{R})^{2} - 2i ({\bar \phi}'_{R}
   + \frac{1}{2} {\bar \phi}'_{L} m_{o} )( \psi'_{L} + \frac{1}{2}
   m'_{o} \psi'_{R} ) + \frac{i}{2} ( {\bar \phi}'_{L} m'^{2}_{o}
   \psi'_{R} ), \label{AZ55compsq}
   \end{eqnarray}
 where we have utilized the fact that $(\Gamma^{\sharp})^{2} = {\bf
 1}_{32 \times 32}$, and thus $\Gamma^{\sharp}$ possesses an inverse
 matrix. 
 The effective action is given by the following path
 integral\footnote{The analogy of this path integral is a 
 following toy model. (i)For the action $S=ax^{2}+bx+c$, the path
 integral is $e^{-W} = \int dx \exp( -(ax^{2}+bx+c)) =
 \int dx  \exp(-a(x+\frac{b}{2a})^{2} ) \exp( -c+ \frac{b^{2}}{4a})
 \propto  \exp(-c+ \frac{b^{2}}{4a})$. And thus the effective action
 is $W = \frac{-b^{2} + 4ac}{4a}$. (ii)The second example is the
 following action, $S=axy+bx+cy$, with $x,y$ being auxiliary
 fields. This integration is performed by $e^{-W} = \int dx dy \exp (
 a(x+ \frac{b}{a})(y+\frac{c}{a}) + \frac{bc}{a} )$, and thus the
 effective action is $W = - \frac{bc}{a}$. This holds even if $x,y$
 are Grassmann odd quantities.}. However, this effective action 
 turns out to vanish:
   \begin{eqnarray}
 & &    e^{-W} = \int dm'_{e} d \psi'_{L} d {\bar \phi}'_{R} dv e^{-S} 
 \nonumber \\
 &\Rightarrow& W = - \frac{1}{4}  tr(  \{ m'^{2}_{o} + i  (\psi'_{R}
 {\bar \phi}'_{L} ) \}^{2} \Gamma^{\sharp} ) - \frac{1}{4}  ( {\bar 
 \phi}'_{L} \psi_{R})^{2}  + \frac{i}{2} ( {\bar \phi}'_{L} m'^{2}_{o}
   \psi'_{R} ) \label{AZ55effectiveac} \\
 & & \hspace{5mm} =  -  \frac{1}{4} tr( m'^{4}_{o} \Gamma^{\sharp} ) +
 \frac{i}{2} ( tr( - m'^{2}_{o} (\psi'_{R} {\bar \phi}'_{L}) 
 \Gamma^{\sharp} ) + {\bar \phi}'_{L} m'^{2}_{o} \psi'_{R} ) +
 \frac{1}{4} ( tr( (\psi'_{R} {\bar \phi}'_{L} )^{2} \Gamma^{\sharp} ) - 
 ({\bar \phi}'_{L} \psi'_{R})^{2} ) = 0. \nonumber
   \end{eqnarray}
  That the first term vanishes is discerned from  the
  anti-commutativity of the matrices $m'_{o}$ and $\Gamma^{\sharp}$;
  namely $m'_{o} \Gamma^{\sharp}= - \Gamma^{\sharp} m'_{o}$. Noting this fact,
  we rewrite this term  as 
       \begin{eqnarray}
         -  \frac{1}{4} tr( m'^{4}_{o} \Gamma^{\sharp} ) 
         \stackrel{\{ \Gamma_{\sharp}, m'_{o} \} = 0}{=} \frac{1}{4}
         tr( m'^{3}_{o} \Gamma^{\sharp} m'_{o} ) \stackrel{\textrm{cyclic}}{=}
         \frac{1}{4} tr( m'^{4}_{o} \Gamma^{\sharp} )  = 0. 
       \end{eqnarray}

  While the effective action in the ten dimensions turns out to
  vanish, we investigate its symmetry. This model is invariant under
  the $gl(1|32,R)$ rotation $\delta M = [A,M]$ where
  $A = \left( \begin{array}{cc} a & \chi \\ i {\bar \epsilon} & b
  \end{array} \right)$. Like the $osp(1|32,R)$ nongauged model, the
  fermionic part of this rotation serves as the supercharge. In this
  sense, the $gl(1|32,R)$ supermatrix model also has $32 + 32 = 64$
  supercharges. We consider the following rescaling of this parameter: 
   \begin{eqnarray}
    A = \left( \begin{array}{cc} a'_{e} + R^{-\frac{3}{4}} a'_{o} &
    \chi'_{L} + R^{-\frac{3}{4}} \chi'_{R} \\ i ( R^{-\frac{3}{4}} {\bar 
    \epsilon}'_{L} + {\bar \epsilon}'_{R}) & b' \end{array} \right).
    \label{AZ55chres} 
   \end{eqnarray}
  For the $gl(1|32,R)$ rotation 
  \begin{eqnarray}
  \delta M = [A,M] =  \left( \begin{array}{cc}
   [a, m + R \Gamma^{\sharp}] + i 
   ( \chi {\bar \phi} - \psi {\bar \epsilon}) & - (m + R
   \Gamma^{\sharp}) \chi + a \psi - b \psi + \chi v \\ i {\bar
   \epsilon} (m + R \Gamma^{\sharp}) - i ( {\bar \phi} a + v {\bar
   \epsilon} ) + i b {\bar \phi} & i ({\bar \epsilon} \psi - {\bar
   \phi} \chi ) + [b, v] \end{array} \right),  \label{AZ55subotr}
  \end{eqnarray}
   each component is transformed as follows:
   \begin{eqnarray}
      \delta m'_{e} &=&  [a_{e}, m'_{e}] + [a_{o}, m'_{o}] + i
     ( \chi'_{L} {\bar \epsilon}'_{R} + \chi'_{R} {\bar \epsilon'_{L}}
     ) - i ( \psi'_{L} {\bar \epsilon}'_{R} + \chi'_{R} {\bar
     \epsilon}'_{L} ), \label{AZ55tr1964me} \\
       \delta m'_{o} &=& \underline{[ a_{o}, \Gamma^{\sharp}]} +
     [a'_{e}, m'_{o}]  + i ( \chi'_{L} {\bar \phi}'_{L} - \psi'_{R}
     {\bar \epsilon}'_{R} ), \label{AZ55tr1964mo} \\ 
      \delta \psi'_{L} &=& - ( m'_{o} \chi'_{R} + m'_{e} \chi_{L}) +
       (a_{o} \psi'_{R} + a_{e} \psi'_{L}) - b' \psi'_{L} + v'
       \chi'_{L}, \label{AZ55tr1964psil} \\
      \delta \psi'_{R} &=& \underline{2 \chi'_{R}} + ( a'_{e} \psi'_{R}  - b
       \psi'_{R} - (m'_{o} \chi'_{L}) ), \label{AZ55tr1964psir} \\
      \delta {\bar \phi}'_{L} &=& \underline{- 2 {\bar \epsilon}'_{L}}
     + ( ({\bar 
         \epsilon}'_{R} m'_{o}) + b' {\bar \phi}'_{L} - {\bar
         \phi}'_{L} a'_{e}  ), \label{AZ55tr1964phil} \\
   \delta {\bar \phi}'_{R} &=& ( {\bar \epsilon}'_{R} m'_{e} +
      {\bar \epsilon}'_{L} m'_{o} ) + b' {\bar \phi}'_{R} - ({\bar
      \phi}'_{L} a'_{o} + {\bar \phi}'_{R} a'_{e} ) - v' {\bar
      \epsilon}'_{R}, \label{AZ55tr1964phir} \\
    \delta v' &=& i ( {\bar \epsilon}'_{R} \psi'_{L} + {\bar
         \epsilon}'_{L} m'_{o} ) - i ({\bar \phi}'_{R} \chi'_{L} +
         {\bar \phi}'_{L} \chi'_{R} ) + [b', v']. \label{AZ55tr1964v}
   \end{eqnarray}
   These results possess two major significances. First is that this
  reconfirms that the effective action (\ref{AZ55effectiveac})
  vanish. The effective
  action is invariant under the transformation of the fields $m'_{o}$,
  $\psi'_{R}$ and ${\bar \phi}'_{L}$. Therefore, the effective action
  makes no difference if we translate these fermions
  arbitrarily. Taking the fields $m'_{o}$, $\psi'_{R}$ and ${\bar
  \phi}'_{L}$ to be all zero, we find that the effective action is $W =
  0$. Therefore, the effective action remains $W=0$ even if we translate
  these fields into non-zero values. 

  The second significance is that this clarifies the structure of the
  supersymmetry of the effective theory. This gauged theory originally
  possesses no trivial translation like the $osp(1|32,R)$ non-gauged 
  model, because this theory is deprived of the symmetry of the
  commutator. However, the Wigner-In{\"o}n{\"u} contraction 
  serves to retrieve the translation in the ten dimensions.
  Actually, the underlined transformation of (\ref{AZ55tr1964mo}),
  (\ref{AZ55tr1964psir}) and (\ref{AZ55tr1964phil}) represents a
  translation independent of the matter field $M$.
  We thus find it natural to extract the structure of the IIB matrix
  model from the fields $m'^{(\pm)}_{o}$, $\psi_{R}'$ and $\phi_{L}$. Their
  supersymmetry transformation is read off from (\ref{AZ55tr1964mo}),
  (\ref{AZ55tr1964psir}) and (\ref{AZ55tr1964phil}) as
  \begin{eqnarray}
       \delta m'^{(+)}_{o} = -  i  \psi'_{R}
      {\bar \epsilon}'_{R}, \hspace{2mm}
       \delta m'^{(-)}_{o} = + i \chi'_{L} {\bar \phi}'_{L},
      \hspace{2mm}  
      \delta \psi'_{R} = 2 \chi'_{R}  - (m'_{o} \chi'_{L}),
      \hspace{2mm}
      \delta {\bar \phi}'_{L} = - 2 {\bar \epsilon}'_{L} + ( {\bar
         \epsilon}'_{R} m'_{o} ). \label{AZ55susy1964} 
   \end{eqnarray}
  This immediately gives the correspondence of the chirality among the 
  bosonic fields and the fermionic fields:
  $(m^{(-)'}_{o}, \phi'_{L})$ and $(m^{(+)'}_{o}, \psi'_{R})$.
  This pairing also resembles the case of the  $osp(1|32,R)$ nongauged model.

  We first scrutinize the former correspondence. In this case, the
  homogeneous and the inhomogeneous supersymmetry is identified as
    \begin{eqnarray}
    \delta^{(1)}_{\chi_{L}} = [Q_{\chi_{L}}, \hspace{1mm} \bullet
     \hspace{1mm} ] = [ 
     \left( \begin{array}{cc}  0 & \chi_{L} \\ 0 & 0 \end{array}
     \right), \hspace{1mm} \bullet \hspace{1mm} ], \hspace{3mm} 
    \delta^{(2)}_{\epsilon_{L}} = [Q_{\epsilon_{L}}, \hspace{1mm}
     \bullet \hspace{1mm} ] = [ 
     \left( \begin{array}{cc} 0 & 0 \\ i {\bar \epsilon}_{L} & 0
     \end{array} \right) , \hspace{1mm} \bullet \hspace{1mm} ]. 
   \end{eqnarray}
  Then, the rank-1 fields are transformed by this homogeneous
  transformation as
   \begin{eqnarray}
 \delta^{(1)}_{\chi_{L}} A'^{(-)}_{\mu} &=& \frac{1}{32} tr(i \chi_{L}
    {\bar \phi}'_{L} \Gamma_{\mu}) - \frac{-1}{32} tr(i \chi_{L} {\bar
    \phi}'_{L} \Gamma_{i \sharp}) =  - \frac{i}{16} {\bar
    \phi}'_{L} \Gamma_{\mu} \chi_{L}.     
   \end{eqnarray}
 The inhomogeneous translation of course do not affect the transformation
 of the vector fields.

  We next explore the commutation relations. Firstly, we discern that 
  the commutator $[\delta^{(1)}_{\chi_{L}}, \delta^{(1)}_{\rho_{L}}]$, 
  as well as $[\delta^{(2)}_{\epsilon_{L}}, \delta^{(2)}_{\eta_{L}}]$,
  manifestly vanish.
    \begin{eqnarray}
   & & [\delta^{(1)}_{\chi_{L}}, \delta^{(1)}_{\rho_{L}}] = 
 [ [ \left( \begin{array}{cc} 0 & \chi_{L} \\ 0 & 0 \end{array}
  \right),   \left( \begin{array}{cc} 0 & \rho_{L} \\ 0 & 0
  \end{array} \right)], \hspace{1mm} \bullet \hspace{1mm} ]  = 0,
 \label{AZ55SUSYcom1} \\ 
   & &  [\delta^{(2)}_{\epsilon_{L}}, \delta^{(2)}_{\eta_{L}}] =
  [ [ \left( \begin{array}{cc} 0 & 0 \\ i {\bar \epsilon}_{L} & 0
  \end{array} \right), \left( \begin{array}{cc} 0 & 0 \\ i {\bar
  \eta}_{L} & 0 \end{array} \right)] , \hspace{1mm} \bullet
  \hspace{1mm} ]= 0, \label{AZ55SUSYcom2} 
   \end{eqnarray}
  This is a great advantage compared with the $osp(1|32,R)$ case, in
  which we are beset by the nonvanishing commutator of the homogeneous
  supersymmetry. 
  The commutation relation $[ \delta^{(1)}_{\chi_{L}},
  \delta^{(2)}_{\epsilon_{L}}]$ is obtained by
   \begin{eqnarray}
    [ \delta^{(1)}_{\chi_{L}}, \delta^{(2)}_{\epsilon_{L}}]
   A'^{(-)}_{\mu} = \frac{iR}{16} {\bar \epsilon}_{L}
   \Gamma_{\mu}  \chi_{L}.  \label{AZ55finalresult1}
    \end{eqnarray}

  The correspondence of the latter case  goes in totally the same way.
  The supercharges are defined by
   \begin{eqnarray}
   \delta^{(1)}_{\epsilon_{R}} = [Q_{\epsilon_{R}}, \hspace{1mm}
   \bullet \hspace{1mm} ] = [ \left( \begin{array}{cc} 0 & 0 \\ i {\bar 
   \epsilon}_{R} & 0 \end{array} \right), \hspace{1mm} \bullet
   \hspace{1mm} ], \hspace{3mm} \delta^{(2)}_{\chi_{R}} =
   [Q_{\chi_{R}}, \hspace{1mm}  \bullet \hspace{1mm} ] =
   [ \left( \begin{array}{cc} 0 & \chi_{R} \\ 0 & 0 \end{array}
   \right), \hspace{1mm} \bullet \hspace{1mm} ].
   \end{eqnarray}
  Then, the supersymmetry transformation of the rank-1 fields is
  obtained by
   \begin{eqnarray}
   \delta^{(1)}_{\epsilon_{R}} A'^{(+)}_{\mu} &=& \frac{1}{32} tr( -i
   \psi'_{R} {\bar \epsilon}_{R} \Gamma_{\mu}) + \frac{-1}{32} tr( -i
   \psi'_{R} {\bar \epsilon}_{R} \Gamma_{i \sharp}) =
   \frac{i}{16} {\bar \epsilon}_{R} \Gamma_{\mu} \psi'_{R}.
   \label{AZ55trmata+}
  \end{eqnarray}
  The commutators $[\delta^{(1)}_{\epsilon_{R}},
  \delta^{(1)}_{\eta_{R}} ]$ and $[\delta^{(2)}_{\chi_{R}},
  \delta^{(2)}_{\rho_{R}} ]$ vanish likewise, and the commutation
  relation $[\delta^{(1)}_{\epsilon_{R}}, \delta^{(2)}_{\chi_{R}}]$
  gives
   \begin{eqnarray}
    [\delta^{(1)}_{\epsilon_{R}}, \delta^{(2)}_{\chi_{R}}]
    A'^{(+)}_{\mu} =
    \frac{iR}{16} {\bar \epsilon}_{R} \Gamma_{\mu} \chi_{R}. 
   \label{AZ55finalresult2} 
   \end{eqnarray}
  
  We see a better correspondence of the supersymmetry with the
  IIB matrix model than the $osp(1|32,R)$ model. However, the
  supersymmetry transformation of the fermion is not balanced by the
  commutator $[A_{\mu}, A_{\nu}]$. This obstacles are shared with the
  $osp(1|32,R)$ case.
  The decisive drawback of this argument is that the ten-dimensional
  effective action totally vanishes. This comes from too large a
  symmetry of the model. However, it may be related to the 
  notion of the topological matrix model\cite{9708039}, in which the IIB
  matrix model is induced ex nihilo.
  The exploration of such a gauged matrix model are interesting from
  various points of view, especially the existence of the local
  Lorentz invariance, and it is worth further investigations. 
  
 \subsection{Curved-space classical solutions of the massive
  $osp(1|32,R)$ nongauged supermatrix model} \label{sec0209057}
 In this section, we review the work\cite{0209057}, in which we
 elaborated on the relation between the supermatrix model and the
 curved-space background. In Section \ref{alternativemodel}, we have
 introduced several generalizations of the IIB matrix model to
 accommodate the curved-space background. The work \cite{0209057}
 focuses on the similarity between the massive IIB matrix model
 (\ref{IIBtmass}) and the $osp(1|32,R)$ supermatrix model with the
 quadratic term. This has led us to expect that the $osp(1|32,R)$
 supermatrix model could also have interesting noncommutative static
 solutions. We delegate the detailed properties of the
 higher-dimensional fuzzy sphere, which plays a main role in
 this analysis, to Section \ref{kimurayhigher}.
 
 \subsubsection{Action}
  Here, we go back to the $osp(1|32,R)$ nongauged supermatrix model
  instead of the gauged matrix model, since our aim is to unravel the
  similarity with the alteration of the IIB matrix model described in
  (\ref{IIBtmass}). We start with the following massive model with
  the cubic interaction\footnote{We warn the readers that the notation 
  of this section is different from that of \cite{0209057}. In this
  paper, we define the path integral of the action $S$ as
   \begin{eqnarray}
     Z = \int dA \cdots e^{-S_{E}}, \nonumber
   \end{eqnarray}
  where $S_{E}$ is defined in the ten-dimensional Euclidean space by
  the Wick rotation of the components $A_{0}, \cdots$ and the gamma
  matrices $\Gamma^{0}$, Therefore, the action is different from that
  of \cite{0209057} by the overall sign.}:
  \begin{eqnarray}
  S &=& 3 \mu Tr \sum_{Q=1}^{33} \left( \left( \sum_{p=1}^{32} M_{p}^{Q}
  M_{Q}^{p} \right) - M_{33}^{Q} M_{Q}^{33} \right) 
    - i  Tr \sum_{Q,R=1}^{33} \left( 
  \left( \sum_{p=1}^{32} M_{P}^{Q} [M_{Q}^{R}, M_{R}^{P}] \right)
  - M_{33}^{Q} [M_{Q}^{R}, M_{R}^{33}] \right) \nonumber \\ 
    &=&  3 \mu Tr (m_{p}^{q} m_{q}^{p} - 2 i {\bar \psi} \psi) 
       - i Tr (m_{p}^{q} [m_{q}^{r}, m_{r}^{p}] - 3i {\bar \psi}_{p} 
               [m_{p}^{q}, \psi_{q}] ). \label{AZaction3}  
  \end{eqnarray}
  Here, $P,Q,R, \cdots$ and $p,q,r, \cdots$ likewise run over
  $1,2,\cdots, 33$ and $1,2,\cdots, 32$. 
  Here, we focus on the nongauged action, and this action has the
  separated $osp(1|32,R)$ rotational symmetry and the $U(N)$ gauge
  symmetry. We reduce the above matrix model to ten dimensions, by
  specializing the tenth space direction $x^{10} = x^{\sharp}$.
  In  Section \ref{osp132r}, we focused on the identification of the
  supersymmetry with that of the IIB matrix model, and we resorted
  to the chiral decomposition. However, we now focus on the
  nontrivial curved-space background, which leads us to reduce the 
  bosonic $32 \times 32$ matrix $m$ to the ten dimensions without the
  chiral decomposition. Namely, $m$ is reduced as
  \begin{eqnarray}
   m = W \Gamma^{\sharp} + A_{\mu} \Gamma^{\mu} + B_{\mu} \Gamma^{\mu \sharp}
     + \frac{1}{2!} C_{\mu_{1} \mu_{2}} \Gamma^{\mu_{1} \mu_{2}}
     + \frac{1}{4!} H_{\mu_{1} \cdots \mu_{4}} \Gamma^{\mu_{1} \cdots
     \mu_{4} \sharp}
     + \frac{1}{5!} Z_{\mu_{1} \cdots \mu_{5}} \Gamma^{\mu_{1} \cdots
     \mu_{5}}. \label{reduction2}
  \end{eqnarray}
  Then, the relevant action (\ref{AZaction3}) is rewritten as
   \begin{eqnarray}
 S &=& 96 \mu Tr \left( W^{2} + A_{\mu} A^{\mu} - B_{\mu} B^{\mu}
   - \frac{1}{2} C_{\mu_{1} \mu_{2}} C^{\mu_{1} \mu_{2}}
   + \frac{1}{4!} H_{\mu_{1} \cdots \mu_{4}} H^{\mu_{1} \cdots
     \mu_{4}}  
      + \frac{1}{5!} Z_{\mu_{1} \cdots \mu_{5}} Z^{\mu_{1} \cdots
     \mu_{5}} -  \frac{i}{16} {\bar \psi} \psi \right) \nonumber \\
    &-& 32 i Tr \Bigg( - 3 C_{\mu_{1} \mu_{2}} [A^{\mu_{1}},
     A^{\mu_{2}}] + 3 C_{\mu_{1} \mu_{2}}  [B^{\mu_{1}}, B^{\mu_{2}}] 
     + 6 W [A_{\mu}, B^{\mu}]
     + C_{\mu_{1} \mu_{2}} [ {C^{\mu_{2}}}_{\mu_{3}}, C^{\mu_{3}
     \mu_{1}}] \nonumber \\
   & & \hspace{12mm} + \frac{1}{4} B_{\mu_{1}} [H_{\mu_{2} \cdots
     \mu_{5}}, Z^{\mu_{1} \cdots \mu_{5}}]
     - \frac{1}{8} C_{\mu_{1} \mu_{2}}
     ( 4 [ {H^{\mu_{1}}}_{\rho_{1} \rho_{2} \rho_{3}}, H^{\mu_{2}
     \rho_{1} \rho_{2} \rho_{3}}]
      + [ {Z^{\mu_{1}}}_{\rho_{1} \cdots \rho_{4}}, Z^{\mu_{2}
     \rho_{1} \cdots \rho_{4}}] ) \nonumber \\
   & & \hspace{12mm} + \frac{3}{(5!)^{2}} \epsilon^{\mu_{1} \cdots
     \mu_{10} \sharp} \left( - W [Z_{\mu_{1} \cdots \mu_{5}},
     Z_{\mu_{6} \cdots \mu_{10}}] + 10 A_{\mu_{1}} [H_{\mu_{2} \cdots
     \mu_{5}}, Z_{\mu_{6} \cdots \mu_{10}}] \right) \nonumber \\
   & & \hspace{12mm} + \frac{200}{(5!)^{3}}
     \epsilon^{\mu_{1} \cdots  \mu_{10} \sharp}
     \left( 5 H_{\mu_{1} \cdots \mu_{4}}  [{Z_{\mu_{5} \mu_{6} \mu_{7}}}^{\rho
     \chi}, Z_{\mu_{8} \mu_{9} \mu_{10} \rho \chi}]
      +10 H_{\mu_{1} \cdots \mu_{4}} [ {H_{\mu_{5} \mu_{6}
     \mu_{7}}}^{\rho}, H_{\mu_{8} \mu_{9} 
     \mu_{10} \rho}]  \right. \nonumber \\
   & & \hspace{35mm} \left.  + 6 {H^{\rho \chi}}_{\mu_{1} \mu_{2}} 
     [ Z_{\mu_{3} \mu_{4} \mu_{5} \rho \chi}, Z_{\mu_{6} \cdots
     \mu_{10}} ] \right) \Bigg)\nonumber \\
   &-& 3 Tr \left( {\bar \psi} \Gamma^{\sharp} [W, \psi]
         + {\bar \psi} \Gamma^{\mu} [A_{\mu}, \psi]
         + {\bar \psi} \Gamma^{\mu \sharp} [B_{\mu}, \psi]
         + \frac{1}{2!} {\bar \psi} \Gamma^{\mu_{1} \mu_{2}}
           [C_{\mu_{1} \mu_{2}}, \psi] +  \right. \nonumber \\
   & & \hspace{10mm} \left. + \frac{1}{4!} {\bar
   \psi} \Gamma^{\mu_{1} \cdots  
     \mu_{4} \sharp} [H_{\mu_{1} \cdots \mu_{4}}, \psi] 
     + \frac{1}{5!} {\bar \psi} \Gamma^{\mu_{1} \cdots \mu_{5}}
      [Z_{\mu_{1} \cdots \mu_{5}}, \psi] \right).
   \end{eqnarray}
  In the purely cubic supermatrix model 
 (without mass term, which has been studied in 
 \cite{0002009,0006137,0102168,0103003}), the rank-2 field 
 $C_{\mu_{1} \mu_{2}}$ possesses a cubic interaction term but has no 
 quadratic term. This has been a severe obstacle to the appearance of 
 a Yang-Mills-like structure in the supermatrix model, because it has been
 impossible to identify $C_{\mu_{1} \mu_{2}}$ with the commutators of
 the rank-1 fields $[B_{\mu_{1}}, B_{\mu_{2}}]$ (or $[A_{\mu_{1}},
 A_{\mu_{2}}]$). In the eleven-dimensional case, this difficulty has
 been overcome  
 in \cite{0201183} through the addition of a mass term, and we thus expect 
 this model to contain the massive IIB matrix model, the bosonic part of which
 has been studied in \cite{0103192} to investigate perturbation theory around
 noncommutative curved-space backgrounds.

 \subsubsection{Resolution of the equations of motion}
   We proceed to search for possible curved-space classical configurations
 solving the equations of motion that follow from the action 
 (\ref{AZaction3}). To get a clearer picture of the problem, we now set 
 the fermions and the positive mass-squared bosonic fields to zero: 
  \begin{eqnarray}
   \psi = W = A_{\mu} = H_{\mu_{1} \cdots \mu_{4}} = Z_{\mu_{1} \cdots
   \mu_{5}} = 0. \label{AZ2trivial}
  \end{eqnarray}
 Since their masses are positive (at least in the spatial directions,
 while the time-like direction of quantum fields is generally unphysical), 
 (\ref{AZ2trivial}) is a stable classical solution. 
 Furthermore, we choose to identify the tachyonic ten-dimensional vector field 
 $B_{\mu}$, rather than the well-defined $A_{\mu}$ with the bosonic fields 
 of the massive IIB matrix model, in order to obtain a possibly stable 
 curved-space classical solution. 
 The classical equations of motion for the remaining tachyonic fields 
  $B_{\mu}$ and $C_{\mu \nu}$ following from (\ref{AZaction3}) are
  \begin{eqnarray}
   B_{\mu} &=& - i \mu^{-1} [B^{\nu}, C_{\mu \nu}], \label{AZeqB+} \\
   C_{\mu \nu} &=& - i \mu^{-1} ( [B_{\mu}, B_{\nu}]
    + [ {C_{\mu}}^{\rho}, C_{\nu \rho}] ). \label{AZeqC+}
  \end{eqnarray}
 Although it is difficult to solve these equations in full generality,
 the equation of motion for $C_{\mu \nu}$ suggests to take 
 $C_{\mu \nu} \propto [B_{\mu}, B_{\nu}]$ for $B_{\mu}$'s satisfying a fairly
 simple commutator algebra. If we look for objects having a clear geometrical 
 interpretation, it is tempting to look for solutions building fuzzy spheres.
 
 \paragraph{$S^{2} \times S^{2} \times S^{2}$ classical solution}
 \hspace{0mm}  

 The simplest tentative solution is the product of three fuzzy 2-spheres 
 with the symmetry $SO(3) \times SO(3) \times SO(3)$. Such a system is
 described by $N \times N$ hermitian matrices building a representation
 of the $so(3) (\sim su(2))$ Lie algebra in the following way:
  \begin{equation}
   [B_{i}, B_{j}]  = i \mu r
   \epsilon_{ijk} B_{k}, \hspace{12mm} B_{1}^{2} + B_{2}^{2} + B_{3}^{2} =
   \mu^{2} r^{2} \frac{N^{2} - 1}{4} {\bf 1}_{N \times N} 
   \textrm{ for } (i,j,k=1,2,3)\label{AZso3so3so3} 
  \end{equation}
 with similar relations for $i,j,k=4,5,6$ and $i,j,k=7,8,9$, trivial 
 commutators for indices that do not belong to the same group of 3,
 and $B_{0} = 0$ ($\epsilon_{ijk}$ is defined as usually).

 This set of fields (\ref{AZso3so3so3}) describes a space formed
 by the Cartesian product of three fuzzy spheres located in the directions
 $(x_{1}, x_{2}, x_{3})$, $(x_{4}, x_{5}, x_{6})$ and $(x_{7}, x_{8},
 x_{9})$, whose radii are all $\mu r \sqrt{N^{2}-1}/2$.
 $(N^{2}-1)/4$ is the quadratic Casimir operator of the $so(3)$ 
 Lie algebra. 
 Note that any positive-integer value of $N$ is possible here, since
 $N$ indexes the dimensions 
 of irreducible representations. For $SO(3)$, the irreps have dimensions
 $N=2j+1$, for all integer values of the spin $j$. However, we can also use 
 spinorial representations with half-integer spins in this case.   
 We have to consider this classical solution instead of the single $S^{2}$ 
 fuzzy sphere
  \begin{eqnarray}
   [B_{i}, B_{j}] = i \mu r \epsilon_{ijk} B_{k} \textrm{ (for 
   $i,j,k =1,2,3$)}, \hspace{3mm} B_{\mu} = 0 \textrm{ (for $\mu=0, 4,
   5, \cdots, 9$)},
  \end{eqnarray}
 because the solution $B_{4} = \cdots = B_{9} = 0$ is unstable in the 
 directions 4 to 9 due to the negative mass-squared\footnote{The classical 
 solution with $B_{0} = 0$ has no problem, because it has a positive mass 
 unlike the other directions of the field $B$.} of the rank-1 fields 
 $B_{\mu}$. Without restricting the generality, we can focus on the first 
 sphere located in the direction $(x_{1}, x_{2}, x_{3})$, since the three 
 fuzzy spheres all share the same equations of motion.
 
 In the framework of fuzzy 2-spheres, we can solve the equations of motion 
 (\ref{AZeqB+}) and (\ref{AZeqC+}) with the following ansatz  for the rank-2 
 field $C_{ij}$:
  \begin{eqnarray}
   C_{ij} = f(r) \epsilon_{ijk} B_{k}, \label{AZeqC+ansatz}
  \end{eqnarray}
  where $f(r)$ is a function depending on the radius parameter $r$.
  Indeed, the equation of motion (\ref{AZeqC+}) reduces then to:
  \begin{eqnarray}
    \epsilon_{ijk} B_{k} (-f(r) + r + r f^{2}(r) ) =
   0. \label{AZeqC+1} 
  \end{eqnarray}
  (\ref{AZeqC+1}) has two solutions: $f_{\pm} (r) = \frac{1 \pm
  \sqrt{1 - 4r^{2}}}{2r}$. When we plug this result in the equation of motion
  for $B$ (\ref{AZeqB+}), this leads to
  \begin{eqnarray}
   B_{i} ( 1- 2 r f_{\pm}(r) ) = 0.
  \end{eqnarray}
  This gives the same condition on the radius parameter $r$ for 
  both $f_{+}(r)$ and $f_{-}(r)$, namely:
  \begin{eqnarray}
    \sqrt{1 - 4r^{2} } = 0. \label{AZeqBC+so3}
  \end{eqnarray}
  Therefore, when we assume the ansatz (\ref{AZeqC+ansatz}), we obtain 
  the classical solution (\ref{AZso3so3so3}) with the radius parameter 
  set to $r = \frac{1}{2}$, which is fortunately real. Indeed, $r^2 \leq 0$ 
  would indicate that the fuzzy sphere solution is unstable. For example, 
  in the IIB massive matrix model described by (\ref{IIBtmass}), the sign of 
  the squared radius of the fuzzy 2-sphere is linked to the sign of the mass 
  term in the action and it would become negative for a correct-sign mass 
  term, which is to be expected, since in that case, the trivial commutative 
  solution becomes the stable vacuum of the theory.  

  We next want to discuss the stability of the 
  $S^{2} \times S^{2} \times S^{2}$ classical solution in more qualitative
  terms\cite{0101102}. To this end, we compare the energy of 
  the trivial commutative solution $B_{\mu} = 0$ with that of the
  fuzzy-sphere solution. The classical energy for $B_{\mu} = 0$ is
  obviously $E_{\tiny{ B_{\mu}=0}} = S_{\tiny{ B_{\mu}=0}} =
  0$.\footnote{Since we now consider a classical solution with $B_{0}
  = 0$ (thus no need  of Wick rotation), the energy is simply equal to the
  classical action in which  we substitute the solution.}  
    
  In the $S^{2} \times S^{2} \times S^{2}$ fuzzy-sphere background, 
  the 2-form field $C_{ij}$ is
  \begin{eqnarray}
   C_{ij} = \epsilon_{ijk} B_{k}.  
  \end{eqnarray}
 Therefore, the total energy is
  \begin{eqnarray}
  E_{\tiny S^{2}}
  &=& S_{\tiny S^2} = - 64 \mu \sum_{\mu = 1}^{9} Tr(B_{\mu}
  B^{\mu} )
  =  - 3 \times 64 \mu \sum_{i=1}^{3}  Tr(B_{i}
  B_{i}) \nonumber \\ 
  &=& - 12 \mu^{3} N (N-1)(N+1). \label{AZ2so3energy} 
  \end{eqnarray}
 This result shows that the $S^{2} \times S^{2} \times S^{2}$ fuzzy-sphere 
 classical solution has a lower energy compared to the trivial commutative
 solution and hence a higher probability. 

 \paragraph{Other curved-space solutions and the fuzzy 8-sphere}
 \hspace{0mm}

  So far, we have considered the simplest curved-space solution $S^{2} 
  \times S^{2} \times S^{2}$. Here, we consider the other curved-space 
  solutions. We focus on the fuzzy 8-sphere solution,
  which has the $SO(9)$ rotational symmetry and is spanned by the
  nine-dimensional space. We assume the following fuzzy sphere
  solution
  \begin{eqnarray}
   B^{(8)}_{0} = 0, \hspace{2mm} B^{(8)}_{p} = \frac{\mu r}{2} G^{(8)}_{p},
  \end{eqnarray}
 where $p$ runs over the space component $1,2,\cdots,9$, and $G^{(8)}_{p}$
 is already defined by (\ref{fspsolution}). We likewise assume the
 ansatz for the rank-2 field $C_{\mu \nu}$ as
  \begin{eqnarray}
   C^{(8)}_{0 p} = 0, \hspace{2mm}
   C^{(8)}_{pq} = -i \mu^{-1} g(r) [B^{(8)}_{p}, B^{(8)}_{q}]. 
  \end{eqnarray}
  Then, the equation of motion (\ref{AZeqC+}) implies
  \begin{eqnarray}
   \frac{-i}{\mu} [B^{(8)}_{p}, B^{(8)}_{q}]
  ( -g(r) + 1 + 7 r^{2} g^{2}(r) ) = 0.
  \end{eqnarray}
 We again have two choices for the function $f(r)$:
  \begin{eqnarray}
   g_{\pm}(r) = \frac{1 \pm \sqrt{1 - 28 r^2}}{14 r^{2}}.
  \end{eqnarray}
 The equation of motion (\ref{AZeqB+}) for the rank-1 field $B^{(8)}_{p}$
 gives 
  \begin{eqnarray}
   B^{(8)}_{p} ( 1 - 8r^{2} g_{\pm} (r) ) = 0.
  \end{eqnarray}
 Now, unlike the case of the $S^{2} \times S^{2} \times S^{2}$ fuzzy
 sphere, $1 - 8 r^{2} g_{-}(r) = 0$ does not have any real positive
 solution for $r$. However, there is exactly one such solution for $1 - 8r^{2}
 g_{+}(r) = 0$, which is $r = \frac{1}{8}$.

 More generally, for an $S^{2k}$ fuzzy sphere $B^{(2k)}_{p} =
 (\frac{\mu r}{2}) G^{(2k)}_{p}$, the same ansatz would give
 \begin{eqnarray}
 g_{\pm}(r) &=& \frac{1 \pm \sqrt{1 - 4(2k-1)r^2}}{2(2k-1)r^{2}} \; ,
 \nonumber\\
 1-2kr^{2} g_{\pm} (r)&=& 0 \; , \textrm{ solvable only for } g_{+} (r) 
 \textrm{ at }  r=\frac{1}{2k}.\nonumber
  \end{eqnarray}
 We discuss the stability of the $S^{8}$ fuzzy-sphere classical
 solution by computing its classical energy. At the classical level,
 we obtain
  \begin{eqnarray}
   E_{\tiny S^{8}} = - \frac{5}{8} \mu^{3} n(n+8)
   N_{4}. \label{AZ2so9energy} 
  \end{eqnarray}  
  $N_{4}$ is given by (\ref{sizefsp}), and we recall that this is
  given by $N_{4} = (n+1)(n+2)(n+3)^{2}(n+4)^{2}(n+5)^{2}(n+6)(n+7) /
  302400$. In contrast with the $S^{2} \times S^{2} \times S^{2}$
  case, $N$ can take here only certain precise  
  values. For example, the smallest non-trivial representation ($n=1$)
  has dimension 16, the following one ($n=2$) 126, then 672, etc...
  The classical energy for the $S^{2} \times S^{2} \times S^{2}$
  fuzzy-sphere solution is of the order ${\cal O}( - \mu^3 n^3 ) = {\cal 
  O}(- \mu^3 N^3)$ while that of the  fuzzy 8-sphere solution 
  is of the order ${\cal O}(- \mu^3 n^{12}) = {\cal
  O}( - \mu^3 N^{\frac{6}{5}})$. Therefore, at large $N$, the 
  $S^{2} \times S^{2} \times S^{2}$ triple fuzzy-sphere solution is 
  energetically favored compared to the $S^{8}$ solution at the equal
  size $N$ of 
  the matrices. The presence of a spherical solution for all $N$ in the 
  $S^{2} \times S^{2} \times S^{2}$ case may indeed be a stabilizing
  factor. On the other hand, at equal value of $n$, whose physical meaning 
  is less clear, the fuzzy 8-sphere solution has lower energy.  

  We have seen that the fuzzy 8-sphere is energetically favored. This
  is in contrast to the matrix model with the Chern-Simons term
  (\ref{IIBfsp}), in which the classical energy is given by
  (\ref{fspclassical}) and thus the fuzzy 8-sphere is positive.
  In this case, the fuzzy 8-sphere is less energetically favored than
  the trivial commutative background. In this sense, this situation has more
  similarity to the IIB matrix model with the tachyonic mass term
  (\ref{IIBtmass}), in  
  which the higher-dimensional fuzzy sphere, as well as the $S^{2}$
  fuzzy sphere, is more energetically favored, as we have seen in
  (\ref{fstmassclassical}). 

  The single fuzzy $q$-spheres for $q \leq 7$ do not 
  constitute a stable classical solution of our model. When the $S^{q}$ sphere
  occupies the direction $x_{1}, x_{2}, \cdots, x_{q+1}$, the solution
  $B^{(q)}_{q+2} = B^{(q)}_{q+3} = \cdots = B^{(q)}_{9} =0$ is
  trivially unstable because of the negative mass-squared. Whereas, 
  the Cartesian product of several fuzzy spheres, such as $S^{2}
  \times S^{5}$, is a possible candidate for a stable classical solution.  

 \subsubsection{Nucleation process of spherical branes}
Starting from a vacuous spacetime, it is interesting to try to guess how
spherical brane configurations could be successively produced through 
a sequence of decays into energetically more favorable meta-stable brane 
systems. The reader may have noticed that we have so far limited ourselves to 
the study of curved branes building irreducible representations of their
symmetry groups. This could seem at first to be an unjustified prejudice,
but it turns out that such configurations are energetically favored
at equal values of $N$. For example, for $SO(3)$, an irreducible representation
${\cal R}_{N}$ of dimension $N$ contributes as
 \begin{equation}
  E_{{\cal R}_{N}} = - 4 \mu^{3} (N^3-N) \nonumber
 \end{equation}
per fuzzy 2-sphere, while a reducible representation ${\cal R}_{N_1} \oplus 
\ldots \oplus {\cal R}_{N_m}$ of equal dimension $N_1+\ldots+N_m=N$ would
contribute as
 \begin{equation}
  E_{{\cal R}_{N_1} \oplus \ldots \oplus {\cal R}_{N_m}} = - 4 \mu^{3} 
  \sum_{i=1}^{m} (N_i^3-N_i).\nonumber
 \end{equation}
This is obviously a less negative number, especially for big values of $N$.
A similar conclusion was reached in \cite{0101102} for the case of a Euclidean
three-dimensional IIB matrix model with a Chern-Simons term and it seems to be 
a fairly general feature of matrix models admitting non-trivial classical 
solutions. This property is particularly clear for low-dimensional branes, 
since the classical energy is of order ${\cal O}(- \mu^3 N^3)$ for the 
$S^{2}$ fuzzy sphere,
but it remains true for any fuzzy $2k$-sphere solution, whose energy
is of order ${\cal O}( - \mu^3 N^{1+4/(k(k+1))})$, which also shows
that low-dimensional configurations are favored. As hinted for in the 
preceding subsection, this latter fact can be physically understood by 
remarking that there are more irreps available for low-dimensional fuzzy 
spheres, which makes it easier for them to grow in radius through 
energetically favorable configurations. A third obvious fact is that 
configurations described by representations of high dimensionality are 
preferred.

Put together, these comparisons give us a possible picture for the brane 
nucleation process in this and similar matrix models. As they appear, 
configurations of all spacetime dimensions described by small representations
will be progressively absorbed by bigger representations to form irreducible 
ones, that will slowly grow in this way to bigger values of $N$.
Parallel to that, branes of higher dimensionalities will tend to decay into a 
bunch of branes of smaller dimensionalities, finally leaving only
2-spheres and noncommutative tori of growing radii. If the size of the
hermitian matrices is left open, as is usually the case in completely
reduced models, where the path integration contains a sum on that
size, no configuration will be truly stable, since the size of the
irreps will grow continuously. 

Of course, this is a relatively qualitative study, which could only be proven
correct by a full quantum statistical study of the model. However, it seems 
to be an interesting proposal for the possible physics of such theories.

\subsubsection{Supersymmetry}
 We next comment on the structure of the supersymmetry.
 The biggest difference with the purely cubic supermatrix model, due to
 the addition of the mass term, is that this model is not invariant 
 under the inhomogeneous supersymmetry
  \begin{eqnarray}
    \delta_{\tiny \textrm{inhomogeneous}} m=0, \hspace{3mm}
    \delta_{\tiny \textrm{inhomogeneous}} \psi = \xi, \label{AZinhomo}
  \end{eqnarray}
 which is a translation of the fermionic field.
 However, this model has 2 homogeneous supersymmetries in ten dimensions, 
 which are part of the $osp(1|32,R)$ symmetry:
   \begin{eqnarray}
    \delta_{\epsilon} M = \left[ \left( \begin{array}{cc} 0 & \epsilon \\ i 
    {\bar \epsilon} & 0 \end{array} \right), 
     \left( \begin{array}{cc} m & \psi \\ i 
    {\bar \psi} & 0 \end{array} \right) \right]
  = \left( \begin{array}{cc} i (\epsilon {\bar \psi} - \psi {\bar
    \epsilon} )   & - m \epsilon \\ i {\bar \epsilon} m & 0
    \end{array} \right), \label{AZhomo}
   \end{eqnarray}
 which transforms the bosonic and fermionic fields as
  \begin{eqnarray}
   \delta_{\epsilon} m = i (\epsilon {\bar \psi} - \psi {\bar
    \epsilon} ), \hspace{3mm}
   \delta_{\epsilon} \psi = - m \epsilon. \label{AZ3SUSY}
  \end{eqnarray}

  In the IIB matrix model, the supersymmetry has to balance between a quartic 
  term $Tr([A_{\mu},A_{\nu}])^2$ and a trilinear contribution 
  $Tr {\bar \psi} \Gamma^{\mu} [A_{\mu}, \psi]$ in the action 
  (\ref{AZM31IKKT}), which implies that the supersymmetry
  transformation of the  
  fermionic field has to be bilinear in the bosonic field. On the other hand, 
  the homogeneous supersymmetries are all linear in the fields in the purely 
  cubic supermatrix model\cite{0102168,0103003}. By incorporating the mass 
  term, we are allowed to integrate out the rank-2 field $C_{\mu_{1} \mu_{2}}$
  by solving the classical equation of motion iteratively as
  in~\cite{0201183}\footnote{The Yang-Mills-like structure of the
  homogeneous supersymmetry transformation on the fermion comes from $-
  \frac{1}{2} C_{\mu \nu} \Gamma^{\mu \nu} \epsilon$, which is a part
  of $\delta_{\epsilon} \psi = - m \epsilon$. The explicit form of the 
  iterative solution of the equations of motion (\ref{AZeqB+}) and
  (\ref{AZeqC+}) is 
 \begin{eqnarray}
    C_{\mu \nu} &=& - i \mu^{-1} [B_{\mu}, B_{\nu}] 
   + i \mu^{-3} [[B_{\mu}, B_{\rho}], [B_{\nu}, B^{\rho}]] \nonumber \\
   & & - i \mu^{-5} [[B_{\mu}, B_{\rho}], [[B_{\nu}, B_{\chi}],[B^{\rho},
   B^{\chi}]] ]
   + i \mu^{-5} [[B_{\nu}, B_{\rho}], [[B_{\mu}, B_{\chi}],[B^{\rho},
   B^{\chi}]] ] 
   + {\cal O}(\mu^{-7}). \nonumber
 \end{eqnarray}
 }.   

 Thanks to this procedure, the homogeneous SUSY transformation for the 
  fermionic field becomes
  \begin{eqnarray}
  \delta_{\epsilon} \psi = \frac{i}{2} [B_{\mu_{1}}, B_{\mu_{2}}]
  \Gamma^{\mu_{1} \mu_{2}} \epsilon + \cdots, \label{AZ2SUSYferm}
  \end{eqnarray}
  while the transformation of the field $B_{\mu}$ is
  \begin{eqnarray}
   \delta_{\epsilon} B_{\mu} = - \frac{1}{32} tr_{32 \times 32}
      ( i (\epsilon {\bar \psi} - \psi {\bar \epsilon}) \Gamma_{\mu
      \sharp} ) = - \frac{i}{16} {\bar \epsilon} \Gamma_{\mu \sharp}
      \psi. \label{AZ2SUSYvec}
   \end{eqnarray}

  In that sense, the mass term is essential to realize the Yang-Mills-like 
  structure for the homogeneous supersymmetries. On the other hand, if we
  want to preserve the homogeneous supersymmetries, we cannot just put 
  a mass term for $C_{\mu_{1} \mu_{2}}$ by hand. The 
  $osp(1|32,R)$ symmetry forces all fields to share the same mass,
  since they all 
  lie in the same multiplet. In particular, we are forced to introduce a 
  mass term for the fermions as well, which breaks the inhomogeneous
  supersymmetries. In other words, it seems difficult to have
  super-Yang-Mills-type structure for both homogeneous and
  inhomogeneous supersymmetries in the context of supermatrix models.
  
  Indeed, in contrast with the purely cubic supermatrix 
  model\cite{0102168,0103003}, which has twice as many supersymmetry
  parameters, 
  the massive supermatrix model has only ${\cal N}=2$ supersymmetries
  in ten dimensions, because it lacks the inhomogeneous supersymmetries.  
  In consequence, we cannot realize the translation of the vector field 
  $A_{\mu}$ as a commutator of two linear combinations of the homogeneous 
  and inhomogeneous supersymmetries (\ref{AZ3SUSY}) as in the IIB matrix model,
  where it leads to the interpretation of the eigenvalues of $A_{\mu}$ as 
  spacetime coordinates. On the contrary,  
   \begin{eqnarray}
   [\delta_{\epsilon}, \delta_{\chi}] m = i [(\epsilon {\bar \chi} -
   \chi {\bar \epsilon}), m], \hspace{3mm}
   [\delta_{\epsilon}, \delta_{\chi}] \psi = i (\epsilon {\bar \chi} - 
   \chi {\bar \epsilon} ) \psi
   \end{eqnarray}
  vanishes up to an $sp(32,R)$ rotation. This problem 
  is a serious obstacle for the identification 
  of the supersymmetry of this model with that of the IIB matrix model. 
  More analysis will be reported elsewhere.

 \subsection{Other related studies}
  We have so far gone over the author's works\cite{0102168,0209057}
  concerning the supermatrix models. We conclude this section by
  touching on other related studies. 
  In \cite{0201183}, Bagnoud, Carlevaro and Bilal discussed the
  $osp(1|32,R)$ supermatrix model in the twelve-dimensional and the 
  eleven-dimensional contexts. Especially, they indicated how to
  perform the infinite momentum frame limit and used the matrix
  version of the T-duality to obtain a supersymmetric matrix quantum
  mechanics. They introduced the quadratic terms as in
  Sec. \ref{sec0209057}, and resorted to the perturbative technique
  that led to an infinite tower of the higher-order interaction among
  the physical fields. On the other hand, the lower-order term
  reproduced the BFSS model with an additional mass term together with 
  interaction terms involving the five-brane degrees of freedom.

  Other attempts for a background-independent matrix models have been
  achieved by the cubic action of the exceptional Jordan
  algebra. Firstly, L. Smolin proposed a matrix model based on the
  exceptional Jordan algebra whose automorphism group is $F_{4}$ (in
  this sense, we refer to this model as "the $F_{4}$ model" here). The 
  exceptional Jordan algebra is related to 
  the octonion, and gives a theory that has a compactification which
  reduces to the matrix string theory\cite{9703030}.
  Its complex extension has been proposed in \cite{0110106}. The
  $E_{6}$ exceptional Lie algebra is known to be the automorphism of
  the complexification. Therefore, the $E_{6}$ matrix model has twice
  as many degrees of freedom as the $F_{4}$ model, and derived the
  effective action of the matrix string type.

 \section{Matrix model with manifest general coordinate invariance}
  \label{manifestgr}
  In the pervious section, we have investigated various aspects of the 
  supermatrix models based on the $osp(1|32,R)$ super Lie algebra.
  This extension of the IIB matrix model poses a lot of interesting
  questions. Especially, in Sec. \ref{sec0102168l}, we have
  investigated the gauged $gl(1|32,R) \otimes gl(R)$ model as a
  candidate for the extension of the IIB matrix model equipped with
  the local Lorentz invariance. The key idea is to render the
  parameter of the $SO(9,1)$ Lorentz transformation dependent on the
  $u(N)$ matrices in a nontrivial manner. This is an essential
  alteration from the IIB matrix model, since the spacetime is
  interpreted as being immersed in the eigenvalues of the bosonic
  matrices $A_{\mu}$. In addition, the $u(1|16,16)$ (and its analytic
  connection $gl(1|32,R)$) supermatrix model
  incorporates the higher-rank tensor field. Especially, the rank-3
  field has a possibility to be identified with the spin connection
  in the quantum field theory on the curved spacetime.

  In this section, we review the author's work\cite{0204078} that
  inherits the above idea. The most serious obstacle of the
  $gl(1|32,R)$ supermatrix model is that the effective action reduced
  to the ten dimensions by the Wigner-In\"{o}n\"{u} contraction
  finally vanishes. This is ascribed to the fact that the element of 
  $gl(1|32,R)$ of all ranks are allocated for both the
  matter fields and the parameter of the local Lorentz transformation.
  These two elements had to be mixed in the eleven-dimensional theory, 
  due to the self-duality condition of the gamma matrices
  (\ref{duality}). Namely, even when the rank-6 fields come in the local
  Lorentz parameter due to the rule (\ref{higherevengauged}), this is
  identified with the rank-5 fields due to the self-duality rule
  (\ref{duality}). By the same token, the rank-7 matter fields coming
  from (\ref{higheroddgauged}) are tantamount to the rank-4 fields.

  In \cite{0204078}, we address the above approach 
  in a different angle without the supermatrix model. We start from the 
  ten-dimensional action from the outset, without the reduction from
  the eleven-dimensional model. Since we start from the
  ten-dimensional field theory, we can separate the matter fields and
  the local Lorentz parameter. We allocate the
  odd-rank fields for the matter fields, and the even-rank fields go
  for the local Lorentz parameter. In addition, we explicitly give an
  expansion of the matrices with the differential operators, to embody 
  the idea of the identification of the matrices and the differential
  operators. In these senses, the work\cite{0204078} gives a clearly 
  different angle from our previous work in Sec. \ref{sec0102168l}. 

 \subsection{Bosonic part of the matrix model}
  This time, we start from the following action, defined in the
  ten-dimensional spacetime from the outset:
  \begin{eqnarray}
   S = Tr_{N \times N} [ tr_{32 \times 32} V(m^{2})]. \label{AZ2action}
  \end{eqnarray}
   The uppercase $Tr$ and the lowercase  $tr$ denote the trace for the
 $N \times N$ and $32 \times 32$ matrices, respectively.
  The indices $\mu, \nu, \cdots$ and $i, j, \cdots$ both run
 over $0,1,\cdots, 9$.The former is allocated for the flat Minkowskian 
 spacetime, and the latter is for the curved spacetime. $V(x)$ is some 
 function determined later, and $m$ consists of the odd-rank fields:
  \begin{eqnarray}
   m = m_{\mu} \Gamma^{\mu} + \frac{i}{3!} m_{\mu_{1} \mu_{2} \mu_{3}} 
   \Gamma^{\mu_{1} \mu_{2} \mu_{3}} 
  - \frac{1}{5!} m_{\mu_{1} \cdots \mu_{5}} \Gamma^{\mu_{1} \cdots
   \mu_{5}} 
  - \frac{i}{7!} m_{\mu_{1} \cdots \mu_{7}} \Gamma^{\mu_{1} \cdots
   \mu_{7}} 
  + \frac{1}{9!} m_{\mu_{1} \cdots \mu_{9}} \Gamma^{\mu_{1} \cdots
   \mu_{9}}. \label{AZ2mex}
  \end{eqnarray}
  Here, $m_{\mu_{1} \cdots \mu_{2n-1}}$ ($n=1,2,3,4,5$) are all
  hermitian matrices, and this satisfies $\Gamma^{0} m^{\dagger}
  \Gamma^{0} = m$. In this sense, our action is shown to be hermitian
  as
  \begin{eqnarray}
   S^{\dagger} = Tr[tr V((m^{\dagger})^{2})]
   = Tr[ tr V((\Gamma^{0} m \Gamma^{0})^{2})]
   = Tr[ tr V(m^{2})] = S. 
  \end{eqnarray}
  Here, we exclude the odd-power of $m$ in the action, because
  $m^{2k+1}$ all consists of the odd-rank gamma matrices and thus
  couple the fermions of the different chirality.

 \subsubsection{Identification of large $N$ matrices with differential
 operators} 
\begin{figure}[htbp]
   \begin{center}
    \scalebox{.7}{\includegraphics{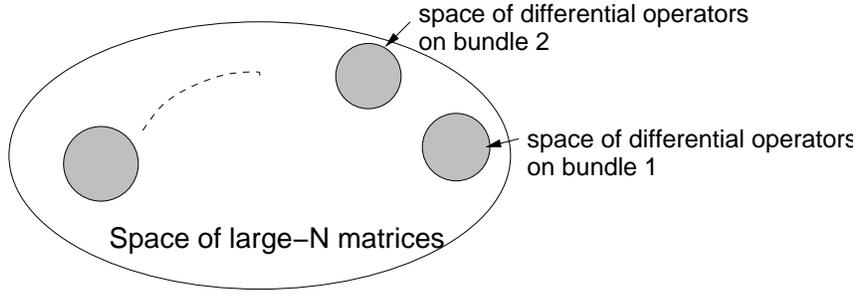} }
   \end{center}
    \caption{The way the spaces of the differential operators are
      embedded in the space of large $N$ matrices.}
   \label{diff-matrix}
  \end{figure}

  We identify the space of the large-$N$ matrices with that of the
  differential operators. By this identification, we can describe the
  differential 
 operators on an arbitrary spin bundle over an arbitrary manifold in the
 continuum limit simultaneously, because they are embedded in the
 space of large $N$ matrices, as is depicted in Fig. \ref{diff-matrix}.
 
\begin{figure}[htbp]
   \begin{center}
    \scalebox{.7}{\includegraphics{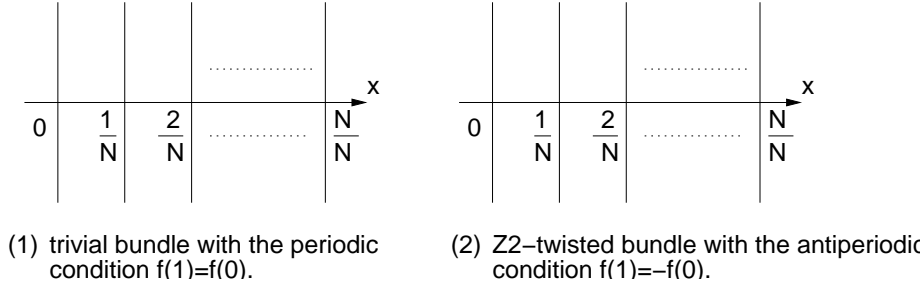} }
   \end{center}
    \caption{Differential operators on (1)the trivial bundle and
      (2)the $Z_{2}$-twisted bundle over $S_{1}$.}  
   \label{diff-matrixex}
  \end{figure}

 To illustrate this idea, we visit several simple examples on the
  differential operators of the scalar fields on two different bundle
  over the $S_{1}$ circle (see Fig. \ref{diff-matrixex}).
   We first consider the trivial bundle with the periodic condition
 $f(1) = f(0)$. We discritize the region $0 \leq x \leq 1$ into small
 slices of spacing $\epsilon = \frac{1}{N}$.
 Then, the differential operator is approximated by the finite
 difference as 
  \begin{eqnarray}
   \partial_{x} f \left( \frac{k}{N} \right) \to \frac{1}{2} \left( 
   \frac{f (\frac{k+1}{N}) - f(\frac{k}{N})}{\epsilon}  
 + \frac{f(\frac{k}{N}) - f(\frac{k-1}{N})}{\epsilon} \right)
 = \frac{N}{2} \left( f \left( \frac{k+1}{N} \right) -  
                      f \left( \frac{k-1}{N} \right) \right).
  \end{eqnarray}
  Due to the periodic condition, this finite difference is expressed
  by the large $N$ matrix as
  \begin{eqnarray}
  \partial_{x} \to  A = \frac{N}{2} \left( \begin{array}{cccccc}
  0  &  1 &   &   &   &  -1 \\
  -1 &  0 & 1 &   &   &     \\
     & -1 & 0 & 1 &   &     \\
     &    &   & \ddots &  &   \\
  1  &    &   &   & -1 & 0 
   \end{array} \right).
     \end{eqnarray}
 We next consider the similar problem with respect to the $Z_{2}$-twisted
 bundle, in which the antiperiodic condition $f(1) = - f(0)$ is
 imposed. Paying attention to this antiperiodicity, we understand that 
 the finite difference in the discritized space is expressed as 
  \begin{eqnarray}
  \partial_{x} \to  A = \frac{N}{2} \left( \begin{array}{cccccc}
  0  &  1 &   &   &   & 1 \\
  -1 &  0 & 1 &   &   &     \\
     & -1 & 0 & 1 &   &     \\
     &    &   & \ddots &  &   \\
  - 1  &    &   &   & -1 & 0 
   \end{array} \right).
  \end{eqnarray}

 In the following, from the big space of
 large $N$ matrices, we pick up a subspace consisting of the
 differential operators over one manifold.
 We  regard $m$ as the differential operators
 over this manifold. Then, we analyze the effective theory of the
 fields appearing in the expansion of the differential 
 operators (the explicit form is given later). 

 The space of the differential operator is infinite-dimensional.
 The trace $Tr$ for this space is generally divergent. However, we can 
 render this trace finite by choosing the function $V(m^{2})$ damping
 rapidly. We would like to identify this matrix $m$ with the Dirac
 operator in the curved spacetime. Clearly, the Dirac operator has a
 dimensionality [(length)$^{-1}$], while $m$ is a dimensionless
 quantity. Then, we introduce a constant $\tau$ with the
 dimensionality [(length)$^{2}$], and express $m$ as
  \begin{eqnarray}
    m &=& \tau^{\frac{1}{2}} D, \textrm{ where } \nonumber \\
   D &=& A_{\mu} \Gamma^{\mu} + \frac{i}{3!} A_{\mu_{1} \mu_{2} \mu_{3}} 
   \Gamma_{\mu_{1} \mu_{2} \mu_{3}}
   - \frac{1}{5!} A_{\mu_{1} \cdots \mu_{5}}
     \Gamma^{\mu_{1} \cdots \mu_{5}}
   - \frac{i}{7!} A_{\mu_{1} \cdots \mu_{7}} \Gamma^{\mu_{1} \cdots
   \mu_{7}} 
   + \frac{1}{9!} A_{\mu_{1} \cdots \mu_{9}} \Gamma^{\mu_{1} \cdots
   \mu_{9}}. \nonumber \\ \label{AZ2expansionD}
  \end{eqnarray}
  Here, $D$ and $A_{\mu_{1} \cdots \mu_{2n-1}}$ are matrices of
  dimensionality [(length)$^{-1}$].  $\tau$ is similar to the Regge
  slope $\alpha'$ in string theory, 
 and is introduced as a reference scale.
 This parameter $\tau$ is not a cut-off parameter.
 $V(m^{2})$ is an exponentially decreasing function, and
 the damping factor is supplied by the action itself.
 Since $V(m^{2})$ is a function of the dimensionless quantity $m$,
 $\tau$ represents a damping scale.
 When we approximate the differential operators by finite $N$
 matrices, an $N$-dependent ultraviolet cut-off naturally appears.
 When we take $N$ to infinity, this cut-off becomes
 infinitely small. But the scale $\tau$ is completely
 independent of this ultraviolet cut-off, and takes a constant value
 even in the large $N$ limit.  Thus, we can fairly take $N$ to
 infinity and identify the large $N$ matrices with the differential
 operators, at least when we investigate the effective theory at tree
 level. In this sense, our model {\it differs} from the induced
  gravity.

  The matrices $A_{\mu_{1} \cdots \mu_{2n-1}}$ are expanded by the
  number of the derivatives. The hermiticity of $A_{\mu_{1} \cdots
  \mu_{2n-1}}$ leads us to expand it by the anti-commutator of the
  derivatives as
  \begin{eqnarray}
   A_{\mu_{1} \cdots \mu_{2n-1}} = a_{\mu_{1} \cdots \mu_{2n-1}} (x)
   + \sum_{k=1}^{\infty} \frac{i^{k}}{2} \{ \partial_{i_{1}}
   \partial_{i_{2}} \cdots \partial_{i_{k}}, {a^{(i_{1} \cdots
   i_{k})}}_{\mu_{1} \cdots \mu_{2n-1}}(x) \}. \label{numberex}
  \end{eqnarray}
  Here, the indices $i_{1}, \cdots, i_{k}$ are symmetric, while the
  indices $\mu_{1}, \cdots, \mu_{2k-1}$ are anti-symmetric. 
  Since we identify $D$ with the extension of the Dirac operator in
  the curved spacetime, we find it natural to identify the
  coefficients $a^{(i)}_{\mu}(x)$ with the vielbein of the background
  metric. A simple dimensional analysis immediately indicates that the 
  coefficients $a^{(i_{1} \cdots i_{k})}_{\mu_{1} \cdots \mu_{2n-1}}(x)$
  has a dimensionality [(length)$^{-1+k}$]. Then, $D$ is identified with 
  the Dirac operator in the curved space as
  \begin{eqnarray}
   D = e^{\frac{1}{2}}(x) \left[ i {e_{\mu}}^{i}(x) \Gamma^{\mu} 
  \left( \partial_{i} +\frac{1}{4} \Gamma^{\nu \rho} \omega_{i \nu
  \rho}(x) \right) \right] e^{-\frac{1}{2}}(x)
  + \textrm{(higher-rank and higher-derivative terms)}. \label{matrixdirac}
  \end{eqnarray}
  
 \subsubsection{Local Lorentz invariance}
 We next clarify that our matrix model is invariant under the local
 Lorentz transformation. Here, the local Lorentz transformation is
 given by
  \begin{eqnarray}
  \delta m &=& [m, \varepsilon], \textrm{ where } \nonumber \\
  \varepsilon &=& -i \varepsilon_{\emptyset} + \frac{1}{2!}
      \Gamma^{\mu_{1} \mu_{2}} \varepsilon_{\mu_{1} \mu_{2}}
   + \frac{i}{4!} \Gamma^{\mu_{1} \cdots \mu_{4}} \varepsilon_{\mu_{1} 
 \cdots \mu_{4}}
   - \frac{1}{6!} \Gamma^{\mu_{1} \cdots \mu_{6}} \varepsilon_{\mu_{1} 
 \cdots \mu_{6}} \nonumber \\
   &-& \frac{i}{8!} \Gamma^{\mu_{1} \cdots \mu_{8}} \varepsilon_{\mu_{1} 
 \cdots \mu_{8}}
   + \frac{1}{10!} \Gamma^{\mu_{1} \cdots \mu_{10}}
 \varepsilon_{\mu_{1} \cdots \mu_{10}}.  \label{locallorentz}
  \end{eqnarray}
  $\varepsilon$ is the parameter of the local Lorentz transformation,
  and it satisfies $\Gamma^{0} \varepsilon^{\dagger} \Gamma^{0} =
  \varepsilon$. The coefficients $\varepsilon_{\emptyset}, \cdots,
  \varepsilon_{\mu_{1} \cdots \mu_{10}}$ are all hermitian matrices.
 The discussion in (\ref{higherevengauged}) immediately indicates that
 the closure of the algebra of the local Lorentz transformation
  necessitates all the even-rank fields. The invariance of the action
  is verified as
  \begin{eqnarray}
   \delta S = Tr [ tr(2V'(m^{2}) m [m,\varepsilon])] =
   0. \label{llinvariance} 
  \end{eqnarray}
  Here, $V'(x)$ denotes $V'(x) = \frac{\partial V(x)}{\partial
  x}$. The vanishing of (\ref{llinvariance}) can be verified by noting 
  the cyclic rule of the trace $Tr$. Since we are now discussing the
  space of the infinite-dimensional space, it is not necessarily
  trivial whether the trace satisfies the cyclic rule. However,
  the cyclic rule holds when the coefficients ${a^{(i_{1} \cdots
  i_{k})}}_{\mu_{1} \cdots \mu_{2n-1}}(x)$ damps rapidly.
 The only nontrivial part for the cyclic rule is the commutator
  between the derivative and the fields
  \begin{eqnarray}
   Tr([\partial_{i}, {a^{(j_{1} \cdots j_{k})}}_{\mu_{1} \cdots
   \mu_{2n-1}} (x) ])
 = \int d^{d} x \langle x | (\partial_{i}{a^{(j_{1} \cdots
   j_{k})}}_{\mu_{1} \cdots \mu_{2n-1}} (x)) | x \rangle
 =   \int d^{d} x (\partial_{i}{a^{(j_{1} \cdots
   j_{k})}}_{\mu_{1} \cdots \mu_{2n-1}} (x)) \langle x | x
   \rangle. \nonumber
  \end{eqnarray}
 However, this surface term does not contribute when the coefficients
  ${a^{(i_{1} \cdots  i_{k})}}_{\mu_{1} \cdots \mu_{2n-1}}(x)$ damp
  rapidly. 

 \subsubsection{Determination of $V(m^{2})$}
  We next determine the function $V(m^{2})$. We require that 
  the action (\ref{AZ2action}) should have the flat metric $m_{0} = i
  \tau^{\frac{1}{2}} \Gamma^{\mu} \partial_{\mu}$ as a classical
  background. However, this turns out to be a draconian constraint on
  $V(x)$. The bosonic part is expressed by the heat-kernel
  expansion. We delegate the details of the heat-kernel expansion to
  Appendix \ref{seeleydewitt}. We find it more convenient to express 
  $V(u)$ in terms of its Laplace transformation
  \begin{eqnarray}
   V(u) = \int^{\infty}_{0} ds g(s) e^{-su}, \label{laplace}
  \end{eqnarray}
 in order to see a more transparent correspondence with the heat kernel.
 Then, the action is now expressed as
  \begin{eqnarray}
  Tr [ tr V(m^{2}) ] = \int^{\infty}_{0} ds g(s) Tr[ tr e^{-s \tau
  D^{2}} ]
   = \int \frac{d^{d}x}{(2 \pi \tau)^{\frac{d}{2}}}
     \left( \sum_{k=-\infty}^{\infty} \left( \int^{\infty}_{0} 
     ds g(s) s^{-\frac{d}{2}+k} \right) \tau^{k} {\cal A}_{k} (x) 
     \right), \label{sdex}
  \end{eqnarray}
  where ${\cal A}_{k}(x)$ are the Seeley-de-Witt coefficients of the
  heat kernel $Tr[ tr (e^{-\tau D^{2}}) ]$, namely
   \begin{eqnarray}
    Tr[ tr(e^{-\tau D^{2}})] &=& \int \frac{d^{d}x}{(2 \pi
    \tau)^{\frac{d}{2}}} \sum_{k=-\infty}^{\infty} \tau^{k} {\cal
    A}_{k}(x). 
   \end{eqnarray}
  Here, $k$ runs all the integers. We can show that the
  half-integer $k$ can be excluded in the following way.
  The coefficients ${\cal A}_{k}(x)$ has a dimensionality
  [(length)$^{-2k}$]. On the other hand the terms $\prod_{k=1}^{n}
   (\partial_{i_{1}} \cdots \partial_{i_{p_{k}}} {a^{(j_{1} \cdots
  j_{l_{k}} ) }}_{\mu_{1} \cdots \mu_{2m_{k}-1}} (x) )$ has a
  dimensionality ${\cal D} = -n + \sum_{i=1}^{n}(-p_{i} +
  l_{i})$. Therefore, this belongs to the coefficients ${\cal
  A}_{- \frac{\cal D}{2} } (x)$. On the other hand, this term has 
  $-n + \sum_{i=1}^{n} (2m_{i} + p_{i} + l_{i})$ indices, which is
  equal to ${\cal D}$ modulo 2. Therefore, when ${\cal D}$ is odd,
  these terms cannot contract the indices, and thus cannot survive in
  the action. If such terms existed, they would no longer be a scalar
  and thus would violate the Lorentz invariance of the action. 
  In this way, we eliminate the contribution of the half-integer $k$.
 
  Then, let us think about the condition for $m_{0}$ to be a classical 
  solution. To this end, the linear terms of all the fluctuations must 
  vanish in the action. Now, only the scalar fields can contribute to
  the action, we have only to consider the linear terms
  ${a_{\mu}}^{(\mu i_{1} i_{1} i_{2} i_{2} \cdots i_{l} i_{l})} (x)$
  (for $l=0,1,2,\cdots$). Since ${a^{(i)}}_{\mu}(x)$ is identified
  with the vielbein, its cancellation means the cancellation of the
  cosmological constant. This amounts to the cancellation of the
  Seeley-de-Witt coefficient ${\cal A}_{0}(x)$.
 We disregard the derivative terms
  $\partial_{i_{1}} \cdots  {a^{(j_{1} \cdots)}}_{\mu_{1} \cdots
  \mu_{2n-1}} (x)$, because these can be eliminated by the partial
  integration. Since the fields ${a_{\mu}}^{(\mu i_{1} i_{1} i_{2}
  i_{2} \cdots i_{l} i_{l})} (x)$ has a dimensionality
  [(length)$^{2l}$], it belongs to the Seeley-de-Witt coefficient
  ${\cal A}_{-l}(x)$. Therefore, we generally demand that the coefficients
  ${\cal A}_{-l}(x)$ for $l=0,1,2,\cdots$ should all cancel.

  However, it turns out that this imposes a demanding condition on the 
  function $V(m^{2})$. The above condition is translated as
  \begin{eqnarray}
   \int^{\infty}_{0} ds g(s) s^{-\frac{d}{2} - n} = 0, \textrm{ for } 
   n=0, 1, 2, \cdots. 
  \end{eqnarray}
  This condition can further be rewritten as
  \begin{eqnarray}
   \int^{\infty}_{0} du V(u) u^{n - 1 + \frac{d}{2}} = \int^{\infty}_{0} du
   \int^{\infty}_{0} ds g(s) e^{-su} u^{n + \frac{d}{2} -1}
   = \Gamma (n + \frac{d}{2}) \int^{\infty}_{0} ds g(s)
   s^{-\frac{d}{2} - n} = 0, \label{conditionv}
  \end{eqnarray}
  for $n=0,1,2,\cdots$. One example for such a function is the
  following:
   \begin{eqnarray}
    V_{0}(u) = \frac{\partial^{\frac{d}{2}-1} (\exp(-u^{\frac{1}{4}})
    \sin (u^{\frac{1}{4}}) )}{\partial u^{\frac{d}{2}-1}}. \label{answev}
   \end{eqnarray}
  We can show that the function (\ref{answev}) satisfies the condition 
  (\ref{conditionv}) as follows\cite{Hausdorff}. 
  Firstly, we note the following integral
  \begin{eqnarray}
   \int^{\infty}_{0} dy y^{m} e^{-a y} = m! a^{-m-1}, \textrm{ where }
   a = \exp( \frac{i \pi}{4} ) = \frac{1+i}{\sqrt{2}}.  
  \end{eqnarray}
  This is a real number when $m-3$ is a multiple of 4. Taking the
  imaginary part of the both hand sides, we obtain
   \begin{eqnarray}
    \int^{\infty}_{0} dy y^{4n+3} \sin(\frac{y}{\sqrt{2}}) \exp( -
    \frac{y}{\sqrt{2}} ) = 0, \textrm{ for } n = 0, 1, 2, \cdots.
   \end{eqnarray}
  We make a substitution $u=\frac{y^{4}}{4}$ and perform the partial
  integration to obtain the solution (\ref{answev})\footnote{This is the 
  famous counter-example against Hausdorff's moment problem when
  we replace the finite integral with the infinite integral.
  When the function $V(u)$ satisfies
  \begin{eqnarray}
    \int^{1}_{0} du u^{n} V(u) = 0, \textrm{ for } n=0, 1, 2,\cdots,
    \nonumber 
  \end{eqnarray}
  The function $V(u)$ must be zero constantly. This problem is called
  "Hausdorff's moment problem"\cite{Hausdorff}. However, this is not
  the case with (\ref{conditionv}) because we are considering the
  infinite integral.}. 

  We then derive the Einstein gravity in the low energy limit, from
  this bosonic part. The linear term of the vielbein
  ${a^{(i)}}_{\mu}(x)$ is eliminated, and thus the graviton is now
  massless since the mass term vanishes due to the general coordinate
  invariance. When we retain the curved-space Dirac operator in
  the expansion (\ref{matrixdirac}), the Seeley-de-Witt expansion is
  now obtained by
   \begin{eqnarray}
    Tr[tr(e^{-\tau D^{2}})] = \int d^{d} x \frac{32}{(2 \pi
    \tau)^{\frac{d}{2}} } e(x) \left( \tau \frac{R(x)}{6} + \cdots
    \right). \label{einsteingravity}
   \end{eqnarray}
  This term should be clearly allocated for the coefficient ${\cal
  A}_{1} (x)$. 

  We next give a qualitative argument for the mass and the kinetic
  terms of the matter fields. While the explicit heat-kernel expansion 
  of (\ref{matrixdirac}) entails an involved calculation, it is easy
  to estimate which coefficient ${\cal A}_{k}(x)$ the mass term and
  the kinetic term belong to via the dimensional analysis.
  Since the coefficients $ {a^{(i_{1} \cdots i_{k})}}_{\mu_{1} \cdots 
   \mu_{2n-1}} (x)$ and ${a^{(j_{1} \cdots j_{l})}}_{\mu_{1} \cdots
   \mu_{2n-1}} (x)$ respectively have the dimensionality
   [(length)$^{-1+k}$] and [(length)$^{-1+l}$], we have
  \begin{eqnarray}
   & & \textrm{mass terms:} {a^{(i_{1} \cdots i_{k})}}_{\mu_{1} \cdots 
   \mu_{2n-1}} (x) {a^{(j_{1} \cdots j_{l})}}_{\mu_{1} \cdots
   \mu_{2n-1}} (x) \in {\cal A}_{1 - \frac{k+l}{2}} (x), \\
  & & \textrm{kinetic terms:} (\partial_{k_{1}}  {a^{(i_{1} \cdots
   i_{k})}}_{\mu_{1} \cdots  \mu_{2n-1}} (x)) 
   (\partial_{k_{2}} {a^{(j_{1} \cdots j_{l})}}_{\mu_{1} \cdots
   \mu_{2n-1}} (x)) \in {\cal A}_{2 - \frac{k+l}{2}} (x).
  \end{eqnarray}
  Here, $k+l$ must be an even number. Firstly, the mass term and the
  kinetic terms of the fields $a_{\mu_{1} \cdots \mu_{2n-1}}(x)$
  belong to ${\cal A}_{1}(x)$ and ${\cal A}_{2}(x)$, respectively.
  The explicit heat-kernel calculation reveals that the relevant mass
  terms are given by
  \begin{eqnarray}
  Tr[tr(e^{-\tau D^{2}})] &=& \int d^{d} x \frac{32}{(2 \pi
  \tau)^{\frac{d}{2}}} \tau \left( \frac{2}{3!} a_{\mu_{1} \cdots
  \mu_{3}}(x) a^{\mu_{1} \cdots \mu_{3}}(x)
  + \frac{4}{5!} a_{\mu_{1} \cdots \mu_{5}}(x) a^{\mu_{1} \cdots
  \mu_{5}}(x) \right. \nonumber \\
  & & \hspace{15mm}
  + \left. \frac{6}{7!} a_{\mu_{1} \cdots \mu_{7}}(x) a^{\mu_{1}
  \cdots \mu_{7}}(x) 
   +  \frac{8}{9!} a_{\mu_{1} \cdots \mu_{9}}(x) a^{\mu_{1} \cdots
  \mu_{9}}(x) \right) + \cdots. \label{hkmass}
  \end{eqnarray}
  We give the derivation of this result in Appendix. \ref{seeleydewitt}.
  Thus, the fields $a_{\mu_{1} \cdots \mu_{2n-1}}(x)$ for $n=2,3,4,5$
  are clearly massive. While (\ref{hkmass}) cancels the mass term
  $a_{\mu}(x) a^{\mu}(x)$, the field $a_{\mu}(x)$ is also presumably
  massive because there is no reason to prohibit the cross terms
  $a_{\mu}(x) {a^{(i_{1} \cdots i_{k})}}_{\mu}(x)$. 

   Next, we consider the fields ${a^{(i)}}_{a_{1} \cdots a_{2n-1}}
 (x)$. Especially the fields
 ${a^{(i)}}_{i a_{1} \cdots a_{2n}} (x)$ (with $n=1, 2, 3, 4$) are
 identified with the anti-symmetric tensor fields in the type IIB
 supergravity. Their mass terms are included in ${\cal A}_{0} (x)$, 
 while the kinetic terms reside in ${\cal A}_{1} (x)$. These 
 fields are therefore regarded as massless. 
 This observation augurs well for the 
 matrix model which reduces to the type IIB supergravity.

 However, it is not clear whether the higher-spin fields
 ${a^{(i_{1} \cdots  i_{k})}}_{a_{1} \cdots a_{2n-1}} (x)$
 ($k=2, 3, \cdots$) are massive, since the mass terms and the
 kinetic terms belong to the coefficients  ${\cal A}_{1-k} (x)$
  and ${\cal A}_{2-k}(x)$ respectively and both of them vanish in the
 action.  

 \subsection{Supersymmetric action}
  In the previous section, we have discerned that the bosonic part of
  the matrix model reduces to the Einstein gravity in the low energy
  limit. Our goal is to establish the extension of the IIB matrix
  model that reduces to the type IIB supergravity in the low energy
  limit. This leads us to scrutinize the fermionic part. We pay
  attention of the correspondence of the supersymmetry.

  We start from the following action:
  \begin{eqnarray}
   S_{S} = Tr[ tr(V(m^{2}))] + Tr {\bar \psi} m \psi. \label{ssaction}
  \end{eqnarray}
  Here, $m$ is defined in the same way as in the bosonic model.
  $\psi$ is a Weyl fermion. However, the crucial difference from the
  IIB matrix model is that we abstain its Majorana property. Namely,
  we abstain its hermiticity.
  It is extremely difficult to build a matrix model that closes with
  respect to the local Lorentz transformation $\delta \psi =
  \frac{1}{4} \Gamma^{\mu \nu} \{ \varepsilon_{\mu \nu} , \psi
  \}$. That is why we abandon the hermiticity and define the local
  Lorentz transformation as $\delta \psi = \frac{1}{4} \Gamma^{\mu
  \nu} \varepsilon_{\mu \nu} \psi$. 
  The fermionic field is also expanded by the number of the
  derivatives as
  \begin{eqnarray}
    \psi = \left( \chi(x) + \sum_{l=1}^{\infty} \chi^{(i_{1} \cdots
    i_{l})} (x) \partial_{i_{1}} \cdots \partial_{i_{l}} \right)
    (e^{-\tau D^{2}})^{\alpha}. \label{expansionferm}
  \end{eqnarray}
  Clearly, the fields $\chi^{(i_{1} \cdots i_{l})} (x)$ has the
  dimensionality [(length)$^{l}$].
  We put the damping factor $(e^{-\tau D^{2}})^{\alpha}$, so that the
  action ${\bar \psi} m \psi$ should be finite. We choose this power
  in accordance with the function (\ref{answev}), which leads us to set
  $\alpha=\frac{1}{4}$. 

 \subsubsection{Local Lorentz invariance}
  We discuss its local Lorentz invariance, which is defined as
   \begin{eqnarray}
    \delta m = [m, \varepsilon], \hspace{2mm}
    \delta \psi = \epsilon \psi, \label{llsuper}
   \end{eqnarray}
  where $\varepsilon$ is defined in (\ref{locallorentz}). ${\bar \psi}$ 
  is transformed as
   \begin{eqnarray}
    \delta {\bar \psi} = \delta ( \psi^{\dagger} (\Gamma^{0}))
                       = - \psi^{\dagger} (\Gamma^{0})^{2}
                       \varepsilon^{\dagger} \Gamma^{0}
                       = - {\bar \psi} \epsilon.
   \end{eqnarray}
   The action is transformed by this local Lorentz transformation as
  \begin{eqnarray}
   \delta S_{S} = 2 Tr[ tr(V_{S}'(m^{2})) [m, \varepsilon]]
                 + Tr[ tr({\bar \psi}[m, \varepsilon] \psi)] = 0.
     \label{llsuper2}
  \end{eqnarray}
  This holds true when the coefficients ${a^{(i_{1} \cdots
  i_{k})}}_{\mu_{1} \cdots \mu_{2n-1}}(x)$ and $\chi^{(i_{1} \cdots
  i_{k})} (x)$ damp rapidly in the infinite distance.

 \subsubsection{Determination of $V_{S}(m^{2})$}
  We next discuss the determination of the function $V_{S}(m^{2})$.
  Here, we do not give a specific form of $V_{S}(m^{2})$ and we just
  list up the conditions for $V_{S}(m^{2})$.

  Firstly, the flat background $m_{0} = i \tau^{\frac{1}{2}}
  \Gamma^{\mu} \partial_{\mu}$ must be a classical solution.
  As we have seen in the bosonic case, this imposes a strict condition
  on the function $V_{S}^{2}(m^{2})$; namely this must satisfy the
  criterion (\ref{conditionv}).

  Secondly, this model must retain the even-rank anti-symmetric tensor
  fields ${a^{(i)}}_{i \mu_{1} \cdots \mu_{2n}}(x)$ $(n=1,2,3,4)$.
  The action $V_{S}(m^{2})$ must be determined so that these fields
  should be massless and thus be bequeathed in the low-energy limit.

  It is not clear whether the function (\ref{answev}) satisfies the
  latter condition, since we have yet to elucidate the mass of the
  higher-spin fields.

 \subsubsection{Supersymmetry}
  We next discuss its supersymmetry. We define the supersymmetric
  transformation as
  \begin{eqnarray}
    \delta_{\epsilon} m = \epsilon {\bar \psi} + \psi {\bar \epsilon},
    \hspace{2mm} 
    \delta_{\epsilon} \psi = 2 V'_{S}(m^{2}) \epsilon. \label{susy-ll}
  \end{eqnarray}
  Here, $V'_{S}(x) = \frac{\partial V_{S}(x)}{\partial x}$. Now, we
  assume that the function $V_{S}(u)$ can be expanded around $u=0$ as
  \begin{eqnarray}
   V_{S}(u) = \sum^{\infty}_{k=1} \frac{a_{2k}}{2k} u^{k}.
  \end{eqnarray}
 Therefore, the following argument does not apply to the function
 (\ref{answev}), because it is not possible to expand this around the
 origin $u=0$.
 The supersymmetry transformation of ${\bar \psi}$ is derived as
  \begin{eqnarray}
   \delta_{\epsilon} {\bar \psi} &=& \delta_{S} (\psi^{\dagger} \Gamma^{0})
 =  - \epsilon^{\dagger} (\Gamma^{0})^{2} (\sum_{k=0}^{\infty} a_{2(k+1)}
 m^{2k})^{\dagger} \Gamma^{0} \nonumber \\
 &=& - \epsilon^{\dagger} \Gamma^{0} (\sum_{k=0}^{\infty} (-1)^{2k-1}
 a_{2(k+1)} (\Gamma^{0} m^{\dagger} \Gamma^{0})^{2k} ) \Gamma^{0} 
 = 2 {\bar \epsilon} V_{S}'(m^{2}). \nonumber \\
  \end{eqnarray}
 Then, the supersymmetry invariance of the action can be easily
 verified as
  \begin{eqnarray}
   \delta_{\epsilon} S_{S}
  = Tr[ tr(2 V'_{S}(m^{2})m (\epsilon {\bar \psi} + \psi {\bar
  \epsilon})
      + 2 {\bar \psi} m V'_{S} (m^{2}) \epsilon
      + 2 {\bar \epsilon} m V'_{S}(m^{2}) \psi ) ] = 0.
  \end{eqnarray}
  We next analyse the commutation relation of the supersymmetry
  transformation. This is given by
  \begin{eqnarray}
  & &  [\delta_{\epsilon}, \delta_{\xi}] m = 2[(\xi {\bar \epsilon} -
    \epsilon {\bar \xi}), V'_{S}(m^{2})], \label{com-bos-ll} \\
  & &  [\delta_{\epsilon}, \delta_{\xi}] \psi = 2 \psi \left(
    {\bar \epsilon} m \frac{V'_{S}(m^{2}) - V'_{S}(0)}{m^{2}} \xi
  - {\bar \xi} m \frac{V'_{S}(m^{2}) - V'_{S}(0)}{m^{2}} \epsilon
    \right), \label{com-ferm-ll}
  \end{eqnarray}
  with the help of the equation of motion
  \begin{eqnarray}
   \frac{\partial S_{S}}{\partial {\bar \psi}} = 2 m \psi = 0,
   \hspace{2mm}
   \frac{\partial S_{S}}{\partial \psi} = 2 {\bar \psi} m = 0.
  \end{eqnarray}
   This can be verified by taking the difference of the two
   transformations, as we did for the IIB matrix model or the
   supermatrix model. For the boson, we compare the following two
   transformation:
  \begin{eqnarray}
     m &\stackrel{\delta_{\xi}}{\to}& m + \xi {\bar \psi} + \psi {\bar \xi} 
     \stackrel{\delta_{\epsilon}}{\to}
     m + (\epsilon + \xi) {\bar \psi} + \psi ({\bar \epsilon} + {\bar
     \xi}) + 2 \xi {\bar \epsilon} V'_{S}(m^{2})
           + 2 V'_{S}(m^{2}) \epsilon {\bar \xi}, \nonumber \\
     m &\stackrel{\delta_{\epsilon}}{\to}& m + \epsilon {\bar \psi}
        + \psi {\bar \epsilon} 
     \stackrel{\delta_{\xi}}{\to}
     m + (\epsilon + \xi) {\bar \psi} + \psi ({\bar \epsilon} + {\bar
     \xi}) + 2 \epsilon {\bar \xi} V'_{S}(m^{2})
           + 2 V'_{S}(m^{2}) \xi {\bar \epsilon}. \nonumber
  \end{eqnarray}
  For the fermion, we compare the following two paths:
  \begin{eqnarray}
   \psi &\stackrel{\delta_{\xi}}{\to}& \psi + 2 V'_{S}(m^{2}) \xi
         \nonumber \\
         &\stackrel{\delta_{\epsilon}}{\to}& 
   \psi + 2 V'_{S}(m^{2}) (\epsilon + \xi)
        + \sum^{\infty}_{k=2} a_{2k} [(\epsilon {\bar \psi} + \psi
         {\bar \epsilon}) m^{2k-3}
         + m (\epsilon {\bar \psi} + \psi {\bar \epsilon}) m^{2k-4}
         + \cdots m^{2k-3} (\epsilon {\bar \psi} + \psi {\bar \epsilon})
         ] \xi, \nonumber \\
   \psi &\stackrel{\delta_{\epsilon}}{\to}& \psi + 2 V'_{S}(m^{2}) \epsilon
        \nonumber \\
         &\stackrel{\delta_{\xi}}{\to}&
   \psi + 2 V'_{S}(m^{2}) (\epsilon + \xi)
        + \sum^{\infty}_{k=2} a_{2k} [(\xi {\bar \psi} + \psi
         {\bar \xi}) m^{2k-3}
         + m (\xi {\bar \psi} + \psi {\bar \xi}) m^{2k-4}
         + \cdots m^{2k-3} (\xi {\bar \psi} + \psi {\bar \xi})
         ] \epsilon. \nonumber
 \end{eqnarray}

  In order to see the structure of the supersymmetry transformation 
  of the ${\cal N}=2$ supersymmetry. Since we now abstain the
  hermiticity of the fermion, we separate the fermions into the 
  real part and the imaginary part as
  \begin{eqnarray}
   \epsilon = \epsilon_{1} + i \epsilon_{2}, \hspace{2mm}
   \xi = \xi_{1} + i \xi_{2},
  \end{eqnarray}
 where $\epsilon_{1}, \epsilon_{2}, \xi_{1}$ and $\xi_{2}$ are
 Majorana-Weyl fermions. Now, we assume that the supersymmetry
 parameters are proportional to the unit matrix for brevity, namely
  \begin{eqnarray}
   \epsilon_{1,2} = c_{\epsilon_{1,2}} {\bf 1}_{N \times N}, \hspace{2mm} 
   \xi_{1,2} = c_{\xi_{1,2}} {\bf 1}_{N \times N}, 
  \end{eqnarray}
  where $c_{\epsilon_{1,2}}$ and $c_{\xi_{1,2}}$ are real
  Grassmann-odd c-numbers.
  Another assumption is that the background metric is flat. 

  Firstly, the transformation of the bosonic field is calculated as
  \begin{eqnarray}
    & & [ \delta_{\epsilon}, \delta_{\xi}] A_{\mu}
   = \frac{1}{16} tr ( [\delta_{\epsilon}, \delta_{\xi}] m \Gamma_{\mu}
   )  = 
   \frac{1}{16} \sum_{k=2}^{\infty} a_{2k} tr ( \xi {\bar \epsilon} m^{2k-2}
   \Gamma_{\mu} - \epsilon {\bar \xi} m^{2k-2} \Gamma_{\mu}
   - m^{2k-2} \xi {\bar \epsilon} \Gamma_{\mu}
   + m^{2k-2} \epsilon {\bar \xi} \Gamma_{\mu} ) \nonumber \\
  & & \hspace{15mm} = 
   \frac{1}{16} \sum_{k=2}^{\infty} a_{2k} ( {\bar \xi} [m^{2k-2},
   \Gamma_{\mu}] \epsilon - {\bar \epsilon} [m^{2k-2}, \Gamma_{\mu}]
   \xi )  \nonumber \\
 & & \hspace{15mm}  =
   \frac{a_{4}}{16}  ( {\bar \xi} [\Gamma^{\nu_{1}} \Gamma^{\nu_{2}},
   \Gamma_{\mu}] \epsilon - {\bar \epsilon} [\Gamma^{\nu_{1}} \Gamma^{\nu_{2}},
   \Gamma_{\mu}] \xi ) A_{\nu_{1}} A_{\nu_{2}} + \cdots \nonumber \\
 & & \hspace{15mm}  = \frac{a_{4}}{16} ( {\bar \xi} \Gamma^{i} \epsilon
    - {\bar \epsilon} \Gamma^{i} \xi ) [A_{i}, A_{\mu}] + \cdots
   = 
   \frac{a_{4}}{8} ( {\bar \xi}_{1} \Gamma^{i} \epsilon_{1}
                    +  {\bar \xi}_{2} \Gamma^{i} \epsilon_{2} )
      [A_{i}, A_{\mu} ] + \cdots, \label{AZ3trans-a}
  \end{eqnarray}
  We focus on the supersymmetry transformation of the vector field
  $a_{\mu}(x)$. The commutator $[A_{i}, A_{\mu}]$ represents the
  translation and the gauge transformation:
  \begin{eqnarray}
    [A_{i}, A_{\mu}] = [i \partial_{i} + a_{i} (x), i \partial_{\mu}
  +  a_{\mu}(x) ] + \cdots 
   = \underbrace{i (\partial_{i} a_{\mu} (x) )}_{\textrm{translation}} 
  \underbrace{ - i (\partial_{\mu} a_{i} (x) )
    + [a_{i} (x), a_{\mu} (x) ]}_{\textrm{gauge transformation}} + \cdots.
  \end{eqnarray}
  Therefore, the commutator of the bosonic fields represents the
  right translation of the vector fields.
  However, this does not hold true of the fermionic fields. The
  commutator is computed as
  \begin{eqnarray}
      [\delta_{\epsilon}, \delta_{\xi}] \psi
  &=&  -  \sum_{k=2}^{n} a_{2k} \psi ( {\bar \xi} m^{2k-3} \epsilon
      - {\bar \epsilon} m^{2k-3} \xi ) + \cdots
  = -  a_{4} ( {\bar \xi} \Gamma^{j} \epsilon - {\bar
   \epsilon} \Gamma^{j} \xi ) \psi A_{j} + \cdots \nonumber \\
  &=& - 2 a_{4} ({\bar \xi}_{1} \Gamma^{j} \epsilon_{1}
      + {\bar \xi}_{2} \Gamma^{j} \epsilon_{2} ) \psi A_{j} + \cdots.
  \nonumber
  \end{eqnarray}
  We have a closer look at the term $\psi A_{j}$:
  \begin{eqnarray}
   \psi A_{j} = i \psi \partial_{j} + \cdots 
              = \left( \chi(x) \partial_{j} + \sum_{i=1}^{\infty}
      \chi^{(i_{1} \cdots \i_{l})} (x) \partial_{i_{1}} \cdots
              \partial_{i_{l}} \partial_{j} \right) (e^{-\tau
              D^{2}})^{\alpha} + \cdots, 
  \end{eqnarray}
  where $\cdots$ denotes the omission of the non-linear terms of the
  fields. Therefore, each fermionic field is subject to the following
  transformation:
  \begin{eqnarray}
   [\psi_{\epsilon}, \psi_{\xi}] \chi(x) = 0 + \cdots, \hspace{2mm}
   [\psi_{\epsilon}, \psi_{\xi}] \chi^{(i_{1} \cdots i_{l+1})} (x) 
 = - 2 a_{4} ({\bar \xi}_{1} \Gamma^{j} \epsilon_{1}
           +  {\bar \xi}_{2} \Gamma^{j} \epsilon_{2}) 
       \chi^{(i_{1} \cdots i_{l}} (x) \delta^{i_{l+1}) j} + \cdots. 
  \end{eqnarray}
  Therefore, the fermionic fields are not subject to the translation
  by the commutator of the supersymmetry. 
 
  However, the argument in this section is only applicable to the
  function $V'_{S}(u)$ that can be expanded around $u=0$, and does not 
  hold true of the function (\ref{answev}). We speculate that we may
  find a right supersymmetry for such functions.

  We surmise that the fermionic fields $\chi(x)$ and $\chi^{(i)}(x)$,
  which are respectively identified with the dilatino and the
  gravitino, would be massless due to the supersymmetry, when the
  even-rank fields ${a^{(i)}}_{i \mu_{1} \cdots \mu_{2n}} (x)$ are
  massless. It is an intriguing future problem to seek such an action
  more rigorously.

  \section{Monte Carlo simulation of the fuzzy sphere solutions}
 We have seen several generalizations of the IIB matrix model to
 accommodate the curved-space classical background in
 Sec. \ref{alternativemodel}. We have reviewed the author's
 work\cite{0209057} in Sec. \ref{sec0209057} about the curved-space
 background in the context of the supermatrix model.

 In this section, we address the stability of the curved-space
 background in the quantum sense, based on the author's
 work\cite{0401038}. It has been a conundrum to discuss
 the stability of the fuzzy sphere in the quantum sense. 
 There have been hitherto several attempts for the
 quantum stability. In \cite{0101102}, the quantum 
 stability has been discussed in terms of the one-loop effective
 action. In \cite{0303120,0307007,0312241}, they performed
 the two-loop perturbative calculation of the Feynman diagram,
 to unravel the stability of the fuzzy sphere and other backgrounds.
 Another attempt is to investigate the quantum stability via the
 Gaussian expansion, and the first-order calculation is given in
 \cite{0303196}. 

 Here, we address the quantum stability of the fuzzy sphere via the
 heat bath algorithm of the Monte Carlo simulation, whose detail we
 delegate to Appendix \ref{simulation}.
 Our approach is totally different from these foregoing
 works\cite{0101102,0303120,0303196,0307007,0312241}, in the sense that we 
 discuss the stability nonperturbatively via the numerical
 simulation. The Monte Carlo simulation
 has played a pivotal role in elucidating the nonperturbative aspects
 of the lattice gauge theory. By the same token, the application of
 the Monte Carlo simulation to the IIB matrix model
 has given us many interesting
 nonperturbative insights. In \cite{9811220}, the bosonic IIB matrix
 model has been scrutinized nonperturbatively via the Monte Carlo
 simulation. This analysis\cite{9811220} was followed by the supersymmetric
 extension\cite{0003208,0005147} via the hybrid Monte Carlo simulation.

 Here, we scrutinize the simplest case for the matrix model with the
 curved-space classical solution. Namely, we concentrate on the
 three-dimensional bosonic IIB matrix model with the Chern-Simons
 term which incorporates the $S^{2}$ fuzzy sphere solution:
  \begin{eqnarray}
    S = N Tr \left( - \frac{1}{4} \sum_{\mu, \nu=1}^{3} [A_{\mu},
   A_{\nu}]^{2} + \frac{2i \alpha}{3} \sum_{\mu, \nu, \rho = 1}^{3} 
   \epsilon_{\mu \nu \rho} A_{\mu} A_{\nu} A_{\rho}
   \right). \label{verydefinition}
  \end{eqnarray}
  The notation is the same as that in Sec. \ref{sec0101102}, except 
  we set $\frac{1}{g^{2}} = N$ in this section. We delegate the
  miscellaneous properties of the $S^{2}$ fuzzy sphere classical solution
  $A_{\mu} = \alpha L_{\mu}$ to Sec. \ref{sec0101102}.
  
 \subsection{Nonperturbative studies of the fuzzy sphere initial
  condition} 
  We start with the Monte Carlo simulation for the initial condition
 of the irreducible representation of the $N \times N$ fuzzy sphere
 classical solution:
  \begin{eqnarray}
   A^{(0)}_{\mu} = \alpha L_{\mu}. \label{fsinitial}
  \end{eqnarray}
 Starting from this initial condition, we see whether or not the fuzzy sphere 
 decays due to the quantum effect of the matrix model.
 Especially, the histogram of the eigenvalues of the Casimir operator 
   \begin{eqnarray}
    Q = A^{2}_{1} + A^{2}_{2} + A^{2}_{3} \label{cas2}
   \end{eqnarray}
 describes the fuzzy sphere's stability visually. In the following, 
 $x$ denotes the eigenvalues of $Q$, and $f(x)$ denotes the eigenvalue
 distribution 
 function normalized as $\int^{\infty}_{0} dx f(x) = 1$. At the initial state,
 the histogram is peaked at $R^{2} = \frac{N^{2}-1}{4} \alpha^{2}$,
 because the Casimir is proportional to the 
 unit matrix for the fuzzy sphere. If the fuzzy sphere is stable in
 the quantum sense, the histogram does not decay greatly from the
 peaked distribution. On the other hand, if the fuzzy sphere is
 unstable, the histogram undergoes a drastic deformation.

 \subsubsection{Nonperturbative stability of the fuzzy sphere}
   We first investigate the $N=16$ and $\alpha=0.5, 1.0, 2.0$ case
    starting from the initial condition (\ref{fsinitial}).
  We see a clear difference between the $\alpha=0.5$ and the
    $\alpha=1.0, 2.0$ case.
 \begin{figure}[htbp]
  \begin{center}
   \scalebox{0.6}{\includegraphics{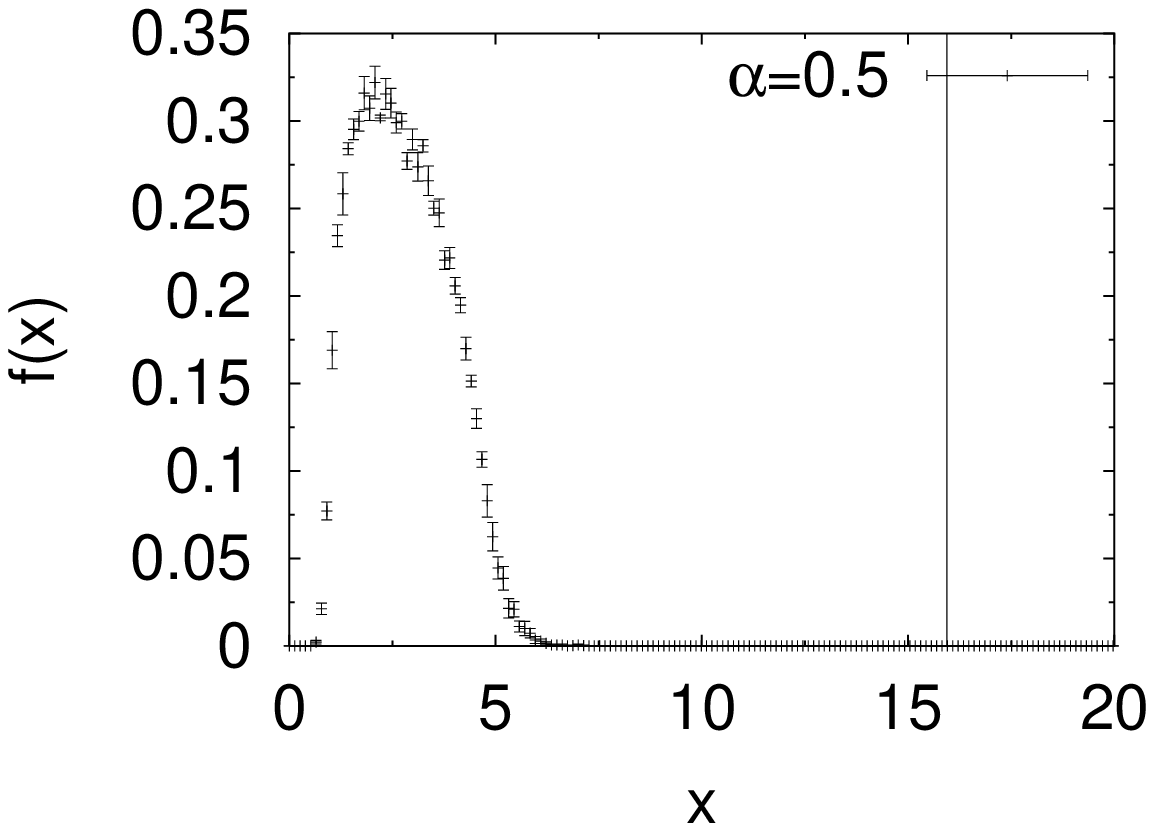}
                  \includegraphics{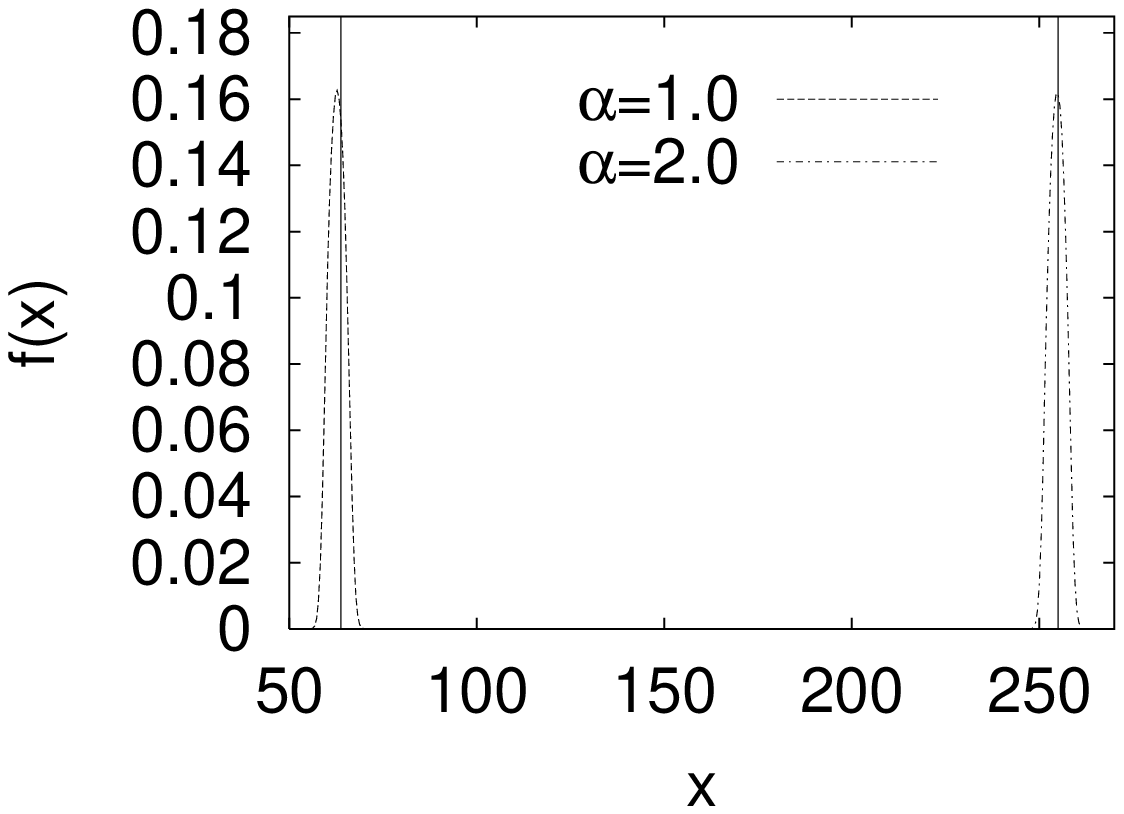}}
  \end{center}
  \caption{The instability at $\alpha=0.5$(left) and the stability at
    $\alpha=1.0, 2.0$(right) of the fuzzy sphere for $N=16$.}
  \label{n16stability}
 \end{figure}

 We see from Fig. \ref{n16stability} the instability of the
fuzzy sphere at $\alpha=0.5$.
The vertical lines in the histogram denote the radius-square of the
fuzzy sphere classical solution $R^{2} = \alpha^{2}
\frac{N^{2}-1}{4}$. Now, this value is $15.9375$ for $\alpha=0.5$, and 
the eigenvalues of $Q$ were originally peaked there. This
fuzzy sphere decays due to the quantum effect and the eigenvalues are
concentrated far away from the original fuzzy sphere configuration.

However, this situation changes for $\alpha=1.0, 2.0$.
Fig. \ref{n16stability} indicates that the eigenvalue distribution
constitutes only one lump in the vicinity of  the classical value of the
radius-square (which is $63.75$ and $255$ for $\alpha=1.0$ and
$\alpha=2.0$ respectively), and that the eigenvalues are not
dissipated otherwise at all. This implies that the eigenvalues remain
constituting a thin sphere shell near its original fuzzy sphere
configuration. In this sense,
we regard the fuzzy sphere classical solution as stable.
This stability is also reconfirmed in Section \ref{evo}, by
demonstrating the dynamical evolution of the single fuzzy sphere
state from the initial condition $A^{(0)}_{\mu} = 0$.

  \subsubsection{First-order phase transition}
  In the previous section, we have discerned that the fuzzy sphere
  solution. In this section, we deepen the above observation and
  demonstrate the first-order transition of the matrix model
  (\ref{verydefinition}).

  Firstly, we ascribe the above stability of the fuzzy sphere to the
  meager quantum effect of the model. 
  The effective action $W$ of this matrix model is given by
   \begin{eqnarray}
  W = - \log \left( \int dA e^{-S} \right). \label{vev}
   \end{eqnarray}
  The effect of the classical fuzzy sphere solution is of the order
  ${\cal O}(\alpha^{4} N^{4})$, as one can easily verify by plugging 
  the fuzzy sphere solution (\ref{fssolution}) into the action
  (\ref{verydefinition}). On the other hand, the effect of the path
  integral measure is of the order ${\cal O}(N^{2})$. 
  This comparison leads to the observation that the matrix model
  receives only a small quantum effect at 
   \begin{eqnarray}
     \alpha \gg {\cal O}(\frac{1}{\sqrt{N}}). \label{powercounting}
   \end{eqnarray}
  
\begin{figure}[htbp]
   \begin{center}
    \scalebox{0.56}{\includegraphics{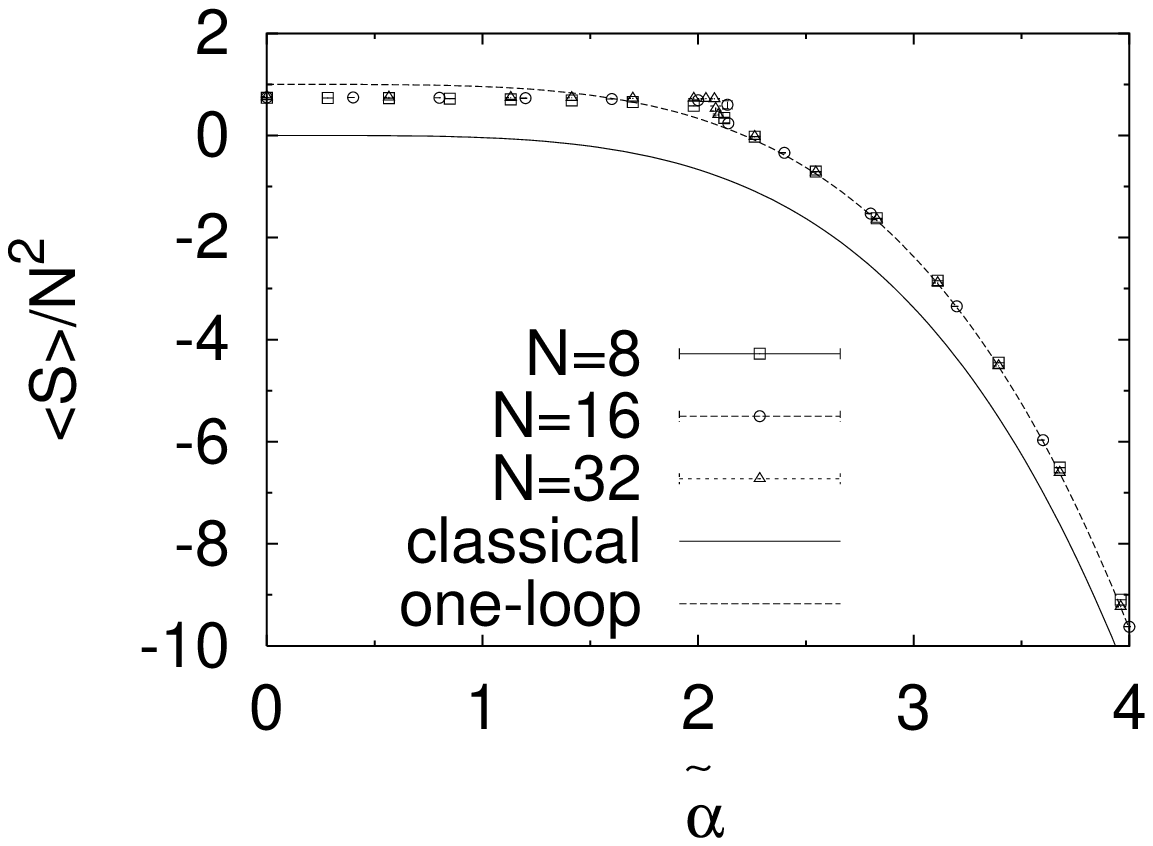}
          \includegraphics{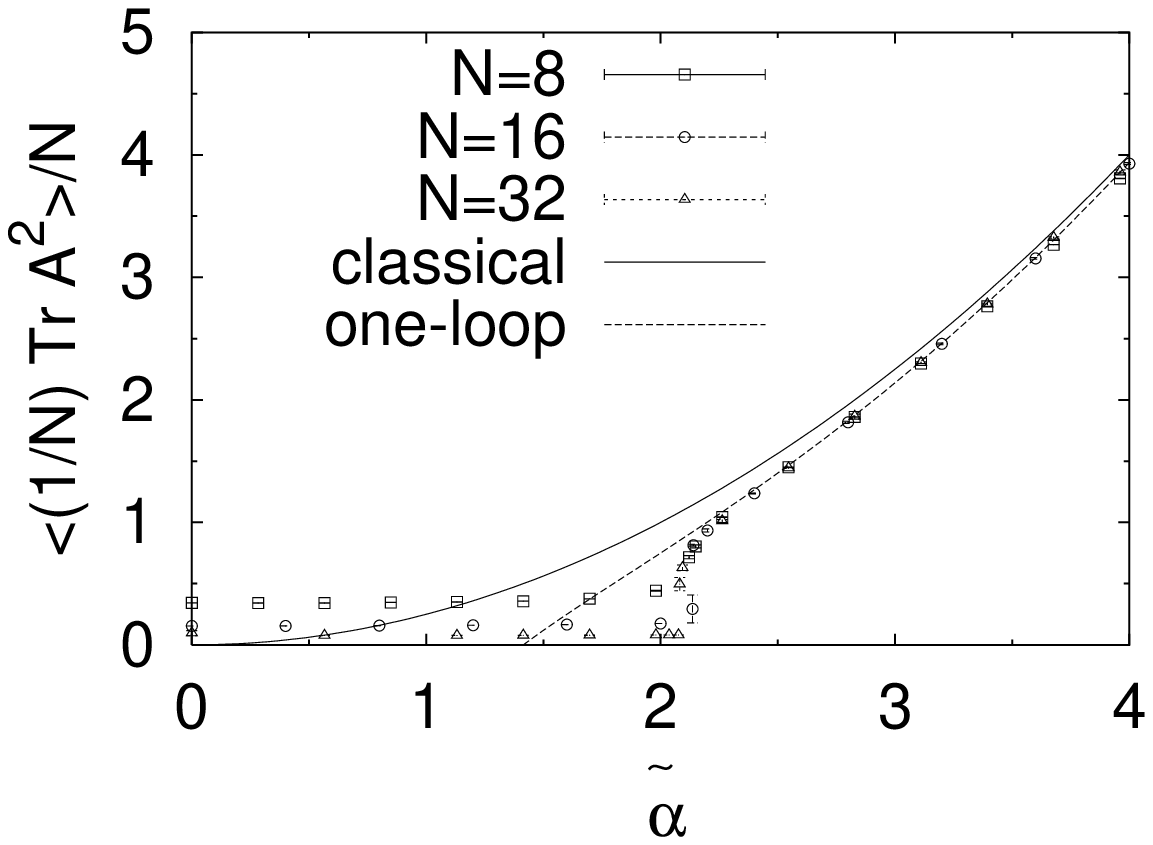} }
    \scalebox{0.56}{\includegraphics{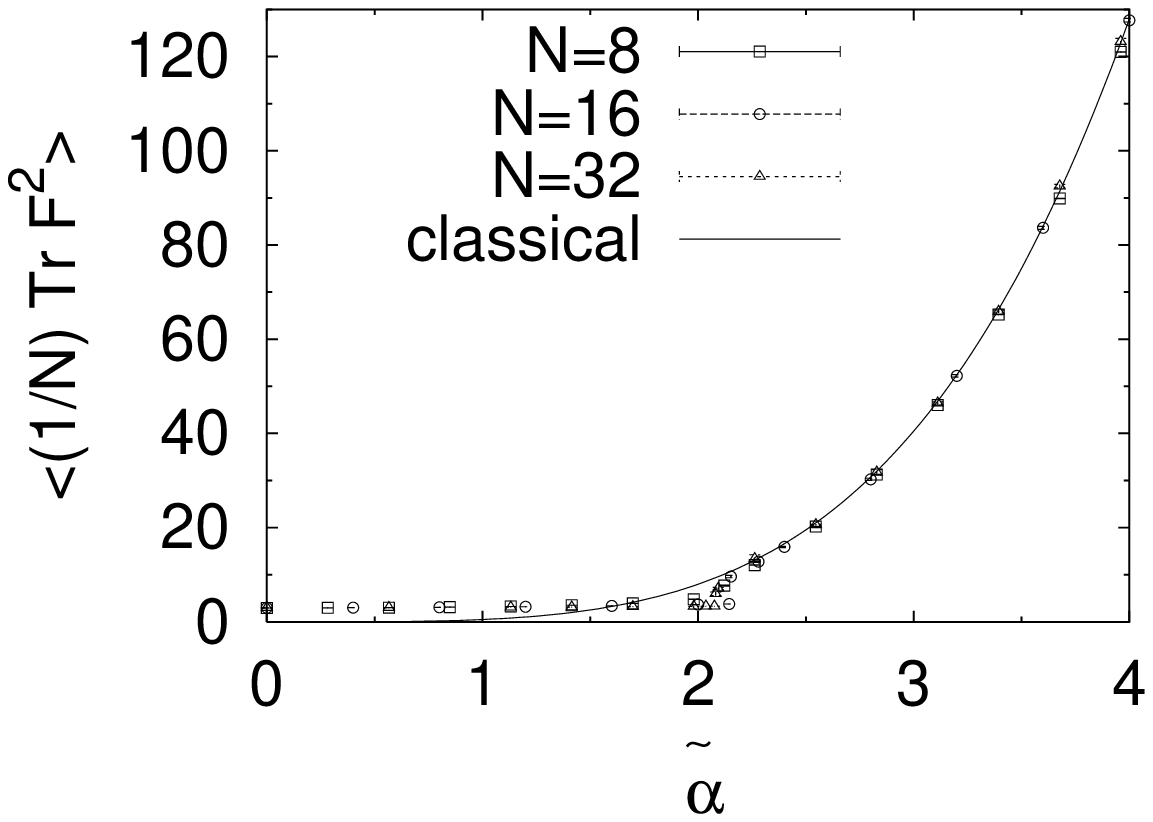}
          \includegraphics{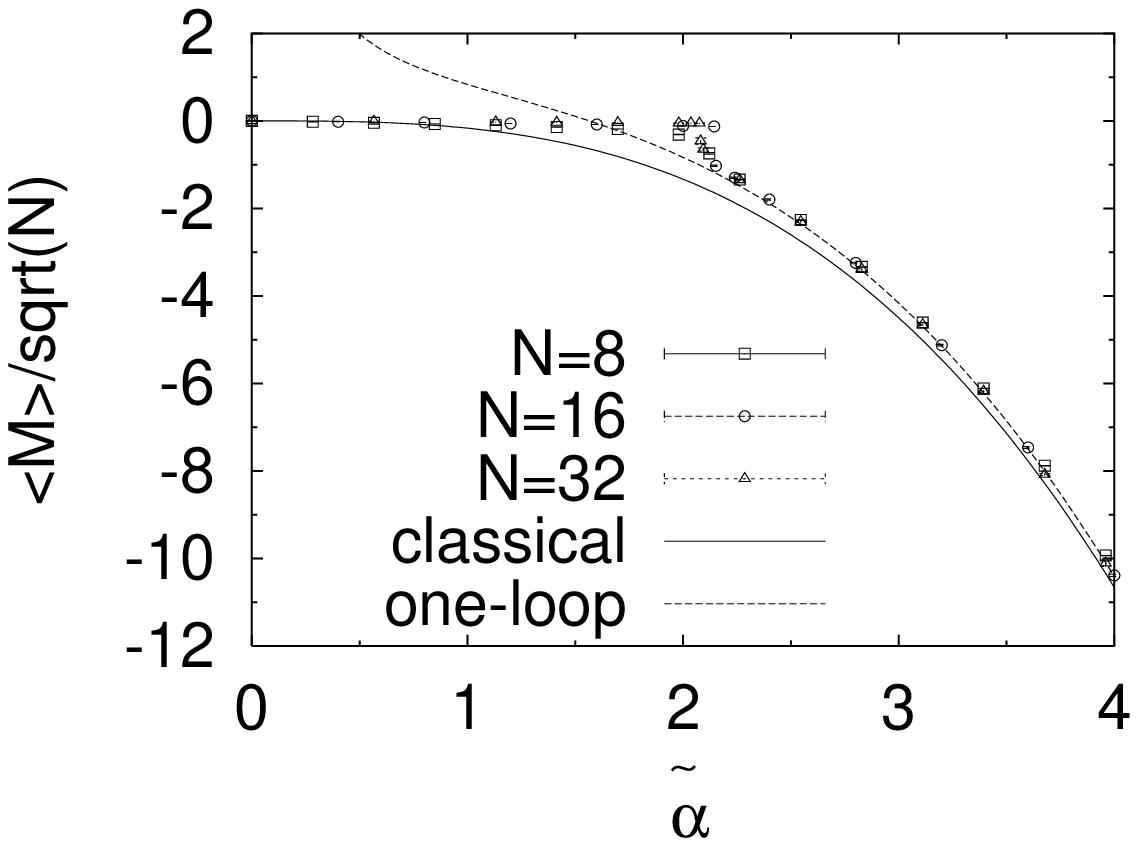} }
    \end{center}
    \caption{$\frac{1}{N^{2}} \langle S \rangle$ (upper left), $\frac{1}{N} 
  \langle \frac{1}{N} Tr A^{2}_{\mu} \rangle$ (upper right), $\langle
  \frac{1}{N} Tr F_{\mu \nu}^{2} \rangle$ (lower left) 
  and $\frac{1}{\sqrt{N}}  \langle M \rangle$ (lower right)
  against ${\tilde \alpha}$, for $N=8,16,32$ with the fuzzy sphere
  initial condition (\ref{fsinitial}).}
  \label{miscFS}
  \end{figure}

 To elaborate on this observation, we next calculate the following
 quantities, for $N=8,16,32$ with the fuzzy sphere 
 initial condition (\ref{fsinitial}). 
  \begin{enumerate}
   \item{The action $\langle S \rangle$: This quantity gives the
       energy of the numerical configuration.}
   \item{The spacetime extent $\langle \frac{1}{N} Tr A^{2}_{\mu}
       \rangle$: This provides us with the information of the
       eigenvalue distribution, and hence the stability of the fuzzy
       sphere.} 
   \item{The Yang-Mills term: $\langle \frac{1}{N} Tr F^{2}_{\mu \nu}
       \rangle$: This quantity, as well as the spacetime extent,
       have been already investigated in the
       bosonic IIB matrix model without the Chern-Simons term (this
       amounts to the $\alpha=0$ case) in \cite{9811220}.
       For the bosonic IIB matrix model, both are shown to behave at
       the order ${\cal O}(1)$.
       It is interesting to compare the behavior of this quantity with
       that of the bosonic IIB matrix model.}
 \item{We derive the following exact result from the Schwinger-Dyson
      equation of the matrix model:
      \begin{eqnarray}
      0 = \int dA \frac{\partial}{\partial A^{a}_{\mu}} (Tr(t^{a}
      A_{\mu}) e^{-S}). \label{SDE}
      \end{eqnarray}
       Here, $t^{a}$ ($a=1, 2, \cdots, N^{2}-1$) are the basis of the
       $SU(N)$ Lie algebra, and the matrices $A_{\mu}$ are expanded as
       $A_{\mu} = \sum_{a=1}^{N^{2}-1} A_{\mu}^{a} t^{a}$. The
       Schwinger-Dyson equation (\ref{SDE}) gives
      \begin{eqnarray}
       \langle K \rangle = \frac{1}{N} \langle 
        Tr( - [A_{\mu}, A_{\nu}]^{2} + 2 i \alpha
        \epsilon_{\mu \nu \rho} A_{\mu} A_{\nu} A_{\rho}) \rangle = 3(1 - 
        \frac{1}{N^{2}} ) = K_{0}, \label{sde-quantity}
      \end{eqnarray} 
     as we prove explicitly in Appendix \ref{IIBsim-app}. 
     While $K$ is given analytically, the numerical computation
     of its vacuum expectation $\langle K \rangle$ is significant for
     the legitimacy of the algorithm.}   
   \item{The Chern-Simons term:
    \begin{eqnarray}
     \langle M \rangle = \langle \frac{2i}{3N}  
     Tr \epsilon_{\mu \nu \rho} A_{\mu} A_{\nu} A_{\rho} \rangle.
     \label{cs-a} 
    \end{eqnarray}
    We have obtained the analytical result $K$, which amounts to
    \begin{eqnarray}
     K = \frac{1}{N} Tr F^{2}_{\mu \nu} + 3 \alpha
     M. \label{sdeqbalance}
    \end{eqnarray}
    While $\langle K \rangle$ itself obviously behaves as ${\cal
        O}(1)$, it is interesting to discern the behavior of each of
        the separate terms $M$, as well as $\frac{1}{N} Tr F^{2}_{\mu \nu}$.}  
   \end{enumerate}
 We postpone the plot of $\langle K \rangle$ later, and plot the other
 four quantities. 
 The above power-counting observation for the quantum effect motivates us
 to plot these quantities against 
 ${\tilde \alpha} = \alpha \sqrt{N}$, instead of $\alpha$.
 For the fuzzy sphere classical solution, they
 respectively scale as
  \begin{eqnarray}
 & & S = {\cal O}(\alpha^{4} N^{4}) = {\cal O}({\tilde
 \alpha}^{4} N^{2}), \hspace{2mm}
 \frac{1}{N} Tr A^{2}_{\mu}  = {\cal O}(\alpha^{2}
 N^{2}) = {\cal O}({\tilde \alpha}^{2} N), \nonumber \\
 & & \frac{1}{N} Tr F_{\mu \nu}^{2} = {\cal
 O}(\alpha^{4} N^{2}) = {\cal O}({\tilde \alpha}^{4}),
 \hspace{2mm}
  M = {\cal O}(\alpha^{3} N^{2}) = {\cal O}({\tilde
 \alpha}^{3} \sqrt{N}). \label{scalemisc}
  \end{eqnarray}
  This leads us to plot the quantities
  $\frac{1}{N^{2}} \langle S \rangle$, $\frac{1}{N} 
  \langle \frac{1}{N} Tr A^{2}_{\mu} \rangle$, $\langle \frac{1}{N} Tr
  F_{\mu \nu}^{2} \rangle$ and $\frac{1}{\sqrt{N}}  \langle M \rangle$ 
  against ${\tilde \alpha}$ in Fig. \ref{miscFS}. 

  This plotting poses a serious consequence.
  This matrix model has the first-order phase transition 
  with the change of the parameter ${\tilde \alpha}$. In the above
  four cases, we see a discontinuity at the critical point 
  \begin{eqnarray}
  {\tilde \alpha} = {\tilde \alpha}^{(l)}_{cr} \sim
  2.1. \label{criticalpointfs} 
  \end{eqnarray}
  We also find that this
  discontinuity is steeper for the larger $N$. This indicates that
  there is a first-order phase transition between the two phases. 

  We call the phase at ${\tilde \alpha} < {\tilde \alpha}^{(l)}_{cr}$ the 
  "Yang-Mills phase". In this phase, the quantum effect of the 
  matrix model becomes strong, and thus the fuzzy sphere classical
  solution is no longer stable. Rather, the behavior for this phase
  has a similarity to that of the bosonic IIB matrix model without the 
  Chern-Simons term. Fig. \ref{miscFS} clearly indicates that this
  model undergoes a smooth transition from $\alpha=0$ (the bosonic IIB
  matrix model itself) to the small $\alpha$. The vacuum expectation
  values of $\langle \frac{1}{N} Tr A^{2}_{\mu} \rangle$ and $\langle
  \frac{1}{N} Tr F_{\mu \nu}^{2} \rangle$ have been analyzed in
  \cite{9811220}, where they are shown to behave at ${\cal O}(1)$. 
  This behavior is inherited in the Yang-Mills phase.
  We elaborate on this phase more closely in Section \ref{Yang-Mills}.

  We call the phase at ${\tilde \alpha} > {\tilde \alpha}^{(l)}_{cr}$ the
  "fuzzy sphere phase", where the quantum effect is so small that the
  classical fuzzy sphere is stable. Now, we discern that the
  fuzzy sphere is stable at $N=16$, $\alpha=1.0, 2.0$, because the
  system is in the fuzzy sphere phase in this case. They are scaled as
  (\ref{scalemisc}), and this behavior is
  totally different from that of the Yang-Mills phase.

  This first-order phase translation is a product of the 
  the Monte Carlo simulation, which could not be seen by the
  perturbative approaches. In this sense, we accentuate that 
  this first-order phase transition is a nonperturbative phenomenon.

\begin{figure}[htbp]
   \begin{center}
    \scalebox{0.6}{\includegraphics{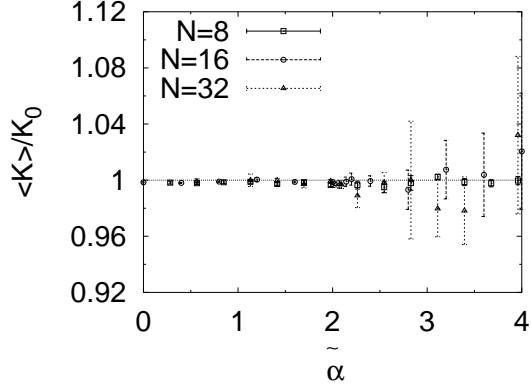}}
   \end{center}
    \caption{The plot of the quantity $\frac{\langle K
 \rangle}{K_{0}}$, for the fuzzy sphere initial
 condition (\ref{fsinitial}).}
  \label{sdeqFS}
  \end{figure}

 We next exhibit the corresponding plot of the
 quantity $\langle K \rangle$ in Fig. \ref{sdeqFS}.  
 In the Yang-Mills phase, the observables $\langle \frac{1}{N} 
 Tr F_{\mu \nu}^{2} \rangle$ and $\langle M \rangle$ behave as ${\cal
 O}(1)$, similarly to the bosonic IIB matrix model.
 The fluctuation of $K$ is thus small in the Yang-Mills phase.

 On the other hand, $\langle \frac{1}{N} 
 Tr F_{\mu \nu}^{2} \rangle$ and $3 \alpha \langle M \rangle$ both behave
 as ${\cal O}({\tilde \alpha}^{4})$ in the fuzzy sphere phase.
 Nevertheless, the quantity 
 $\langle K \rangle = \langle (\frac{1}{N} Tr F_{\mu \nu}^{2} + 3
 \alpha M ) \rangle$ is of the order ${\cal O}(1)$, as is analytically 
 verified. This is ascribed to the huge cancellation between these two 
 terms in the fuzzy sphere phase. 
 Therefore, the fluctuation of $\langle K \rangle$ is naturally bigger
 in the fuzzy sphere phase than in the bosonic Yang-Mills phase. 

 \subsubsection{One-loop dominance in the fuzzy sphere phase}
  The behavior of these four quantities in the fuzzy sphere phase
  gives an even more striking consequence. The numerical value of
  these four quantities coincides their one-loop results. 
  The one-loop vacuum expectation values are calculated in the
  large-$N$ limit as
   \begin{eqnarray}
  \frac{\langle S \rangle}{N^{2}} = - \frac{1}{24} {\tilde
    \alpha}^{4} + 1, \hspace{2mm}
    \frac{1}{N} \langle \frac{1}{N} Tr A_{\mu}^{2} \rangle
   = \frac{{\tilde \alpha}^{2}}{4} - \frac{1}{{\tilde \alpha}^{2}},
    \hspace{2mm}
   \langle \frac{1}{N} Tr F_{\mu \nu}^{2} \rangle = \frac{{\tilde
  \alpha}^{2}}{2}, \hspace{2mm}
   \frac{1}{\sqrt{N}} \langle M \rangle = - \frac{{\tilde
  \alpha}^{3}}{6} + \frac{1}{{\tilde \alpha}}. \label{one-loopmisc}
   \end{eqnarray}
 We delegate the derivation of these results to the appendices of
 \cite{0401038}.  The first term comes from the classical effect,
 while the one-loop correction comes in the second term of each
 quantity. The dotted lines in the graphs denote these
 one-loop result of these quantities\footnote{For $\langle \frac{1}{N} Tr
 F_{\mu \nu}^{2} \rangle$, we omit the dotted line, because it is not
 subject to the one-loop correction.}. 
 The coincidence of the numerical values of these observables with the 
 one-loop result suggests the higher-loop effects are suppressed in
 the large-$N$ limit and hence that the one-loop effect is dominant in 
 the matrix model (\ref{verydefinition}).

 \subsubsection{Width of the eigenvalue distribution}
\begin{figure}[htbp]
   \scalebox{0.56}{\includegraphics{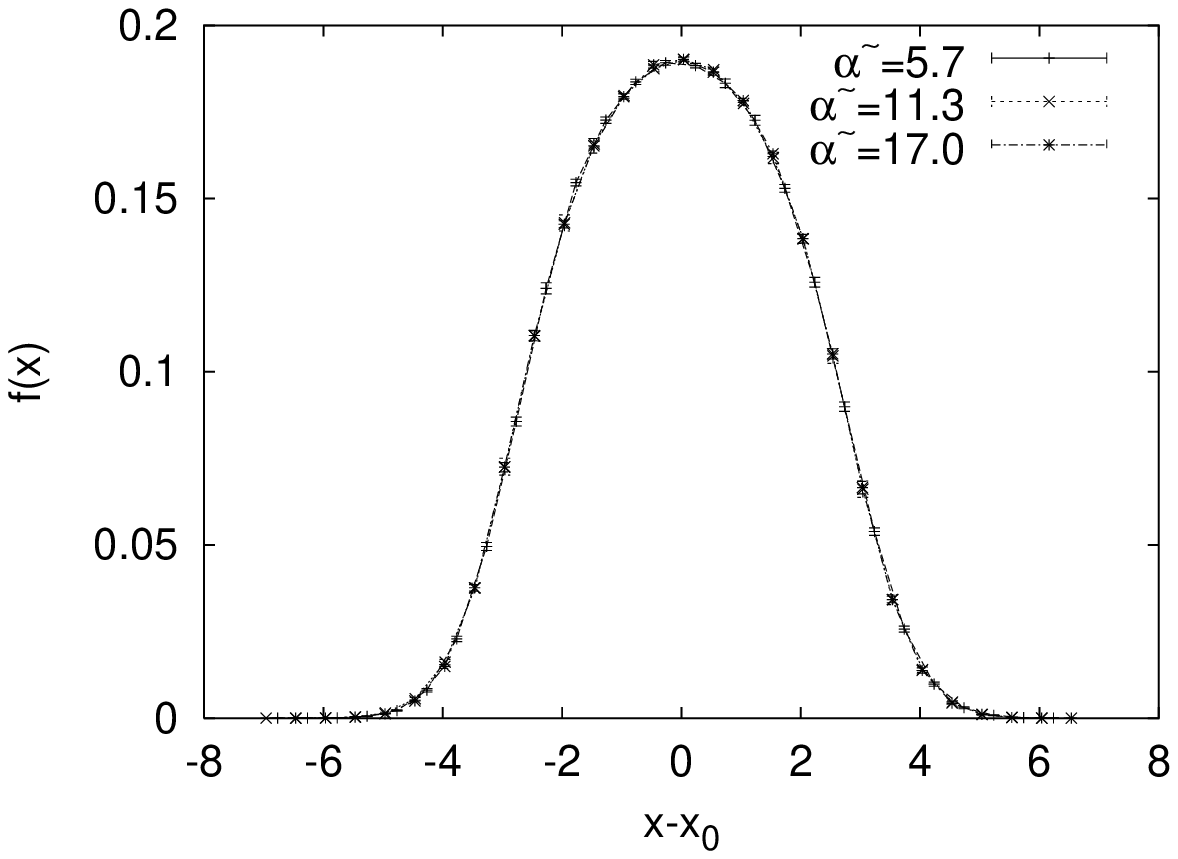}
                   \includegraphics{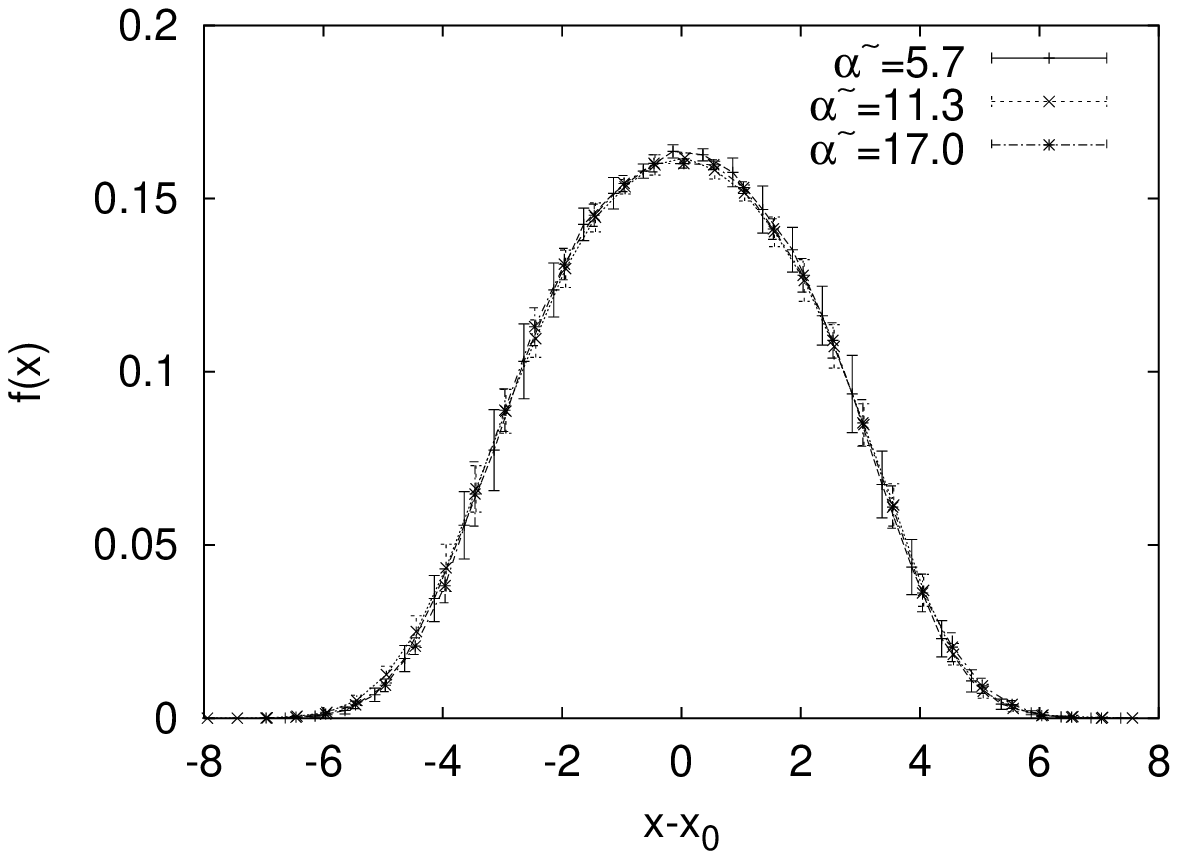}}
   \scalebox{0.56}{\includegraphics{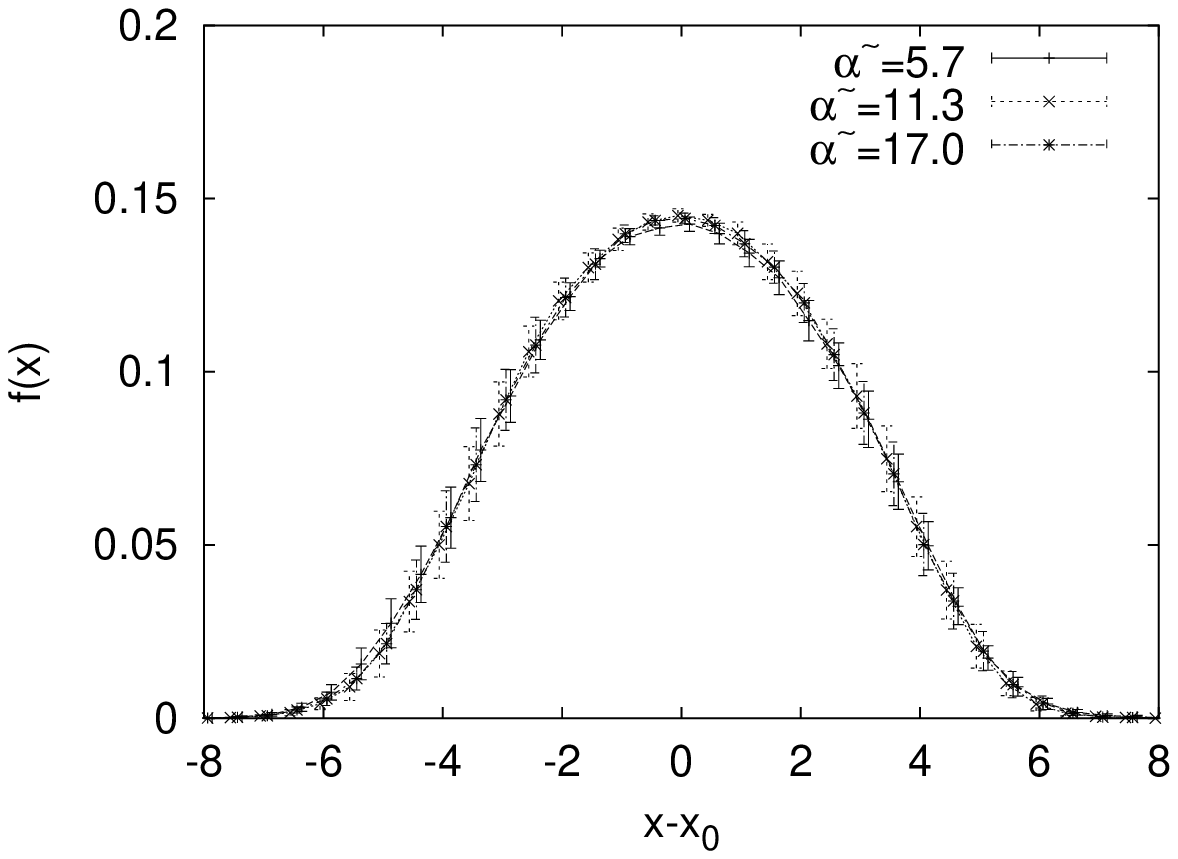}
                   \includegraphics{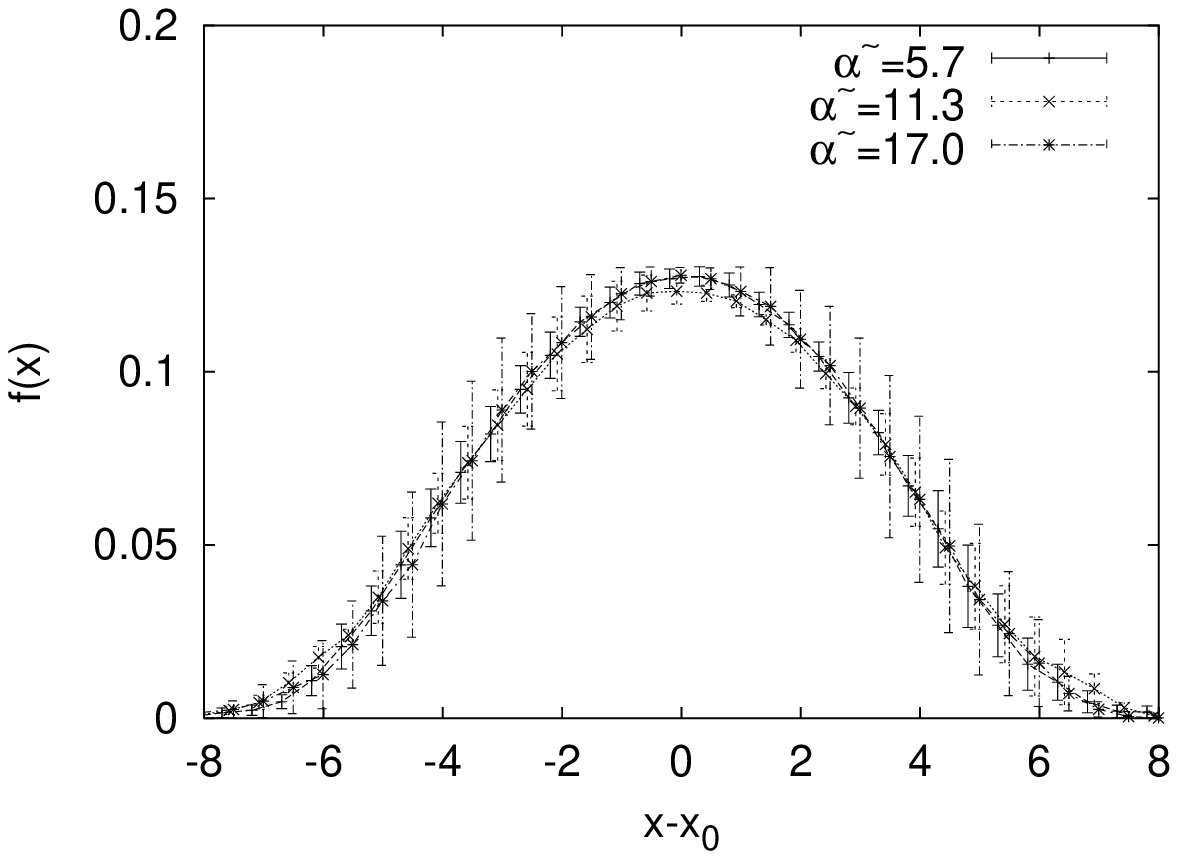}}
    \caption{Eigenvalue distribution of $Q$ around its mean, for
         $N=8$ (upper left), $N=16$ (upper right), $N=32$ (lower left) 
         and $N=64$ (lower right), respectively.}
  \label{hists}
  \end{figure}

 We next discuss the dependence of the width of the lump in the
 histogram of the Casimir $Q$ on the parameter $\alpha$ and $N$. 
 We define the width $\sigma$ as
  \begin{eqnarray}
   \sigma^{2} = \int^{\infty}_{0} dx x^{2} f(x) - 
   \left( \int^{\infty}_{0} dx x f(x) \right)^{2} 
    = \langle \frac{1}{N} Tr (A_{\mu}^{2})^{2} \rangle
    - \langle \frac{1}{N} Tr A_{\mu}^{2} \rangle^{2}
    = \langle \frac{1}{N} \sum_{i=1}^{N} \lambda_{i}^{2} \rangle
    - \langle \frac{1}{N} \sum_{i=1}^{N} \lambda_{i} \rangle^{2}.
    \nonumber \\ \label{width}
  \end{eqnarray}
 Here, $\lambda_{1}, \lambda_{2}, \cdots, \lambda_{N}$ are the
 eigenvalues of $Q$. The above three expressions are all tautological.
  For the convenience of comparing the
histograms for different $N$ and $\alpha$, we plot the eigenvalue
distribution against the  $(x-x_0)$ in Fig. \ref{hists},
instead of the usual argument $x$, where $x_0(N,{\tilde \alpha})$ are
 the mean of the observed eigenvalues for a particular $N$ and
 ${\tilde \alpha}$ and are functions of $N$ and ${\tilde \alpha}$.

  The overlapping of the different $\alpha$ histograms for a particular
$N$ shows that the width of the
eigenvalue distribution does not depend on $\alpha$.
We also observe that the width of the distributions for different $N$
has a linear dependence on  $\log N$ in Fig. \ref{width}.
The width $\sigma$ turns out to behave as
  \begin{eqnarray}
    \sigma^{2} \sim 2 \log N - 0.84.
  \end{eqnarray}
 This is in agreement with the one-loop calculation $\sigma^{2} = 2
 \log N$, which is also derived in the appendices of \cite{0401038}.
 \begin{figure}[htbp]
   \begin{center}
    \scalebox{0.55}{\includegraphics{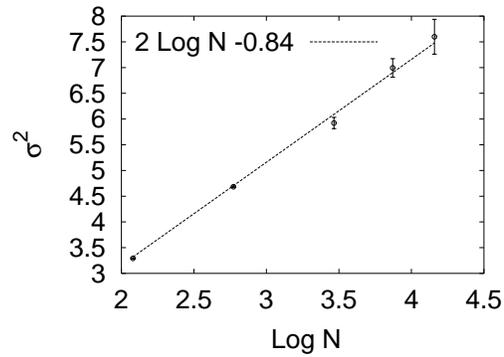}}
   \end{center}
    \caption{The square of the width of the distribution $\sigma^{2}$ 
      against $\log N$.}
  \label{width}
  \end{figure}

  \subsection{Analysis of the Yang-Mills phase}
 We have so far investigated the numerical simulation starting from
 the fuzzy sphere initial condition (\ref{fsinitial}). We have found
 that there is a phase transition between the Yang-Mills phase and
 the fuzzy sphere phase at ${\tilde \alpha}^{(l)}_{cr} \sim 2.1$
 (namely, $\alpha^{(l)}_{cr} \sim \frac{2.1}{\sqrt{N}}$).
 In this section, we elaborate on the behavior of the Yang-Mills
 phase, and we launch the Monte Carlo simulation from the initial
 condition 
  \begin{eqnarray}
     A^{(0)}_{\mu} = 0. \label{zeroinitial}
  \end{eqnarray}

 \subsubsection{Eigenvalue distribution in the Yang-Mills phase}
 \label{Yang-Mills} 
\begin{figure}[htbp]
    \begin{center}
   \scalebox{0.55}{\includegraphics{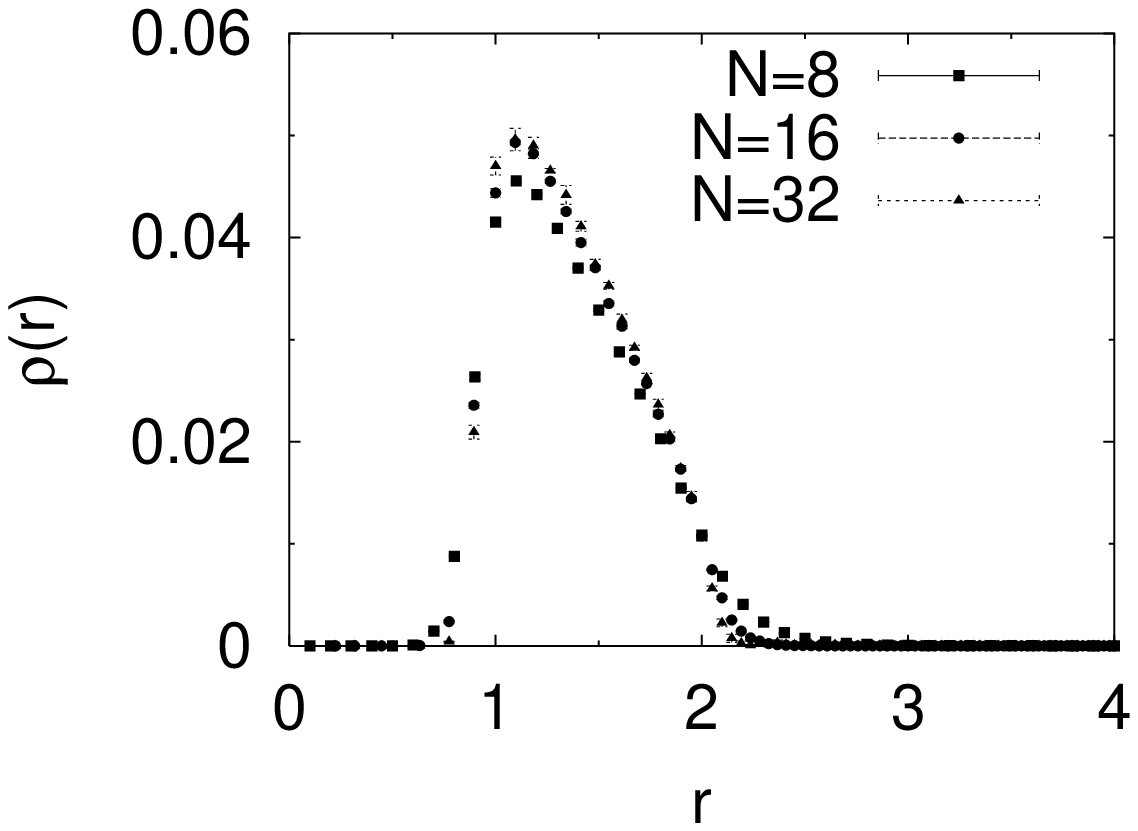}
           \includegraphics{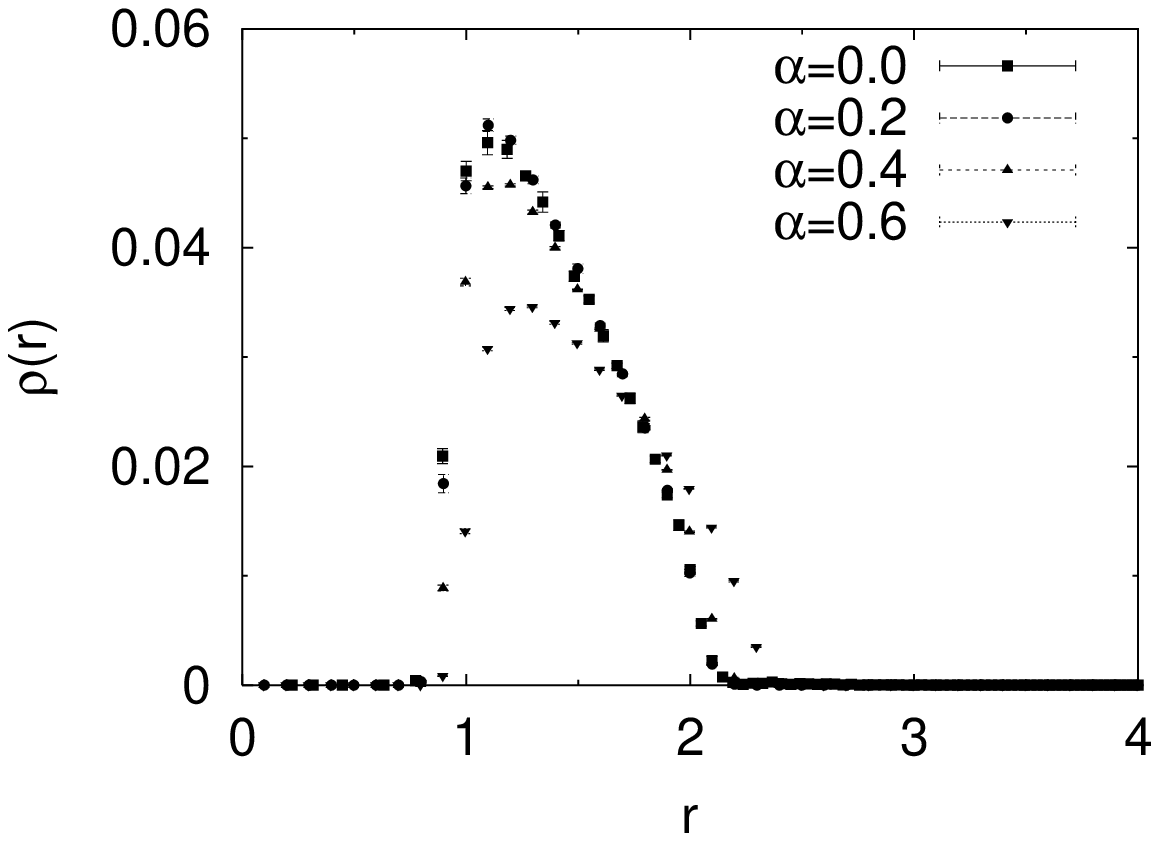}}
    \end{center}
    \caption{The plots of the eigenvalue density against the radius
 $r$, for $N=8,16,32$ with $\alpha=0.0$ to indicate the $N$
 dependence(left), and $\alpha=0.0, 0.2, 0.4, 0.6$ and $N=32$ to
 indicate the $\alpha$ dependence(right).}
    \label{hist-anYM}
   \end{figure}

 We start with discussing the eigenvalue distribution of the Casimir
 $Q$, for $N=8,16,32$ with $\alpha=0.0$ to indicate the $N$ dependence 
 and $\alpha = 0.0, 0.2, 0.4, 0.6$ with $N=32$ to indicate the
 $\alpha$ dependence. 
 Here, we plot the eigenvalue density $\rho(r)$ against $r =
 \sqrt{(\textrm{eigenvalues})}$ in Fig. \ref{hist-anYM}.
 This $r$ trivially represents the
 distance from the origin. $\rho(r)$ is defined as
  \begin{eqnarray}
   \rho(r) = \frac{\textrm{(distribution of
   $\sqrt{(\textrm{eigenvalues})}$ of $Q$)}}{4 \pi r^{2}},
  \end{eqnarray}
  so that this may represent the eigenvalue density. $\rho(r)$ is
  normalized so that $\int_{0}^{\infty} dr 4 \pi r^{2} \rho(r) =1$.
  
     The histogram for $\alpha=0.0$ is devoid of the eigenvalue in the
  vicinity of the origin, and the histogram soars around the radius 0.8.
  In this sense, there is an ultraviolet cutoff in the eigenvalue
  distribution. This phenomenon is interpreted in the following way.
  If a certain state is an eigenstate of the Casimir $Q$ with
  almost zero eigenvalues, this implies that the same state is almost
  an eigenstate of $A_{1}$, $A_{2}$ and $A_{3}$ individually with
  almost zero eigenvalues. However, this violates the uncertainty
  principle, because a generic configuration of $A_{\mu}$ do not
  commute with each other. In the bosonic IIB matrix
  model\cite{9811220}, the vacuum expectation value $\langle
  \frac{1}{N} Tr F_{\mu \nu}^{2} \rangle$ is of the order
  ${\cal O}(1)$. Therefore, the scale of the uncertainty $[A_{\mu},
  A_{\nu}]$ is also of the order ${\cal O}(1)$. The above histogram is
  consistent with this consideration.

  The histogram for $\alpha=0.0, 0.2, 0.4, 0.6$ with $N=32$
  demonstrates that the eigenvalue distribution  undergoes a gradual
  change as we vary $\alpha$ within the Yang-Mills phase.  
  
 \subsubsection{Phase transition and the hysteresis}
\begin{figure}[htbp]
    \begin{center}
      \scalebox{1.0}{\includegraphics{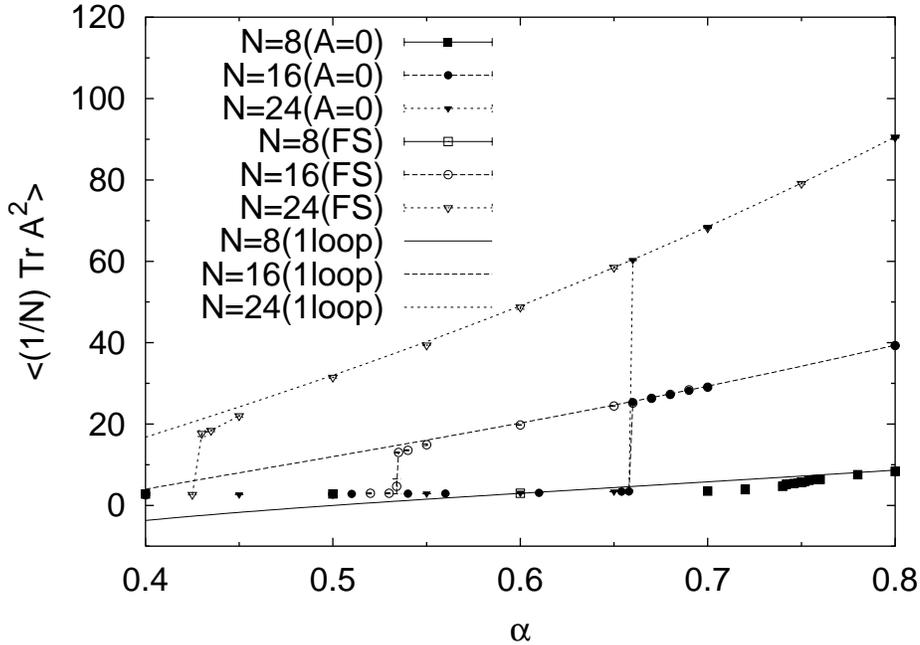}}
    \end{center}
    \caption{The plot of $\langle \frac{1}{N} Tr A^{2}_{\mu} \rangle$
   against $\alpha$, for $N=8,16,24$.}
  \label{hiscycle}
  \end{figure}

  We next investigate the phase transition of the simulation initiated 
  from (\ref{zeroinitial}). We find that the critical point differs
  in the initial condition (\ref{zeroinitial}) from that for the
  fuzzy sphere initial condition (\ref{fsinitial}). When we initiate
  the simulation from $A^{(0)}_{\mu} = 0$, the critical point is found at 
  \begin{eqnarray}
   \alpha^{(u)}_{cr} \sim 0.66. \label{criticalpointzero}
  \end{eqnarray}
   We plot $\langle \frac{1}{N} Tr A^{2}_{\mu} \rangle$ for $N=8,16,24$
  with the two initial condition 
  (\ref{fsinitial}) and (\ref{zeroinitial}) in Fig. \ref{hiscycle}. 
  The important difference in the Yang-Mills phase is that the
  critical point $\alpha^{(u)}_{cr}$ is independent of $N$.
  Since the critical point $\alpha^{(u)}_{cr}$ is larger
  than that for the fuzzy sphere initial condition $\alpha^{(l)}_{cr}
  \sim \frac{2.1}{\sqrt{N}}$ for the sufficiently large $N$, we call
  the critical point for the Yang-Mills (fuzzy sphere) phase 'the
  upper (lower) critical point' respectively.
  The above hysteresis cycle is realized for the sufficiently large
  $N$. For the $N=8$ case, the lower critical point is approximately
  $\alpha^{(l)}_{cr} \sim \frac{2.1}{\sqrt{8}} = 0.742 \cdots$, which
  is larger than $\alpha^{(u)}_{cr}$. In this case, the hysteresis
  disappears, as is discerned from Fig. \ref{hiscycle}. On the other
  hand, this is not the case with $N=16, 24$ in which the hysteresis
  cycle is realized.
   
  \subsection{Metastability of the multi-sphere solutions}  
   In this section, we investigate the multi-fuzzy-sphere solution of
  the matrix model (\ref{verydefinition}). It accommodates the
  following multi-fuzzy-sphere solution
  \begin{eqnarray}
   A_{\mu} = \alpha \times \textrm{diag} (L^{(n_{1})}_{\mu},
   L^{(n_{2})}_{\mu}, \cdots, L^{(n_{k})}_{\mu}). \label{multifs}
  \end{eqnarray}
  Here, $L^{(n_{a})}_{\mu}$ is the $n_{a}$-dimensional irreducible
  representation of the $SU(2)$ Lie algebra. The whole size of the
  matrices $A_{\mu}$ is $N = \sum_{a=1}^{k} n_{a}$.  
  For this multi-fuzzy-sphere solution, the eigenvalue distribution of $Q$ is
  peaked at 
  \begin{eqnarray}
   r^{2}_{a} = \frac{\alpha^{2}}{4} (n^{2}_{a} - 1), \hspace{3mm}
   (a=1,2,\cdots,k). \label{multipeak}
  \end{eqnarray}
  The classical value of the action is easily calculated as
   \begin{eqnarray}
    S = - \frac{\alpha^{4} N}{24} \sum_{a=1}^{k} (n_{a}^{3} -
    n_{a}). \label{multiaction}
   \end{eqnarray}
  The value of the action (\ref{multiaction}) for the
  multi-fuzzy-sphere condition is bigger than that for the single
  fuzzy sphere state (namely, $S = - \frac{\alpha^{4} N^{2}(N^{2}
  -1)}{24}$). Therefore, the multi-fuzzy-sphere state can be realized
  as a metastable state. In the following, we report the results
  of the simulations to unravel this metastability.

  \subsubsection{Evolution of the multi-fuzzy-sphere solution} \label{evo}
\begin{figure}[htbp]
   \begin{center}
    \scalebox{0.65}{\includegraphics{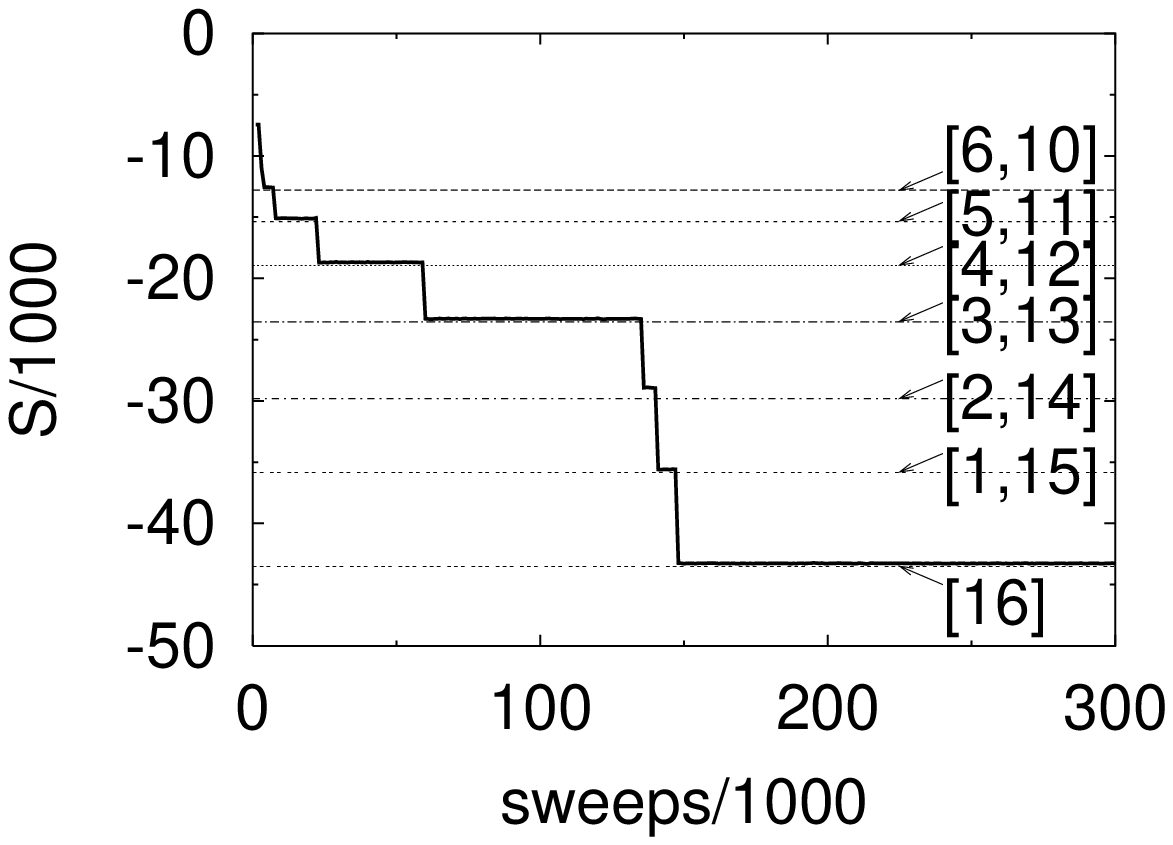}
                   \includegraphics{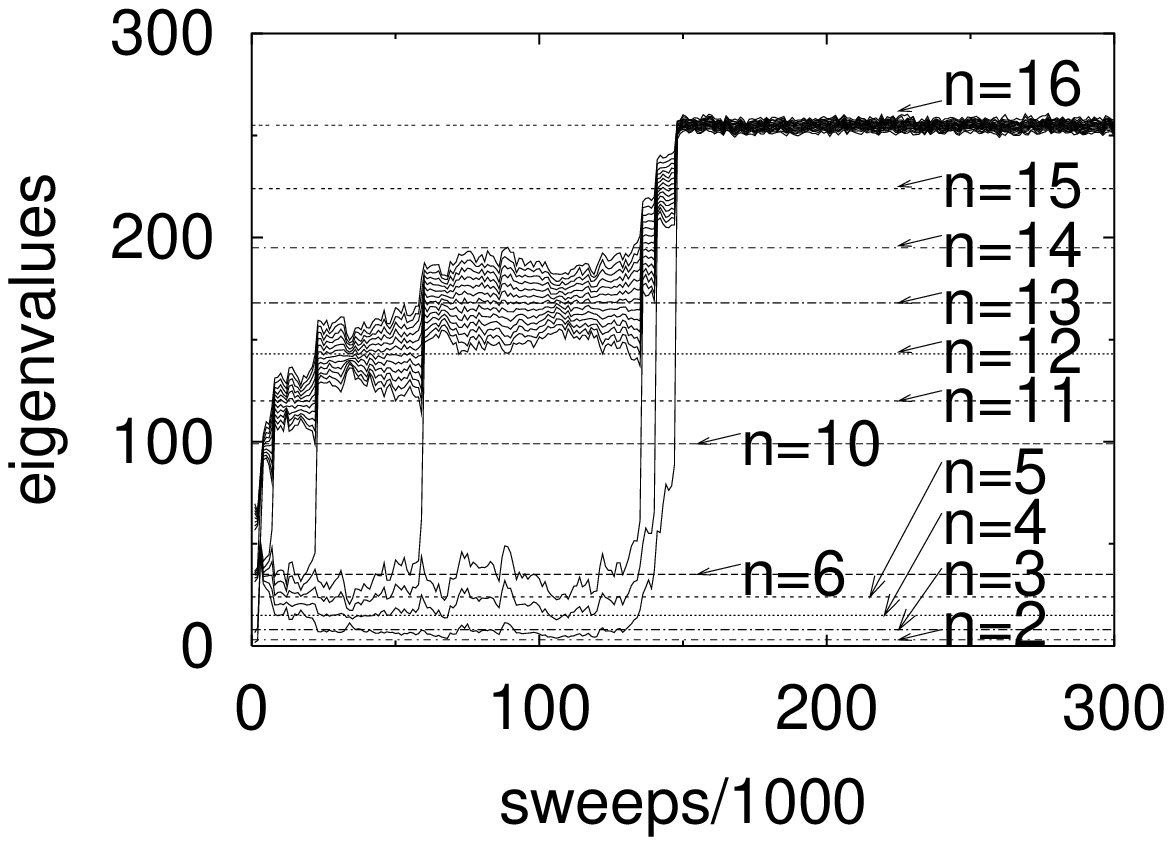} }
   \end{center}
    \caption{The history of the vacuum expectation value of the action 
  $\langle S \rangle$ (left), and the eigenvalues of $Q$ (right)
  against the sweeping time, for $N=16$, $\alpha=2.0$.}
  \label{n16a2}
  \end{figure}

  Firstly, we report the simulation with the initial condition 
  (\ref{zeroinitial}), and observe how the fuzzy sphere state
  evolves. Here, we focus on the large-$\alpha$ case in the fuzzy
  sphere phase  $N=16$ and $\alpha=2.0$.
  We plot the history of the vacuum expectation value of the action 
  $\langle S \rangle$, and the eigenvalues of $Q$ 
  against the sweeping time in Fig. \ref{n16a2}.
  This indicates the process in which the multi-fuzzy-sphere state is
  generated dynamically. Firstly, we start to see a tiny plateau in
  the graph of $\langle S \rangle$ corresponding to the
  multi-fuzzy-sphere state 
  $A_{\mu} = \alpha \left( \begin{array}{cc} L^{(6)}_{\mu} & 0 \\ 0 &
  L^{(10)}_{\mu} \end{array} \right)$. When we further undergo the
  sweeps, this multi-fuzzy-sphere state falls into the lower-energy
  state $A_{\mu} = \alpha \left( \begin{array}{cc} L^{(5)}_{\mu} & 0 \\ 0 &
  L^{(11)}_{\mu} \end{array} \right)$. In this way, the state falls
  off to the lower-energy state $A_{\mu} = \alpha \left(
  \begin{array}{cc} L^{(4,3,2,1)}_{\mu} & 0 \\ 0 & 
  L^{(12,13,14,15)}_{\mu} \end{array} \right)$.

  Finally the system arrives at the one-fuzzy-sphere 
  irreducible representation $A_{\mu} = \alpha L^{(16)}_{\mu}$.
  Namely, the one-fuzzy-sphere state is dynamically generated in the
  course of the thermalization.
  When we further undergo the sweeps, the system is stuck in this
  one-fuzzy-sphere representation, and never falls off to the other state
  any more. This observation reinforces the stability of the
  one-fuzzy-sphere state in the fuzzy sphere phase.

%   There remain two questions for Fig. \ref{n16a2}. First is the behavior
%  of the eigenvalues in the smaller representation $\alpha
%  L^{(6,5,4,3,2,1)}_{\mu}$. Fig. \ref{n16a2} indicates that these
%  eigenvalues do not reside in the value predicted classically. 
%  The second question concerns the eigenvalues in the larger
%  representation $\alpha L^{(10,11,12,\cdots,16)}_{\mu}$.
%  There is a distinct difference with respect to the width of the
%  eigenvalues between the representation $\alpha L^{(16)}_{\mu}$ and
%  the others $\alpha L^{(10,11,12,\cdots,15)}_{\mu}$. It is
%  interesting to consider what differentiates the representation 
%  $\alpha L^{(16)}_{\mu}$ in this sense.
%  These two riddles are presumably ascribed to the quantum effect.

 \subsubsection{Metastability of the fuzzy sphere solution}
\begin{figure}[htbp]
   \begin{center}
  \scalebox{0.44}{\includegraphics{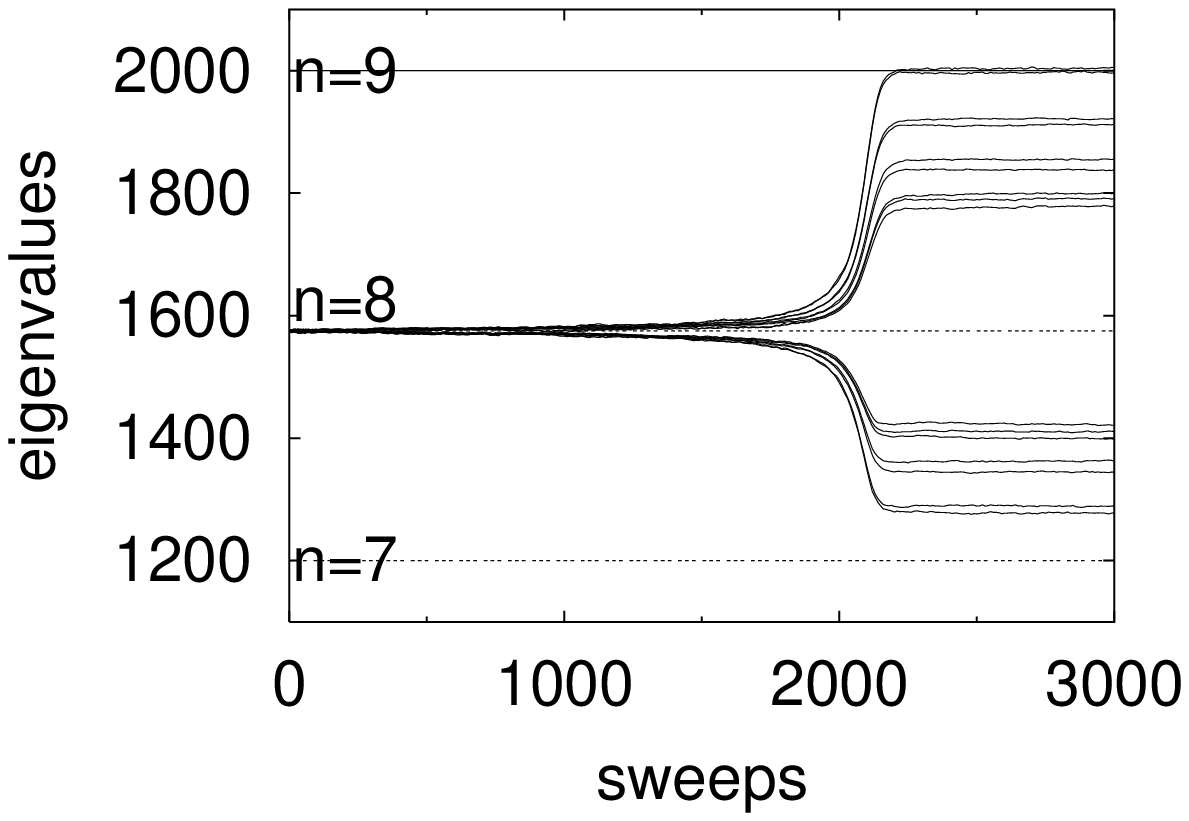}
                  \includegraphics{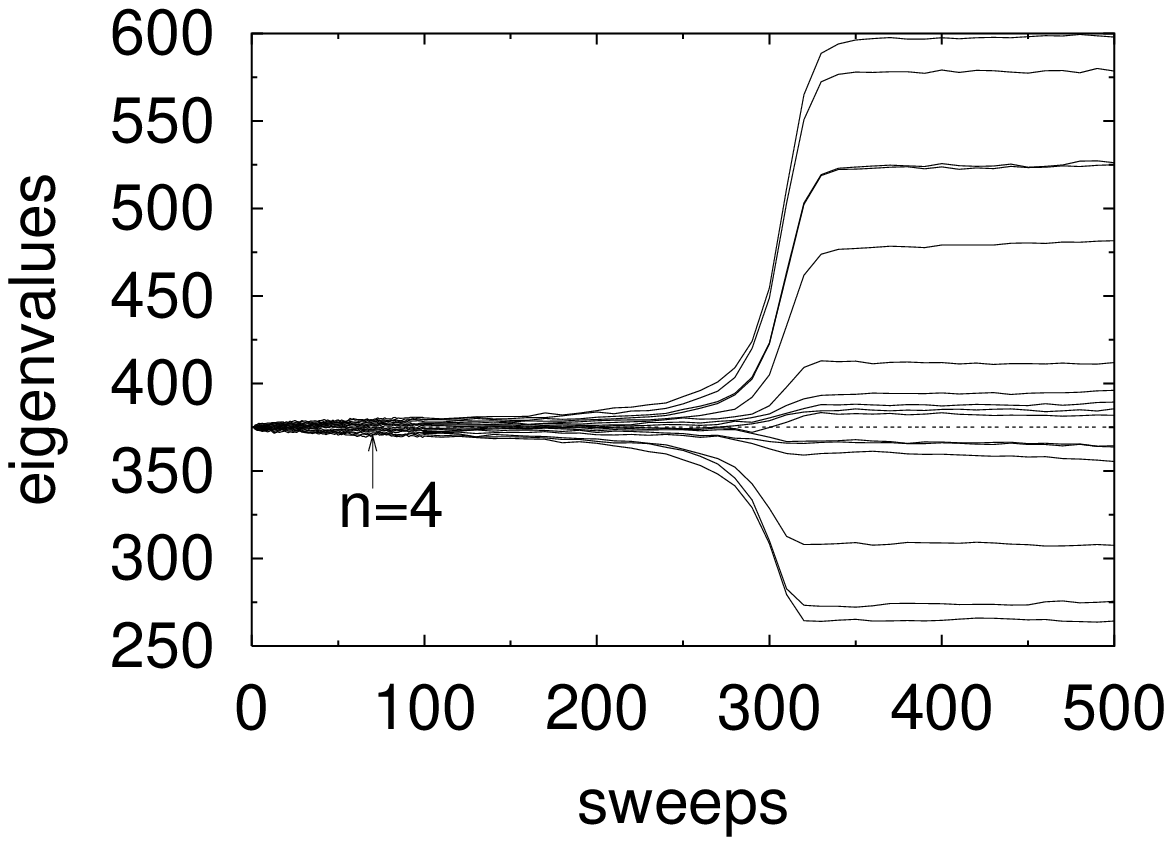}
                  \includegraphics{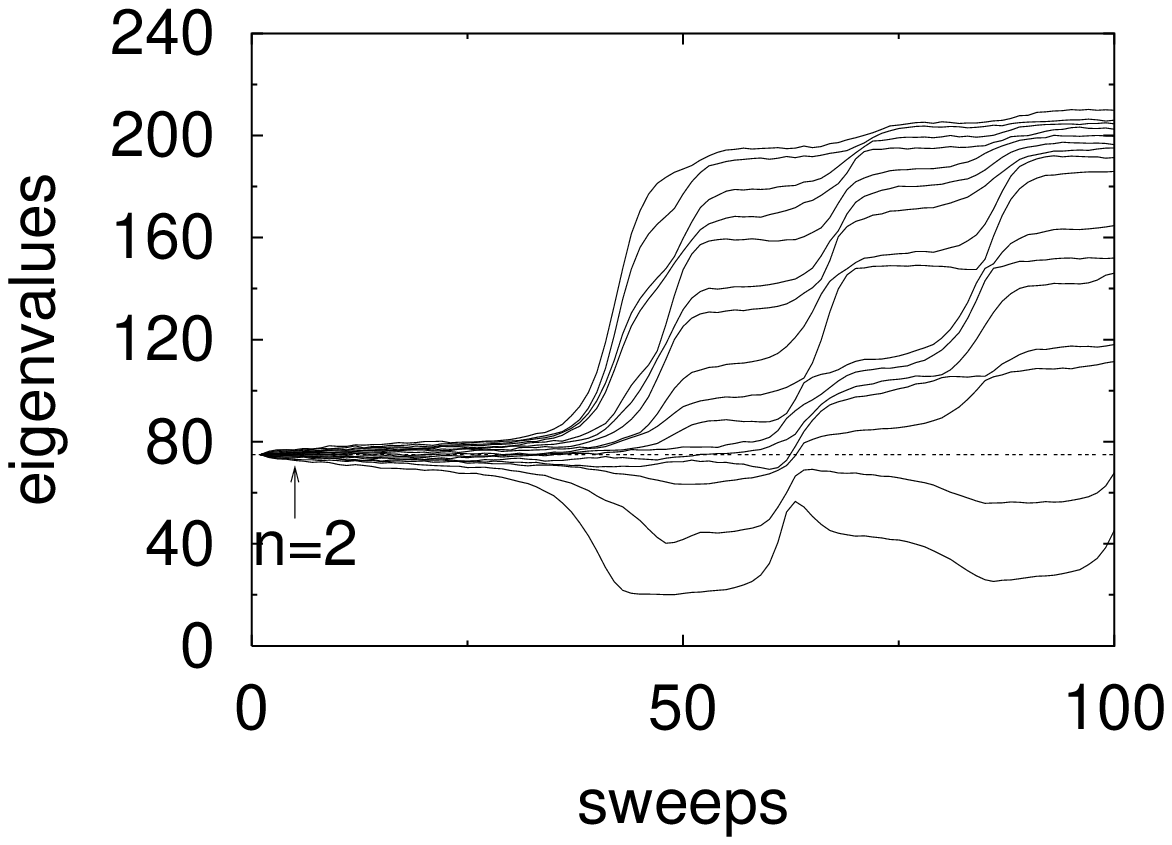}}
    \end{center}
    \caption{The eigenvalue distribution of the Casimir $Q$, as the
      multi-fuzzy-sphere solution (\ref{kini}) starts to decay, for
      $N=16$, $\alpha=10.0$ and $k=2$ (left), $k=4$ (middle), $k=8$ (right).}
  \label{decay248}
  \end{figure}

 In this section, we obtain an insight into the metastability 
of the fuzzy sphere state, by focusing on the dependence on 
$k$, $\alpha$ and $N$. 
We adopt the initial condition of the 
$n=n_{1}=n_{2}= \cdots n_{k} = \frac{N}{k}$ case 
in (\ref{multifs}) for brevity.
Namely, we initiate the simulation from the multi-fuzzy-sphere 
state
\begin{eqnarray}
A^{(0)}_{\mu} = \alpha L^{(n)}_{\mu} \otimes {\bf 1}_{k \times k}.
\label{kini} 
\end{eqnarray}
We have observed in the previous section that the 
multi-fuzzy-sphere state falls into the lower-energy state 
in the course of the thermalization of the system. Firstly, we observe 
how the multi-fuzzy-sphere classical solution (\ref{kini}) decays, for 
$N=16$, $\alpha=10.0$ and $k=2,4,8$ in Fig. \ref{decay248}.
 This indicates that the multi-fuzzy-sphere state (\ref{kini})
 decays faster for larger $k$ (namely, for the smaller representation
 of the size $n=\frac{N}{k}$). This is a natural result, and we
 elaborate on this dependence on $k$ later.
% The second significance is that the eigenvalues tend to be paired
% after the decay. We clearly observe this phenomenon especially for
% $k=2$. This phenomenon may be related to the dynamical
% generation of the gauge group, in the sense that Iso and
% Kawai\cite{9903217} proposed that the cluster distribution of $m$
% eigenvalues gives the $U(m)$ gauge group. It is intriguing to 
% elaborate on this viewpoint in the future.

\begin{figure}[htbp]
   \begin{center}
    \scalebox{0.56}{\includegraphics{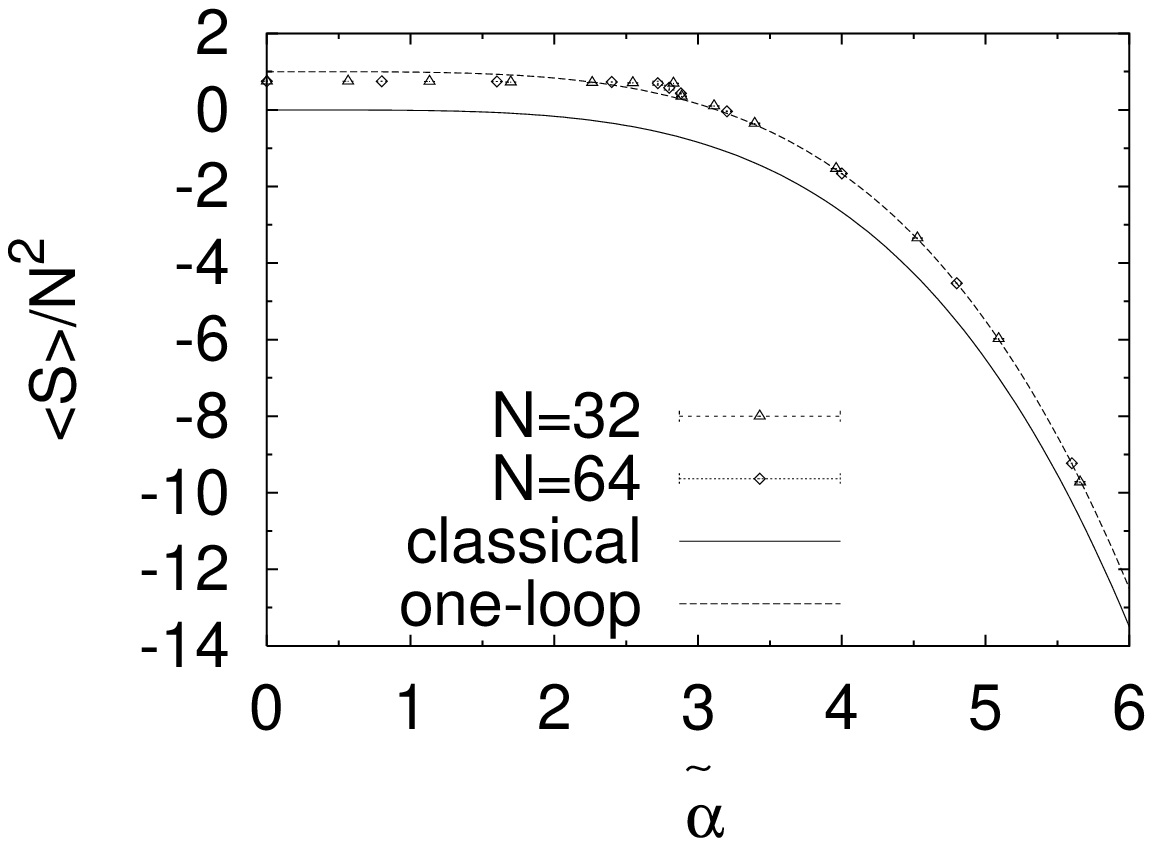}
                   \includegraphics{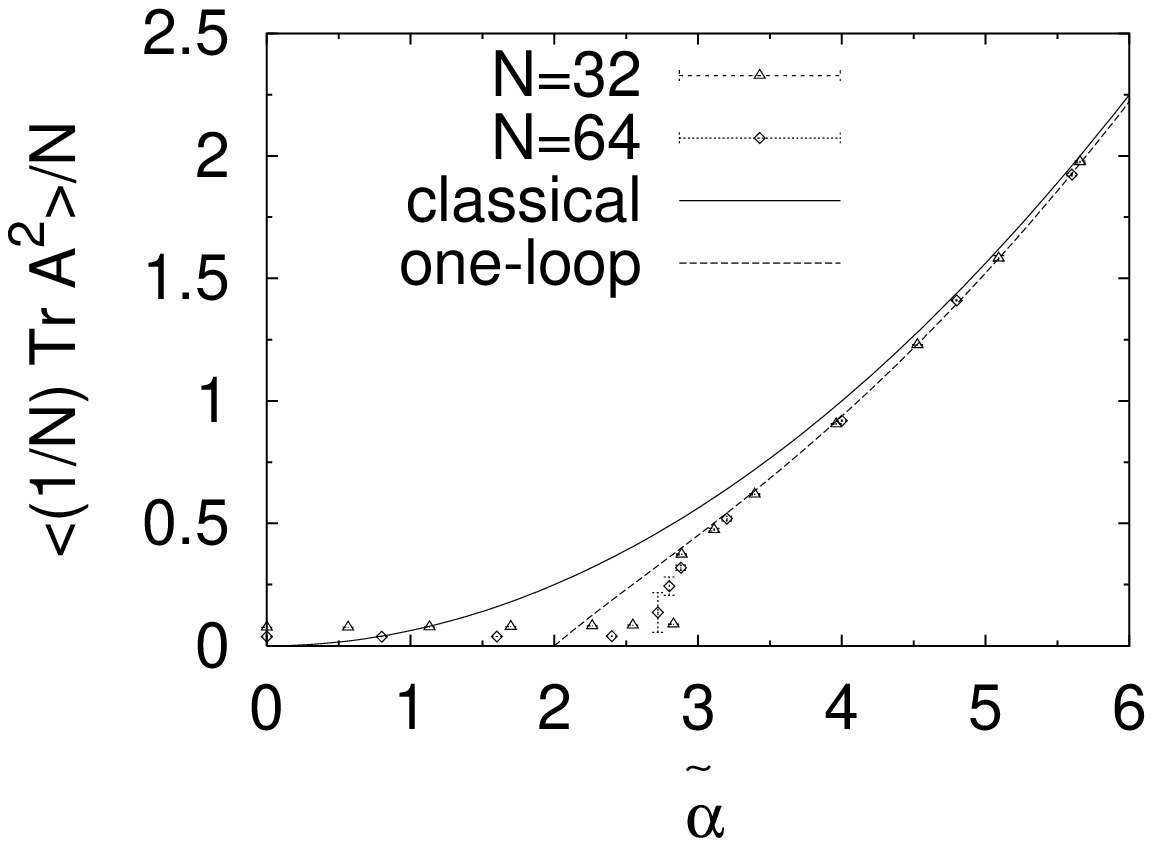}}
    \scalebox{0.56}{\includegraphics{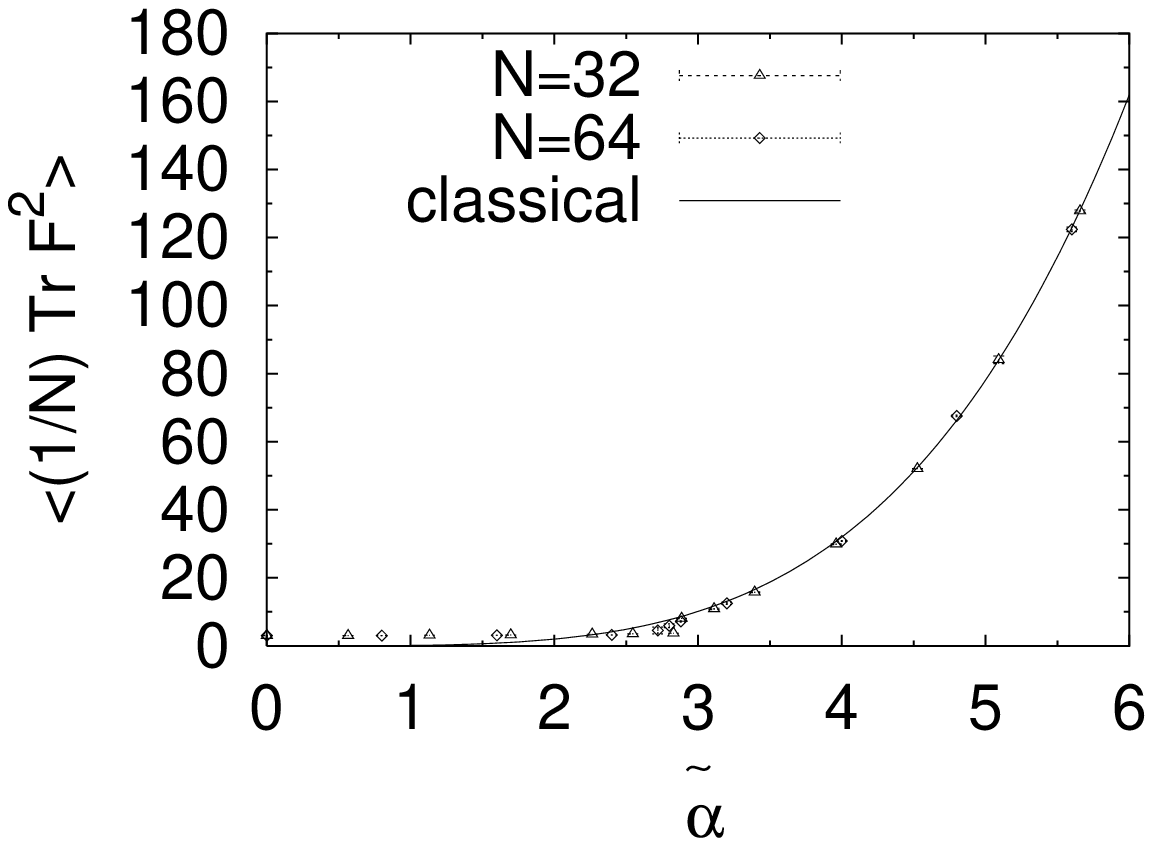}
                   \includegraphics{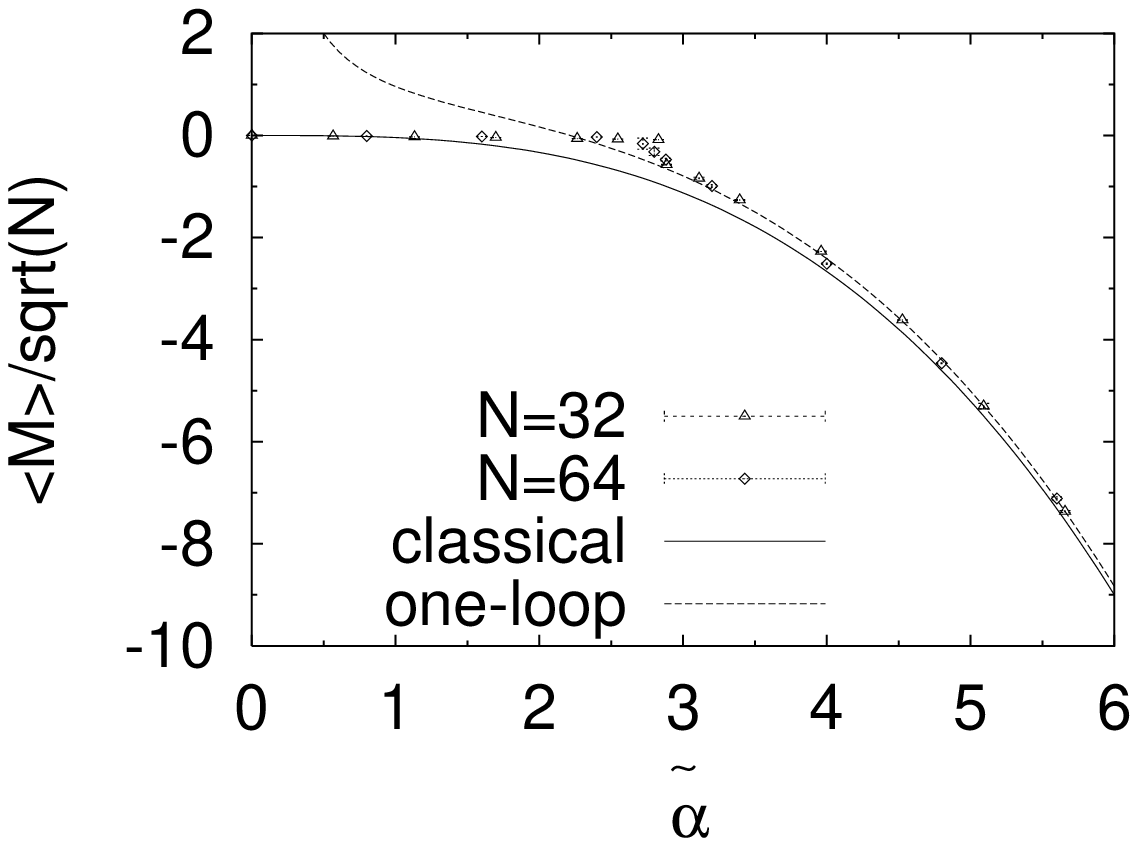}}
   \end{center}
\caption{$\frac{1}{N^{2}} \langle S \rangle$ (upper left), $\frac{1}{N} 
  \langle \frac{1}{N} Tr A^{2}_{\mu} \rangle$ (upper right), $\langle
  \frac{1}{N} Tr F_{\mu \nu}^{2} \rangle$ (lower left) 
  and $\frac{1}{\sqrt{N}}  \langle M \rangle$ (lower right) against
  ${\tilde \alpha}$, for $N=32,64$ for the multi fuzzy sphere  state
  (\ref{kini}).} 
 \label{miscFSk}
 \end{figure}

 Nextly, we show the first-order phase transition and the one-loop
 dominance of the $k=2$ reducible representation. 
 The observables $\frac{\langle S
 \rangle}{N^{2}}$, $\frac{1}{N} \langle \frac{1}{N} Tr A^{2}_{\mu}
 \rangle$, $\langle \frac{1}{N} Tr F_{\mu \nu}^{2} \rangle$
 and $\frac{1}{\sqrt{N}} \langle M \rangle$ are calculated in the
 appendices of \cite{0401038} as
  \begin{eqnarray}
    \frac{\langle S \rangle}{N^{2}} = - \frac{{\tilde
    \alpha}^{4}}{24k^{2}} + 1, \hspace{2mm}
    \frac{1}{N} \langle \frac{1}{N} Tr A_{\mu}^{2} \rangle
   = \frac{{\tilde \alpha}^{2}}{4k^{2}} - \frac{1}{{\tilde \alpha}^{2}},
    \hspace{2mm}
   \langle \frac{1}{N} Tr (F_{\mu \nu})^{2} \rangle 
  = \frac{{\tilde
  \alpha}^{2}}{2k^{2}}, \hspace{2mm}
   \frac{1}{\sqrt{N}} \langle M \rangle = - \frac{{\tilde
  \alpha}^{3}}{6k^{2}} + \frac{1}{{\tilde \alpha}}. \label{one-loopk}
  \end{eqnarray}
 for the one-loop effect around the multi fuzzy sphere (\ref{multifs})
 with $n = n_{1} = n_{2} = \cdots = n_{k} = \frac{N}{k}$.
 We launch the simulation from the initial condition (\ref{kini})
 for $k=2$.  We plot these four observables in Fig. \ref{miscFSk}.

  We have seen in Fig. \ref{decay248} that the multi fuzzy sphere
 decays in the course of the thermalization. 
 Here we undergo a sufficiently short sweep in the fuzzy sphere
 phase, such that the multi fuzzy sphere (\ref{kini}) may not decay.
 In the $k=2$ multi fuzzy sphere, we find two consequenses
 similar to the $k=1$ irreducible representation.
 Firstly, we note that there is a first-order phase transition.
 This time, the critical point lies at 
  \begin{eqnarray}
   {\tilde \alpha}^{(l)k=2}_{cr} \sim 2.8. \label{criticalpointfsk}
  \end{eqnarray}
 Secondly, the $k=2$ multi fuzzy sphere has the one-loop dominance in
 the fuzzy sphere phase in the large-$N$ limit.

 We can understand the critical point (\ref{criticalpointfsk}) from
 the one-loop dominance. 
 In the fuzzy sphere phase, the spacetime extent $\langle \frac{1}{N}
 Tr A^{2}_{\mu} \rangle$ behaves at the one-loop level as $\frac{1}{N} \langle
 \frac{1}{N} Tr A^{2}_{\mu} \rangle = \frac{{\tilde \alpha}^{2}}{4k^{2}} -
 \frac{1}{{\tilde \alpha}^{2}}$ in the fuzzy sphere phase. This is
 positive only when  
 ${\tilde \alpha} > \sqrt{2k}$. In this sense, we obtain the lower
 bound on the critical point from the one-loop dominance
 as ${\tilde \alpha}^{(l)k}_{cr} \sim {\cal 
 O}(\sqrt{k})$. The critical point (\ref{criticalpointfs})
 for $k=1$ and (\ref{criticalpointfsk}) for $k=2$ are consistent with
 this observation.

\begin{figure}[htbp]
   \begin{center}
    \scalebox{1.0}{\includegraphics{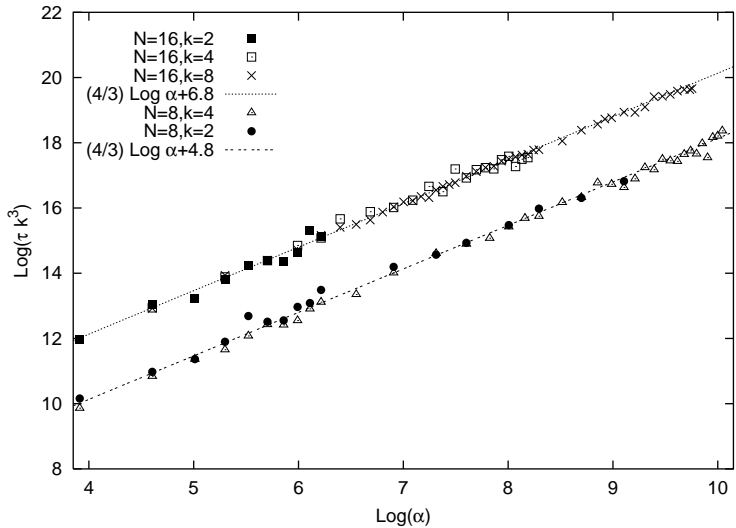}}
   \end{center}
    \caption{The plot of  $\log( \tau k^{3} )$
      against $\log \alpha$ for $N=8,16$.}
  \label{life16}
  \end{figure}

 We next discuss the sweeping time $\tau$ that takes for the initial
   condition (\ref{kini}) to decay more qualitatively.
To this end, we plot $\log( \tau  k^{3} )$
against $\log \alpha$ for $N=8,16$ in Fig. \ref{life16}. This
   indicates that the sweeping time for the multi-fuzzy-sphere state
   to decay depends on $\alpha$ and $k$ as  
 \begin{eqnarray}
  \tau \sim \alpha^{\frac{4}{3}} k^{-3}.
 \end{eqnarray}
 Fig. \ref{life16} indicates that this power law is independent of the
 size of the matrix $N$. 

  This is in contrast to the intuitive observation that the decay
  probability $P$ obeys 
  $P \sim e^{-S} = \exp (- N^{4} \alpha^{4} k^{-2} )$; 
  namely the sweeping time for the decay is $\frac{1}{P}
  \sim \exp \left( N^{4} \alpha^{4} k^{-2} \right)$. 
  It is interesting to ruminate on the reasoning for the deviation
  from this intuitive observation.
 
 \subsection{Miscellaneous future directions}
 We conclude this section by listing the future approaches to this
 work. There are a lot of interesting future works in this direction.
 Firstly, this toy model is expected to serve for the dynamical
 generation of the gauge group. As we have reviewed in Section
 \ref{sec0101102}, the expansion around the reducible representation
 (\ref{multifs}) for $n_{1} = \cdots n_{k} = \frac{N}{k}$ gives the
 noncommutative Yang-Mills theory with the $U(k)$ gauge group.
 Iso and Kawai\cite{9903217} suggested that the dynamical generation of 
 the gauge group is ascribed to the cluster distribution of the
 eigenvalues. By examining the eigenvalue distribution of this model,
 we may gain insight into the dynamical generation of the gauge group.

 Secondly, it is also exciting to extend our analysis to the
 supersymmetric case. For the simplest case, we start with the
 four-dimensional supersymmetric model with the three-dimensional
 Chern-Simons term\footnote{Austing and Wheater\cite{0310170}
 corroborated that the three-dimensional action (\ref{IIBfs}) has a
 divergent path-integral. 
 Therefore, the numerical simulation of the three-dimensional model
 is not possible.}. It is intriguing to investigate how the
 supersymmetry affects the results we have obtained for the bosonic
 case.

 Thirdly, the extension to the higher-dimensional extension is also an 
 intriguing issue. We can extend our analysis to the
 higher-dimensional fuzzy sphere, as we have reviewed in Section
 \ref{kimurayhigher}. There has been hitherto no works which discuss the
 quantum stability of the higher-dimensional fuzzy sphere $S^{2k}$,
 since it is much more mathematically involved than the simplest
 $S^{2}$ fuzzy sphere. On the other hand, it is easy to extend the
 algorithm for the matrix model with the three-dimensional
 Chern-Simons term to the higher dimension. We expect that the Monte
 Carlo simulation may open the door to the study of the
 higher-dimensional fuzzy sphere.

 The fourth future direction is to extend our analysis to the general
 homogeneous space, which is proposed by Kitazawa in \cite{0207115}.
 By investigating several homogeneous spaces (such as $CP^{2} =
 SU(3)/U(2)$), we may be able to understand which homogeneous space is 
 the most favored in the quantum sense. We expect that this direction
 may give us a clue to the compactification of the spacetime.

 In this way, the Monte Carlo simulation we have described here gives
 us a lot of exciting prospects.

 \section{Conclusion}
 In this thesis, we have reviewed the author's
 works\cite{0102168,0204078,0209057,0401038} about the relation
 between the gravitational interaction and the matrix model.
 Now, it is believed that the 'Theory of Everything', which unifies
 all the interactions in the nature, is realized by the superstring
 theory. The large-$N$ reduced model is now regarded as the promising
 candidate for the nonperturbative formulation of the superstring
 theory. If we build
 the nonperturbative formulation of the superstring theory, 
 this will be the right framework to unify all interactions.
 Then, we will be able to solve all the riddles of the Standard Model,
 such as the dimensionality of our spacetime, the generation of the
 quark, the gauge group and the eighteen parameters of the Standard Model, 
 from the parameterless theory.

 In order to arrive at this ultimate goal, it is an important step 
 to see the correspondence between the matrix model and the
 gravitational interaction. In the future, it is indispensable to
 elucidate the more manifest correspondence with the gravitational
 interaction. For example, it is interesting to pursue the matrix
 model that is equipped with the local Lorentz symmetry and reduces to
 the supergravity  in the low-energy limit, by inheriting the idea we
 described in Section 4. If we find such a matrix model, it will be
 certainly a better extension of the IIB matrix model, in the sense
 that it has clearer relation to the gravitational interaction.
 In addition, it is also interesting how the matrix model realizes the 
 curved-space background.  The curved spacetime is also an essential
 feature of the general relativity. So far, it is known that the IIB
 matrix model, per se, realizes the curved-space background through the
 condensation of the graviton. In addition, several alterations of the 
 IIB matrix model have been advocated, so that they may be equipped
 with the curved-space manifold. On the other hand, the alterations of 
 the IIB matrix model is merely limited to the simple manifolds;
 such as the $S^{2k}$ fuzzy sphere, fuzzy torus or $CP^{2}$
 manifold. It is exciting to pursue how the matrix model realizes the
 general curved spacetime more manifestly.
 
  We finally mention the relation between the IIB matrix model and its 
  miscellaneous extensions. In the quantum field theory, some different
  models which share the same symmetry are equivalent in the continuum
  limit, known as universality.
  We expect that the similar mechanism may hold true of 
  the matrix models and hence that various matrix models may
  have the same large $N$ limit. Thus, there is a possibility that these 
  new extensions are equivalent to the IIB matrix model.
  We can expect that both the IIB matrix model and the new extensions are
  equally authentic constructive definition of superstring theory in
  the large $N$ limit. It is an intriguing future work to elaborate on 
  this conjecture of the universality.

 Mankind has yet to grasp what is the true 'Theory of Everything'.
 We must exert more effort to arrive at the answer to this ultimate 
 and most difficult question of the elementary particle physics.

%%%%%%%%%%%%%%%%%%%%%%%%%%%%%%%%%%%%%%%%%%%%%%%%%%%%%%%%%%%%%%%%%%%
%%%%%%%%%%%%%%%%%%%%%%%%%%%%%%%%%%%%%%%%%%%%%%%%%%%%%%%%%%%%%%%%%%%
%%%%%%%%%%%%%%%%%%%%%%%%%%%%%%%%%%%%%%%%%%%%%%%%%%%%%%%%%%%%%%%%%%%
%%%%%%%%%%%%%%%%%%%%%%%%%%%%%%%%%%%%%%%%%%%%%%%%%%%%%%%%%%%%%%%%%%%

 \paragraph{Acknowledgment} \hspace{0mm} 

  It is my great pleasure to acknowledge the collaboration and the
  advice to complete the present work.
  
  I would like to express my sincere gratitude to
  my immediate supervisor Prof. Hikaru Kawai. Without his guidance, I
  could not have completed these works.

  I would also like to express my thanks to Dr. Subrata Bal,
  Dr. Maxime Bagnoud, Prof. Satoshi Iso, Prof. Hikaru Kawai,
  Dr. Keiichi Nagao, Prof. Jun Nishimura and Dr. Yuhi Ohwashi for the
  collaboration. 
  Especially, Prof. Nishimura's pedagogical lecture at KEK invited me
  to the numerical simulation of the large-$N$ reduced model.

  I also thank Prof. Hajime Aoki, 
  Dr. Shoichi Kawamoto, Dr. Yusuke Kimura, Prof. Yoshihisa Kitazawa,
  Dr. Tsunehide Kuroki, Mr. Takeshi Morita, Dr. Shin Nakamura,
  Dr. Shun'ichi Shinohara, Prof. Tsukasa Tada,
  Mr. Yastoshi Takayama, Dr. Dan Tomino and Prof. Asato Tsuchiya for
  valuable discussion. 

  I thank the Yukawa Institute for Theoretical Physics at Kyoto
  University. Discussions during the YITP-W-02-11 and YITP-W-03-09 
  on "YONUPA Summer School", YITP-W-01-04, YITP-W-02-04 and
  YITP-W-03-07 on "Quantum Field Theory", YITP-W-02-15 on "YITP School
  on Lattice Field Theory" and YITP-W-02-18 and YITP-W-03-19 on
  "Quantum Field Theories: Fundamental Problems and Applications"  
  were useful to complete the works in this thesis. In addition, the
  numerical computations in the work\cite{0401038} were in part
  performed at the Yukawa Institute Computer Facility.  

 These works are supported in part by Grant-in-Aid for
 Scientific Research from Ministry of Education, Culture, Sports,
 Science and Technology of Japan $\#$01282.

%%%%%%%%%%%%%%%%%%%%%%%%%%%%%%%%%%%%%%%%%%%%%%%%%%%%%%%%%%%%%%%%%%%%
%%%%%%%%%%%%%%%%%%%%%%%%%%%%%%%%%%%%%%%%%%%%%%%%%%%%%%%%%%%%%%%%%%%%
%%%%%%%%%%%%%%%%%%%%%%%%%%%%%%%%%%%%%%%%%%%%%%%%%%%%%%%%%%%%%%%%%%%%
%%%%%%%%%%%%%%%%%%%%%%%%%%%%%%%%%%%%%%%%%%%%%%%%%%%%%%%%%%%%%%%%%%%%
%%%%%%%%%%%%%%%%%%%%%%%%%%%%%%%%%%%%%%%%%%%%%%%%%%%%%%%%%%%%%%%%%%%%

\appendix
 \section{Notation} \label{notation}
  \subsection{Definition and properties of the gamma matrices and the
  fermions} 
  Basically, we follow the same notation as for \cite{0103003}.
  Throughout this thesis, the metric of the Minkowskian spacetime is
  taken so that the time component should be minus and the space
  components should be plus. Namely, the metric is defined as
   \begin{eqnarray}
    \eta^{\mu \nu} = \textrm{diag}(-1,+1, \cdots, +1). \label{minkowski}
   \end{eqnarray}
  The gamma matrices are defined so that it may comply with the
  following Clifford algebra:
   \begin{eqnarray}
    \{ \Gamma^{\mu}, \Gamma^{\nu} \} = 2 \eta^{\mu \nu} {\bf
    1}_{32 \times 32}. \label{clifford} 
   \end{eqnarray}
  Here, the gamma matrices $\Gamma^{\mu}$ are the $32 \times 32$ real
  matrices and the indices $\mu, \nu, \cdots$ run over $0,1,\cdots,
  9$.  The explicit components are given by 
  \begin{eqnarray}
   \Gamma^{0} = {\bf 1}_{16 \times 16} \otimes (-i \sigma_{2})
              = \left( \begin{array}{cc} 0 & - {\bf 1}_{16 \times 16}
              \\ {\bf 1}_{16 \times 16} & 0 \end{array} \right), 
   \hspace{2mm}
     \Gamma^{p} = \gamma^{p} \otimes \sigma_{3}
              = \left( \begin{array}{cc} \gamma^{p} & 0
              \\ 0 & \gamma^{p} \end{array} \right). \nonumber
  \end{eqnarray}
   Here, $\gamma^{p}$ are the nine-dimensional (and thus $16 \times
   16$) gamma matrices, and obey the Clifford algebra for the
   nine-dimensional Euclidean spacetime
   $\{ \gamma^{p}, \gamma^{q} \} = 2 \delta^{pq} {\bf 1}_{16 \times
   16}$, where $p, q, \cdots$ run over $1,2,\cdots, 9$.

   The transpose of the gamma matrices are given by 
   \begin{eqnarray}
     ^{T}(\Gamma^{0}) = - \Gamma^{0}, \hspace{2mm}
     ^{T}(\Gamma^{p}) = + \Gamma^{p} \hspace{2mm}
     (p=1,2,\cdots,9). \label{gammatranspose}. 
   \end{eqnarray}
   The chirality matrices $\Gamma^{10} (=\Gamma^{\sharp})$ are defined by
   \begin{eqnarray}
    \Gamma^{\sharp} = \Gamma^{0} \Gamma^{1} \Gamma^{2} \cdots
    \Gamma^{9} = {\bf 1}_{16 \times 16} \otimes \sigma_{1}
    =  \left( \begin{array}{cc} 0 &  {\bf 1}_{16 \times 16}
              \\ {\bf 1}_{16 \times 16} & 0 \end{array} \right).
   \label{chiralitydef}
   \end{eqnarray}
  For the indices $A,B, \cdots = 0,1,\cdots, 10$ in the eleven-dimensional 
  Minkowski spacetime, the gamma matrices
  comply with the Clifford algebra $\{ \Gamma^{A}, \Gamma^{B} \} = 2 
  \eta^{AB} {\bf 1}_{32 \times 32}$. 

  Next, we define the charge conjugation matrix $C$. $C$ should satisfy
  the conditions
   \begin{eqnarray}
     C \Gamma^{\mu} = - ^{T}(\Gamma^{\mu}) C, \hspace{2mm}
     C + ^{T}C = 0. \label{chargeconjugation}
   \end{eqnarray}
  In our case, $C$ is identical to $\Gamma^{0}$. 

  The anti-symmetrized gamma matrices are abbreviated as
   \begin{eqnarray}
    \Gamma^{A_{1} \cdots A_{k}} = \frac{1}{k!} \sum_{\sigma \in
    {\cal S}_{n}} \textrm{sgn}(\sigma) \Gamma^{A_{\sigma(1)}}
    \Gamma^{A_{\sigma(2)}} \cdots \Gamma^{A_{\sigma(k)}}.
    \label{antisymmetry}  
   \end{eqnarray}
  Here, ${\cal S}_{n}$ is the set of the permutation of the integers
  $1,2,\cdots, k$. 

  \subsubsection{Chirality of the Weyl fermion} 
  We summarize the notation for the chirality of the fermion.
  The left-hand (right-hand) fermion is defined as
   \begin{eqnarray}
     \psi_{L} = \frac{1+\Gamma^{\sharp}}{2} \psi, \hspace{2mm}
     \psi_{R} = \frac{1-\Gamma^{\sharp}}{2} \psi. \label{chiralfermion}
   \end{eqnarray}
  The left-handed (right-handed) fermions satisfy the following
  identities for the ten-dimensional gamma matrices:
  \begin{eqnarray}
   {\bar \chi}_{L} \Gamma^{\mu_{1} \cdots \mu_{2k}} \epsilon_{L}
 = {\bar \chi}_{R} \Gamma^{\mu_{1} \cdots \mu_{2k}} \epsilon_{R} = 0,
 \hspace{2mm} 
   {\bar \chi}_{L} \Gamma^{\mu_{1} \cdots \mu_{2k+1}} \epsilon_{R}
 = {\bar \chi}_{R} \Gamma^{\mu_{1} \cdots \mu_{2k+1}} \epsilon_{L} =
 0. \label{leftright}
  \end{eqnarray}
  Here, we prove only ${\bar \chi}_{L} \Gamma^{\mu_{1} \cdots
  \mu_{2k}} \epsilon_{L} =0$, so that the rest can be verified
  likewise.
  \begin{eqnarray}
       {\bar \chi}_{L} \Gamma^{\mu_{1} \cdots \mu_{2k}} \epsilon_{L}
   &=&  {^{T} \chi} {^{T} (\frac{1 + \Gamma^{\sharp}}{2})} \Gamma^{0} 
       \Gamma^{\mu_{1} \cdots \mu_{2k}} \frac{1 + \Gamma^{\sharp}}{2}
       \epsilon = {^{T} \chi} \frac{1 + \Gamma^{\sharp}}{2} \frac{1 +
       (-1)^{2k+1} \Gamma^{\sharp}}{2} \Gamma^{0} \Gamma^{\mu_{1} \cdots
       \mu_{2k}} \epsilon \nonumber \\
    &=&  {^{T} \chi} \frac{1 + \Gamma^{\sharp}}{2} \frac{1 -
       \Gamma^{\sharp}}{2} \Gamma^{0} \Gamma^{\mu_{1} \cdots \mu_{2k}}
       \epsilon = 0.
  \end{eqnarray}

 \subsubsection{Epsilon tensors and the duality of the matrices}
  Here, we define the anti-symmetric epsilon tensors as
  \begin{eqnarray}
   \epsilon_{01 \cdots k} = 1, \hspace{2mm} (\textrm{so that }
   \epsilon^{01 \cdots k} = -1). \label{epsilon}
  \end{eqnarray}
  This definition leads the duality relations of the gamma matrices
  \begin{eqnarray}
   \Gamma^{A_{0} \cdots A_{k}} =
   \frac{(-1)^{\frac{k(k+1)}{2}}}{(10-k)!} \epsilon^{A_{0} \cdots
   A_{9} A_{\sharp}} \Gamma_{A_{k+1} \cdots A_{\sharp}}. \label{duality}
  \end{eqnarray}

  \subsubsection{Multiplication law of the gamma matrices}
  The anti-symmetrized gamma matrices obey the following formula of
  the product.
  \begin{eqnarray} \Gamma^{A_{1} \cdots A_{m}} \Gamma^{B_{1}
    \cdots B_{n}} &=& \Gamma^{A_{1} \cdots A_{m} B_{1} \cdots
    B_{n}} + {(-1)^{m-1}} {_{m}C_{1}}{_{n}C_{1}}
    {\eta^{[A_{1}}}^{[B_{1}} {\Gamma^{A_{2} \cdots
    A_{m}]}}^{B_{2} \cdots B_{n}]} \nonumber \\
   &+& {(-1)^{(m-1)+(m-2)}} {_{m}C_{2}}{ _{n}C_{2}} 2!
    {\eta^{[A_{1}}}^{[B_{1}} {\eta^{A_{2}}}^{B_{2}}
    {\Gamma^{A_{3} \cdots A_{m}]}}^{B_{3} \cdots B_{n}]}
    \label{AZproduct} \\
  &+& {(-1)^{(m-1)+(m-2)+(m-3)}} {_{m}C_{3}}{ _{n}C_{3}} 3!
    {\eta^{[A_{1}}}^{[B_{1}} {\eta^{A_{2}}}^{B_{2}}
    {\eta^{A_{3}}}^{B_{3}} {\Gamma^{A_{4} \cdots
    A_{m}]}}^{B_{4} \cdots B_{n}]} + \cdots.  \nonumber
     \end{eqnarray}   
 
 \subsubsection{Flipping property of the Majorana fermions} 
   Here, we introduce a useful formula for the flipping of the Majorana 
  fermion.  The gamma matrices has the following property:
   \begin{eqnarray}
    \Gamma^{0} (^{T} \Gamma^{A}) \Gamma^{0} = \Gamma^{A}. 
   \end{eqnarray}
  This leads to the following relation:
   \begin{eqnarray}
    \Gamma^{0} ({^{T}\Gamma^{A_{1} \cdots A_{k}}}) \Gamma^{0}
    &=&
    (-1)^{k-1} (\Gamma^{0} ({^{T} \Gamma^{A_{k}}})  \Gamma^{0} )
    \cdots (\Gamma^{0} ({^{T} \Gamma^{A_{1}}})  \Gamma^{0} ) 
     =
    (-1)^{k-1} \Gamma^{A_{k} \cdots A_{1}} \nonumber \\
    &=& (-1)^{k-1}
    (-1)^{\frac{k(k-1)}{2}} \Gamma^{A_{1} \cdots A_{k}} =
    (-1)^{\frac{(k+2)(k-1)}{2}} \Gamma^{A_{1} \cdots A_{k}}.
    \label{flipgamma}  
   \end{eqnarray}
  Then, the Majorana fermions are flipped as
   \begin{eqnarray}
    {\bar \chi} \Gamma^{A_{1} \cdots A_{k}} \epsilon
 &=&  ^{T}({\bar \chi} \Gamma^{A_{1} \cdots A_{k}} \epsilon)
 = - ^{T}\epsilon (^{T}\Gamma^{A_{1} \cdots A_{k}}) (^{T}\Gamma^{0})
      \chi
 = - ^{T}\epsilon (\Gamma^{0})^{2} (^{T}\Gamma^{A_{1} \cdots A_{k}})
      (\Gamma^{0}) \chi \nonumber \\
 &=& (-1)^{\frac{k(k+1)}{2}}  {\bar \epsilon} \Gamma^{A_{1} \cdots
 A_{k}} \chi. \label{flippingproperty}
   \end{eqnarray}
 Namely, the sign is minus for $k=1,2,5$, and plus for $k=0,3,4$.

 \subsubsection{Proof of the Fierz identity (\ref{AZM31fierz}) and
   (\ref{fierz3dim})} \label{sec-fierz}
  We give a proof of the formula of the Fierz identity
  (\ref{AZM31fierz}) and (\ref{fierz3dim}), which contributes to the
  proof of the supersymmetry algebra of the matrix model. 

  We start with the ten-dimensional formula (\ref{AZM31fierz}). 
  The fermions $\epsilon_{1}$ and $\epsilon_{2}$ are Majorana-Weyl
  fermions, and effectively act on the $16 \times 16$ space of the
  gamma matrices. The left hand side is written as
   \begin{eqnarray}
  {\bar \epsilon}_{1} \Gamma_{\nu} \psi \Gamma^{\mu \nu} \epsilon_{2} 
  =  ({\bar \epsilon}_{1})^{\alpha} {(\Gamma_{\nu})_{\alpha}}^{\beta} 
     \psi_{\beta} {(\Gamma^{\mu \nu})_{\gamma}}^{\delta} \epsilon_{2
  \delta} 
  = - {(\Gamma^{\mu \nu})_{\gamma}}^{\delta} {({\bar \epsilon}_{1}
  \epsilon_{2})_{\delta}}^{\alpha} {(\Gamma_{\nu})_{\alpha}}^{\beta}
  \psi_{\beta}. \label{AZMAf2}
   \end{eqnarray} 
   Here, the minus sign emerges because we have flipped the order of
  the Grassmann odd fermions. The matrix ${({\bar \epsilon}_{1}
  \epsilon_{2})_{\delta}}^{\alpha}$ can be decomposed as
   \begin{eqnarray}
   {({\bar \epsilon}_{1} \epsilon_{2})_{\delta}}^{\alpha} &=& 
    \frac{1}{16} ({\bar \epsilon}_{1} \epsilon_{2}) {\bf 1}
  + \frac{1}{16} ({\bar \epsilon}_{1} \Gamma_{\mu} \epsilon_{2})
    \Gamma^{\mu} 
  - \frac{1}{16 \times 2!} ({\bar \epsilon}_{1} \Gamma_{\mu_{1} \mu_{2}}
    \epsilon_{2}) \Gamma^{\mu_{1} \mu_{2}} 
  -  \frac{1}{16 \times 3!} ({\bar \epsilon}_{1} \Gamma_{\mu_{1} \mu_{2}
    \mu_{3}} \epsilon_{2}) \Gamma^{\mu_{1} \mu_{2} \mu_{3}} \nonumber \\ 
  &+&  \frac{1}{16 \times 4!} ({\bar \epsilon}_{1} \Gamma_{\mu_{1}
    \cdots \mu_{4}}  \epsilon_{2}) \Gamma^{\mu_{1} \cdots \mu_{4}} 
  +  \frac{1}{16 \times 5!} ({\bar \epsilon}_{1} \Gamma_{\mu_{1}
    \cdots \mu_{5}} 
    \epsilon_{2}) \Gamma^{\mu_{1} \cdots \mu_{5}}. \label{AZMAf}
  \end{eqnarray}
  Since the fermions $\epsilon_{1}$ and $\epsilon_{2}$ have the same
  chirality, the terms of the even rank vanish due to the relation
  (\ref{chiralfermion}). We ignore the rank-3 term because we are
  interested in the difference $ {\bar \epsilon}_{1} \Gamma_{\nu} \psi
  \Gamma^{\mu \nu} \epsilon_{2}
       -  {\bar \epsilon}_{2} \Gamma_{\nu} \psi \Gamma^{\mu \nu}
       \epsilon_{1}$, and the rank-3 term does not contribute due to
  the relation (\ref{flippingproperty}). This leads us to focus only
  on the rank-1,5 terms. The relevant gamma matrices are computed as
    \begin{eqnarray}
   \Gamma^{\mu \nu} \Gamma_{\rho} \Gamma_{\nu} &=& \Gamma^{\mu\nu}
   (\Gamma_{\rho\nu} + \eta_{\rho\nu})  
   = (- {\eta_{\nu}}^{\nu} {\Gamma^{\mu}}_{\rho} + 2 {\Gamma^{\mu}}_{\rho}
     - {\eta_{\nu}}^{\nu} {\eta^{\mu}}_{\rho} + {\eta^{\mu}}_{\rho} ) +
   {\Gamma^{\mu}}_{\rho}
   \stackrel{{\eta^{\nu}}_{\nu} = 10}{=} 7 {\Gamma_{\rho}}^{\mu} - 9
   {\eta^{\mu}}_{\rho} \nonumber \\ 
   &=& 7 ( \Gamma_{\rho} \Gamma^{\mu} - {\eta^{\mu}}_{\rho} ) -  9
   {\eta^{\mu}}_{\rho} 
   = 7 \Gamma_{\rho} \Gamma^{\mu} - 16 {\eta^{\mu}}_{\rho},
   \label{AZMArank1} \\  \Gamma^{\mu\nu} \Gamma_{\rho_{1} \cdots
   \rho_{5}} \Gamma_{\nu}  
   &=& \Gamma^{\mu\nu} ( - \Gamma_{\nu \rho_{1} \cdots \rho_{5}} + 5
   \eta_{\nu [\rho_{1}} \Gamma_{\rho_{2} \cdots \rho_{5}]} ) \nonumber
   \\ 
  &=& - ( - \eta^{\mu \nu} \Gamma_{\nu \rho_{1} \cdots \rho_{5}} +
   {\eta^{\nu}}_{\nu} {\Gamma^{\mu}}_{\rho_{1} \cdots \rho_{5}}
   - 5 \eta_{\nu [\rho_{1}} {\Gamma^{\mu \nu}}_{\rho_{2} \cdots \rho_{5}]} )
      + 5 ( {\Gamma^{\mu}}_{\rho_{1} \cdots \rho_{5}} 
        - {\eta^{\mu}}_{[ \rho_{2}} \Gamma_{\rho_{1} \rho_{3} \rho_{4}
   \rho_{5}]} )  \nonumber \\  
  &=& - {\Gamma_{\rho_{1} \cdots \rho_{5}}}^{\mu} + 5 {\eta^{\mu}}_{[\rho_{1}}
   \Gamma_{\rho_{2} \cdots \rho_{5}]}
   = \Gamma_{\rho_{1} \cdots \rho_{5}} \Gamma^{\mu}. \label{AZMArank5}
  \end{eqnarray}
  When we substitute (\ref{AZMAf}), (\ref{AZMArank1}) and (\ref{AZMArank5})
  into (\ref{AZMAf2}), the computation of the Fierz transformation
  goes as follows:
  \begin{eqnarray}
 {\bar \epsilon}_{1} \Gamma_{\nu} \psi \Gamma^{\mu \nu} \epsilon_{2} 
  &=&  - {(\Gamma^{\mu \nu})_{\gamma}}^{\delta} {({\bar \epsilon}_{1}
  \epsilon_{2})_{\delta}}^{\alpha} {(\Gamma_{\nu})_{\alpha}}^{\beta}
  \psi_{\beta} \nonumber \\
  &=& - \frac{1}{16} ({\bar \epsilon}_{1} \Gamma_{\rho} \epsilon_{2}) 
    (\Gamma^{\mu \nu} \Gamma^{\rho} \Gamma_{\nu}) \psi
    - \frac{1}{16 \times 5!} ({\bar \epsilon}_{1} \Gamma_{\rho_{1} \cdots
  \rho_{5}} \epsilon_{2}) (\Gamma^{\mu \nu} \Gamma^{\rho_{1} \cdots \rho_{5}}
  \Gamma_{\nu} ) \psi  +  (\textrm{rank 3 term}) \nonumber \\
  &=&     {\bar \epsilon_{1}} \Gamma^{\mu} \epsilon_{2} \psi - \frac{7}{16}
      {\bar \epsilon_{1}} \Gamma^{\rho} \epsilon_{2} \Gamma_{\rho}
      \Gamma^{\mu} \psi - \frac{1}{16 \times 5!} {\bar \epsilon_{1}}
      \Gamma^{\rho_{1} \cdots \rho_{5}} \epsilon_{2} \Gamma_{\rho_{1} \cdots
      \rho_{5}} \Gamma^{\mu} \psi + (\textrm{rank 3 term}). \label{AZMAQED}
  \end{eqnarray}

  We next go to the proof of (\ref{fierz3dim}) for the
  three-dimensional spacetime. This time, the fermions are Majorana
  ones. The proof goes in the same way as for (\ref{AZM31fierz}).
  Here, we note that the $2 \times 2$ matrices $M$ are decomposed by the
  Paulian matrices as $ M = \frac{1}{2} (Tr M) {\bf 1} + \frac{1}{2}
  (Tr M \sigma_{\mu}) \sigma_{\mu}.$ 
  Then, the Fierz decomposition is verified as follows:
  \begin{eqnarray}
   {\bar \epsilon}_{1} \sigma_{\mu} \psi \sigma_{\mu \nu}
  \epsilon_{2} 
   =  ({\bar \epsilon}_{1})^{\alpha} {(\sigma_{\mu})_{\alpha}}^{\beta} 
      \psi_{\beta} {(\sigma_{\mu \nu})_{\gamma}}^{\delta}
  (\epsilon_{2})_{\delta}
   =  - {(\sigma_{\mu \nu})_{\gamma}}^{\delta} {({\bar \epsilon}_{1}
  \epsilon_{2})_{\delta}}^{\alpha} {(\sigma_{\mu})_{\alpha}}^{\beta}
  \psi_{\beta}
  \sim - \frac{1}{2} ({\bar \epsilon}_{1} \sigma_{\chi} \epsilon_{2}) 
      (\sigma_{\mu \nu} \sigma_{\chi} \sigma_{\mu}) \psi. \nonumber \\
   \label{fierz3dimint1}
  \end{eqnarray}
  Here, we drop the contribution of the rank-0 term $({\bar
  \epsilon}_{1} {\bf 1} \epsilon_{2})$, because we are interested in
  the difference ${\bar \epsilon}_{1} \sigma_{\mu} \psi \sigma_{\mu \nu}
  \epsilon_{2} - {\bar \epsilon}_{2} \sigma_{\mu} \psi \sigma_{\mu \nu}
  \epsilon_{1}$ and the rank-0 term cancels. Likewise, the latter
  formula is rewritten as
   \begin{eqnarray}
   {\bar \epsilon}_{1} \sigma_{\mu} \psi \sigma_{\mu}
  \epsilon_{2} 
   =  ({\bar \epsilon}_{1})^{\alpha} {(\sigma_{\mu})_{\alpha}}^{\beta} 
      \psi_{\beta} {(\sigma_{\mu})_{\gamma}}^{\delta}
  (\epsilon_{2})_{\delta}
   =  - {(\sigma_{\mu})_{\gamma}}^{\delta} {({\bar \epsilon}_{1}
  \epsilon_{2})_{\delta}}^{\alpha} {(\sigma_{\mu})_{\alpha}}^{\beta}
  \psi_{\beta}
  \sim - \frac{1}{2} ({\bar \epsilon}_{1} \sigma_{\chi} \epsilon_{2}) 
      (\sigma_{\mu} \sigma_{\chi} \sigma_{\mu}) \psi. \nonumber
  \\ \label{fierz3dimint2} 
  \end{eqnarray}
 The relevant Paulian matrices are calculated as
  \begin{eqnarray}
   \sigma_{\mu \nu} \sigma_{\chi} \sigma_{\mu}
    &=& i \epsilon_{\mu \nu \rho} \sigma_{\rho}
      (\delta_{\chi \mu} + i \epsilon_{\chi \mu \eta} \sigma_{\eta})
    = i  \epsilon_{\chi \nu \rho} \sigma_{\rho}
      - (\delta_{\nu \eta} \delta_{\rho \chi} - \delta_{\nu \chi}
    \delta_{\rho \eta} ) (\delta_{\rho \eta} + i \epsilon_{\rho \eta
    \mu} \sigma_{\mu}) \nonumber \\
   &=&  i  \epsilon_{\chi \nu \rho} \sigma_{\rho}
      -  i  \epsilon_{\chi \nu \rho} \sigma_{\rho}
      - \delta_{\nu \chi} + \delta_{\nu \chi} \underbrace{\delta_{\rho \eta}
    \delta_{\rho \eta}}_{=3} = 2 \delta_{\nu \chi}, \label{fierz3dimint3} \\
 \sigma_{\mu} \sigma_{\chi} \sigma_{\mu} 
  &=& \sigma_{\mu} (\delta_{\chi \mu} + i \epsilon_{\chi \mu \eta}
    \sigma_{\eta}) 
   + \sigma_{\chi} + i \epsilon_{\chi \mu \eta} (i \epsilon_{\mu \eta
    \rho} \sigma_{\rho} + \delta_{\mu \eta})
   =  \sigma_{\chi} - (\delta_{\eta \eta} \delta_{\chi \rho} -
    \delta_{\eta \chi} \delta_{\eta \rho}) \sigma_{\rho} \nonumber \\
   &=& \sigma_{\chi} (1 - 3 + 1) = - \sigma_{\chi}. \label{fierz3dimint4}
  \end{eqnarray}
  Plugging (\ref{fierz3dimint3}) and (\ref{fierz3dimint4}) into
  (\ref{fierz3dimint1}) and (\ref{fierz3dimint2}), we complete the
  proof of the Fierz identity (\ref{fierz3dim}). 

 \subsubsection{Proof of the Fierz identity (\ref{cycleferm})} 
  \label{sec-fierz2}
 We next verify the Fierz identity (\ref{cycleferm}). This formula
 plays an important role in the supersymmetry  of the super
 Yang-Mills theory or the Green-Schwarz action of the superstring.
 This holds only for the following cases:
  \begin{enumerate}
   \item{$d=3$, and $\psi_{1,2,3}$ are Majorana.}
   \item{$d=4$, and $\psi_{1,2,3}$ are Majorana or Weyl.}
   \item{$d=6$, and $\psi_{1,2,3}$ are Weyl.}
   \item{$d=10$, and $\psi_{1,2,3}$ are Majorana-Weyl.}
  \end{enumerate}
 To substantiate this relation, we carry
 out the following decomposition.
  \begin{eqnarray}
  & & ({\bar \epsilon} \Gamma^{\mu} \psi_{1})({\bar \psi}_{2} \Gamma_{\mu} 
  \psi_{3})
 + ({\bar \epsilon} \Gamma^{\mu} \psi_{2})({\bar \psi}_{3} \Gamma_{\mu} 
  \psi_{1})
 + ({\bar \epsilon} \Gamma^{\mu} \psi_{3})({\bar \psi}_{1} \Gamma_{\mu} 
  \psi_{2}) \nonumber \\
 &=& (\epsilon)_{\alpha} (\Gamma^{0} \Gamma^{\mu})_{\alpha \beta}
  (\psi_{1})_{\beta} (\psi_{2})_{\gamma} (\Gamma^{0}
  \Gamma_{\mu})_{\gamma \delta} (\psi_{3})_{\delta}
 + (\epsilon)_{\alpha} (\Gamma^{0} \Gamma^{\mu})_{\alpha \gamma}
  (\psi_{2})_{\gamma} (\psi_{3})_{\delta} (\Gamma^{0}
  \Gamma_{\mu})_{\delta \beta} (\psi_{1})_{\beta} \nonumber \\
 &+& (\epsilon)_{\alpha} (\Gamma^{0} \Gamma^{\mu})_{\alpha \delta}
  (\psi_{3})_{\delta} (\psi_{1})_{\beta} (\Gamma^{0}
  \Gamma_{\mu})_{\beta \gamma} (\psi_{2})_{\gamma}.
  \end{eqnarray}
 Namely, our job reduces to verifying the vanishing of the following
 quantity:  
  \begin{eqnarray}
  (\Gamma^{0} \Gamma^{\mu})_{\alpha \beta} (\Gamma^{0}
  \Gamma_{\mu})_{\gamma \delta}
  +  (\Gamma^{0} \Gamma^{\mu})_{\alpha \gamma} (\Gamma^{0}
  \Gamma_{\mu})_{\delta \beta}
  +  (\Gamma^{0} \Gamma^{\mu})_{\alpha \delta} (\Gamma^{0}
  \Gamma_{\mu})_{\beta \gamma}. \label{cycleinterm2}
  \end{eqnarray}
  We multiply $(\chi_{1})_{\gamma}$ and $(\chi_{2})_{\delta}$
  with (\ref{cycleinterm2}) to obtain\footnote{The flipping property
  ${\bar \chi}_{1} \Gamma_{\rho_{1} \cdots \rho_{k}} \chi_{2}
  = (-1)^{\frac{k(k+1)}{2}} {\bar \chi}_{2} \Gamma_{\rho_{1} \cdots
  \rho_{k}} \chi_{1}$ holds true {\it only for the Majorana fermion},
  because ${\bar \chi} = {^{T}\chi} \Gamma^{0}$ holds only for the
  Majorana fermions. However, the property
  ${^{T}\chi}_{1} \Gamma^{0} \Gamma_{\rho_{1} \cdots \rho_{k}} \chi_{2}
  = (-1)^{\frac{k(k+1)}{2}} {^{T}\chi}_{2} \Gamma^{0} \Gamma_{\rho_{1} \cdots
  \rho_{k}} \chi_{1}$ holds true without assuming the Majorana fermion.}
  \begin{eqnarray}
   (\Gamma^{\mu})_{\alpha \beta} ({^{T}\chi}_{1} \Gamma_{0} 
      \Gamma_{\mu} \chi_{2})
 + (\Gamma^{\mu} \chi_{1})_{\alpha} ({^{T}\chi}_{2} \Gamma^{0}
 \Gamma_{\mu})_{\beta} 
 - (\Gamma^{\mu} \chi_{2})_{\alpha} ({^{T}\chi}_{1} \Gamma^{0}
      \Gamma_{\mu})_{\beta}. 
   \label{goalll}
  \end{eqnarray}
  We verify the vanishing of (\ref{goalll}) by decomposing them by the 
  gamma matrices. For example, the $16 \times 16$ matrices for the
  ten-dimensional case are decomposed in the same way as
  (\ref{AZMAf}). This leads us to verify the vanishing for the
  components of each rank.

 For $(\Gamma^{\mu})_{\alpha \beta}$ of the first term, it is trivial
 that only the rank-1 components contribute.
 For the second and third terms which entail the
 fermions $\chi_{1,2}$, we can limit the argument at most to the rank-5 
 components. We note that the rank-0,3,4 components do not contribute
 because of the anti-symmetry of $\Gamma_{0} \Gamma_{\rho_{1} \cdots
 \rho_{k}}$ (for $k=0,3,4$). 
 The even-rank components do not contribute for the Weyl fermion
 either, because of the Weyl projection.
 We list up which rank we should consider in the following:
  \begin{enumerate}
   \item{For $d=3$ Majorana case, only the rank-1 matrices may contribute.}
   \item{For $d=4$ Majorana case, only the rank-1,2 components may
       survive. For the Weyl case, only the rank-1 may survive.}
   \item{For the $d=6$ Weyl case, we consider the rank-1
       components.}
   \item{For the $d=10$ Majorana-Weyl case, we consider the rank-1,5
       components.} 
  \end{enumerate}
 
  For the rank-1 case, we multiply $(\Gamma_{\rho})_{\beta \alpha}$,
  and we find that this quantity vanishes:
  \begin{eqnarray}
  Tr(\Gamma^{\mu} \Gamma_{\rho})
  ({^{T}\chi}_{1} \Gamma_{0} \Gamma_{\mu} \chi_{2})
  - ({^{T}\chi_{2}} \Gamma_{0} \Gamma_{\mu}
  \Gamma_{\rho} \Gamma^{\mu} \chi_{1})
  + ({^{T}\chi_{1}} \Gamma_{0} \Gamma_{\mu} \Gamma_{\rho} \Gamma^{\mu}
  \chi_{2}) 
  =  (-{\cal P} + 2(d-2)) ({^{T}\chi}_{1} \Gamma_{0} \Gamma_{\rho}
  \chi_{2})  = 0. \nonumber
  \end{eqnarray}
  ${\cal P}$ is the size of the gamma matrices, and $d$ is the
  spacetime dimensionality. Here, we utilize the following formula for 
  the general rank-$k$ gamma matrices $\Gamma_{\rho_{1} \cdots \rho_{k}}$:
   \begin{eqnarray}
     \Gamma_{\mu} \Gamma_{\rho_{1} \cdots \rho_{k}} \Gamma^{\mu}
 &=&    (\Gamma_{\mu \rho_{1} \cdots \rho_{k}} + k \eta_{\mu [\rho_{1}} 
      \Gamma_{\rho_{2} \cdots \rho_{k}]}) \Gamma^{\mu} 
 = (-1)^{k-1} (- \eta^{\mu}_{\mu} \Gamma_{\rho_{1} \cdots \rho_{k}}
     + k \eta^{\mu}_{[\rho_{1}} \Gamma_{\mu \rho_{2} \cdots \rho_{k}]})
     + k  \eta_{\mu [\rho_{1}} \Gamma_{\rho_{2} \cdots \rho_{k}] \mu}
 \nonumber \\ 
 &=& (-1)^{k} (d-2k) \Gamma_{\rho_{1} \cdots \rho_{k}}. \label{gammaform1}
   \end{eqnarray}
  The rank-1 terms cancel for $d = 3,4,6,10$,
  where ${\cal P} = 2(d-2) = 2,4,8,16$ respectively. 

  Next, we go on to the generic rank-$k$ ($k=2,3,\cdots$) case. 
  We multiply $(\Gamma_{\rho_{1} \cdots \rho_{k}})_{\beta \alpha}$ to obtain
  \begin{eqnarray}
   - ({^{T}\chi}_{2} \Gamma_{0} \Gamma_{\mu} \Gamma_{\rho_{1} \cdots \rho_{k}}
   \Gamma^{\mu} \chi_{1})
  + ({^{T}\chi}_{1} \Gamma_{0} \Gamma_{\mu} \Gamma_{\rho_{1} \cdots \rho_{k}}
   \Gamma^{\mu} \chi_{2})
 = 2 ({^{T}\chi}_{1} \Gamma_{0} \Gamma_{\mu} \Gamma_{\rho_{1} \cdots \rho_{k}}
   \Gamma^{\mu} \chi_{2}).
  \end{eqnarray}
  Due to the formula of the gamma matrices (\ref{gammaform1}), these
  contributions actually vanish for $k=2,5$ for $d=4,10$ respectively.

 \subsection{Supermatrices} \label{AZMA091441}
 This section is devoted to introducing the definitions of the notion
 of supermatrices. In treating supermatrices, there are many points we 
 should be meticulous about, because what holds true of ordinary matrices 
 is not applicable to the supermatrices. 

 \subsubsection{Transpose} 
  We first introduce a notion of the transpose, emphasizing on the
 difference from the ordinary matrices. In considering such objects,
 it is extremely important to settle the starting point, because the
 other notions are defined so that they are consistent with this
 starting point. 
 
 \paragraph{Transpose of Vector} \hspace{0mm} 
 
 The guiding principle in considering the transpose of the
 supermatrices is {\it the transpose of the vector}. 
  \begin{itemize}
   \item{ The guiding principle is that {\it the transpose of a
     vector} is defined as 
     \begin{eqnarray}
         { \left( \begin{array}{c}  x_{1} \\ \vdots  \\  x_{n}  \end{array}
         \right)}^{T} = ( x_{1}  ,\cdots , x_{n}  ). 
     \end{eqnarray}      
     We denote $\{ x_{\mu} \}$ as the  components of $v$, and these
     components mean  {\sf both bosons and fermions}.}
   \item{ We define the vector as $v = \left( \begin{array}{c}  \eta
     \\ b  \end{array} \right) $ where $\eta$ and $b$ are fermionic and
     bosonic {\it real}  fields respectively. }
  \end{itemize} 

  \paragraph{Transpose of Supermatrices} \hspace{0mm} 
 
  The transpose of supermatrices must be defined so that the
  definition is consistent with the transpose of a vector. Therefore the 
  transpose of a supermatrix must satisfy 
   \begin{eqnarray}
     {^{T}(Mv)} = {^{T}v} {^{T}M},
   \end{eqnarray}
  where $M$ is a supermatrix and $v$ is a vector. Following this rule, 
  the transpose of a supermatrix is defined as follows:
  \begin{eqnarray}
     \textrm{For } M =  \left( \begin{array}{cc} a & \beta \\ \gamma 
    & d \end{array} \right) \hspace{2mm} and  \hspace{2mm} v = 
    \left( \begin{array}{c}   \eta \\ b 
    \end{array} \right)  ,\hspace{4mm} {^{T}{M}} =  \left
    ( \begin{array}{cc} {^{T}{a}} & - {^{T}{\gamma}} \\ {^{T}{\beta}}
    & {^{T}{d}} \end{array} \right). \label{defsupertrace}
  \end{eqnarray}
   \begin{itemize}
    \item{ $a$ and $d$ are bosonic (i.e. Grassmann even) $m \times
        m$ and $n \times n$ matrices, respectively.}
    \item{ $\beta$ and $\gamma$ are $m \times n$ and $n \times m$
        fermionic (i.e. Grassmann odd) matrices, respectively.} 
    \item{ $\eta$ ($b$) denote the upper $m$ (lower $n$) 
        bosonic(fermionic) components of the supervector respectively.}
   \end{itemize}  

 {\sf (Proof)  This can be verified using the very definition of the
 transpose.  
  \begin{eqnarray} Mv =   \left( \begin{array}{c}   a\eta + \beta b  \\
       \gamma \eta + d b  \end{array} \right).
  \end{eqnarray}
  Then the  transpose of this vector is by definition 
   \begin{eqnarray} 
    {^{T}(Mv)} = ( {^{T}\eta} {^{T}a} + {^{T}b} {^{T}\beta} , - {^{T}\eta}
    {^{T}\gamma} + {^{T}b} {^{T}d} ) = {^{T}v} {^{T}M}.
    \end{eqnarray}
   The point is that {\sf the sign of } $\gamma \eta$ {\sf has changed
   because these are Grassmann odd}.   Noting this fact, we can read
   off the result (\ref{defsupertrace}). (Q.E.D.) }\\

  We have one caution about the transpose of the supermatrix. {\sf
  The transpose of the transpose does not give an original matrix.}
  This 'anomalous' property can be immediately read off from the
  definition of the supermatrix (\ref{defsupertrace}):
  \begin{eqnarray}
   {^{T} ({^{T} \left( \begin{array}{cc} a & \beta \\ \gamma  & d
   \end{array} \right)} ) } = {^{T} \left( \begin{array}{cc} {^{T}{a}}
   & - {^{T}{\gamma}} \\ {^{T}{\beta}} & {^{T}{d}} \end{array} \right) 
    } = \left( \begin{array}{cc} a & - \beta \\ - \gamma
    & d \end{array} \right).
  \end{eqnarray}

  \paragraph{Transpose of Transverse Vector} \hspace{0mm} 

  We have seen an important fact that the transpose of the transpose
  of a supermatrix does not give the original supermatrix. In fact,
  the same holds true of the transpose of the transpose of a vector. 
   Conclusion coming first, the definition is 
    \begin{eqnarray}
     {^{T} y} = {^{T} ( \eta ,b)} = \left( \begin{array}{c} - {^{T} \eta} \\
     {^{T}b} \end{array} \right).
    \end{eqnarray} 
   We confirm that this is actually a well-defined settlement. This
   notion is defined so that 
    \begin{eqnarray}
      {^{T} (yM)} = {^{T}M } {^{T}y},
    \end{eqnarray}    
   where $y$ is a transverse vector $y = (\eta ,b)$ and $M$ is a
   supermatrix $M = 
   \left( \begin{array}{cc} a & \beta \\ \gamma & d \end{array}
   \right)$.  We
   compute both the L.H.S and the R.H.S and verify that they actually
   match if we follow the above definition.
   \begin{itemize}
    \item{L.H.S. :  ${^{T} (yM) } = {^{T} ( \eta a + b \gamma , \eta
         \beta + bd)} =  
         \left( \begin{array}{c} - {^{T} (\eta a)} - {^{T} (b \gamma)} \\
         {^{T}( \eta \beta)} +  
         {^{T} (bd)} \end{array} \right) = \left( \begin{array}{c}
         - {^{T}a} {^{T} \eta} - {^{T} \gamma} {^{T}b} \\ -
         {^{T}\beta} {^{T} \eta} + 
         {^{T}d} {^{T}b} \end{array} \right)$. \\
         We have used the fact that in the transpose of $fF$, we must
         multiply $-1$ because a fermion jumps over another fermion.}
    \item{ R.H.S. : ${^{T}M} {^{T}y} = \left( \begin{array}{cc} {^{T}a} 
            & - {^{T}\gamma} \\ {^{T} \beta} & {^{T} d} \end{array} \right) 
            \left( \begin{array}{c} - {^{T}\eta } \\ {^{T}b}
            \end{array} \right) = \left( \begin{array}{c} - {^{T}a} 
            {^{T} \eta} - {^{T}\gamma} {^{T}b} \\ - {^{T}\beta} {^{T}
            \eta} + {^{T}\gamma}  
            {^{T}b} \end{array} \right)$. }
   \end{itemize} 
  Thus we have verified that the above definition of the transpose is
  consistent with the condition ${^{T} (yM)} = {^{T}M } {^{T}y}$. 
  As we have mentioned before, the transpose of the transpose of a
  vector does not give the original vector, because 
   \begin{eqnarray}
    {^{T} ( {^{T} \left( \begin{array}{c} \eta \\ b \end{array} \right)}
    )} = {^{T} ( {^{T} \eta}, {^{T}b} )} = \left( \begin{array}{c}
    -\eta \\ b \end{array} \right).
   \end{eqnarray}

  \subsubsection{Hermitian Conjugate}
  We introduce a notion of hermitian conjugate of the supermatrix. This
  notion is much simpler than the transpose, and we do not see 
  anomalous properties as emerged in the transpose. The starting
  point of this notion is 
    \begin{itemize}
     \item{ For a fermionic {\it number} $\alpha, \beta$, the complex
         conjugate is $(\alpha \beta)^{\dagger} = (\beta)^{\dagger}
         (\alpha)^{\dagger}$.}
    \item{ For a vector $v = \left( \begin{array}{c} \eta \\ b
          \end{array} \right)$, the complex conjugate is 
        $ \left( \begin{array}{c} \eta \\ b \end{array}
         \right)^{\dagger} = ( \eta^{\dagger}, b^{\dagger})$. }
    \end{itemize} 

  Under this definition, the hermitian conjugate of a supermatrix is
  defined as 
   \begin{eqnarray}
    \textrm{For } M = \left( \begin{array}{cc} a & \beta \\ \gamma &
    d \end{array} \right), \hspace{4mm}  M^{\dagger} = \left
    ( \begin{array}{cc} a^{\dagger} & \gamma^{\dagger} \\ \beta^{\dagger} & 
    d^{\dagger} \end{array} \right).
   \end{eqnarray} 
 {\sf
 (Proof) The guiding principle to determine the hermitian conjugate of 
 a supermatrix is the condition 
  \begin{eqnarray}
   ( M v)^{\dagger} = (v^{\dagger}) (M^{\dagger}).
  \end{eqnarray}
  For $M = \left( \begin{array}{cc} a & \beta \\ \gamma &
    d \end{array} \right)$ and $v = \left( \begin{array}{c} \eta \\ b
    \end{array} \right)$, $(Mv)^{\dagger}$ is computed to be,
    utilizing the definition for the vector, 
     \begin{eqnarray}
        (Mv)^{\dagger} = \left( \begin{array}{c} a \eta + \beta b \\
        \gamma \eta + d b \end{array} \right)^{\dagger} = 
        ( (a \eta)^{\dagger} + (\beta b)^{\dagger} , (\gamma
        \eta)^{\dagger} + (d b)^{\dagger} ) = ( {\eta}^{\dagger}
        a^{\dagger} + b^{\dagger} {\beta}^{\dagger} ,
        {\eta}^{\dagger} {\gamma}^{\dagger} + b^{\dagger} d^{\dagger} ). 
     \label{AZMA134718} 
    \end{eqnarray}
    The explicit form of the hermitian conjugate of a supermatrix can
    be read off from (\ref{AZMA134718}), and this completes the
    proof. (Q.E.D.)}

    The definition of the hermitian conjugate of a transverse vector
    is now straightforward. This is given by 
    \begin{eqnarray}
       y^{\dagger} = ( \eta , b)^{\dagger} = \left( \begin{array}{c}
       \eta^{\dagger} \\ b^{\dagger} \end{array} \right).  
    \end{eqnarray} 
    It is straightforward to verify that this definition is consistent
    with the guiding principle 
     \begin{eqnarray}
       (y M)^{\dagger} = M^{\dagger} y^{\dagger},
     \end{eqnarray}
   because $(l.h.s.) = (r.h.s.) = \left( \begin{array}{c} a^{\dagger}
   \eta^{\dagger} + \gamma^{\dagger} b^{\dagger} \\ \beta^{\dagger}
   \eta^{\dagger} + d^{\dagger} b^{\dagger} \end{array} \right)$.

  \subsubsection{Complex Conjugate} \label{AZCaho}
  We define a notion of complex conjugate for a supermatrix. The
  guiding principle to define the complex conjugate is to require the
  matrices and the vectors to satisfy the condition 
   \begin{eqnarray}
       (Mv)^{\ast} = M^{\ast} v^{\ast}. \label{AZMA1guidingcc}
   \end{eqnarray}
  In order to satisfy this condition, we define the complex conjugate
  of the vectors and the matrices as follows\footnote{  Be careful
  about the fact that ${^{T}(M^{\dagger})}$ {\it is 
  different from } $({^{T} M})^{\dagger} = M^{\ast}$. They are
  computed to be ${^{T}(M^{\dagger})} = \left( \begin{array}{cc}
  a^{\ast} & - \beta^{\ast} \\ \gamma^{\ast} & d^{\ast} \end{array}
  \right)$, ${^{T}(v^{\dagger})} = \left( \begin{array}{c}
  - \eta^{\ast} \\  b^{\ast} \end{array} \right) $ and ${^{T}
  (y^{\dagger}) } =  (\eta^{\ast}, b^{\ast})$ and these are {\it
  not} the complex conjugate of the vectors or the matrices.}. 
   \begin{eqnarray}
      v^{\ast} \stackrel{def}{=} ({^{T} v})^{\dagger}, \hspace{4mm}
      M^{\ast} \stackrel{def}{=} ({^{T} M})^{\dagger}. \label{AZMA3defcc}
   \end{eqnarray} 
  It is straightforward to verify that this definition is consistent
  with the guiding principle (\ref{AZMA1guidingcc}).
   \begin{eqnarray}
    (Mv)^{\ast} = (({^{T}v})({^{T}M}))^{\dagger} = ({^{T}
    M})^{\dagger} ({^{T} v})^{\dagger} = M^{\ast} v^{\ast}.
   \end{eqnarray}
  Combining the results obtained in the previous section, the explicit 
  form of the complex conjugate of the vectors and the matrices are
    \begin{eqnarray}
      M^{\ast} = \left( \begin{array}{cc} a & \beta \\ \gamma & d
      \end{array} \right)^{\ast} = \left( \begin{array}{cc}
        a^{\ast} & \beta^{\ast} \\ - \gamma^{\ast} & d^{\ast}
        \end{array}  \right), \hspace{2mm} 
      v^{\ast} = \left( \begin{array}{c} \eta \\ b \end{array}
        \right)^{\ast} = \left( \begin{array}{c} \eta^{\ast} \\
        b^{\ast} \end{array} \right), \hspace{2mm}
      y^{\ast} = ( \eta, b)^{\ast}  = ( - \eta^{\ast}, b^{\ast}).  
    \label{AZMA1misccc} 
   \end{eqnarray} 
  This is clearly consistent with the guiding principle $(Mv)^{\ast}
  = M^{\ast} v^{\ast}$ because $(l.h.s.) = (r.h.s.) = 
  \left( \begin{array}{c} a^{\ast} \eta^{\ast} + \beta^{\ast} b^{\ast} \\
   - \gamma^{\ast} \eta^{\ast} + d^{\ast} b^{\ast} \end{array}
  \right) $. \\ 

   We have following properties which relates the transpose, hermitian
  conjugate and the complex conjugate. \\

  (Prop) (1)${^{T}M} = (M^{\ast})^{\dagger}$, (2)$M^{\dagger} =
  {^{T} (M^{\ast})}$.  \\

  {\sf
  (Proof) These properties can be verified by noting that the
  hermitian conjugate of the hermitian conjugate gives back the
  original quantity, which can be readily verified by definition.
  \begin{enumerate}
   \item{ $(M^{\ast})^{\dagger} = (({^{T} M})^{\dagger} )^{\dagger} = 
       {^{T} M}$.}
   \item{ ${^{T} (M^{\ast})} \stackrel{(1)}{=}
       ((M^{\ast})^{\ast})^{\dagger} = M^{\dagger}$. In the last
       equality, we have utilized the fact that, for a supermatrix,
       $(M^{\ast})^{\ast} = M$, which can be readily verified from
       the explicit form of the complex conjugate
       (\ref{AZMA1misccc}).  }
  \end{enumerate}
  This completes the proof of the above properties. (Q.E.D.)
  }\\

  Now we are ready to answer the question : {\it what do we mean by 'a
  supermatrix is real ?'}. In considering physics, we must
  take into account the reality condition. We utilize supermatrices in 
  the context of expressing the action of superstring theory, the
  action must be real, and we are required to solidify the definition
  of the reality of a supermatrix. \\

   {\sf (Def) A supermatrix $M$ is real
  $\stackrel{def}{\Leftrightarrow}$ $M$ is a mapping from {\it a real
  vector} to {\it a real vector}}.  \\

   This statement is equivalent to, for the above definition of
  complex conjugate, 
    \begin{eqnarray}
      M^{\ast} = M. \label{AZMA3defreality}
    \end{eqnarray}
  This can be verified by noting the starting guiding principle that
  $M$ is designed to satisfy $(Mv)^{\ast} = 
  M^{\ast} v^{\ast}$. If $M^{\ast} = M$ is satisfied,
  $(Mv)^{\ast}$ is a real vector if $v$ is real, because
   \begin{eqnarray}
     (Mv)^{\ast} = M^{\ast} v^{\ast} \stackrel{v \textrm{ is
     real}}{=} M^{\ast} v \stackrel{M^{\ast} = M}{=} M v.
   \end{eqnarray}
    The relationship  (\ref{AZMA3defreality}) tells us the conditions
  for the components of $M$ to satisfy. Noting the explicit form of
  complex conjugate (\ref{AZMA1misccc}), we can derive $a^{\ast} =
  a$, $d^{\ast} =  d$ , $\beta^{\ast} = \beta$ and $\gamma^{\ast} =
  - \gamma$,  id est, 
  \begin{itemize}
   \item{$a, \beta, d$ should be real.}
   \item{$\gamma$ should be pure imaginary.}
  \end{itemize}

 \section{Calculation of the Seeley-de-Witt coefficients}
   \label{seeleydewitt} 
  This appendix is devoted to introducing the basic technique of
  computing the Seeley-de-Witt coefficients in the heat kernel
  expansion. There are a number of ways to compute the coefficients,
  and here we focus on the calculation based on the
  Campbell-Baker-Hausdorff formula. We consider the general elliptic
  differential operator
 \begin{eqnarray}
  D^{2} = - \left( g^{ij}(x) \frac{d}{dx^{i}} \frac{d}{dx^{j}} +
  A^{i} (x) \frac{d}{dx^{i}} + B(x) \right). \label{elliptic}
 \end{eqnarray}
  In this section, we consider the $d$-dimensional spacetime, and the
  indices $i,j, \cdots$ run over $i,j, \cdots = 1,2, \cdots, d$.
  These indices are allocated for the curved spacetime.

  We consider the trace of the large $N$ matrices
 in terms of the heat kernel. The trace of the operators
 are expressed using the complete system as
  \begin{eqnarray}
    Tr e^{-\tau D^{2}} = \int d^{d} x \langle x | e^{-\tau D^{2}} | x
    \rangle, 
  \end{eqnarray}
 where the bracket $|x \rangle$ and $\langle x |$ satisfies $\sum_{x} | 
 x \rangle \langle x |=1$. However, it is difficult to consider the
 trace of a general operator, and we regard the operator as the sum of 
 the first term of (\ref{elliptic}) and the perturbation around it. This
 is a famous procedure, and the perturbation is expressed in terms of
 the Seeley-de-Witt coefficients. 

  It is well known that the Green function is computed to be
  \begin{eqnarray}
  \langle x | \exp  \left( \tau g^{ij}(y) \frac{d}{d x^{i}}
   \frac{d}{dx^{j}} \right) | y \rangle 
  = \frac{e(y)}{(2 \pi \tau)^{\frac{d}{2}} } \exp \left( -
   \frac{(x-y)^{i} (x-y)^{j} g_{ij}(y) }{4 \tau} \right).  
  \end{eqnarray} 
  Therefore, its trace is easily derived as
  \begin{eqnarray}
   Tr \left( \exp  \left( \tau g^{ij}(y) \frac{d}{d x^{i}}
   \frac{d}{dx^{j}} \right) \right)
  = \int d^{d} x  \langle x | \exp  \left( \tau g^{ij}(y) \frac{d}{d x^{i}}
   \frac{d}{dx^{j}} \right) | x \rangle 
  = \int d^{d} x \frac{e(x)}{(2 \pi \tau)^{\frac{d}{2}}}. \label{hklap}
  \end{eqnarray}
  On the other hand, the heat kernel expansion of the general elliptic 
  operator (\ref{elliptic}) is expanded as
  \begin{eqnarray}
   Tr(e^{-\tau D^{2}}) = \int d^{d} x \langle x | e^{-\tau D^{2}} | x
   \rangle  = \int d^{d} x   \frac{e(x)}{(2 \pi \tau)^{\frac{d}{2}}}
    (a_{0} + \tau a_{1} + \tau^{2} a_{2} + \cdots ). \label{hkelliptic}
  \end{eqnarray}
  These coefficients $a_{0}, a_{1}, a_{2}, \cdots$ are called "the
  Seeley-de-Witt coefficients". (\ref{hklap}) trivially gives
  $a_{0}=1$. However, the other coefficients give nontrivial results.
  In the following, we derive the first order $a_{1}$ for the general
  elliptic operator (\ref{elliptic}). 

  To this end, we divide $-\tau D^{2}$ as $- \tau D^{2} = X+Y$, where
    \begin{eqnarray}
 & & X = \tau \left( g^{ij} (y) \frac{d}{dx^{i}} \frac{d}{dx^{j}} \right),
 \label{AZ32X}  \\
 & & Y = \tau \left( (g^{ij} (x) - g^{ij}(y) ) \frac{d}{dx^{i}}
 \frac{d}{dx^{j}} +  A^{i} (x) \frac{d}{dx^{i}} + B(x) \right).
   \end{eqnarray}
  From now on, we compute the exponential $\exp(X+Y)$ via the
  Campbell-Baker-Hausdorff formula, while the computation is rather
  involved:
  \begin{eqnarray}
    e^{A} e^{B} = \exp \left( A + B + \frac{1}{2}[A,B] + \frac{1}{12} ( [A,
  [A, B]] + [B, [B, A]] ) + \cdots \right).  
  \end{eqnarray}

  Since we know that $\langle x | e^{X} |y \rangle = \frac{e(y)}{(2 \pi
 \tau)^{\frac{d}{2} } } \exp \left( - \frac{1}{4 \tau} (x-y)^{i}
 (x-y)^{j} g_{ij}(y) \right)$, the quantity in question is computed as
  \begin{eqnarray}
   e^{X+Y} e^{-X} &=& \exp \left( Y + \frac{1}{2} [X,Y] + \frac{1}{12} 
  ( [X+Y, [X+Y, -X]] + [ -X, [-X, X+Y]] )  + \cdots \right) \nonumber
  \\
  &=& \exp \left( Y + \frac{1}{2} [X,Y] + \frac{1}{12} ( 2 [X, [X, Y]] 
  - [Y, [Y, X]] ) + \cdots \right) \nonumber \\
  &=& 1 + Y + \frac{1}{2} [X,Y] + \frac{1}{6} [X, [X, Y]] +
  \frac{1}{12} [Y, [X, Y]] + \cdots \nonumber \\
  & & +  \frac{1}{2} (  Y + \frac{1}{2} [X,Y] + \frac{1}{6} [X, [X, Y]] +
  \frac{1}{12} [Y, [X, Y]] + \cdots )^{2} + \cdots \nonumber \\
  &=& 1 + Y + \frac{1}{2} [X,Y] + \frac{1}{6} [ X, [X, Y]] 
   + \frac{1}{2} Y^{2} + \frac{1}{8} [X,Y]^{2} 
   + \frac{1}{3} Y [ X, Y] + \frac{1}{6} [X,Y] Y + \cdots. \nonumber
  \\  \label{cbh}
  \end{eqnarray}
 Before we enter the computation of the quantity $\langle x | e^{X+Y}
 | y \rangle$, we summarize the formula of the differentiation of
 $e^{X}$: 
  \begin{eqnarray}
   \frac{d e^{X}}{d x^{i}} &=& - \frac{1}{2 \tau} (x-y)^{j} g_{ij}(y)
   e^{X}, \label{dell-1} \\
   \frac{d^{2} e^{X} }{dx^{i_{1}} dx^{i_{2}} } &=& 
  \left( - \frac{1}{2 \tau} g_{i_{1} i_{2}} (y) + \frac{1}{4 \tau^{2}} 
   (x-y)^{l_{1}} (x-y)^{l_{2}} g_{i_{1} l_{1}} (y) g_{i_{2} l_{2}} (y) 
   \right) e^{X}, \label{dell-2} \\
   \frac{d^{3} e^{X} }{dx^{i_{1}} dx^{i_{2}} dx^{i_{3}} } &=& \left( 
   \frac{1}{4 \tau^{2}} (x-y)^{l} ( g_{i_{1} i_{2}} (y) g_{i_{3} l}
   (y) + g_{i_{2} i_{3}} (y) g_{i_{1} l} (y) + g_{i_{3} i_{1}} (y)
   g_{i_{2} l} (y) ) \right.  \nonumber \\
  & &  \hspace{-30mm} - \left. \frac{1}{8 \tau^{3}} (x-y)^{l_{1}}
   (x-y)^{l_{2}} 
   (x-y)^{l_{3}} g_{i_{1} l_{1}} (y) g_{i_{2} l_{2}} (y) g_{i_{3}
   l_{3}} (y) \right) e^{X}, \label{dell-3} \\
   \frac{d^{4} e^{X} }{dx^{i_{1}} dx^{i_{2}} dx^{i_{3}} dx^{i_{4}} }
   &=& \left(  \frac{1}{4 \tau^{2}} ( g_{i_{1} i_{2}} (y) g_{i_{3} i_{4} }
   (y) + g_{i_{2} i_{3}} (y) g_{i_{4} i_{1} } (y) + g_{i_{1} i_{3}} (y)
   g_{i_{2} i_{4} } (y) ) \right.  \nonumber \\ 
   & & \hspace{-30mm} - \frac{1}{8 \tau^{3}} (x-y)^{l_{1}}
   (x-y)^{l_{2}} ( g_{i_{1} i_{2}}(y) g_{i_{3} l_{1}}(y) g_{i_{4}
   l_{2}} (y) + g_{i_{2} i_{3}}(y) g_{i_{1} l_{1}}(y) g_{i_{4} l_{2}}
   (y) + g_{i_{1} i_{3}}(y) g_{i_{2} l_{1}}(y) g_{i_{4} l_{2}} (y)
   \nonumber \\
   & & \hspace{-20mm} + g_{i_{1} i_{4}}(y) g_{i_{2} l_{1}}(y) g_{i_{3} 
   l_{2}} (y) + g_{i_{2} i_{4}}(y) g_{i_{1} l_{1}}(y) g_{i_{3}
   l_{2}}(y) + g_{i_{3} i_{4}}(y) g_{i_{1} l_{1}}(y) g_{i_{2}
   l_{2}}(y) ) \nonumber \\
   & & \hspace{-30mm} \left. + \frac{1}{16 \tau^{4}} (x-y)^{l_{1}}
   (x-y)^{l_{2}} (x-y)^{l_{3}} (x-y)^{l_{4}} g_{i_{1} l_{1}}(y)
   g_{i_{2} l_{2}}(y) g_{i_{3} l_{3}}(y) g_{i_{4} l_{4}}(y) \right)
   e^{X}.  \label{AZ32Dproject}
  \end{eqnarray}

 \paragraph{ Computation of $Ye^{X}$} \hspace{0mm} 

 We start with the computation of the easiest case:
   \begin{eqnarray}
   Ye^{X} &=& \tau \left( ( g^{ij}(x) - g^{ij}(y) ) \frac{d}{dx^{i}}
   \frac{d}{dx^{j}} + A^{i} (x) \frac{d}{dx^{i}} + B(x) \right) e^{X}
   \nonumber \\
  &=& \tau \left( (g^{ij}(x) - g^{ij}(y)) ( - \frac{1}{2} g_{ij}(y) +
   \frac{1}{4 \tau} (x-y)^{l_{1}} (x-y)^{l_{2}} g_{i l_{1}}(y)
   g_{j l_{2}} (y) ) \right. \nonumber \\
  & & +  \left.  B(x) - \frac{1}{2} A^{i} (x-y)^{j} g_{ij}(y) 
  \right) e^{X}. \nonumber \\ 
  \end{eqnarray} 
  Therefore, the trace is obtained by
  \begin{eqnarray}
   Tr (Y e^{X} ) = \int d^{d} x \langle x | Y e^{X} | x \rangle 
   = \int d^{d} x \frac{\tau e(x)}{(2 \pi \tau)^{\frac{d}{2}} }
   B(x). \label{AZ32type1} 
  \end{eqnarray} 

 \paragraph{ Computation of $\frac{1}{2} [X,Y] e^{X}$} \hspace{0mm} 

   We next go on to a bit more complicated case, and we compute the
   operator $[X,Y]$ itself:
  \begin{eqnarray}
  [X,Y] &=& \tau^{2} \left( g^{i_{1} i_{2}}(y) \frac{d}{dx^{i_{1}}}
  \frac{d}{dx^{i_{2}} } \right) \times 
  \left( (g^{j_{1} j_{2}}(x) - g^{j_{1} j_{2}}(y) )
  \frac{d}{dx^{j_{1}}} \frac{d}{dx^{j_{2}}} + A^{j}(x)
  \frac{d}{dx^{j}} + B(x) \right) \nonumber \\
  &-& \tau^{2} \left( (g^{j_{1} j_{2}}(x) - g^{j_{1} j_{2}}(y) )
  \frac{d}{dx^{j_{1}}} \frac{d}{dx^{j_{2}}} + A^{j}(x)
  \frac{d}{dx^{j}} + B(x) \right) \times  \left( g^{i_{1} i_{2}}(y)
  \frac{d}{dx^{i_{1}}} \frac{d}{dx^{i_{2}} } \right) \nonumber \\
  &=& \tau^{2} \left( 2g^{i_{1} i_{2}}(y) (\frac{d  g^{j_{1}
  j_{2}}(x)}{dx^{i_{1}}} ) \frac{d^{3}}{dx^{i_{2}} dx^{j_{1}} dx^{j_{2}}} 
  + g^{i_{1} i_{2}}(y) ( \frac{d^{2}  g^{j_{1} j_{2}}(x)}{dx^{i_{1}}
  dx^{i_{2}}} )   \frac{d^{2}}{dx^{j_{1}} dx^{j_{2}}} \right. \nonumber \\
  & & + 2 g^{i_{1} i_{2}}(y) (\frac{d A^{j}(x) }{dx^{i_{1}}} )
  \frac{d^{2}}{dx^{i_{2}} dx^{j}} + g^{i_{1} i_{2}} (y)
  ( \frac{d A^{j}(x)}{dx^{i_{1}} dx^{i_{2}}} ) \frac{d}{dx^{j}}
  \nonumber \\
  & & \left.
   + 2g^{i_{1} i_{2}}(y) (\frac{d B(x)}{dx^{i_{1}}} )
  \frac{d}{dx^{i_{2}}} + g^{i_{1} i_{2}}(y) (\frac{d^{2} B(x)}{dx^{i_{1}}
  dx^{i_{2}} } ) \right). 
  \end{eqnarray}
   Therefore, the trace is computed to be, with the help of the
   formulae (\ref{AZ32Dproject}), 
  \begin{eqnarray}
 & &   Tr ( \frac{1}{2} [X,Y] e^{X} ) = \int d^{d} x \langle x |
 \frac{1}{2} [X,Y] e^{X} | x \rangle \nonumber \\
 &=& \int d^{d} x \frac{e(x)}{(2 \pi \tau)^{\frac{d}{2}}} \left\{ \tau
 \left( -  \frac{1}{4} g^{i_{1} i_{2}}(x) g_{j_{1} j_{2}}(x)
 ( \frac{d^{2} g^{j_{1} j_{2}}(x) }{dx^{i_{1}} dx^{i_{2}}} ) -
 \frac{1}{2} (\frac{d A^{i}(x)}{dx^{i}} ) \right)  +
 \frac{\tau^{2}}{2} g^{i_{1}  i_{2}}(x) (\frac{d^{2} B(x) }{dx^{i_{1}}
 dx^{i_{2}}})   \right\}. \nonumber \\ \label{AZ32type2} 
  \end{eqnarray}

 \paragraph{ Computation of $\frac{1}{6}[X, [X,Y]] e^{X}$}
 \hspace{0mm}

 We compute the operator $[X, [X, Y]]$ as 
 \begin{eqnarray}
  & & [X, [X, Y]] = \tau^{3}  \left( 4 g^{i_{1} i_{2}}(y) g^{k_{1}
  k_{2}}(y) ( \frac{d^{2} g^{j_{1} j_{2}}(x) }{dx^{i_{1}} dx^{k_{1}}} )
  \frac{d^{4}}{dx^{i_{2}} dx^{k_{2}} dx^{j_{1}} dx^{j_{2}} }
  \right. \nonumber \\
  & & + 4 g^{i_{1} i_{2}}(y) g^{k_{1} k_{2}}(y) ( \frac{d^{3} g^{j_{1}
  j_{2}}(x)}{d  x^{i_{1}} dx^{i_{2}} dx^{k_{1}} } ) 
  \frac{d^{3}}{dx^{k_{2}} dx^{j_{1}} dx^{j_{2}}}  \nonumber \\
 & & + g^{i_{1} i_{2}}(y) g^{k_{1} k_{2}}(y) ( \frac{d^{4} g^{j_{1}
  j_{2}}(x)}{dx^{i_{1}}  dx^{i_{2}} dx^{k_{1}} dx^{k_{2}} } ) 
  \frac{d^{2}}{dx^{j_{1}} dx^{j_{2}}} 
   + 4 g^{i_{1} i_{2}}(y) g^{k_{1} k_{2}}(y)
  (\frac{d^{2} A^{j}(x) }{dx^{i_{1}} dx^{k_{1}} } ) 
  \frac{d^{3}}{dx^{i_{2}} dx^{k_{2}} dx^{j}} \nonumber \\
 & & + 4 g^{i_{1} i_{2}}(y) g^{k_{1} k_{2}}(y)
  ( \frac{d^{3} A^{j}(x) }{dx^{i_{1}} dx^{i_{2}} dx^{k_{1}}} )
  \frac{d^{2}}{dx^{i_{2}} dx^{j}} 
  + g^{i_{1} i_{2}}(y) g^{k_{1} k_{2}}(y) ( \frac{d^{4}
  A^{j}(x)}{dx^{i_{1}}  dx^{i_{2}} dx^{k_{1}} dx^{k_{2}} } ) \frac{d}{dx^{j}}
  \nonumber   \\
 & & +  4 g^{i_{1} i_{2}}(y) g^{k_{1} k_{2}}(y)
  ( \frac{d^{2} B(x)}{dx^{i_{1}} dx^{k_{1}}} ) \frac{d^{2}}{dx^{i_{2}}
  dx^{k_{2}} } 
 + 4 g^{i_{1} i_{2}}(y) g^{k_{1} k_{2}}(y)
  ( \frac{d^{3} B(x)}{dx^{i_{1}} dx^{i_{2}} dx^{k_{1}}} )
  \frac{d}{dx^{k_{2}}} \nonumber \\
 & & \left. + g^{i_{1} i_{2}}(y) g^{k_{1} k_{2}}(y) ( \frac{d^{4} B(x) 
  }{dx^{i_{1}} dx^{i_{2}} dx^{k_{1}} dx^{k_{2}}} ) \right). 
 \end{eqnarray}
 Therefore, the trace is computed as
  \begin{eqnarray}
 & &  Tr ( \frac{1}{6} [X, [X, Y]]e^{X} ) = \int d^{d} x \langle x |
   \frac{1}{6} [X, [X, Y]] e^{X} | x \rangle \nonumber \\ 
 &=& \int d^{d} x \frac{e(x)}{(2 \pi \tau)^{\frac{d}{2}}} 
   \left\{ \tau \left( \frac{1}{6} g^{i_{1} i_{2}}(x) g_{j_{1} j_{2}}(x)
   ( \frac{d^{2} g^{j_{1} j_{2}}(x) }{dx^{i_{1}} dx^{i_{2}}}  )
   + \frac{1}{3}  (\frac{d^{2} g^{ij}(x) }{dx^{i} dx^{j}} ) \right)
   \right. \nonumber \\
 & & - \tau^{2} \left( \frac{1}{12} g^{i_{1} i_{2}}(x) g^{j_{1}
   j_{2}}(x) g^{k_{1} k_{2}}(x) ( \frac{d^{4} g^{j_{1} j_{2}}
   (x)}{dx^{i_{1}} dx^{i_{2}} dx^{k_{1}} dx^{k_{2}}}  )
   \right. \nonumber \\
 & & \left.  + \frac{1}{3} g^{i_{1} i_{2}}(x)
   ( \frac{d^{3} A^{j}(x) }{dx^{i_{1} i_{2} j}} )  
   + \frac{1}{3} g^{i_{1} i_{2}}(x) (\frac{d^{2} B(x)}{dx^{i_{1}}
   dx^{i_{2}}} ) \right) 
   + \left. \frac{\tau^{3}}{6} ( g^{i_{1} i_{2}}(x) g^{j_{1}
   j_{2}}(x) )( \frac{d^{4} B(x)}{dx^{i_{1}} dx^{i_{2}} dx^{j_{1}}
   dx^{j_{2}} } ) \right\}. \nonumber \\ 
  \end{eqnarray} 

  \paragraph{ Computation of $\frac{1}{2} Y^{2} e^{X}$}
 \hspace{0mm} 

 The next job is the computation of the term $\frac{1}{2} Y^{2}$: 
  \begin{eqnarray}
  Y^{2} &=& \tau^{2} \left( (g^{i_{1} i_{2}}(x) - g^{i_{1} i_{2}}(y))
  \frac{d^{2}}{dx^{i_{1}} dx^{i_{2}}} + A^{i}(x) \frac{d}{dx^{i}} +
  B(x) \right) \nonumber \\
   & & \times
    \left( (g^{j_{1} j_{2}}(x) - g^{j_{1} j_{2}}(y))
  \frac{d^{2}}{dx^{j_{1}} dx^{j_{2}}} + A^{j}(x) \frac{d}{dx^{j}} +
  B(x) \right) \nonumber \\
  &=& \tau^{2} \left( (g^{i_{1} i_{2}}(x) - g^{i_{1} i_{2}}(y) )
  ( g^{j_{1} j_{2}}(x) - g^{j_{1} j_{2}}(y) ) \frac{d^{4}}{dx^{i_{1}}
  dx^{i_{2}} dx^{j_{1}} dx^{j_{2}}} \right. \nonumber \\
  & & + 2 (g^{i_{1} i_{2}}(x) - g^{i_{1} i_{2}}(y) )
  ( \frac{d  g^{j_{1} j_{2}}(x)}{dx^{i_{1}}}  ) \frac{d^{3}}{dx^{i_{2}}
  dx^{j_{1}} dx^{j_{2}}} 
   + (g^{i_{1} i_{2}}(x) - g^{i_{1} i_{2}}(y) )
  ( \frac{d^{2}  g^{j_{1} j_{2}}(x)}{dx^{i_{1}} dx^{i_{2}}} )
  \frac{d^{2}}{dx^{j_{1}} dx^{j_{2}}} \nonumber \\
 & & + 2 (g^{i_{1} i_{2}}(x) - g^{i_{1} i_{2}}(y)) A^{j}(x) 
    \frac{d^{3}}{d x^{i_{1}} dx^{i_{2}} dx^{j}} 
    + 2 (g^{i_{1} i_{2}}(x) - g^{i_{1} i_{2}}(y))
  ( \frac{d A^{j}(x)}{dx^{i_{1}}} ) \frac{d^{2}}{dx^{i_{2}} dx^{j}}
  \nonumber \\
 & & + (g^{i_{1} i_{2}}(x) - g^{i_{1} i_{2}}(y) )
  (\frac{d^{2} A^{j}(x)}{dx^{i_{1}} dx^{i_{2}} } ) \frac{d}{dx^{j}}
   +  (g^{i_{1} i_{2}}(x) - g^{i_{1} i_{2}}(y) ) B(x)
  \frac{d^{2}}{dx^{i_{1}} dx^{i_{2}}} \nonumber \\
 & &   + 2 (g^{i_{1} i_{2}}(x) - g^{i_{1} i_{2}}(y) )
  ( \frac{d B(x)}{dx^{i_{1}}} ) \frac{d}{dx^{i_{2}}} 
  + (g^{i_{1} i_{2}}(x) - g^{i_{1} i_{2}}(y) )
  (\frac{d^{2} B(x)}{dx^{i_{1}} dx^{i_{2}} }  ) \nonumber \\
 & & + A^{i}(x) ( \frac{d  g^{j_{1} j_{2}}(x)}{dx^{i}} )
  \frac{d^{2}}{dx^{j_{1}} dx^{j_{2}}} 
   + A^{i}(x) A^{j}(x) \frac{d^{2}}{dx^{i} dx^{j}} 
   + A^{i} (x) B(x) \frac{d}{dx^{i}} + A^{i}(x) (\frac{d
  B(x)}{dx^{i}}) \nonumber \\ 
 & & \left. + (g^{i_{1} i_{2}}(x) - g^{i_{1} i_{2}}(y) ) B(x)
  \frac{d^{2}}{dx^{j_{1}} dx^{j_{2}}} + B(x) A^{i}(x) \frac{d}{dx^{i}} 
  + B(x) B(x) \right).  
  \end{eqnarray}
 The trace is thus 
  \begin{eqnarray}
  & &  Tr ( \frac{1}{2} Y^{2} e^{X} ) = \int d^{d} x \langle x |
   \frac{1}{2} Y^{2} e^{X} | x \rangle \nonumber \\
  &=& \int d^{d} x \frac{e(x)}{(2 \pi \tau)^{\frac{d}{2}}} \left\{ 
  \tau \left( - \frac{1}{4} A^{i}(x) g_{j_{1} j_{2}}(x)
   (\frac{d  g^{j_{1} j_{2}}(x)}{dx^{i}}  ) - \frac{1}{4} A^{i}(x)
   A^{j}(x) g_{ij}(x) \right) \right. \nonumber \\
  & & \hspace{10mm} \left.  + \tau^{2} \left( \frac{1}{2} A^{i}(x) (\frac{d
   B(x)}{dx^{i}} ) + \frac{1}{2} B(x) B(x) \right)  \right\}. 
  \end{eqnarray}
 
  \paragraph{ Computation of $\frac{1}{8} [X,Y]^{2} e^{X}$}
 \hspace{0mm} 

 We next compute the commutator $[X,Y]^{2}$. From now on, the 
 computation becomes more complicated than before, and we give only
 the trace:
  \begin{eqnarray}
 & &  Tr ( \frac{1}{8} [X,Y]^{2} e^{X} ) = \int d^{d} x \langle x |
   \frac{1}{8} [X,Y]^{2} e^{X} | x \rangle \nonumber \\
 &=& \int d^{d} x \frac{e(x)}{(2 \pi \tau)^{\frac{d}{2}}}
  \left\{ \tau \left(- \frac{1}{16} g^{ik}(x) g_{j_{1} j_{2}}(x)
   g_{l_{1} l_{2}}(x)  ( \frac{d g^{j_{1} j_{2}}(x)}{dx^{i}} )
   ( \frac{d g^{l_{1} l_{2}}(x)}{dx^{k}} ) \right. \right. \nonumber \\
 & & \hspace{5mm} - \frac{1}{8} g^{ik}(x) g_{j_{1} l_{1}}(x) g_{j_{2}
   l_{2}}(x) ( \frac{d g^{j_{1} j_{2}}(x)}{dx^{i}} ) ( \frac{d
   g^{l_{1} l_{2}}(x)}{dx^{k}} ) 
   - \frac{1}{4} ( \frac{d g^{j_{1} j_{2}}(x)}{dx^{j_{1}}} ) ( \frac{d 
   g^{l_{1} l_{2}}(x) }{dx^{j_{2}}} ) g_{l_{1} l_{2}}(x) \nonumber \\ 
 & & \hspace{5mm} \left. \left. - \frac{1}{4} g_{j_{2} l_{2}}(x)
   ( \frac{d g^{l_{1} 
   l_{2}} (x)}{dx^{j_{1}}} )( \frac{d g^{j_{1} j_{2}}(x)}{dx^{l_{1}}}
   ) - \frac{1}{4} g_{ij}(x) (\frac{d g^{ip}(x)}{dx^{p}} )( \frac{d
   g^{jq}(x) }{dx^{q}} ) \right) + {\cal O}(\tau^{2}) \right\}.
  \end{eqnarray}

  \paragraph{ Computation of $\frac{1}{3} Y [X,Y] e^{X}$}
 \hspace{0mm} 
  \begin{eqnarray}
 & &  Tr ( \frac{1}{3} Y [X,Y] e^{X} ) = \int d^{d} x \langle x | 
   \frac{1}{3} Y [X,Y] e^{X} | x \rangle \nonumber \\
 &=& \int d^{d} x \frac{e(x)}{( 2 \pi \tau)^{\frac{d}{2}}} \left\{ 
  \tau \left( \frac{1}{6} A^{i}(x) (\frac{d g^{j_{1}
   j_{2}}(x)}{dx^{i}} ) g_{j_{1} j_{2}}(x) + \frac{1}{3} A^{i}(x) g_{ij} 
   (x) ( \frac{d g^{j_{1} j_{2}}(x)}{dx^{j_{2}}} ) \right)
   \right. \nonumber \\
 & &  - \tau^{2} \left( \frac{1}{6}  g^{k_{1} k_{2}}(x)
   g_{ij}(x) A^{i}(x) (\frac{d^{2} A^{j}(x)}{dx^{k_{1}} dx^{k_{2}}} )
   + \frac{1}{3} A^{i}(x) (\frac{d B(x)}{dx^{i}}) \right. \nonumber \\
 & & \hspace{5mm} + \frac{1}{6} g^{k_{1} k_{2}}(x) g_{j_{1} j_{2}}(x)
   ( \frac{d^{2} g^{j_{1} j_{2}}(x)}{dx^{k_{1}} dx^{k_{2}}} ) B(x) 
   + \frac{1}{6} g^{k_{1} k_{2}}(x) g_{j_{1} j_{2}}(x) A^{i}(x)
   (\frac{d^{3} g^{j_{1} j_{2}}(x) }{dx^{i} dx^{k_{1}} dx^{k_{2}}} ) 
  \nonumber \\
 & & \hspace{5mm} \left. + \frac{1}{3} A^{i}(x) (\frac{d^{2}
   A^{j}(x)}{dx^{i} dx^{j}}) + \frac{1}{3} B(x) (\frac{d
   A^{i}(x)}{dx^{i}} ) \right) \nonumber \\
 & &  \left. + \frac{\tau^{3}}{3} B(x) g^{k_{1} k_{2}}(x)
   (\frac{d^{2} B(x)}{dx^{k_{1}} dx^{k^{2}}} ) \right\}.
  \end{eqnarray}

  \paragraph{ Computation of $\frac{1}{6} [X,Y] Y e^{X}$}
 \hspace{0mm} 
  \begin{eqnarray}
 & &  Tr ( \frac{1}{6} [X,Y] Y e^{X} ) = \int d^{d} x \langle x | 
   \frac{1}{6} [X,Y] Y e^{X} | x \rangle \nonumber \\
 &=& \int d^{d} x \frac{e(x)}{( 2 \pi \tau)^{\frac{d}{2}}} \left\{ 
  \tau \left( \frac{1}{12} g^{k_{1} k_{2}}(x) g_{i_{1} i_{2}}(x)
   g_{j_{1} j_{2}}(x) (\frac{d g^{i_{1} i_{2}}(x)}{dx^{k_{1}}} )
   ( \frac{d g^{j_{1} j_{2}}(x)}{dx^{k_{2}}} )
   \right. \right. \nonumber \\
 & &  + \frac{1}{6} g^{k_{1} k_{2}} (x) g_{i_{1} j_{1}}(x) g_{i_{2}
   j_{2}}(x)  (\frac{d g^{i_{1} i_{2}}(x)}{dx^{k_{1}}} )
   ( \frac{d g^{j_{1} j_{2}}(x)}{dx^{k_{2}}} ) \nonumber \\
 & & + \frac{1}{6} g_{i_{1} i_{2}}(x) (\frac{d g^{i_{1}
   i_{2}}(x)}{dx^{j_{1}}} ) ( \frac{d g^{j_{1} j_{2}}(x)}{dx^{j_{2}}}
   ) + \frac{1}{3} g_{i_{2} j_{2}}(x) (\frac{d g^{j_{1}
   j_{2}}(x)}{dx^{i_{1}}} )( \frac{d g^{i_{1} i_{2}}(x)}{dx^{j_{1}}} ) 
   \nonumber \\
 & & + \frac{1}{12} g_{j_{1} j_{2}}(x) \left. \left.(\frac{d g^{j_{1} 
   j_{2}}(x)}{dx^{i}} ) A^{i}(x) + \frac{1}{6} (\frac{d g^{j_{1}
   j_{2}}(x)}{dx^{j_{1}} } ) g_{j_{2} i}(x) A^{i}(x)  \right) + {\cal
   O}(\tau^{2}) \right\}.
  \end{eqnarray}

  Summing up these terms altogether, we obtain the general form of the 
  first-order coefficient $a_{1}(x)$ as
  \begin{eqnarray}
   a_{1}(x) &=& B(x) - \frac{1}{2} (\frac{d A^{i}(x)}{dx^{i}})
    + \frac{1}{3} (\frac{d^{2} g^{ij}(x)}{dx^{i} dx^{j}}) 
    - \frac{1}{12} g^{i_{1} i_{2}}(x) g_{j_{1} j_{2}}(x) (\frac{d^{2}
    g^{j_{1} j_{2}}(x)}{dx^{i_{1}} dx^{i_{2}}} ) \nonumber \\
    &+& \frac{1}{12} g_{i_{2} j_{2}}(x) (\frac{d g^{j_{1}
    j_{2}}(x)}{dx^{i_{1}}} )(\frac{d g^{i_{1} i_{2}}(x)}{dx^{j_{1}}} ) 
    - \frac{1}{4} A^{i}(x) A^{j}(x) g_{ij}(x) 
     + \frac{1}{2} A^{i}(x) g_{i j_{1}} (x) (\frac{d g^{j_{1}
    j_{2}}(x)}{dx^{j_{2}} }) \nonumber \\
    &+& \frac{1}{48} g^{k_{1} k_{2}}(x) g_{i_{1} i_{2}}(x) g_{j_{1}
    j_{2}}(x) (\frac{d g^{i_{1} i_{2}}(x) }{dx^{k_{1}}} )(\frac{d
    g^{j_{1} j_{2}}(x)}{dx^{k_{2}}} ) \nonumber \\ 
    &+&  \frac{1}{24} g^{k_{1} k_{2}}(x) g_{i_{1} j_{1}}(x) g_{i_{2}
    j_{2}}(x) (\frac{d g^{i_{1} i_{2}}(x) }{dx^{k_{1}}} )( \frac{d
    g^{j_{1} j_{2}}(x)}{dx^{k_{2}}} ) \nonumber \\
   &-& \frac{1}{12} g_{i_{1} i_{2}}(x) (\frac{d g^{i_{1}
    i_{2}}(x)}{dx^{j_{1}}} )(\frac{d g^{j_{1}
    j_{2}}(x)}{dx^{j_{2}}}) - \frac{1}{4} g_{ij}(x) (\frac{d
    g^{ip}(x)}{dx^{p}} ) ( \frac{d g^{jq}(x)}{dx^{q}} ). \label{AZ32a1}
  \end{eqnarray}

 Especially, the Seeley-de-Witt coefficient of the Laplace Beltrami
 operator
  \begin{eqnarray}
   \Delta(x) &=& \frac{1}{\sqrt{g(x)}} \left( \frac{d}{dx^{i}} \sqrt{g(x)}
   g^{ij} (x) \frac{d}{dx^{j}} \right) \nonumber \\
  &=& g^{ij}(x) \frac{d}{dx^{i}} \frac{d}{dx^{j}}
       + \left( \left( \frac{d g^{ij}(x)}{dx^{j}} \right)
  - \frac{1}{2} g^{ij}(x) \left( \frac{d}{dx^{j}} g^{kl}(x) \right)
   g_{kl}(x) \right) \frac{d}{dx^{i}} 
    \label{laplacebeltrami}
  \end{eqnarray}
  is important on the practical ground. This amounts to $A^{i}(x) 
  =  \left( \frac{d g^{ij}(x)}{dx^{j}} \right)
  - \frac{1}{2} g^{ij}(x) \left( \frac{d}{dx^{j}} g^{kl}(x) \right)
   g_{kl}(x)$ and $B(x)=0$, in the general form (\ref{elliptic}).
  Plugging this into the general formula of $a_{1}$ (\ref{AZ32a1}), we 
  obtain the following important result:
  \begin{eqnarray}
   a_{1} (x) = \frac{R(x)}{6}, \label{seeley-de-witt-laplace}
  \end{eqnarray}
  where $R(x)$ is the Ricci scalar of the spacetime metric.
  This result is used in the context of the superstring theory in
  deriving the Weyl anomaly and the corresponding Liouville action.

 \paragraph{Derivation of (\ref{hkmass})} \hspace{0mm} 

  With the above idea in mind, let us derive the result (\ref{hkmass}) 
  for the simplest example of the calculation in our case.
  We expand the differential operator $- \tau D^{2}$ given by
  (\ref{AZ2expansionD}) around $X = \tau \partial_{\mu} \partial^{\mu}$. 
  Namely, we express $- \tau D^{2}$ as $- \tau D^{2} = X+Y$, and
  perform a perturbative expansion around $X$. The heat kernel for $X$ 
  is given by $ \langle x | e^{X} |y \rangle
  = \int \frac{d^{d}x}{(2 \pi \tau)^{\frac{d}{2}}}
     \exp \left( - \frac{(x-y)_{\mu} (x-y)^{\mu}}{4 \tau} \right)$. 
  Here, we identify the field
  ${a^{(i)}}_{\mu} (x)$ with the vielbein, and we expand this around
  the flat background metric as ${a^{(i)}}_{\mu} (x)=
  ({a^{(i)}}_{\mu})_{0} + {h^{i}}_{\mu} (x)$, where $({a^{(i)}}_{\mu})_{0} 
  ({a^{(j)}}_{\mu})_{0} = \eta^{ij}$.
  Then, $D$ is more explicitly expressed as
  \begin{eqnarray}
  D &=& \Gamma^{\mu} \left( a_{\mu}(x) + i \partial_{\mu}
   + \frac{i}{2} (\partial_{i} {h^{i}}_{\mu} (x)) + i {h^{i}}_{\mu} (x)
  \partial_{i} + \cdots \right) 
  + \frac{i}{3!} \Gamma^{\mu_{1} \mu_{2} \mu_{3}} 
     (a_{\mu_{1} \mu_{2} \mu_{3}} (x) + \cdots) \nonumber \\
  &-& \frac{1}{5!} \Gamma^{\mu_{1} \cdots \mu_{5}} (a_{\mu_{1} \cdots
   \mu_{5}} (x) + \cdots)   
  - \frac{i}{7!} \Gamma^{\mu_{1} \cdots \mu_{7}} (a_{\mu_{1} \cdots
   \mu_{7}} (x) + \cdots) 
  + \frac{1}{9!} \Gamma^{\mu_{1} \cdots \mu_{9}} (a_{\mu_{1} \cdots
   \mu_{9}} (x) + \cdots).  \nonumber \\
  \end{eqnarray}

  Now, we are interested in the mass term of the vector fields 
  $a_{\mu_{1} \cdots \mu_{2n-1}} (x)$.
 In the Campbell-Baker-Hausdorff formula (\ref{cbh}), only the terms
 $Y$ and $\frac{1}{2} Y^{2}$ contribute to their mass terms.
 Now, we have a close look at these two contribution.

 For $Y$, only the elements of the rank-0 gamma matrices contribute to
 the heat kernel, because the trace $tr$ for the $32 \times 32$ gamma
 matrices projects out the higher-rank gamma matrices.
 Therefore, the contribution to the mass term is clearly
  \begin{eqnarray}
 \langle x| tr Y e^{X} | x \rangle  &=&  \int \frac{d^{d}x}{(2 \pi
           \tau)^{\frac{d}{2}}} 32 \tau \left(
           - a_{\mu}(x) a^{\mu} (x) 
           - \frac{1}{3!} a_{\mu_{1} \mu_{2} \mu_{3}} (x) a^{\mu_{1}
           \mu_{2} \mu_{3}} (x)
           - \frac{1}{5!} a_{\mu_{1} \cdots \mu_{5}} (x) a^{\mu_{1}
           \cdots \mu_{5}} (x) \right. \nonumber \\
     & & \hspace{10mm} \left. 
     - \frac{1}{7!} a_{\mu_{1} \cdots \mu_{7}} (x) a^{\mu_{1}
           \cdots \mu_{7}} (x)
           - \frac{1}{9!} a_{\mu_{1} \cdots \mu_{9}} (x) a^{\mu_{1}
           \cdots \mu_{9}} (x) \right), \label{y-cont}
  \end{eqnarray}
  where the constant $32$ comes from the trace $tr {\bf 1}_{32 \times 32}$. 
  We next compute the effect of $\frac{1}{2} Y^{2}$. This time, the
  higher-rank terms of $Y$ contribute, because the product of the
  gamma matrices survives. The relevant terms are as follows:
  \begin{eqnarray}
   Y &=& \tau \left( 2i a_{\mu}(x) \partial_{\mu}
       - \frac{2}{2!} \Gamma^{\nu_{1} \nu_{2}} a_{\mu \nu_{1} \nu_{2}} 
   (x) \partial_{\mu}
       - \frac{2i}{4!} \Gamma^{\nu_{1} \cdots \nu_{4}} a_{\mu \nu_{1}
   \cdots \nu_{4}} (x) \partial_{\mu}
       + \frac{2}{6!} \Gamma^{\nu_{1} \cdots \nu_{6}} a_{\mu \nu_{1}
   \cdots \nu_{6}} (x) \partial_{\mu} \right. \nonumber \\
       &+& \left. \frac{2i}{8!} \Gamma^{\nu_{1} \cdots \nu_{8}} a_{\mu \nu_{1}
   \cdots \nu_{8}} (x) \partial_{\mu} \right) + \cdots, \nonumber
  \end{eqnarray}
  where $\cdots$ denotes the omission of the irrelevant terms. This
  gives the contribution 
  \begin{eqnarray}
  \langle x | \frac{1}{2} tr Y^{2} e^{X} | x \rangle &=&  
  \langle x | \frac{\tau^{2}}{2} \times 32 \times \left( -4 a^{\mu}(x) a^{\nu}
  (x) - \frac{4}{2!} {a^{\mu}}_{\rho_{1} \rho_{2}} (x) a^{\nu \rho_{1} 
  \rho_{2}} (x) 
  - \frac{4}{4!} {a^{\mu}}_{\rho_{1} \cdots \rho_{4}} (x) a^{\nu \rho_{1} 
  \cdots \rho_{4}} (x) \right. \nonumber \\
 & & \left. \hspace{10mm}
  -  \frac{4}{6!} {a^{\mu}}_{\rho_{1} \cdots \rho_{6}} (x) a^{\nu \rho_{1} 
  \cdots \rho_{6}} (x)
  - \frac{4}{8!} {a^{\mu}}_{\rho_{1} \cdots \rho_{8}} (x) a^{\nu \rho_{1} 
  \cdots \rho_{8}} (x) \right) \partial_{\mu} \partial_{\nu} e^{X} | x 
 \rangle \nonumber \\
 &=& \int \frac{d^{d}x}{(2 \pi \tau)^{\frac{d}{2}}} 32 \tau \left( 
           a_{\mu}(x) a^{\mu} (x)   
         + \frac{1}{2!} a_{\mu_{1} \mu_{2} \mu_{3}} (x) a^{\mu_{1}
           \mu_{2} \mu_{3}} (x)
           + \frac{1}{4!} a_{\mu_{1} \cdots \mu_{5}} (x) a^{\mu_{1}
           \cdots \mu_{5}} (x) \right. \nonumber \\
      & & \left. \hspace{10mm}  
      + \frac{1}{6!} a_{\mu_{1} \cdots \mu_{7}} (x) a^{\mu_{1}
           \cdots \mu_{7}} (x) 
           + \frac{1}{8!} a_{\mu_{1} \cdots \mu_{9}} (x) a^{\mu_{1}
           \cdots \mu_{9}} (x) \right), \label{y2-cont}
  \end{eqnarray}
  where we have utilized the formula (\ref{dell-2}) for $g_{i_{1}
  i_{2}}(y) = \eta_{i_{1} i_{2}}$. We obtain the mass term (\ref{hkmass}) in
  question by the sum of (\ref{y-cont}) and (\ref{y2-cont}).

 \section{Basic knowledge for the numerical simulation of the
   matrix model} \label{simulation}
 This appendix is devoted to introducing the basic knowledge of the
 Monte Carlo simulation of the matrix model. While the following ideas 
 can be readily applied to the QCD, the programming for the matrix
 model is much simpler than for the quantum field theory.

 \subsection{Review of the Monte Carlo simulation} \label{sec-mc}
  The Monte Carlo simulation plays an extremely important role in the
  numerical simulation. In many cases, the analytical computation may
  be too complicated to handle. Let us think about the four-dimensional 
  10,000-site lattice gauge theory. This system has 40,000
  sites. Therefore, even for the simplest $Z_{2}$ gauge theory, we
  have to compute the sum of the $2^{40,000} \sim 1.58 \times
  10^{12041}$  configurations. The analytical computation of 
  the partition function is almost impossible and this leads us to
  resort to the numerical treatment.

  To put it simply, the fundamental idea of the Monte Carlo simulation 
  is to produce the equilibrium ensemble artificially. 
  In the case of the matrix model, the state $C$ is identified with
  the element of the $N \times N$ matrices; namely
  \begin{eqnarray}
   C = \{ (A_{\mu})_{ij}, \cdots, \}.
  \end{eqnarray}
  For the quantum field theory, $C$ of course corresponds to the gauge 
  field configuration $U_{\mu}^{ij}(x)$. Since the sequential integral 
  is represented as $\int d A_{\mu} = \sum_{C} \cdots$,
  the partition function is rewritten as $Z = \sum_{C} \exp(-S(C))$. 
  In the Monte Carlo simulation, we derive the configuration $C_{k}$
  that complies with the Boltzmann probability
  \begin{eqnarray}
     P_{\textrm{bol}} (C_{k}) = Z^{-1} \exp(-S(C_{k})). \label{boltzmann}
  \end{eqnarray}
  
  To this end, we consider the following Markov process. Namely, the
  probability to transform from the configuration $C_{k-1}$ to the
  new one $C_{k}$ depends only on the state $C_{k-1}$ and
  $C_{k}$. This probability never depends on the previous history
  $C_{1}, C_{2}, \cdots, C_{k-2}$. In this sense, this probability is
  denoted as
  \begin{eqnarray}
  P(C_{k-1} \to C_{k}) = P(C_{k-1}, C_{k}), \label{markov}
  \end{eqnarray}
 since the probability is the function of only $C_{k-1}$
 and $C_{k}$. The following "detailed balance condition" is a vital
 constraint on the Boltzmann distribution:
  \begin{eqnarray}
   e^{-S(C)} P(C,C') = e^{-S(C')} P(C',C). \label{detailedbalance}
  \end{eqnarray}
 For the transformation complying with the detailed balance condition, 
 we have the following two crucial properties:
  \begin{enumerate}
   \item{An equilibrium sequence of the state transforms into another
       equilibrium state.}
   \item{A nonequilibrium sequence approaches to the equilibrium state.}
  \end{enumerate}
  It is easy to prove the first statement. While it is trivial, we
  recall the following properties of the probabilities.
  Firstly, $\sum_{C} P(C,C') = \sum_{C'}P(C,C') = 1$. Secondly, when
  $E'$ is obtained from $E$ through the Markov process and $P'$ is the 
  probability for the ensemble $E'$, we have
  $P'(C) = \sum_{C'} P(C,C')P(C')$. 
  The first property is verified by taking the sum of the both hand
  sides of (\ref{detailedbalance}) with respect to $C$. Now, the
  previous state $P_{\textrm{bol}}(C)$ complies with the Boltzmann
  distribution $P_{\textrm{bol}}(C) = Z^{-1} e^{-S(C)}$. Then, we have
  \begin{eqnarray}
   P'(C') = \sum_{C} P_{\textrm{bol}}(C) P(C,C') 
          = \sum_{C} P_{\textrm{bol}}(C') P(C',C) = Z^{-1}e^{-S(C')}.
  \end{eqnarray}
  Thus, the newly emerging state also complies with the Boltzmann
  distribution.

  The second property is substantiated by defining the distance of the
  two ensembles $E$ and $E'$ as
   \begin{eqnarray}
    |E-E'| = \sum_{C} |P(C)-P'(C)|. 
   \end{eqnarray}
  Then, the distance from the Boltzmann distribution is given by
  \begin{eqnarray}
    |E'-E_{\textrm{bol}}| &=& \sum_{C}|P'(C)-P_{\textrm{bol}}(C)|
  = \sum_{C} \left| \sum_{C'} P(C,C') (P(C') - P_{\textrm{bol}}(C'))
  \right| \nonumber \\
  &\leq& \sum_{C} \sum_{C'} P(C,C') |P(C') - P_{\textrm{bol}}(C')|
   = \sum_{C'} |P(C') - P_{\textrm{bol}}(C')| = |E-E_{\textrm{bol}}|.
  \end{eqnarray}
  This implies that the distance from the Boltzmann state is smaller
  as we iterate the Markov process obeying the detailed balance
  condition. 

  There are two major algorithms that satisfy the detailed balance
  condition. One is the heat bath algorithm. The heat bath algorithm
  is used when we have a clear correspondence between the uniform random
  number and the configuration of the fields. In this case, the
  probability to transform to the configuration $C'$ is given by
  \begin{eqnarray}
   P(C,C') \propto e^{-S(C')}. \label{heatbath}
  \end{eqnarray}
  In this process, the probability $P(C,C')$ depends only on the new
  configuration, and does not even depend on the previous state. It is
  trivial that this satisfies the detailed balance condition.
  In analyzing the matrix model, we resort to the heat bath
  algorithm. We postpone the explicit example to the subsequent
  sections for the matrix model simulation.

  The other is the Metropolis algorithm. Firstly, we choose the
  configuration $C'$ at random. Then, we compute the difference
  $\Delta S = S(C') - S(C)$. We adopt the new state at the probability 
  $\textrm{min}(1,e^{-\Delta S})$. Namely, the probability is given by
   \begin{eqnarray}
    P(C,C') \to \left\{ \begin{array}{ll} 1 & (\textrm{if }
    S(C)>S(C')), \\
    e^{-\Delta S} = e^{-S(C')} e^{S(C)} & (\textrm{if }
    S(C)<S(C') ). \end{array} \right. \label{metropolis}
   \end{eqnarray}
  We take a uniform random
  number $r \in [0,1]$. When $r \leq e^{-\Delta S}$ we adopt the new
  configuration $C'$, and otherwise we jettison the new configuration
  $C'$ and keep the status quo.
  It is easy to verify that the Metropolis algorithm also satisfies
  the detailed balance condition, by noting that the probability for
  the inverse transformation is given by
   \begin{eqnarray}
    P(C',C) \to \left\{ \begin{array}{ll} e^{\Delta S} = e^{S(C')}
    e^{-S(C)} & (\textrm{if } S(C)>S(C')), \\
    1 & (\textrm{if }S(C)<S(C') ). \end{array} \right. \label{metropolis2} 
   \end{eqnarray}
  The advantage of the Metropolis algorithm is that it can be applied
  to more various systems. When it is impossible to build the heat bath 
  algorithm, the Metropolis algorithm is here to stay. A good example
  is the supersymmetric matrix model. When the model has fermions,
  we often use the hybrid Monte Carlo simulation. It is impossible to
  handle the Grassmann numbers numerically. Therefore, if we are to
  treat the model equipped with the fermion, we must integrate out the 
  fermion analytically. Then, we evaluate the Pfaffian composed of the
  bosonic fields with the numerical method. 

 \subsection{Simulation of the quadratic matrix model}
 We first start with the simplest example of the matrix model. While
 it is a very simple toy model, we scrutinize this case in full detail 
 because this is the heart of the heat bath algorithm of the more
 complicated matrix models. We deal with the following quadratic
 matrix model: 
  \begin{eqnarray}
   S = \frac{N}{2} Tr \phi^{2}.
  \end{eqnarray}
 Here, $\phi$ is an $N \times N$ hermitian matrix, and this model
 is invariant under the $U(N)$ unitary transformation.
 Its path integral is given by 
  \begin{eqnarray}
    Z = \int d^{N^{2}} \phi e^{-S} = \int d^{N^{2}} \phi \exp( -
    \frac{N}{2} Tr \phi^{2}). \label{pathintone}
   \end{eqnarray}
 Analytically, the propagator of this matrix model is evaluated as
 follows\footnote{The result
 (\ref{propone}) is applicable to the case $\phi \in U(N)$. Here, we
 touch on the propagator for the $SU(N)$ case. This can be derived
 from (\ref{propone}) by subtracting the trace part. Namely, we
 replace $\phi_{ii}$ with $\phi'_{ii} = \phi_{ii} - \frac{1}{N}
 \sum_{i=1}^{N} \phi_{ii}$.  The propagator $\langle \phi_{ii}'
 \phi_{ll}' \rangle$ is computed as
  \begin{eqnarray}
   \langle \phi'_{ii} \phi'_{ll} \rangle
 = \langle \phi_{ii} \phi_{ll} \rangle
 - 2 \frac{1}{N} \sum_{j=1}^{N} \langle \phi_{ii} \phi_{jj} \rangle
 + \frac{1}{N^{2}} \sum_{j,k=1}^{N} \langle \phi_{jj} \phi_{kk}
 \rangle
 = \frac{1}{N} \left( \delta_{il} - \frac{1}{N} \right). \nonumber
  \end{eqnarray}
 Therefore, the Feynman rule is replaced for $SU(N)$ case by
  \begin{eqnarray}
   \langle \phi_{ij} \phi_{kl} \rangle = \frac{1}{N}
   \left( \delta_{il} \delta_{jk} - \frac{1}{N} \delta_{ij}
   \delta_{kl} \right). \nonumber
  \end{eqnarray}
  }:
   \begin{eqnarray}
    \langle \phi_{ij} \phi_{kl} \rangle = \frac{1}{Z} \int d^{N^{2}} \phi
    \phi_{ij} \phi_{kl} \exp( - \frac{N}{2} Tr \phi^{2}) = \frac{1}{N} 
    \delta_{il}
    \delta_{jk}. \label{propone}
   \end{eqnarray}
 The proof of this propagator goes as follows.
 We note that, due to the hermiticity of $\phi$, the trace is written as 
   \begin{eqnarray}
   \frac{N}{2} Tr \phi^{2} = \frac{N}{2} \sum_{i,j=1}^{N} \phi_{ij} \phi_{ji}
  = N \sum_{1 \leq i<j \leq N} \phi_{ij} \phi^{\ast}_{ij} 
  + \frac{N}{2} \sum_{i=1}^{N} \phi_{ii} \phi_{ii}. 
  \end{eqnarray}
  We separate $\phi_{ij}$ into the real and the imaginary part as
  \begin{eqnarray}
    \phi_{ij} = \frac{X_{ij} + i Y_{ij}}{\sqrt{2N}}
    (=\phi_{ji}^{\ast}). 
  \end{eqnarray}
 Here, $X_{ij}$ and $Y_{ij}$ are real c-number. The diagonal parts are
 real numbers, and they are rewritten using the real numbers $W_{i}$ as
  \begin{eqnarray}
   \phi_{ii} = \frac{W_{i}}{\sqrt{N}}. 
  \end{eqnarray}
 Then, the quadratic term  is written as
 \begin{eqnarray}
   \frac{N}{2} Tr \phi^{2} = \frac{1}{2} \left( \sum_{i=1}^{N} W_{i}^{2}
     + \sum_{1\leq i<j \leq N} (X_{ij}^{2} + Y_{ij}^{2}) \right).
  \label{heatbathkey}
 \end{eqnarray}
  The derivation of the propagator reduces to the simple Gaussian
  integral:
  \begin{eqnarray}
   \frac{1}{a} = \frac{\int^{+\infty}_{-\infty} dx x^{2}
   \exp(-\frac{ax^{2}}{2})}{\int^{+\infty}_{-\infty} dx
   \exp(-\frac{ax^{2}}{2})}. 
  \end{eqnarray}
  \begin{itemize}
   \item{$\langle \phi_{ii} \phi_{ll} \rangle = \frac{1}{N} \langle
       W_{i} W_{l} \rangle$ survives only for $i=l$.}
   \item{For $\langle \phi_{ij} \phi_{kl} \rangle$ ($i \neq j$), we note the 
       following two results. Firstly, $\langle \phi_{ij} \phi_{ij}\rangle$
       is shown to vanish as 
    \begin{eqnarray}
      \langle \phi_{ij} \phi_{ij}\rangle = \frac{1}{2N} \langle
          (\underbrace{X_{ij} X_{ij} - Y_{ij}
          Y_{ij}}_{\textrm{cancelled}} 
           + 2i \underbrace{X_{ij} Y_{ij}}_{(*)}) \rangle =
          \frac{1-1}{2N}=0.
    \end{eqnarray}
     Here, (*) does not contribute ab initio, since this is a linear
          term of each $X_{ij}$ and $Y_{ij}$. 
     Secondly, we note that 
     \begin{eqnarray}
      \langle \phi_{ij} \phi_{ji} \rangle =
          \frac{1}{2N} \langle 
          (X_{ij} X_{ij} + Y_{ij} Y_{ij}) \rangle = \frac{1}{N}.
     \end{eqnarray}
          survives (namely when $i=l, j=k$).
    }
  \end{itemize}
  This completes the proof of the Feynman rule (\ref{propone}).

  Using the Feynman rule (\ref{propone}), we can easily derive the
  following vacuum expectation values: 
  \begin{eqnarray}
   \langle \frac{1}{N} Tr \phi^{2} \rangle =1, \hspace{2mm}
   \langle \frac{1}{N} Tr \phi^{4} \rangle = 2 + \frac{1}{N^{2}},
   \hspace{2mm} 
   \langle (\frac{1}{N} Tr \phi^{2})^{2} \rangle = 1 +
   \frac{2}{N^{2}}. \label{propanal}
  \end{eqnarray}

 We next go into the numerical treatment of this matrix model, by
 means of the heat bath algorithm. The
 rewriting of the action (\ref{heatbathkey}) plays a crucial role in
 the analysis. The advantage is that all $W_{i}$, $X_{ij}$ and
 $Y_{ij}$ are decoupled. Therefore, in the heat bath algorithm, we can
 update each of these real numbers independently.
  Recall that, in the heat bath algorithm (\ref{heatbath}), the
 probability for the new configuration depends only on the new
 configuration. Now, the partition function is rewritten as 
  \begin{eqnarray}
    Z = \int \prod_{i=1}^{N} dW_{i} \prod_{1 \leq i < j \leq N}
      dX_{ij} dY_{ij}  \exp \left( - \frac{1}{2}
      \sum_{i=1}^{N} W_{i}^{2} - \frac{1}{2} \sum_{1 \leq i<j \leq N} 
      (X_{ij}^{2} + Y_{ij}^{2}) \right). \label{pionematrix}
   \end{eqnarray}
 Thus, for example, the new configuration of $W_{i}$ complies with the
 following probability distribution.  
  \begin{eqnarray}
   P(W_{i}) = \frac{1}{\sqrt{2 \pi}} \exp( - \frac{W_{i}^{2}}{2}),
  \end{eqnarray}
 where $P(W_{i})$ is normalized so that $\int^{+\infty}_{-\infty}
 dW_{i} P(W_{i}) = 1$. Namely, the updating of the quantities $W_{i},
 X_{ij}, Y_{ij}$ all reduces to the Gaussian distribution.
 The process of updating the matrix elements can be divided into the
 following three steps.

 \paragraph{Generation of the uniform random number} \hspace{0mm} 

 Firstly, we introduce the numerical method to generate the sequence
 of the pseudo-random number. A simple way is the "congruence
 method". We generate the sequence of the random 
 numbers $\{ z_{i} \}$ by the following recursive formula:
  \begin{eqnarray}
    z_{i+1} = a z_{i} + c \hspace{2mm} ( \bmod \hspace{1mm} 2^{31}-1
    ). \label{congruence} 
  \end{eqnarray}
 The initial term $z_{1}$ should be given by hand as a random seed. 
 When we normalize this sequence as $\{ \frac{z_{i}}{2^{31}-1} \}$, we 
 obtain a uniform pseudo-random number [0,1]. Especially, it is known
 that we obtain a sensible pseudo-random sequence by taking $a=5^{11}$ and
 $c=0$. The drawback of the congruence method is the price of its
 simplicity. This random sequence has too short a cycle. However, this 
 is not so problematic so long as we undergo the simulation of the
 matrix model, while this is a serious problem for the QCD simulation.

 \paragraph{Generation of the Gaussian distribution} \hspace{0mm} 

 We next introduce the way to generate the Gaussian distribution from
 the uniform random numbers, generated in the previous step.
 We take two independent uniform random numbers $x,y \in [0,1]$.
 Then, we introduce the quantity $r$ as
  \begin{eqnarray}
   r = \sqrt{-a^{2} \log x^{2}}. \label{interr}
  \end{eqnarray}
 The probability distribution of $r$ is given by
  \begin{eqnarray}
   P(r) dr = P(x) \frac{dx}{dr} dr = \frac{2r}{a^{2}} \exp ( -
   (\frac{r}{a})^{2} ).
  \end{eqnarray}
 We next introduce the quantities $X,Y$ as
  \begin{eqnarray}
   X = r \cos(2 \pi y), \hspace{2mm} Y = r \sin(2 \pi y). \label{gaussres}
  \end{eqnarray} 
  We discern that this complies with the Gaussian distribution as
  \begin{eqnarray}
   P(r) dr dy \propto \exp \left( -\frac{1}{a^{2}} (X^{2} + Y^{2})
   \right) dXdY. \label{gaussdis}
  \end{eqnarray}
  Especially when $a=1$, $X$ and $Y$ obey the normal Gaussian
  distribution.

 \paragraph{Updating of the matrix elements} \hspace{0mm} 

 Now that we introduce the random number complying the Gaussian
 distribution, we update each of the matrix elements $W_{i}$, $X_{ij}$ 
 and $Y_{ij}$ by the Gaussian distribution. When we finish updating
 all of them once, this means that we have completed one sweep.
 Then, we reiterate the sweeps until the system is sufficiently
 thermalized.

 \subsection{Simulation of the quartic matrix model}
 We next investigate the $\phi^{4}$ matrix model. This model has a lot 
 of lessons in treating the IIB matrix model and its extensions.
 The action is defined by
  \begin{eqnarray}
  S = N \left( \frac{1}{2} Tr \phi^{2} - \frac{g}{4} Tr
  \phi^{4} \right). \label{phito4} 
  \end{eqnarray}
  In order to perform the heat bath algorithm of this matrix model, we
  introduce the auxiliary field $Q$ as
  \begin{eqnarray}
  {\tilde S} = \frac{N}{2} Tr \phi^{2} + \frac{N}{2} Tr Q^{2} -
  \alpha N Tr(Q \phi^{2}). \label{phito4aux}
  \end{eqnarray}
 Here, $\alpha = \sqrt{\frac{g}{2}}$ and the auxiliary field $Q$ is an $N
 \times N$ hermitian matrix. It is easy to verify that the action
 (\ref{phito4aux}) reduces to (\ref{phito4}) by integrating out $Q$:
 \begin{eqnarray}
 {\tilde S} = \frac{N}{2} Tr (Q - \alpha \phi^{2})^{2} + S. \nonumber
 \end{eqnarray}
 The idea to update the quantities $Q$ and $\phi$ is similar to that
 of the quadratic case, and we proceed rather quickly. We rewrite the
 elements of $Q$ as
  \begin{eqnarray}
  Q_{ii} = \frac{W_{i}}{\sqrt{N}} + \alpha (\phi^{2})_{ii}, \hspace{2mm}
  Q_{ij} = \frac{X_{ij} + i Y_{ij}}{\sqrt{2N}} + \alpha
  (\phi^{2})_{ij}. \label{updateq4}
  \end{eqnarray}
 In this way, the action ${\tilde S}$ is written as
  \begin{eqnarray}
  {\tilde S} = \frac{1}{2} \left( \sum_{i=1}^{N} W_{i}^{2}
       + \sum_{1 \leq i < j \leq N} (X_{ij}^{2} + Y_{ij}^{2}) \right) 
       + S. \nonumber
  \end{eqnarray}
 In the following, $W_{i}$ $(i=1, \cdots, N)$, $X_{ij}$ and $Y_{ij}$ $(1
 \leq i<j \leq N)$ denote the real c-numbers that obey the normal
 Gaussian distribution. 

 We next update the elements of the matrix $\phi$. To this end, we
 extract the dependence of ${\tilde S}$ on each component
 $\phi_{ij}$. Firstly, the dependence on the diagonal part $\phi_{ii}$ 
 is extracted as follows. In the following (up to
 (\ref{updatearhoij})), we sometimes do not take a
 summation with respect to the duplicate indices. 
  \begin{eqnarray}
   {\tilde S} &=& \frac{N}{2} (\phi_{ii})^{2} - \alpha N Q_{ii}
   (\phi_{ii})^{2} - \alpha N \phi_{ii} \sum_{j \neq i}(\phi_{ij}
   Q_{ji} + Q_{ij} \phi_{ji}) + \cdots \nonumber \\
 &=& \frac{Nc_{i}}{2} (\phi_{ii} - \frac{h_{i}}{c_{i}})^{2} -
   \frac{N}{2} \frac{(h_{i})^{2}}{c_{i}} + \cdots, \textrm{ where }
   \nonumber  \\
  c_{i} &=& 1 - 2 \alpha Q_{ii}, \hspace{2mm} 
     h_{i} = \alpha \sum_{j\neq i} (\phi_{ij} Q_{ji} + Q_{ij}
     \phi_{ji}). \label{updatephiii}
  \end{eqnarray}
 The $\cdots$ denotes the terms independent of $\phi_{ii}$. We
 omit them because these do not concern the updating of $\phi_{ii}$
 Therefore, the diagonal parts $\phi_{ii}$ are updated as
  \begin{eqnarray}
   \phi_{ii} = \frac{W_{i}}{\sqrt{N c_{i}}} + \frac{h_{i}}{c_{i}}.
   \label{updatephiii2} 
  \end{eqnarray} 

 We next go on to the updating of $\phi_{ij}$, whose dependence is
 extracted as
  \begin{eqnarray}
   {\tilde S} &=& N |\phi_{ij}|^{2} - \alpha N |\phi_{ij}|^{2}  
   (Q_{ii}+Q_{jj})
   - \left[ \alpha N \phi_{ij} \left( \sum_{k\neq i} \phi_{jk} Q_{ki} +
   \sum_{k \neq j} Q_{jk} \phi_{ki} \right) + \textrm{c.c.} \right] +
   \cdots 
   \nonumber \\
    &=& N c_{ij} \left| \phi_{ij} - \frac{h_{ij}}{c_{ij}} \right|^{2}
     - N \frac{|h_{ij}|^{2}}{c_{ij}} + \cdots, \textrm{ where }
   \label{updatephiij}   \\
    c_{ij} &=& 1 - \alpha(Q_{ii} + Q_{jj}), \hspace{2mm}
    h_{ij} = \alpha \left( \sum_{k \neq j} \phi_{ik} Q_{kj} +
    \sum_{k\neq i} Q_{jk} \phi_{kj} \right). \nonumber  
  \end{eqnarray}
  This leads us to update the nondiagonal parts $\phi_{ij}$ as
  \begin{eqnarray}
   \phi_{ij} = \frac{X_{ij} + i Y_{ij}}{\sqrt{2 N c_{ij}}} +
   \frac{h_{ij}}{c_{ij}}.
  \end{eqnarray}

 This completes the algorithm of updating the matrix elements. 
 However, we must recall that this matrix model is unbounded
 below. The eigenvalues near the origin is stable only for the
 large-$N$ limit, and only {\it metastable} for the finite $N$. 
   \begin{figure}[htbp]
   \begin{center}
    \scalebox{0.4}{\includegraphics{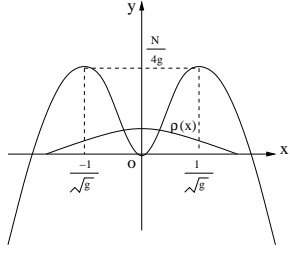} }
   \end{center}
   \caption{The metastability of the origin for finite $N$. The
     height of the potential barrier is $\frac{N}{4g}$.} 
  \label{BIPZeps}
  \end{figure}
  The function $V(x) = N (\frac{1}{2} x^{2} - \frac{g}{4}
  x^{4})$ has extrema at $x=\pm \frac{1}{\sqrt{g}}$, at which
  $V(\pm \frac{1}{\sqrt{g}}) = \frac{N}{4g}$. Therefore, the origin is 
  barriered by the potential with the height $\frac{N}{4g}$. For the
  large-$N$ limit, this barrier prevents the eigenvalues from
  dissipating outside the potential, and the system is kept finite.
  In \cite{BIPZ}, the following vacuum expectation value is computed
  analytically in the large-$N$ limit for $0 \leq g \leq \frac{1}{12}$:
   \begin{eqnarray}
   \lim_{N \to \infty} \langle \frac{1}{N} Tr \phi^{2} \rangle =
   \frac{a^{2}(4-a^{2})}{3}, \textrm{ where } a^{2} =
   \frac{2}{1+\sqrt{1-12g}}. \label{BIPZres}
   \end{eqnarray}
  Here, the parameter $a$ gives the support of the eigenvalue
  distribution. Namely, the eigenvalues are distributed only in the
  region $[-2a,+2a]$. More accurately, the eigenvalue distribution is
  also derived as 
  \begin{eqnarray}
   \rho(\lambda) = \frac{1}{\pi} \sqrt{4a^{2}-\lambda^{2}} \left(-
   \frac{1}{2}g \lambda^{2} - ga^{2} + \frac{1}{2} \right), \label{BIPZdis}
  \end{eqnarray}
  where $\rho(\lambda)$ is normalized as $\int^{+\infty}_{-\infty}
  \rho(\lambda) d \lambda = 1$. We delegate the detailed derivation to
  \cite{BIPZ}. 

  However, for the finite $N$, the height of the potential barrier is
  also finite, thus we see the divergence in the course of updating
  the eigenvalues.  
   \begin{figure}[htbp]
   \begin{center}
    \scalebox{0.4}{\includegraphics{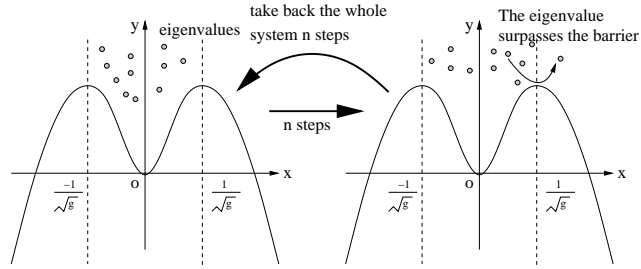} }
   \end{center}
   \caption{The reguralization trick utilized in
     \cite{kawaiokamoto}. When any one of the eigenvalue surpasses the 
     potential barrier, we take back the whole system several steps.} 
  \label{ko}
  \end{figure} 
  Namely, when the eigenvalues surpass the potential
  barrier, we see an avalanche of the eigenvalues rolling off and off
  the potential. This is what happens in the course of the divergence
  of the system. In order to evade the divergence, we need some
  regularization. Here, we utilize the same regularization trick as for 
  the research of the Weingarten model\cite{kawaiokamoto}. Namely,
  when any one of the eigenvalues surpasses the potential barrier
  $\frac{1}{\sqrt{g}}$, we take back the whole system several $n$ steps.

  In explicitly writing our code, we store the history of several
  previous steps for the matrix elements of $\phi$. When the
  eigenvalues happen to cordon the potential barrier, we recall the
  record of the history, and take back several steps (practically,
  around $n=3,4$ steps). In this way, we evade the divergence for finite $N$.
  Here, we plot the numerical result based on the heat bath algorithm
  for the quantity $\langle \frac{1}{N} Tr \phi^{2} \rangle$ for
  $N=32$. 
   \begin{figure}[htbp]
   \begin{center}
    \scalebox{0.7}{\includegraphics{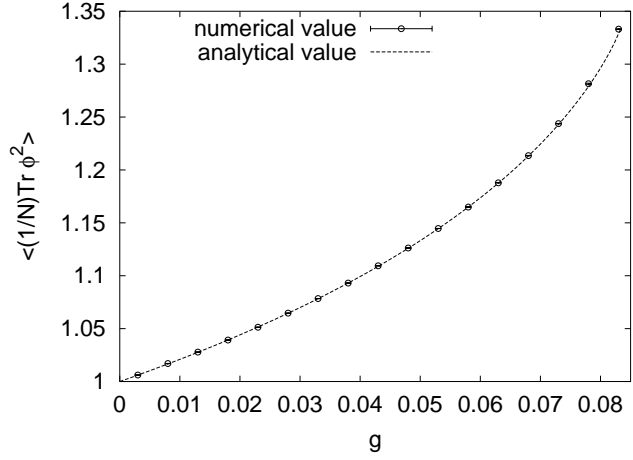} }
   \end{center}
   \caption{The plot of $\langle \frac{1}{N} Tr \phi^{2} \rangle$ for
     the $\phi^{4}$ matrix model for $N=32$.} 
  \label{one-matrix}
  \end{figure}

 \subsection{Simulation of the bosonic IIB matrix model with the
   Chern-Simons term} \label{IIBsim-app}
  We next introduce the algorithm for the IIB matrix model with the
  Chern-Simons term. In this section, we define the action as
  \begin{eqnarray}
    S =  - \frac{N}{4} Tr [A_{\mu}, A_{\nu}]^{2} - 
           \frac{\lambda}{2k+1} N \epsilon_{\mu_{1} \cdots \mu_{2k+1}}
           Tr A_{\mu_{1}} A_{\mu_{2}} \cdots
           A_{\mu_{2k+1}}. \label{IIBfsp2} 
  \end{eqnarray}
 Namely, we take the $g$ in the action (\ref{IIBfsp}) to be
 $\frac{1}{g^{2}} = N$. When we set $\lambda=0$, this is of course
 identical to the bosonic IIB matrix model, which has been scrutinized 
 in \cite{9811220}. There is not so big a difference from the 
 bosonic IIB matrix model, because the Chern-Simons term is nothing
 but a linear term with respect to each $A_{\mu}$.

 We define this matrix model in the  $d=(2k+1)$-dimensional Euclidean
 space, and the indices $\mu, \nu, \cdots$ run over $1,2, \cdots,
 d=(2k+1)$. The following argument is applicable to the bosonic IIB
 matrix model without the Chern-Simons term (namely when $\lambda=0$)
 for an arbitrary dimensions $d \geq 3$(the even $d$ is also
 acceptable)\footnote{The path integral of the two-dimensional bosonic
 IIB matrix model (without the Chern-Simons term) 
 \begin{eqnarray}
  S = - \frac{N}{4} \sum_{\mu,\nu=1}^{2} Tr[A_{\mu}, A_{\nu}]^{2}
    = - \frac{N}{2} Tr [A_{1}, A_{2}]^{2} \nonumber
 \end{eqnarray}
 is easily shown to diverge\cite{0101071} for any size of the matrices
 $N$. When we diagonalize $A_{1}$ as $A_{1} = 
 \textrm{diag} (\lambda_{1}, \lambda_{2}, \cdots \lambda_{N})$, the diagonal
 parts of $A_{2}$ clearly do not contribute to the action. However,
 in the measure of the path integral
  \begin{eqnarray}
  Z = \int d A_{1} d A_{2} e^{-S}, \nonumber
  \end{eqnarray}
 the integral for the diagonal parts of $A_{2}$ runs vacuously. This
 causes the divergence of the path integral.}.
 For the bosonic IIB matrix model without the Chern-Simons term, the
 readers are to ignore the Chern-Simons term in the following argument. 

 The action (\ref{IIBfsp2}) is also analyzed via the heat bath
 algorithm. Firstly, we note that the quartic commutator in
 (\ref{IIBfsp2}) is rewritten as
  \begin{eqnarray}
 & &   - \frac{N}{4} \sum_{\mu,\nu=1}^{d} Tr [A_{\mu}, A_{\nu}]^{2} 
  = - \frac{N}{2} \sum_{1 \leq \mu < \nu \leq d} Tr [A_{\mu}, A_{\nu}]^{2} 
  = N  \sum_{1 \leq \mu < \nu \leq d} [ Tr(A_{\mu}^{2} A_{\nu}^{2})
    - Tr (A_{\mu} A_{\nu} A_{\mu} A_{\nu})] \nonumber \\
 &=& - \frac{N}{2} \sum_{1 \leq \mu < \nu \leq d} Tr G_{\mu \nu}^{2} +
  2N \sum_{1 \leq \mu < \nu \leq d} Tr (A_{\mu}^{2} A_{\nu}^{2} ),
 \end{eqnarray} 
 where $G_{\mu \nu} = \{ A_{\mu}, A_{\nu} \}$, and these are hermitian 
 matrices because these are anti-commutators of $A_{\mu}$.
 This leads us to introduce the auxiliary fields $Q_{\mu \nu}$ as
  \begin{eqnarray}
   {\tilde S} = N \sum_{1 \leq \mu < \nu \leq d} \left( \frac{1}{2} Tr 
   Q_{\mu \nu}^{2} - Tr (Q_{\mu \nu} G_{\mu \nu}) + 2 Tr(A_{\mu}^{2}
   A_{\nu}^{2})  - \frac{\lambda}{2k+1}
    N \epsilon_{\mu_{1} \cdots \mu_{2k+1}} Tr A_{\mu_{1}}
         A_{\mu_{2}} \cdots A_{\mu_{2k+1}} \right). \nonumber \\
  \label{IIBfspaux}
  \end{eqnarray}
 Here, $Q_{\mu \nu}$ are hermitian matrices, and satisfy $Q_{\mu \nu}
 = Q_{\nu \mu}$. $Q_{\mu \nu}$ is defined only for $\mu \neq \nu$.
 Of course, the action (\ref{IIBfspaux}) is equivalent to
 (\ref{IIBfsp2}) after we integrate out $Q_{\mu \nu}$:
  \begin{eqnarray}
   {\tilde S} = \frac{N}{2} \sum_{1 \leq \mu < \nu \leq d} Tr (Q_{\mu
   \nu} - G_{\mu \nu})^{2} + S. \nonumber
  \end{eqnarray}
  Therefore, the matrices $Q_{\mu \nu}$ are updated as
  \begin{eqnarray}
   (Q_{\mu \nu})_{ii} = \frac{W_{i}}{\sqrt{N}} + (G_{\mu \nu})_{ii},
   \hspace{2mm} 
   (Q_{\mu \nu})_{ij} = \frac{X_{ij} + i Y_{ij}}{\sqrt{2N}} + (G_{\mu
   \nu})_{ij}. \label{updateq}
  \end{eqnarray}

 We next update the matrices $A_{\rho}$. To this end, we extract the
 dependence of the action (\ref{IIBfspaux}) on each component of
 $(A_{\rho})_{ij}$. To this end, we rewrite the action
 (\ref{IIBfspaux}) as
  \begin{eqnarray}
   {\tilde S} &=& -N \sum_{\rho < \nu} Tr Q_{\rho \nu}
   (A_{\rho} A_{\nu} + A_{\nu} A_{\rho} ) 
   - N \sum_{\mu < \rho} Tr Q_{\mu \rho}
   (A_{\mu} A_{\rho} + A_{\rho} A_{\mu} ) \nonumber \\
  &+& 2N \sum_{\rho < \nu} Tr (A_{\rho}^{2} A_{\nu}^{2})
  + 2N \sum_{\mu < \rho} Tr (A_{\mu}^{2} A_{\rho}^{2}) \nonumber 
   \\
  &-& \frac{\lambda}{2k+1} N \sum_{\nu_{i} \neq \rho}
   \epsilon_{\nu_{1} \cdots \nu_{2k}  \rho} Tr  A_{\nu_{1}} \cdots
   A_{\nu_{2k}} A_{\rho} + \cdots \nonumber \\ 
  &=&  -N \sum_{\mu \neq \rho} Tr Q_{\mu \rho}
   (A_{\mu} A_{\rho} + A_{\rho} A_{\mu} )
   + 2N \sum_{\mu \neq \rho} Tr (A_{\mu}^{2} A_{\rho}^{2})
   - \lambda N  \sum_{\nu_{i} \neq \rho} 
  \epsilon_{\nu_{1} \cdots \nu_{2k}} Tr A_{\nu_{1}} \cdots
   A_{\nu_{2k}} A_{\rho} + \cdots \nonumber \\
  &=& -N Tr (T_{\rho} A_{\rho}) + 2N Tr(S_{\rho} A_{\rho}^{2}) + \cdots,
   \textrm{ where } \label{IIBfsp3} \\
  S_{\rho} &=& \sum_{\mu \neq \rho} A_{\mu}^{2}, \hspace{2mm}
  T_{\rho} = \sum_{\mu \neq \rho} (A_{\mu} Q_{\rho \mu} + Q_{\rho \mu} 
   A_{\mu}) + \lambda \sum_{\nu_{i} \neq \rho} 
  \epsilon_{\nu_{1} \cdots \nu_{2k}} Tr A_{\nu_{1}} \cdots
   A_{\nu_{2k}}. \label{notationst}
  \end{eqnarray}
  The difference between the pure bosonic IIB matrix model and the
  model with the Chern-Simons term only comes in the definition of
  $T_{\rho}$. Otherwise, both cases go totally in the same way.

  Using the rewriting (\ref{IIBfsp3}), we first update the diagonal
  part $(A_{\rho})_{ii}$. (\ref{IIBfsp3}) is further rewritten as
  \begin{eqnarray}
  {\tilde S} &=& -N (T_{\rho})_{ii} (A_{\rho})_{ii} + 2N
  (S_{\rho})_{ii} (A_{\rho})_{ii}^{2}
  + 2N \sum_{j \neq i} [(S_{\rho})_{ji} (A_{\rho})_{ii}
  (A_{\rho})_{ij} + (S_{\rho})_{ij} (A_{\rho})_{ji} (A_{\rho})_{ii}]
  + \cdots \nonumber \\
  &=& 2N (S_{\rho})_{ii} (A_{\rho})_{ii}^{2} - 4 N h_{i} + \cdots
   = 2N (S_{\rho})_{ii} \left( (A_{\rho})_{ii} -
  \frac{h_{i}}{(S_{\rho})_{ii}} \right) - 2N
  \frac{h_{i}^{2}}{(S_{\rho})_{ii}} + \cdots, \label{IIBfsp4}
   \textrm{ where } \\ 
  h_{i} &=& \frac{1}{4} \left( (T_{\rho})_{ii} -2 \sum_{j\neq i}
  [(S_{\rho})_{ji} (A_{\rho})_{ij} + (S_{\rho})_{ij}
  (A_{\rho})_{ji}] \right). \nonumber
  \end{eqnarray}
 Therefore, the diagonal parts $(A_{\rho})_{ii}$ are updated as
  \begin{eqnarray}
   (A_{\rho})_{ii} = \frac{W_{i}}{\sqrt{4N (S_{\rho})_{ii}}} +
   \frac{h_{i}}{(S_{\rho})_{ii}}. \label{updatearhoii}
  \end{eqnarray}

 We next update the nondiagonal parts $(A_{\rho})_{ij}$, whose
 dependence is now extracted as
  \begin{eqnarray}
   {\tilde S} &=& -N [(T_{\rho})_{ji} (A_{\rho})_{ij} +
   (T_{\rho})_{ij} (A_{\rho})_{ji} ]
  + 2N [(S_{\rho})_{ii} + (S_{\rho})_{jj}] |(A_{\rho})_{ij}|^{2}
   \nonumber \\
  &+& 2N \left[ \sum_{k \neq j} (S_{\rho})_{jk} (A_{\rho})_{ki}
   (A_{\rho})_{ij} + \sum_{k \neq i} (S_{\rho})_{ik} (A_{\rho})_{kj}
   (A_{\rho})_{ji} \right. \nonumber \\
  & & \left.  + \sum_{k \neq j} (S_{\rho})_{kj} (A_{\rho})_{ji}
   (A_{\rho})_{ik} + \sum_{k \neq i} (S_{\rho})_{ki} (A_{\rho})_{ij}
   (A_{\rho})_{jk} \right] + \cdots \nonumber \\
  &=& 2N c_{ij} |(A_{\rho})_{ij}|^{2} - 2N (A_{\rho})_{ij} h_{ji}
    - 2N (A_{\rho})_{ji} h_{ij} + \cdots \nonumber \\
   &=& 2N c_{ij} \left| (A_{\rho})_{ij} - \frac{h_{ij}}{c_{ij}}
   \right|^{2} - 2N \frac{|h_{ij}|^{2}}{c_{ij}} + \cdots, \label{IIBfsp5}
   \textrm{ where } \\
  c_{ij} &=& (S_{\rho})_{ii} + (S_{\rho})_{jj}, \hspace{2mm}
  h_{ij}  = \frac{1}{2} (T_{\rho})_{ij} - \left( 
         \sum_{k \neq i} (S_{\rho})_{ik} (A_{\rho})_{kj}
       + \sum_{k \neq j} (S_{\rho})_{kj} (A_{\rho})_{ik} \right). \nonumber 
  \end{eqnarray}
 Therefore, the nondiagonal part $(A_{\rho})_{ij}$ is updated as
  \begin{eqnarray}
    (A_{\rho})_{ij} = \frac{X_{ij} + i Y_{ij}}{\sqrt{4Nc_{ij}}} +
    \frac{h_{ij}}{c_{ij}}, \label{updatearhoij}
  \end{eqnarray}

 This completes the algorithm for the heat bath algorithm for the
 bosonic IIB matrix model with the Chern-Simons term.
 In order to corroborate the legitimacy of the code, it is useful to
 exploit the following relation coming from the Schwinger-Dyson
 equation
  \begin{eqnarray}
  0 = \int d^{d} A \sum_{a=1}^{N^{2}-1} \sum_{\mu=1}^{d}
  \frac{\partial}{\partial A_{\mu}^{a}} [Tr(t^{a} A_{\mu})
  e^{-S}]. \label{SDE-app} 
  \end{eqnarray}
 Here, $t^{a}$ is the basis of the $SU(N)$ Lie algebra and satisfies
 the following identities:
  \begin{eqnarray}
   Tr(t^{a} t^{b}) = \delta^{ab}, \hspace{2mm}
   \sum_{a=1}^{N^{2}-1} (t^{a})_{ij} (t^{a})_{kl} = \delta_{il}
   \delta_{jk} - \frac{1}{N} \delta_{ij} \delta_{kl}. \label{identitysun}
  \end{eqnarray}
  The matrices $A_{\mu}$ are expanded in terms of $t^{a}$ as
  $A_{\mu} = \sum_{a=1}^{N^{2}-1} A_{\mu}^{a} t^{a}$.
  This leads to the following identities:
  \begin{eqnarray}
    \sum_{a=1}^{N^{2}-1} Tr(t^{a}A) Tr(t^{a}B)
  &=& \sum_{a=1}^{N^{2}-1} A_{ji} B_{lk} (t^{a})_{ij} (t^{a})_{kl} 
  = A_{ji} B_{lk} (\delta_{il}
   \delta_{jk} - \frac{1}{N} \delta_{ij} \delta_{kl}) \nonumber \\
  &=& Tr(AB) - \frac{1}{N} TrA TrB = TrAB. \label{identitysun2}
   \end{eqnarray}
  At the last equality, we utilize the tracelessness of the $SU(N)$
  Lie algebra (namely, $TrA = TrB = 0$). 

  Recalling the fact that
  \begin{eqnarray}
   \frac{\partial S}{\partial A_{\mu}^{a}} 
 = -N Tr(t^{a} [A_{\nu},[A_{\mu}, A_{\nu}]]) - \lambda d N
 \epsilon_{\mu \nu_{1} \cdots \nu_{d-1}} Tr(t^{a} A_{\nu_{1}}
 A_{\nu_{2}} \cdots A_{\nu_{d-1}}), \nonumber
  \end{eqnarray}
 we rewrite the Schwinger-Dyson equation (\ref{SDE-app}) as
 \begin{eqnarray}
  0 &=& \int d^{d} A \sum_{a=1}^{N^{2}-1} [ Tr(t^{a}t^{a}) d e^{-S}
       + N Tr(t^{a}A_{\mu}) Tr (t^{a} [A_{\nu},[A_{\mu}, A_{\nu}]])
       e^{-S} + \nonumber \\
   & & \hspace{20mm} +  \lambda dN \epsilon_{\mu \nu_{1} \cdots
       \nu_{d-1}} Tr(t^{a}A_{\mu}) Tr(t^{a} A_{\nu_{1}} A_{\nu_{2}}
       \cdots A_{\nu_{d-1}}) e^{-S} ] \nonumber \\ 
    &=& \int d^{d} A \left[ d(N^{2}-1) + N Tr \left( [A_{\mu},
       A_{\nu}]^{2} + \lambda \epsilon_{\mu_{1} \cdots \mu_{d}} 
       A_{\mu_{1}} A_{\mu_{2}} \cdots A_{\mu_{d}} \right) \right]
       e^{-S} \nonumber \\
    &=& \left( \int d^{d} A e^{-S} \right) \times \left( d (N^{2}-1) +
       \frac{N \int d^{d} A Tr \left( [A_{\mu}, 
       A_{\nu}]^{2} + \lambda \epsilon_{\mu_{1} \cdots \mu_{d}} 
       A_{\mu_{1}} A_{\mu_{2}} \cdots A_{\mu_{d}} \right) e^{-S}}{ \int d^{d}
       A e^{-S}} \right).  \nonumber \\ 
 \end{eqnarray}
  This implies that the vacuum expectation value of the following
  quantity is analytically computed as
  \begin{eqnarray}
  - \frac{1}{N} \langle Tr[A_{\mu}, A_{\nu}]^{2} \rangle
  - \frac{\lambda}{N} \langle \epsilon_{\mu_{1} \cdots \mu_{d}}
    Tr A_{\mu_{1}} A_{\mu_{2}} \cdots A_{\mu_{d}} \rangle
 = d (1-\frac{1}{N^{2}}). \label{SDEartifact}
  \end{eqnarray}

 \subsection{Jackknife (binning) method}
  Finally, we introduce the jackknife method to estimate the errorbar
  of the quantities obtained by the numerical simulation.
  Now, we have a sample of independent measurements of a primary
  quantity $A$. The measured samples are
  \begin{eqnarray}
   A_{1}, A_{2}, \cdots, A_{N}.
  \end{eqnarray}
 The sample average is simply defined as
  \begin{eqnarray}
   {\bar A} = \frac{1}{N} \sum_{s=1}^{N} A_{s}.
  \end{eqnarray}
 Next, we define the {\it jackknife average}. 
 We exclude {\it one} of the sets $\{ A_{n} \}$ in taking the average:
  \begin{eqnarray}
   A_{(J)s} = \frac{1}{N-1} \sum_{r \neq s} A_{r}.
  \end{eqnarray}
 The variance of the jackknife estimators can be obtained as
  \begin{eqnarray}
    \sigma_{(J){\bar A}}^{2} = \frac{N-1}{N} \sum_{s=1}^{N} (A_{(J)s} - {\bar
    A})^{2}. 
  \end{eqnarray}
  Note that the average ${\bar A_{(J)s}}$ is trivially equivalent to
  the average ${\bar A_{s}}$. We can verify $\sigma_{(J){\bar A}}
  = \sigma_{\bar A}$ in the following way.
  \begin{eqnarray}
   \sigma_{(J){\bar A}}^{2} &=& \frac{N-1}{N} \sum_{s=1}^{N}
   \left( {\bar A}^{2} - \frac{2}{(N-1)} {\bar A} (N {\bar A} - A_{s})
    + \frac{1}{(N-1)^{2}} (N {\bar A} -A_{s})^{2} \right). \nonumber \\
  &=& \frac{N-1}{N} \left[ N {\bar A}^{2} - \frac{2N}{(N-1)} (N {\bar A}^{2} -
   {\bar A}^{2}) + \frac{1}{(N-1)^{2}} \left( (N^{3} {\bar A}^{2}
     - 2 N^{2} {\bar A}^{2}) + \sum_{s=1}^{N}
   (A_{s}^{2}) \right)  \right] \nonumber \\
  &=& {\bar A}^{2} \left( - (N-1)  + \frac{N(N-2)}{(N-1)} \right)
    + \frac{1}{(N-1)N} \sum_{s=1}^{N} (A_{s}^{2}) \nonumber \\
  &=& \frac{1}{(N-1)} \left( \frac{1}{N} \sum_{s=1}^{N} (A_{s}^{2})
    - {\bar A}^{2} \right) = \sigma_{\bar A}^{2}.   
  \end{eqnarray}
  Now, $\sigma_{\bar A}$ is an {\it unbiased estimator} of the {\it
  variance}:
  \begin{eqnarray}
   E \left( \frac{1}{N-1} \sum_{s=1}^{N} (A_{i} - {\bar A})^{2}
   \right)
 = \frac{1}{N-1} \sum_{s=1}^{N} \frac{N-1}{N} \sigma^{2} = \sigma^{2}.
  \end{eqnarray}

 Next, we explain the jackknife binning. This method is utilized when
 $A_{s}$'s have correlations. Especially, the configurations generated 
 by the Markov process are not statistically independent. This poses
 the problem of the "autocorrelation". In coping with this problem,
 we use the jackknife binning. We consider the following bins:
  \begin{eqnarray}
   \underbrace{\underbrace{\spadesuit \spadesuit \spadesuit
   \spadesuit}_{\textrm{one bin with $n=\frac{N}{M}$ elements}}, 
 \underbrace{\spadesuit \spadesuit \spadesuit
   \spadesuit}_{\textrm{one bin}}, \cdots, 
 \underbrace{\spadesuit \spadesuit \spadesuit
   \spadesuit}_{\textrm{one bin}}}_{M \textrm{ bins}}.
  \end{eqnarray}
 Suppose one bin contains $n$ elements, and there are $M$ bins.
 Then, we take the average for {\it each bin} as follows:
  \begin{eqnarray}
  A_{(J)b} = \frac{1}{N - n} \sum_{s \notin \textrm{(Bin)}_{b}} A_{s},
  \end{eqnarray}
 where $b$ runs over $1, 2, \cdots, M$. We take the variance for these 
 $A_{(J)b}$:
  \begin{eqnarray}
   \sigma_{(Bin)}^{2} (n) = \frac{M-1}{M} \sum_{b=1}^{M} (A_{(J)b} -
   {\bar A_{(J)b}} )^{2}.
  \end{eqnarray}
 Note that ${\bar A_{(J)b}} = {\bar A}$. This variance depends on the
 number of the elements in one bin $n$, and this variance increases with
 $n$  and we finally see the plateau.

 To illustrate this process, let us take one simple example. We have
 the data 
  \begin{eqnarray}
  A_{1} =1, \hspace{2mm} A_{2} = 2, \hspace{2mm} \cdots, \hspace{2mm}
  A_{20} = 20,
  \end{eqnarray}
 and we construct the bin as
  \begin{eqnarray}
   \{ A_{1}, \cdots, A_{4} \}, \{A_{5}, \cdots, A_{8} \}, \cdots, 
   \{ A_{17}, \cdots, A_{20} \}.
  \end{eqnarray}
  In this case, $N=20$, $n=4$ and $M=5$. Namely, we define the average 
  for each bin as
  \begin{eqnarray}
  A_{(J)b=1} &=& \frac{(5+6+7+8) + (9+10+11+12) + (13+14+15+16) + 
    (17+18+19+20)}{20-4} = 12.5,
  \nonumber \\
  A_{(J)b=2} &=& \frac{(1+2+3+4) + (9+10+11+12) + (13+14+15+16) +
  (17+18+19+20)}{20-4} = 11.5, \nonumber \\
  A_{(J)b=3} &=& \frac{(1+2+3+4) + (5+6+7+8) + (13+14+15+16) +
  (17+18+19+20)}{20-4} = 10.5, \nonumber \\
  A_{(J)b=4} &=& \frac{(1+2+3+4) + (5+6+7+8) + (9+10+11+12) +
  (17+18+19+20)}{20-4} = 9.5, \nonumber \\
  A_{(J)b=5} &=& \frac{(1+2+3+4) + (5+6+7+8) + (9+10+11+12) +
  (13+14+15+16)}{20-4} = 8.5. \nonumber 
    \end{eqnarray}
  \begin{figure}[htbp]
   \begin{center}
    \scalebox{0.4}{\includegraphics{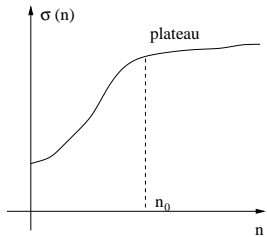}}
   \end{center}
  \caption{The plateau emerges as we increase the bin size.}
  \label{binplateau}
  \end{figure}
 The average is of course ${\bar A_{(J)b}} = 10.5$. The variance is
 computed as
  \begin{eqnarray}
   \sigma_{(Bin)}^{2} &=& \frac{5-1}{5} [(12.5-10.5)^{2} +
   (11.5-10.5)^{2} 
   + (10.5-10.5)^{2} + (9.5-10.5)^{2} + (8.5 -
   10.5)^{2} ] \nonumber \\
  &=& \frac{4}{5} \times 10 = 8. 
  \end{eqnarray}
  The process in taking the plateau is too delicate to delegate to a
 computer, and we usually judge the plateau by hand.

\end{document}